%% file: main_thesis.tex
\documentclass[english, PhD, oneside, noexaminfo]{uaqthesis}

\usepackage[utf8]{inputenx}
\usepackage{indentfirst}
\usepackage{microtype}
\usepackage{amsfonts}
\usepackage[english]{babel}
\usepackage{xcolor}
\usepackage{wrapfig}
\usepackage{lettrine}
\usepackage[nottoc, notlof, notlot]{tocbibind}
\usepackage{hyperref}
\usepackage{tikz}
\usepackage{tikz-feynman}
\usepackage{circuitikz}
\usepackage{wasysym}
\usepackage{cancel}
\usepackage{xcolor}
\usepackage{subcaption}
\usepackage{soul}
\usetikzlibrary {circuits}
\hypersetup{
			colorlinks=true,
			linkcolor=red,
                        linktoc=page,
			anchorcolor=black,
			citecolor=green,
			urlcolor=blue,
			pdftitle={FedericoVirzi_PhD_Thesis},
			pdfauthor={Federico Virzi},
			pdfkeywords={thesis, univaq, l'aquila, university}
 }

\usepackage{indentfirst}
\usepackage{todonotes}
\usepackage{graphicx, rotating, lipsum}
\usepackage{float}
\usepackage{csquotes}
\usepackage{amsmath}
\usepackage{empheq} % braces around aligned/align systems with per-line labels
\usepackage{numprint}
\usepackage{comment}
\usepackage{pgfplots}
\usepackage{pgfplotstable}
\usepackage{lineno}
\usepackage[toc]{appendix}

\usepackage{inconsolata}
\usepackage{color}
\usepackage{listings}

\graphicspath{{./images/}}
\pgfplotsset{compat=1.14}

\definecolor{pgreen}{rgb}{0,0.5,0} % prop keyword
\definecolor{pred}{rgb}{0.9,0,0} % prop comments
\definecolor{javared}{rgb}{0.6,0,0} % for strings
\definecolor{javagreen}{rgb}{0.25,0.5,0.35} % comments
\definecolor{javapurple}{rgb}{0.5,0,0.35} % keywords
\definecolor{javadocblue}{rgb}{0.25,0.35,0.75} % javadoc

\definecolor{forestgreen}{rgb}{0.0, 0.27, 0.13}

\pgfplotstableset{
    every head row/.style={before row=\toprule,after row=\midrule},
    every last row/.style={after row=\bottomrule},
    fixed,precision=2,
}

\title{Real-time Electron Tagging by Cyclotron Radiation Emission Spectroscopy in the PTOLEMY experiment}
\author{Federico Virzi}
\IDnumber{289016\\SSD FIS/01}
\cycle{XXXVIII}
\course{Astroparticle Physics}
\courseorganizer{Department of Physical and Chemical Science}
\submitdate{April 2026}
\copyyear{2026}
\examdate{21 July 2026}
\advisor{Prof. Alfredo Davide Ferella}
\coadvisor{Prof. Marcello Messina}
\authoremail{virzi.federico96@gmail.com}
\begin{document}

\frontmatter
\maketitle
\dedication{ }

 \begin{abstract}
The PTOLEMY project aims at measuring the neutrino mass and, in the future, the Cosmic Neutrino Background from Tritium endpoint measurement. The Radio-Frequency (RF) Tracker is a sub-detector with the goal of measuring in real time the electrons by exploiting the Cyclotron Radiation Emission Spectroscopy. In this work, an overview of the PTOLEMY project is given together with the design, simulation, construction and commissioning of the RF Tracker prototype: the Electron Trap. The measurements and data analysis of the Electron Trap are presented, establishing the primary path for the development of the RF Tracker.
 \end{abstract}

\tableofcontents

\mainmatter

\input{sections/1Introduction}

\input{sections/2PTOLEMY_Project}
\input{sections/3RF_theory}
\input{sections/4Electron_Trap}
\input{sections/5Electron_Trap_Results}

\input{sections/6Conclusions}

%\begin{appendices}
%\input{sections/7appendix}
%\end{appendices}

\backmatter
\cleardoublepage
\phantomsection

\bibliographystyle{plain}
\bibliography{refs}

\end{document}

%% file: sections/1Introduction.tex
\chapter{Introduction} \label{chap:Introduction}

Cosmology is nowadays one of the most active fields of physics, in particular the beginning of the Universe is probably the best way to study particles related to new emerging theories. However, the feasibility of experiments in this field is always problematic, in particular if we consider the very early phases of the Universe. This work matters to the goal of the PTOLEMY project, an experiment that is a bridge between modern cosmology and particle physics aiming at probing the existence of the Cosmic Neutrino Background, a relic radiation decoupled $1s$ after the Big Bang. \\
An introduction on the Standard Cosmological Model, the $\Lambda$-CDM, is given in Chapter (\ref{chap:Introduction}) together with the state of the art of neutrino mass measurement. In Chapter (\ref{chap:PTOLEMY_Project}) the PTOLEMY project is described in detail paying particular attention to the "RF Tracker", a part of the experiment that exploits electron spectroscopy with the cyclotron radiation technique (CRES). Consequently, a detailed description about electron cyclotron radiation is given in Chapter (\ref{chap:RF_theory}) with all the theoretical details about radio-frequency features; finally the "Electron Trap", which is a test setup for the PTOLEMY "RF tracker" and which is the core of this work, is described in Chapter (\ref{chap:Electron_Trap}) and the measurements done at the PTOLEMY Gran Sasso Laboratory are discussed in Chapter (\ref{chap:Electron_Trap_Measurements}).

\newpage
\section{The Cosmological Standard Model:\\ A brief thermal history of the Universe}

The most advanced cosmological model is the $\Lambda$-CDM, which predicts the evolution of the Universe in the scenario of a Hot Big Bang. In this Section the basic concepts of the Universe evolution are given and a brief thermal evolution of the Universe is described. In particular a simplified model of the Universe genesis is given by considering the Universe expansion as adiabatic expansion of a gas, which state variables like densities will be taken from statistical mechanics considerations. The description of the Universe can be done by studying the \emph{Friedman Equation} (FE), which is essentially the energy conservation equation of the expanding Universe. The FE will be solved in this simplified model by using the \emph{Fluid equation}, in this way it will be possible to introduce the main process that is useful for the purposes of this work, that is the \emph{Neutrino decoupling} \cite{grupen}.

\subsection{An expanding Universe and the Friedman equation} \label{sec:friedman_eq}

In the cosmological standard model, the beginning of the Universe has started with a Hot Big Bang with the following expansion of the Universe. It is possible to model this expansion by introducing the scale factor \emph{R}

\begin{equation*}
    R(t)
\end{equation*}

that represents the expansion of a length and depends on the cosmological time $t$. A basic assumption of $\Lambda$-CDM is that this expansion must be \emph{isotropic} and \emph{homogeneous}, in particular two objects that are located at certain positions $R_1(t)$ and $R_2(t)$, the ratio 

\begin{equation*}
    a=\frac{R_1(t)}{R_2(t)}
\end{equation*}

must be constant over time \cite{grupen}. So, if we want to see the evolution over time of the distances:

\begin{equation*}
    \dot{R_1}(t)=a\dot{R_2}(t)=H(t)R_1(t)
\end{equation*}

where H(t) is the \emph{Hubble parameter}, or also called \emph{expansion rate}, the general expression for H(t) is shown in Equation (\ref{eq:hubble_parameter}). 

\begin{equation}
    H(t)=\frac{1}{R(t)}\frac{dR(t)}{dt}
    \label{eq:hubble_parameter}
\end{equation}

The Universe expansion can be modeled as a free expansion of a gas, for which we can define the temperature, the density (density of mass, energy, and number) and pressure. This primordial gas is composed of different types of particles, where each one is evolving differently over time depending on the particle properties like, for example, the \emph{interaction rate} $\Gamma \sim n\langle \sigma v\rangle$, the amplitude of the interaction, and the particle mass. An important remark is that actually in the $\Lambda$-CDM the Universe temperature $T(t)$ is actually considered as the photon temperature at the cosmological time $t$, but for the purposes of the following calculations, this is not necessary. From mechanical statistics considerations \cite{grupen} it is possible to define the energy density $\rho$, the number density $n$ and the average energy $\langle E\rangle$ of each type of particle, taking into account whether that particle is relativistic $T\gg m$ or is non relativistic $T\ll m$.
%In the following is shown that this difference is important to distinguish the decoupling behavior.\\

\subsubsection{Relativistic case}
The energy density of a relativistic particle is

\begin{empheq}[left=\empheqlbrace]{align}
    \rho &= \frac{\pi^2}{30}gT^4 \qquad \text{(bosons)}
    \label{eq:energy_densitiy_bosons} \\
    \rho &= \frac{7}{8}\frac{\pi^2}{30}gT^4 \qquad \text{(fermions)}
    \label{eq:energy_densitiy_fermions}
\end{empheq}

where $T$ is the Temperature and $g$ are the degrees of freedom of that particle. In a similar way it is possible to define the number density,

\begin{empheq}[left=\empheqlbrace]{align}
    n &= \frac{\zeta(3)}{\pi^2}gT^3 \qquad \text{(bosons)}
    \label{eq:number_densitiy_bosons} \\
    n &= \frac{3}{4}\frac{\zeta(3)}{\pi^2}gT^3 \qquad \text{(fermions)}
    \label{eq:number_densitiy_fermions}
\end{empheq}

and then the average energy $\langle E\rangle=\rho/n$

\begin{empheq}[left=\empheqlbrace]{align}
    \langle E\rangle &= \frac{\pi^2}{30\zeta(3)}T \sim 2.7~T \qquad \text{(bosons)}
    \label{eq:average_energy_bosons} \\
    \langle E\rangle &= \rho/n = \frac{7}{6}\frac{\pi^2}{30\zeta(3)}~T \sim 3.1~T \qquad \text{(fermions)}
    \label{eq:average_energy_fermions}
\end{empheq}

It is possible to note that at the very beginning of the Universe, when the temperature was very high, the Universe was dominated by radiation and the difference between bosons and fermions was negligible.

\subsubsection{Non Relativistic case}

In the case where $T\ll m$ and by neglecting the chemical potential \cite{grupen}, the energy density is 

\begin{equation}
    \rho=mn 
\end{equation}

and the number density is given by the Boltzmann distribution

\begin{equation}
n=g \left( \frac{mT}{2\pi} \right) ^{3/2}e^{-m/T}
\label{eq:number_density_non_rel}
\end{equation}

From Equation (\ref{eq:number_density_non_rel}) it is possible to note that the particle density is exponentially suppressed by a term $\sim m/T$. The average energy becomes

\begin{equation}
    \langle E \rangle=m+\frac{3}{2}T
\end{equation}

It is now possible to define the total energy density $\rho_{tot}$, which from now on will be referred to as $\rho$,

\begin{equation}
    \rho=\rho_{tot}=\frac{\pi^2}{30}g_{*}T^4
\end{equation}

where $g_*$ are the effective degrees of freedom of the Universe at a certain time. The effective number of degrees of freedom, as can be seen in Equation (\ref{eq:g_star}) depends on the particle temperature $T_k$ where $k=i,j$ for bosons and fermions and, respectively, $T \equiv T_{\gamma}$.

\begin{equation}
    g_{*}=\sum_i g_i \left( \frac{T_i}{T} \right) ^2+ \frac{7}{8}\sum_j g_j \left( \frac{T_j}{T} \right) ^4
    \label{eq:g_star}
\end{equation}

\subsubsection{Fluid equation}

\begin{figure}[h!]
    \centering
    \includegraphics[scale=0.4]{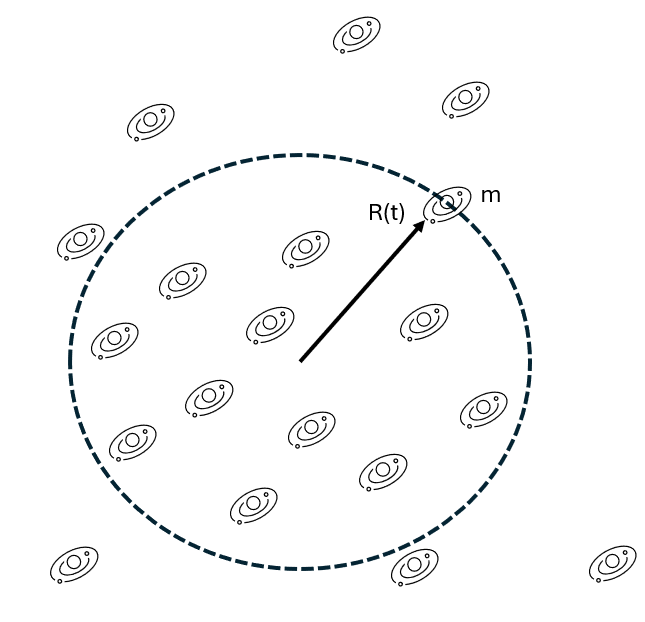}
    \caption{Artistic drawing of mass distribution of the expanding Universe.}
    \label{fig:test_galaxy}
\end{figure}

Now that we have the general formula for $\rho$, we can obtain the fluid equation required to solve the Friedman equation. To obtain the fluid equation we can use the equation of state of an ideal gas. As a first step it is useful to define a test galactic object with mass $m$ and at distance $R(t)$ with respect to a fixed observer, as shown in Figure (\ref{fig:test_galaxy}).

To understand the energy conservation inside the sphere of radius $R(t)$, i.e. the Friedman equation, it is useful to see the time variation of the internal energy.\\
The general form of the internal energy $U(t)$ is 

\begin{equation*}
    dU(t)=T(t)dS-P(t)dV(t)
\end{equation*}

where $dS=0$ because the entropy of a free expanding gas is zero. The total time derivative is

\begin{equation}
    \frac{dU(t)}{dt}+P(t)\frac{dV(t)}{dt}=0
    \label{eq:fluid_eq1}
\end{equation}

and because we are considering a test sphere, the Energy and Volume time derivative is

\begin{equation}
\begin{cases}

    \frac{dU(t)}{dt}=\frac{\partial U(t)}{\partial R(t)} \dot{R}(t) +\frac{\partial U(t)}{\partial \rho (t)} \dot{\rho}(t)=3R^2(t) \rho (t) \dot{R}(t)+R^3(t)\dot{\rho}(t)\\
    
    \frac{dV(t)}{dt}=\frac{dR^3(t)}{dt}=3R^2(t)\dot{R}(t)

\end{cases}
\label{eq:fluid_eq2}
\end{equation}

that means, by using (\ref{eq:fluid_eq2}) inside Equation (\ref{eq:fluid_eq1}),

\begin{equation}
    \dot{\rho}(t)+\frac{3\dot{R}(t)}{R(t)}(\rho(t)+P(t))=0
    \label{eq:fluid_equation}
\end{equation}

At the beginning of the Universe expansion, $P=\rho/3$ \cite{grupen}, which means that the fluid equation becomes 

\begin{equation*}
    \dot{\rho} (t) + 4 \rho (t) \frac{\dot{R}(t)}{R(t)}=0
\end{equation*}

where it is possible to note that 

\begin{equation*}
    \frac{1}{R^4(t)} \frac{d}{dt} \left( \rho(t)R^4(t) \right) =0
\end{equation*}

that gives the dependence of the energy density with respect to the scale length that is shown in Equation (\ref{eq:rho_rad}).

\begin{equation}
\rho(t)=const \frac{1}{R^4(t)}  
\label{eq:rho_rad}
\end{equation}

If we consider when the Universe was dominated by non relativistic matter, $P=0$ and in the same way we obtain  that 

\begin{equation}
    \rho(t)=const \frac{1}{R^3(t)}  
\label{eq:rho_non_rel}
\end{equation}

%considering  from mechanical statistics

%\begin{equation}
%    \dot{\rho}+\frac{4\rho \dot{R}}{R}=0
%\end{equation}

\subsubsection{The Friedman Equation}
In the following, the Friedman equation is derived from Newtonian gravity considerations. Taking back the example from Figure (\ref{fig:test_galaxy}), the part of the Universe inside the sphere of radius $R(t)$ has an average mass (in general energy) density $\rho(t)$. The total mass $M$ is 

\begin{equation*}
    M(t)=\frac{4}{3}R^3(t)\rho (t)
\end{equation*}

and the gravitational potential of the test object is described by Equation (\ref{eq:test_galaxy_pot}).

\begin{equation}
    V(t)=-\frac{GmM(t)}{R(t)}=-\frac{4}{3}\pi G m R^2(t) \rho (t)
    \label{eq:test_galaxy_pot}
\end{equation}

The kinetic energy of the test object is $K(t)=\frac{1}{2}m\dot{R}^2(t)$ so the total energy E of the test object is described by Equation (\ref{eq:E_test_galaxy}).

\begin{equation}
\begin{split}
    E(t)=K(t)+V(t)=\frac{1}{2}m\dot{R}^2(t)-\frac{4}{3}\pi G m R^2(t) \rho (t)= \\
    =\frac{1}{2}mR^2(t) \left( \frac{\dot{R}^2(t)}{R^2(t)}-\frac{8\pi}{3}G\rho (t) \right)
    \label{eq:E_test_galaxy}
    \end{split}
\end{equation}

It is possible to define the \emph{curvature parameter} $k$ as

\begin{equation}
    k=-\frac{2E}{m}
\end{equation}

The energy conservation equation in the Newtonian case can be rewritten as 

\begin{equation}
    \left( \frac{\dot{R}(t)}{R(t)} \right)^2 + \frac{k}{R^2(t)} = \frac{8\pi}{3} G \rho (t)
\end{equation}

It is possible now to include the cosmological vacuum term, that is given from the General relativity and from the \emph{Robertson-Walker metric}, we obtain the Friedman Equation that is shown in Equation (\ref{eq:friedman_equation})

\begin{equation}
    \left( \frac{\dot{R}(t)}{R(t)} \right)^2 = \frac{8\pi}{3} G \rho (t) - \frac{k}{R^2(t)} +\frac{\Lambda}{3}
    \label{eq:friedman_equation}
\end{equation}

At the early stages of the expansion of the Universe, the energy density was described by Equation (\ref{eq:rho_rad}). When this dependence is included in the Friedman Equation, we see that the curvature and vacuum term can be neglected, and so the Friedman Equation becomes

\begin{equation*}
    \dot{R}(t) \sim \frac{1}{R(t)}
\end{equation*}

that gives as solution 

\begin{equation}
    H(t) = \frac{1}{2} \frac{1}{t}
    \label{eq:H_rel}
\end{equation}

while in the case of the Universe dominated by non relativistic matter, which is described by (\ref{eq:rho_non_rel}) gives solutions

\begin{equation}
    H(t)=\frac{2}{3} \frac{1}{t}
    \label{eq:H_non_rel}
\end{equation}

Those two solutions are going to be useful in order to understand the expansion rate when the Universe was dominated by a different composition of the primordial plasma.

%\begin{equation}
%    \rho_{cr}=\frac{3}{8}\frac{H_0^2}{\pi G}
%\end{equation}

\subsection{Thermal history of the Universe}
\label{sec:thermal_history}

In this Section a brief thermal history of the expanding Universe is given. The main concepts that are needed to develop a simple model of the Universe expansion have been discussed in the Section (\ref{sec:friedman_eq}), in particular the Universe expansion can be modeled as a free gas expansion with a temperature that changes over time as

\begin{equation}
    T(t)=\frac{T_0}{R(t)}
    \label{eq:temperature_hubble}
\end{equation}

While the Universe radius $R(t)$ expands, the photon temperature decreases and so also the average energy of the primordial plasma.  This phenomenon is particularly interesting because each particle reaction has an interaction rate $\Gamma$, which changes over time as the Universe cooled down, that and when $\Gamma$ becomes comparable to the photon average energy $\langle E\rangle$ makes the particle formation possible.

\subsubsection{The Primordial Universe}

The energy scale of the beginning of the Universe can be estimated by the comparison between the \emph{Schwarzschild radius} and the \emph{Compton length} because 

\begin{equation*}
    R_S=2\frac{mg}{c^2}
\end{equation*}

\begin{equation*}
    \lambda_C=\frac{h}{mc}
\end{equation*}

in particular
\begin{equation}
    \frac{1}{2\pi}\lambda_C=\frac{1}{2}R_S
\end{equation}

gives an estimation of the \emph{Planck mass} that can be converted in the \emph{Planck energy}

\begin{equation}
    E_{Pl}=\sqrt{\frac{\hbar c}{G}} \sim 1.2 \times 10^{19} ~GeV
\end{equation}

that is the energy scale in the early stages of the Universe expansion. The length and time scale are instead 

\begin{equation}
    l_{Pl}\sim 1.6 \times 10^{-35}~ m
\end{equation}

and 

\begin{equation}
    t_{Pl}\sim 5.4 \times 10^{-44} ~s
\end{equation}

In the Planck phase of the Universe, the energy density was so high that the Universe was radiation dominated, expanding as described in Equation (\ref{eq:H_rel}), and all forces were unified. The coupling constants $\alpha_s$, $\alpha_{ew}$ and $\alpha_{em}$ got separated during the GUT phase, in which the Temperature dropped to about $T=\mathcal{O}(10^{16}~ GeV)$. After a large and fast expansion, also called \emph{Inflation}, the Universe cooled down to $T=\mathcal{O}(10^2~GeV)$ entering the \emph{Electroweak} era, in which leptons began massive.\\
Almost ten microseconds after the Big Bang, the Temperature was such that the strong coupling constant $\alpha_s$ was much larger than the others, in this way hadrons formation has been possible.\\
A resume about those first stages of the Universe expansion is shown in Table (\ref{tab:Primordial_Universe}).

\begin{table}[h!]
    \centering
    \begin{tabular}{|c|ccc|}
    \hline
    Cosmological era& $T[GeV]$ & $\rho[GeV^4]$ & $t[s]$\\
    \hline
    Planck&$10^{19}$&$10^{78}$&$10^{-45}$\\
    G.U.T.&$10^{16}$&$10^{66}$&$10^{-39}$\\
    E.W.&$10^{2}$&$10^{20}$&$10^{-11}$\\
    Q.C.D&$0.2$&$10^{-2}$&$10^{-5}$\\
    \hline
    
    \end{tabular}
    \caption{Temperature, energy density and time scale of the Universe expansion at different era. }
    \label{tab:Primordial_Universe}
\end{table}

\subsubsection{The relic radiations: C$\nu$B and CMB}

After the Q.C.D phase, the Universe was cold enough such that some reactions began to be suppressed, resulting in the formation of light nuclei, entering in the so called \emph{Big Bang Nucleosinthesis era} (BBN) when the temperature reached almost two times the electron mass. At the beginning of the BBN the temperature was $T \gg 2m_e \simeq 1~MeV$ and electrons, neutrinos, photons, and neutrons were in thermal equilibrium with cosmic free streaming by reactions such as 

\begin{equation}
    e^+e^- \leftrightarrow \nu \bar{\nu}
    \label{eq:pair_nu}
\end{equation}

\begin{equation}
    \nu_e n \leftrightarrow e^-p
    \label{eq:n_p}
\end{equation}

\begin{equation}
    e^+e^- \leftrightarrow \gamma \gamma 
    \label{eq:pair_gamma}
\end{equation}

When the temperature reaches $T\sim 1~MeV$ the muonic and tauonic neutrinos start to decouple and when the temperature drops to almost $T  \simeq 0.7~MeV$ the production of protons from neutrons from Equation (\ref{eq:n_p}) starts to become suppressed. It is possible to estimate that temperature by the comparison between the reaction rate $\Gamma$ and the expansion rate $H$, giving as result the so called \emph{freeze-out} temperature $T_f$ which can be estimated to be the temperature of the universe at almost $t\sim 1~s$ after the Big Bang. When the Universe was at temperature $T_f=0.7~MeV$ also the electronic neutrino stops to interact with the neutron to produce protons, and so all the neutrino flavours were decoupled, bringing to the so called \emph{Neutrino Decoupling}, from that time those neutrinos rarely interact with the matter and this neutrino radiation becomes relic particles. Those relic neutrinos emitted one second after the Big Bang are called \emph{Cosmic Neutrino Background}.\\
After Neutrino decoupling the reaction (\ref{eq:pair_gamma}) was still ongoing and so the photon temperature was decreasing less rapidly than the relic neutrinos, then it is possible to relate the relic neutrino temperature $T_{\nu}$ to the photon temperature $T_{\gamma}$. In particular, by using the thermodynamical and statistical equations described before, we obtain 

\begin{equation}
    T_{\nu}= \left( \frac{4}{11}  \right) ^{1/3} T_{\gamma}
    \label{eq:T_nu_T_gamma}
\end{equation}

%\begin{figure}[h!]
 %   \centering
  %  \includegraphics[scale=0.5]{imgs/timeline_universe.png}
   % \caption{Timeline of the Universe expansion.}
    %\label{fig:timeline_universe}
%\end{figure}

So far, all the cosmological era that has been described were related to a radiation dominated universe. However, at a certain point light nuclei start to form. The breaking point has been when the temperature was such that was possible the formation of hydrogen atoms, bringing the cosmological plasma to a neutral gas. This phenomenon is called \emph{Recombination} and can be estimated to be happened at around 380000 years after the Big Bang.\\
A consequence of the recombination is that most of the electrons were no longer available to produce photons by pair annihilation, and so also photons became decoupled from the matter, forming the so called \emph{Cosmic Microwave Background} (CMB). From this point the Universe becomes matter dominated and starts to expand at the rate described in Equation (\ref{eq:H_non_rel}). The CMB has been directly observed during the last century and is considered the one of the most important confirmations of the $\Lambda$-CDM; in particular the today CMB temperature, $T_{\gamma}^0$ has been measured to be

\begin{equation}
    T_{\gamma}^0=2.73~K
\end{equation}

With this information, it is possible to reconstruct, as shown by Equation (\ref{eq:T_nu_T_gamma}), the temperature of the C$\nu$B today, corresponding to 

\begin{equation}
    T_{\nu}^0=1.95~K
\end{equation}

At this temperature corresponds an average kinetic energy $ K_{\nu} \simeq \langle E_{\nu} \rangle$ that can be obtained by Equation (\ref{eq:average_energy_fermions})\footnote{It is important to note that the temperature must be multiplied by the Boltzmann constant.} of about 

%\begin{equation}
 %   K_{\nu}=k_BT_{\nu}^0=8.6 \times 10^{-5} \frac{eV}{K} \cdot 1.95K \sim 0.1 meV
%\end{equation}

\begin{equation}
    K_{\nu} \simeq 3.1 k_BT_{\nu}^0=3.1 \cdot 8.6 \times 10^{-5} \frac{eV}{K} \cdot 1.95~K \sim 0.5 ~meV
\end{equation}

by using the number density of relativistic fermions of Equation (\ref{eq:number_densitiy_fermions}) it is possible to estimate the number density of the C$\nu$B at today temperature, It is worth noting that even if today the relic neutrinos are non relativistic, at the decoupling time they were more energetic and so the estimate of the number density requires the relativistic formula. In particular it is possible to obtain that\footnote{This is obtained by using the Fermi-Dirac distribution, but there exist also other models for this density number.} 

\begin{equation}
    n_{\nu}^0=56~cm^{-3}
\end{equation}

per neutrino flavour, not considering the antineutrinos.\\
In conclusion, in this Section a thermal universe history has been described with the goal of introducing the main concepts for an introduction for C$\nu$B. A more complete overview requires more considerations like CP violation, the \emph{Sakharov conditions}, \emph{Light Nuclei abundance}, the role of \emph{dark matter}, the properties of the CMB like the anisotropies and \emph{Baryon acoustic oscillations}, which are, however, outside the goal of this work.

\subsection{Cosmic Neutrino Background direct detection} \label{sec:CNB_detection}

In Section (\ref{sec:thermal_history}) the basic steps of the thermal expansion of the Universe and the relic radiations have been introduced following the cosmological standard model. After having described the model, it is important to search for direct evidence that, however, are not much about the early stages and if are present are are strongly model dependent. The most important direct experimental evidence of the $\Lambda$-CDM is the CMB and its anisotropies in the spectrum. This discovery was emitted, as mentioned in the last Section, to about 380000 years after the Big Bang, so nowadays is actually missing a strong and model independent  direct detection of an early-stages radiation. An important step could be the detection of the cosmic neutrino background: that has not yet been directly observed; however, in 1962 S. Weinberg has found a method to directly detect C$\nu$B by using Tritium \cite{weinberg}. This method was adapted to a massive neutrino framework in 2007 by A.G. Cocco, G. Mangano and M. Messina \cite{probing_low}, creating the basis of the PTOLEMY experiment. In this Section the reaction that can be used for the C$\nu$B capture is presented.\\
To summarize, the experimental problem is the detection of neutrino radiation that interacts by weak interaction, with the average kinetic energy and number density that are resumed in (\ref{eq:CnuB_properties})

\begin{equation}
    \begin{cases}
        K_{\nu}=0.5 ~meV\\
        n_{\nu}^0=56 ~cm^{-3}
    \end{cases}
    \label{eq:CnuB_properties}
\end{equation}

\begin{figure}[h!]
    \centering
    \includegraphics[scale=0.45]{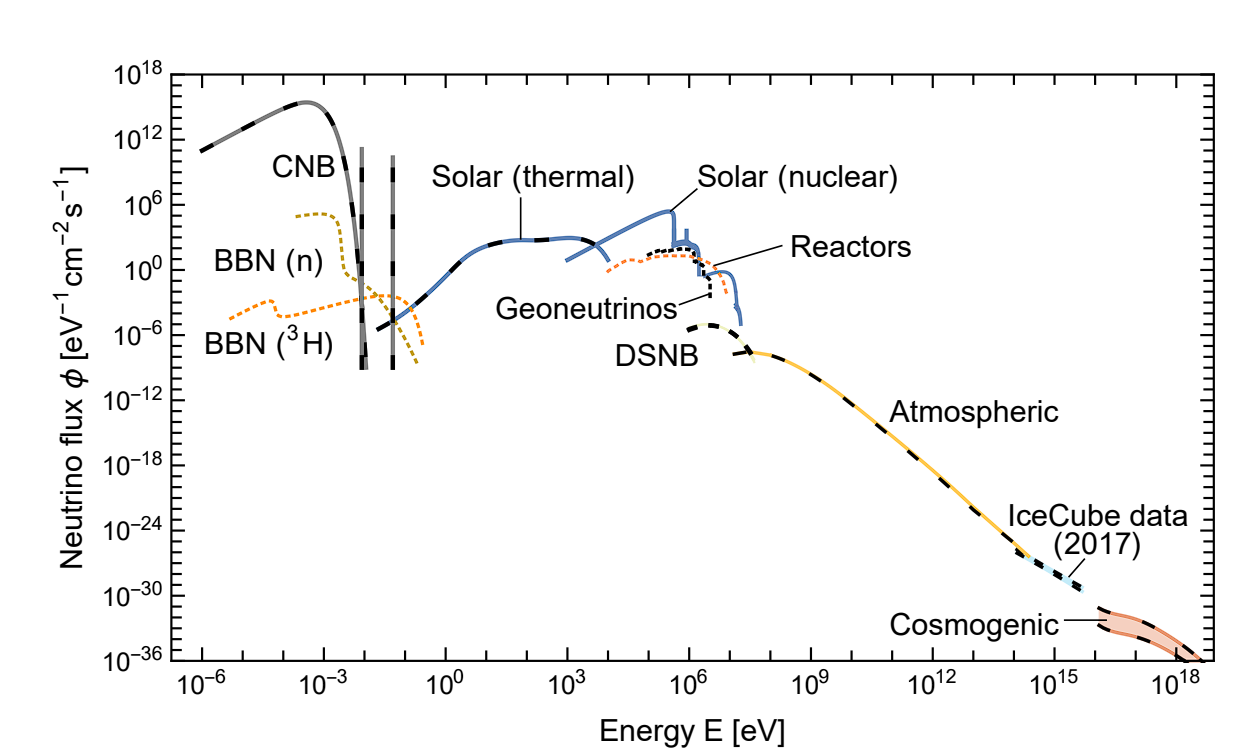}
    \caption{Neutrino spectrum \cite{neutrino_spectrum} for different neutrino sources. Full, dotted and dashed lines represents the case when only neutrinos, antineutrinos or both are produced.}
    \label{fig:neutrino_spectrum}
\end{figure}

It is worth noting that the number density has been calculated by using the Fermi-Dirac distribution as mentioned in Section (\ref{sec:friedman_eq}) but actually, not only exist different model for a correct computation of the number density, but also the $56~cm^{-3}$ are actually calculated in the more conservative scenario, in particular not considering the fact that the neutrino is massive and so the C$\nu$B could clusterize around the centre of the galaxy, increasing the neutrino density by a factor $1-10$ at the Earth position \cite{wong}.\\
In Figure (\ref{fig:neutrino_spectrum}) is shown the energy spectrum for neutrinos from different sources, in particular it is possible to note that at energies $E<0.01~eV$ the C$\nu$B is the only neutrino source and also has the largest flux among all sources.

A possibility for direct detection of C$\nu$B is to exploit \emph{Neutrino capture in Beta decaying nuclei}, or NCB, which is shown in Equation (\ref{eq:NCB}) in the case of Tritium $^3H$.

\begin{equation}
    \nu_e + ^3_1H \rightarrow ^3_2He^+ + e^-
    \label{eq:NCB}
\end{equation}

The important thing of this reaction is that it is non-threshold, making it a suitable choice for the C$\nu$B direct detection.\\
Neutrinos rarely interact with matter and low energy neutrinos are even more elusive because of their small cross section, however, in \cite{probing_low} it has been described that actually the real quantity for describing the interaction rate of low energy neutrinos is the cross section times the neutrino velocity. The neutrino capture rate, $\lambda_{\nu}$, can be written as shown in Equation (\ref{eq:nu_rate}), in which $f$ is a distribution function that can be a simple Fermi-Dirac (FD), a  Navarro-Frenk-White (NFW) distribution \cite{Navarro}, or a distribution obtained from the Milky Way (MW) profile.

\begin{equation}
    \lambda_{\nu}=\int \sigma_{NCB} \cdot v_{\nu} \cdot f(p_{\nu}) \frac{d^3p_{\nu}}{(2\pi)^3}
    \label{eq:nu_rate}
\end{equation}

The problem is that Tritium is a beta unstable element and the final state of beta decay is similar to that of neutrino capture but occurs with a higher rate, as shown in Equation (\ref{eq:beta_rate}), in which $G_{\beta}$, $F$, and $C_{\beta}$ are factors that depend on the Tritium atom.

\begin{equation}
    \lambda_{\beta}=\frac{G^2_{\beta}}{2\pi^3}\int p_eE_e F C_{\beta} E_{\nu} p_{\nu} dE_e
    \label{eq:beta_rate}
\end{equation}

Those electrons from beta decay represent the background process for a direct detection of the C$\nu$B whose spectrum is given by a typical Kurie function with a  $Q_{\beta}=18.6~keV$. The signature from the cosmic neutrino background is instead, in first approximation, a peak at energy $Q_{\beta}+2m_{\nu}$. Considering that the cross section times neutrino velocity can be written as 

\begin{equation}
    \sigma_{NCB}v_{\nu}=\frac{2\pi^2\text{ln2}}{A\cdot T_{1/2}}
\end{equation}

It is possible to calculate the ratio between those two reactions that gets a disadvantage ratio of several orders of magnitudes.

\begin{equation}
    \lambda_{\nu}(^3H)=0.66\times10^{-23}\lambda_{\beta}(^3H)
\end{equation}

However, it is still possible, by using the Formula (\ref{eq:rate})

\begin{equation}
    R=\lambda_{\nu}\frac{N_AM[g]}{A}
    \label{eq:rate}
\end{equation}

to have a sizable number a macroscopic number of neutrino captures for 100~g of Tritium. In Table (\ref{tab:capture_event_rate}) the event rate for three different neutrino density models is shown as a function of the neutrino mass.

\begin{table}[h!]
    \centering
    \begin{tabular}{|c|ccc|}
    \hline
    $m_{\nu} (eV)$ & FD ($events \cdot yr^{-1}$) & NFW ($events \cdot yr^{-1}$) & MW ($events \cdot yr^{-1}$)\\
    \hline
    0.3 & 7.5 & 23 & 33\\
    0.15 & 7.5 & 10 & 12\\
    \hline
    
    \end{tabular}
    \caption{C$\nu$B capture rate for different neutrino mass and clustering models.}
    \label{tab:capture_event_rate}
\end{table}

\newpage
\begin{figure}[h!]
    \centering
    \includegraphics[scale=0.6]{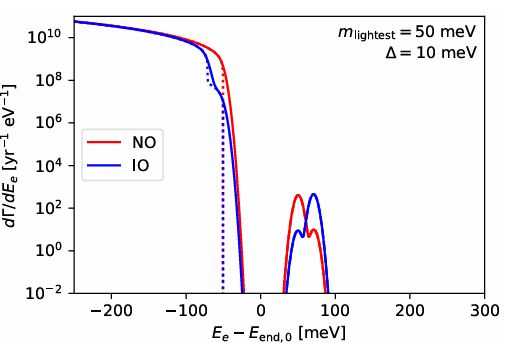}
    \caption{Tritium Kurie plot with C$\nu$B peak at $2m_{\nu}=2m_{lightest}$ from the endpoint ($E_{end,0}$ is the endpoint in the massless neutrino case). "IO" and "NO" are the inverted and normal ordering cases introduced in Section (\ref{sec:neutrino_mass}). Dotted lines are representing the background without energy resolution convolution \cite{neutrino_physics}.}
    \label{fig:tritium_kurie_plot_toy}
\end{figure}

The resultant spectrum is shown in Figure (\ref{fig:tritium_kurie_plot_toy}).\\
For the relic neutrino detection there is also another advantage that comes from the non relativistic energy of the C$\nu$B, in particular if the neutrino is a Majorana particle, there degree of freedom $g_{\nu}$ at the neutrino decoupling is factor 2 larger than in the case of Dirac neutrino. An effect going in the same direction can be a possible coherent effect with the Tritium target which still needs to be investigated.

\newpage
\section{The Neutrino mass} \label{sec:neutrino_mass}

In the last Section has been introduced the problem of the cosmic neutrino background detection by using Tritium atoms as target, it has been shown also that the experimental capability of that measurement depends on the mass of the neutrino. In this Section a brief introduction about the neutrino mass is given \cite{pdg}. 

\subsection{Introduction to Neutrino mass models}

In the Standard Model of particle physics, charged fermions acquire mass by the \emph{Yukawa coupling} of the fermion field  $Y_{ij}$ with the \emph{Higgs} scalar field $\phi$, as shown in Equation (\ref{eq:lagr_higgs}).

\begin{equation}
    \mathcal{L}=Y_{ij}^l \bar{L}_{l,i} \phi L_{r,j}+ h.c.
    \label{eq:lagr_higgs}
\end{equation}

This interaction is written with a term in the standard model Lagrangian such that during the broken symmetry phase, the fermion acquires a mass that depends on the vacuum expectation value of $\phi$. However in this scenario the neutrinos cannot be massive, otherwise this could break the gauge symmetry, this means that to introduce a neutrino mass term, it is necessary to extend the Standard Model.\\
The neutrino mass term can be included in two different theoretical scenarios related by the nature of the neutrino depending on whether it is a \emph{Dirac} or \emph{Majorana} particle. This is done by introducing in the SM Lagrangian a non known a priori number of \emph{sterile} neutrinos and including a term, $\mathcal{L}_D$ for the Dirac neutrino or and $\mathcal{L}_M$ for the Majorana neutrino.

\begin{equation}
    \mathcal{L}=M_{D,ij} \bar{\nu}_{s,i} \nu_{L,j} +h.c.
\end{equation}

\begin{equation}
    \mathcal{L}=\frac{1}{2}M_{N,ij} \bar{\nu}_{s,i} \nu_{s,j}^c + h.c.
\end{equation}
Those two additional terms can be summed in such a way that it is possible to write a neutrino mass Lagrangian $\mathcal{L}_{M_{\nu}}$

\begin{equation}
    \mathcal{L}_{M_{\nu}}=\frac{1}{2}(\bar{\vec{\nu_L^c}},\bar{\vec{\nu_s}})\left(
    \begin{matrix}
        0&M_D^T\\
        M_D&M_N
    \end{matrix}
    \right)\left(
\begin{matrix}
    \vec{\nu}_L\\
    \vec{\nu}_s^c
\end{matrix}
    \right)+h.c.=\bar{\vec{\nu}^c}M_{\nu}\vec{\nu}+h.c.
\end{equation}

where $M_D$ is a matrix $m\times 3$ and $M_{N}$ is $m \times m$.\\
If the neutrino is a Dirac particle, it is possible to obtain the Dirac mass Lagrangian by imposing $M_N=0$, in this way the neutrino mass Lagrangian term can be written as

\begin{equation}
    \mathcal{L}=\sum_{k=1}^{3}m_k\bar{\nu}_{D,k}\nu_{D,k}
\end{equation}

where $\nu_{D,k}$ is the Dirac neutrino field $\nu_{D,k} = \nu_{D,L,k}+\nu_{D,R,k} $.
If the neutrino is a Majorana particle, the neutrino mass Lagrangian is 
 
\begin{equation}
    \mathcal{L}=\frac{1}{2}\sum_{k=1}^{3+m}m_k\bar{\nu}_{M,k}\nu_{M,k}
\end{equation}

where the mass eigenstates are

\begin{equation}
    \nu_{M,k}= \left(V^{\nu \dagger } \vec{\nu}  \right)_k + \left(V^{\nu \dagger } \vec{\nu}  \right)_k^c
\end{equation}

From those eigenstate it is possible to note, as mentioned in the last Section, that the Majorana neutrino and antineutrino are the same thing.\\

\subsubsection{Neutrino Oscillation}

Nowadays we know that neutrino is a massive particle. This was established by \emph{neutrino oscillation} measurement where flavour oscillation takes place during the propagation of those particles between the source and detection point. This result was definitively confirmed in the 90s. We know that the neutrino can be written in a 3 states wavefunctions, in particular it can be written in the flavour basis

\begin{equation}
\nu_{\alpha}=\left(
    \begin{matrix}
        \nu_e\\
        \nu_{\mu}\\
        \nu_{\tau}
    \end{matrix}
    \right)
\end{equation}

or in the mass basis

\begin{equation}
\nu_{i}=\left(
    \begin{matrix}
        \nu_1\\
        \nu_2\\
        \nu_3
    \end{matrix}
    \right)
\end{equation}

It is possible to describe the neutrino states in the flavour basis from the mass eigenstates by using the \emph{Pontecorvo-Maki-Nakagawa-Sakata} matrix $U$ (also known as PMNS matrix), as shown in Equation (\ref{eq:nu_flavor_mass}).

\begin{equation}
    |\nu_{\alpha}(t)\rangle=\sum_{i=1}^nU_{\alpha i}^*|\nu_{i}(t)\rangle
    \label{eq:nu_flavor_mass}
\end{equation}

The PMNS is shown in Equation (\ref{eq:PMNS})

\begin{equation}
    U=\left(
\begin{matrix}
    1&0&0\\
    0&c_{23}&s_{23}\\
    0&-s_{23}&c_{23}
\end{matrix}
    \right)
    \cdot
    \left(
\begin{matrix}
    c_{13}&0&s_{13}e^{-i\delta_{CP}}\\
    0&1&0\\
    -s_{13}e^{i\delta_{CP}}&0&c_{13}
\end{matrix}
   \right)
    \cdot
    \left(
\begin{matrix}
    c_{12}&s_{12}&0\\
    -s_{12}&c_{12}&0\\
    0&0&1
\end{matrix}
    \right)
    \label{eq:PMNS}
\end{equation}

where $c_{ij}$ and $s_{ij}$ are the cosine and sine of the angle $\theta_{ij}$ that is a parameter of the oscillation and $\delta_{CP}$ is a phase term due to CP violation.\\
The neutrino oscillation is the possibility for a neutrino of flavour $\alpha$ to appear in flavour $\beta$ after the neutrino drift. This phenomenon has a probability that can be calculated by Equation (\ref{eq:Pab}) and depends on the length of the neutrino drift $L$, the energy of the neutrino $E$ and the difference of the mass eigenstates $\Delta m^2_{ij}$, which means that by measuring the probability of neutrino oscillation from one flavour to another it is possible to measure the difference of the masses eigenstates.

\begin{equation}
    P_{\alpha\beta}=sin^2(2\theta)sin^2 \left( \frac{\Delta m^2L}{4E} \right)
        \label{eq:Pab}
\end{equation}

It is important to point out that the oscillation probability in Equation (\ref{eq:Pab}) has been calculated by using the two flavour approximation, it is possible to extend it to all three mass eigenstates, however for the purposes of this introduction it is just important to understand that from the probability it is possible to measure the difference of the squared of masses eigenvalues.\\
During the last decades, several neutrino oscillation experiments have been performed and all $\Delta m^2_{ij}$ have been measured. However, there is still today an open problem that is the fact that we do not know the \emph{mass ordering} of the mass eigenstates.

\subsection{Direct Neutrino mass measurements}

Neutrino oscillation experiments have measured the difference of masses of $\nu_{i}$ states, however, are not capable of measuring the absolute mass scale of the neutrino mass, which is still an open problem. From the experimental point of view it is important to point out what means "neutrino mass", in particular, so far there are two main things that are possible to measure (or to put limits): the \emph{effective neutrino mass} $m_{\beta}$ and the sum of the masses $m_{\nu}$, as is shown in Equation (\ref{eq:nu_masses}).

\begin{equation}
    \begin{cases}
        m_{\beta}= \sqrt{\sum_i|U_{e,i}|^2m^2_i} \\
        m_{\nu}=\sum_i m_i
    \end{cases}
    \label{eq:nu_masses}
\end{equation}

The effective neutrino mass is typically measured in endpoint experiments, like PTOLEMY, while the sum of the masses it is reconstructed by cosmological observations. It is useful to mention the possibility to obtain information about neutrino mass from \emph{Neutrinoless double beta decay} experiments, in which it could be possible to measure the effective Majorana mass $m_{\beta \beta}$ \cite{serge}.

\subsubsection{Cosmology}

The cosmic neutrino background is considered to be the most abundant neutrino source in the Universe. Even though neutrinos have a very small mass, they are subject to gravitational force; this means that during the Universe expansion and after the galaxy formation, they have had the time to cluster depending by the gravitational potential of the astrophysical objects during the cosmological evolution. This clusterization process is called \emph{Large Structure formation} and is a useful tool for measuring $m_{\nu}$. However, the evaluation of the neutrino overdensity depends on the model we adopt \cite{direct_neutrino_mass}. The sum of the masses eigenstates has been measured by Planck, limiting $m_{\nu}$ to

\begin{equation}
    m_{\nu}=\sum_im_i<120~meV
\end{equation}

\subsubsection{Endpoint experiments}

When beta unstable elements emit electrons, they are emitted at different energies and as mentioned in Section (\ref{sec:CNB_detection}) the electron energy spectrum follows the Kurie function of the element. The key point for the effective mass measurement is to take into account the effect of the mass $m_{\beta}$ in the Kurie plot, this is done by considering the formula for the energy spectrum as reported in Equation (\ref{eq:Kurie_plot}).

%\begin{equation}
 %   \frac{dN}{dK}=\frac{G_F^2}{2\pi^3}|V_{ud}|^2|M|^2Fp_e(K+m_e)(K_{end,0}-K)\left( \sum_{i=1}^3|U_{e,i}|^2\sqrt{(K_{end,0}-K)^2-m_i^2} \right)
  %  \label{eq:Kurie_plot}
%\end{equation}

\begin{equation}
    \frac{dN}{dK}=\frac{G_F^2}{2\pi^3}|V_{ud}|^2|M|^2Fp_e(K+m_e)(K_{end,0}-K) \sqrt{(K_{end,0}-K)^2-m_{\beta}^2} 
    \label{eq:Kurie_plot}
\end{equation}

where $K_{end,0}$ is the endpoint energy in the case of a massless neutrino, $|V_{ud}|^2$ is the \emph{Cabibbo angle} and $M$ is the nuclear matrix element.\\
An important element of this formula is that in the proximity of the endpoint, the distortion due to the factor $(K_{end,0}-K)^2-m_{\beta}^2$ becomes non negligible. By performing a fit of the number of decays per unit of energy, it is possible to measure the value of the effective neutrino mass. The experiments that exploit this method for mass measurement can be divided, depending on their experimental technique, into three main categories:

\begin{itemize}
    \item Electrostatic filter experiments
    \item Calorimetric experiments
    \item Radio-frequency radiation experiments
\end{itemize}

\begin{figure}[h!]
    \centering

    \begin{tikzpicture}

    \draw (-6,-5.2)--(6,-5.2)--(0,5.2)--(-6,-5.2);

    \node[] at (0,5.5){Electrostatic Filter};
    \node[] at (-6.3,-5.5){Cyclotron radiation};
    \node[] at (5.8,-5.5){Calorimetry};

    \node[] at (0,-1){PTOLEMY};
    \node[] at (-5,-5){Project-8};
    \node[] at (5,-4.8){Holmes};
    \node[] at (4.9,-4.4){Echo};
    \node[] at (0,3.7){KATRIN};
    \node[] at (-4.1,-4.2){QTNM};
    
    \end{tikzpicture}
    
    \caption{World effort for direct neutrino mass measurements. 
    On the vertex, moving clockwise, experiments which exploit electrostatic selection technology (KATRIN), calorimetric (ECHO \cite{echo}, HOLMES) and cyclotron radiation detection (PROJECT-8, QTNM \cite{QTNM}). In the middle PTOLEMY that aims at combine the features of all three technologies.}
    \label{fig:Nu_mass_experiments}
\end{figure}
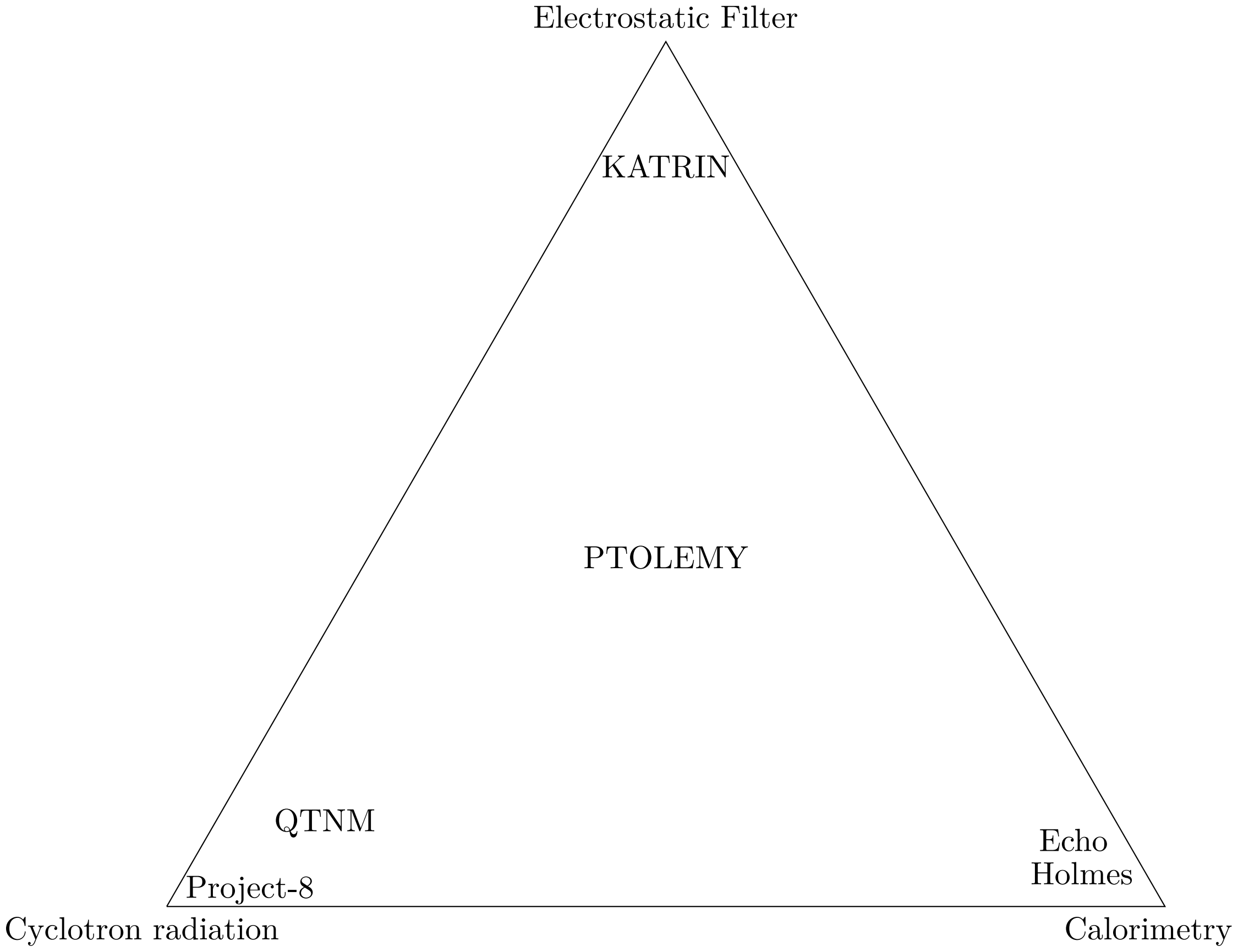

The endpoint neutrino mass experiments can be modeled in a diagram as shown in Figure (\ref{fig:Nu_mass_experiments}). 

\newpage
Each of these experiments has particular experimental limitations and strong points. Taking as example the concept of the KATRIN experiment \cite{KATRIN}, which consists in exploiting a magnetic bottle, is simple and well established in terms of electron trajectory but has the severe limitation of the tritium gas source that cannot be increased in mass (gas pressure) and the spectrometer is not scalable, essentially limiting the detector for further significant increase of performances, which are however expected in 2030 with the KATRIN++ phase. The experiments that instead exploit quantum calorimetric sensors, taking as an example HOLMES \cite{holmes}, have the experimental limitation that the goal neutrino mass sensitivity, which depends by the energy resolution on the holmium endpoint and by the capability of recording large event number, is much higher than the current KATRIN limit. The most advanced project of the radio-frequency (RF) experiments is PROJECT-8 which has proven the experimental concept in two phases, proving the possibility to achieve a very good energy resolution of electron energy with the CRES technique \cite{p8}, but has also shown some limiting factors for a neutrino mass measurement like the amount of Tritium needed to run the experiment and the error on the reconstructed energy, which is given by the additional frequency emitted by the electron. This additional frequency depends on the intensity of the magnetic field in the emission position\footnote{In PROJECT-8 this is represented by the difference between a \emph{shallow-trap} and \emph{deep-trap} configuration, while in the case of the Electron Trap this effect will be called "potential-well effect" and it is described in Section (\ref{sec:montecarlo}).} \cite{p8}; a more detailed description of this kind of limiting factors will be given in Chapter (\ref{chap:Conclusions}) with a comparison with the Electron Trap setup. All of those experiments have in common the non scalability of the target or the difficulty in increasing the energy resolution.\\
The most updated limit in the effective neutrino mass is given by KATRIN experiment \cite{katrin_limit} and is

\begin{equation}
    m_{\beta}<450~meV \hspace{1cm} (90\% \text{ C.L.})
\end{equation}

\newpage
\section{Electron motion in Magnetic and Electric fields} \label{sec:e_motion}

Before entering into the description of the PTOLEMY experiment, it is useful to introduce the electron dynamic in magnetic and electric fields. In fact, the whole PTOLEMY detector concept relies on exploiting the features of trajectories of charged particles in the magnetic field in all steps.

%\subsection{Electron motion in Uniform $\Vec{B}$ field}
\subsection{Uniform $\Vec{B}$}
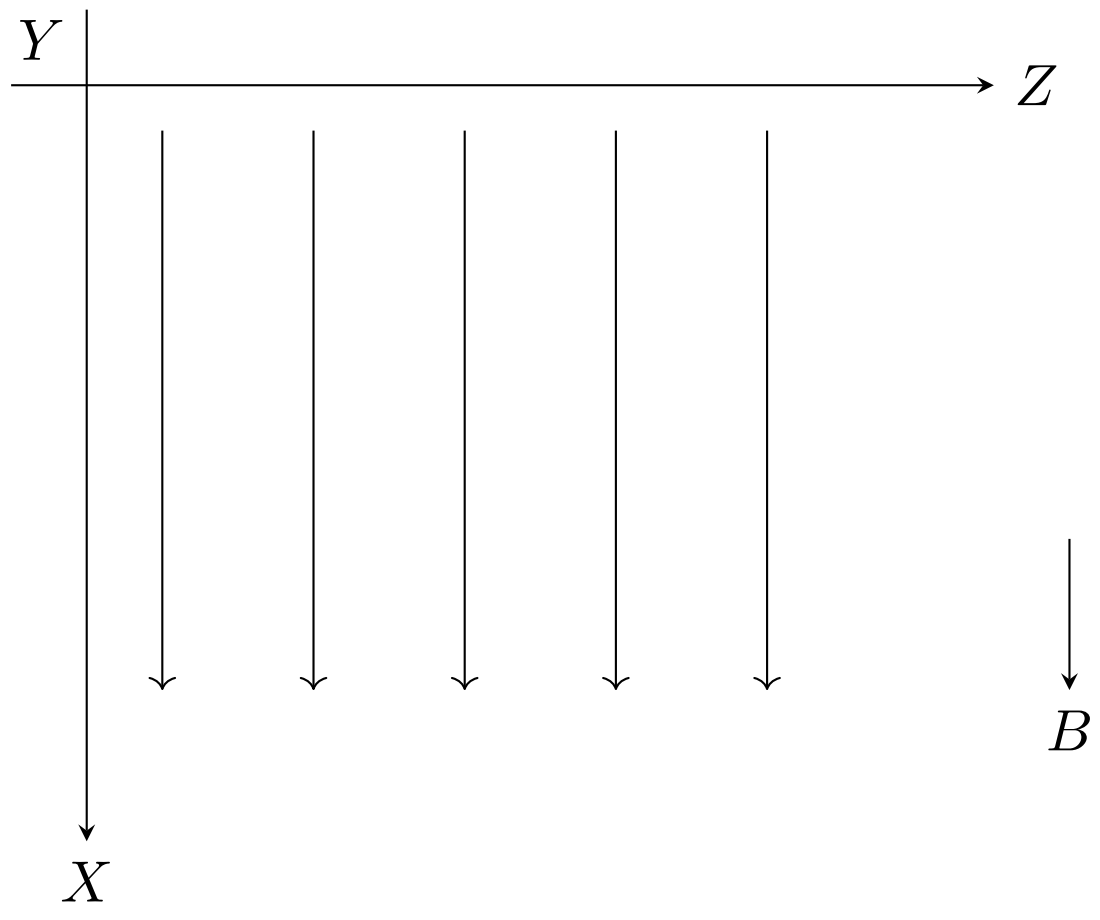
\begin{figure}[h!]
	\centering
	\begin{tikzpicture}
		%axis
		\draw[-stealth](0,0.5)--(0,-5) node[below]{$X$};
		\draw[-stealth](-0.5,0)--(6,0) node[right]{$Z$};
		\node[] at (-0.3,0.3){$Y$};
		
		%fields
		\draw[->](0.5,-0.3)--(0.5,-4);
		\draw[->](1.5,-0.3)--(1.5,-4);
		\draw[->](2.5,-0.3)--(2.5,-4);
		\draw[->](3.5,-0.3)--(3.5,-4);
		\draw[->](4.5,-0.3)--(4.5,-4);
		
		%legend
		\draw[-stealth](6.5,-3)--(6.5,-4) node[below]{$B$};
		
	\end{tikzpicture}
	\caption{Uniform magnetic field $B$, the $Y$ axis is directed from the scheme to the viewer.}
	\label{fig:UniformB_drawing}
\end{figure}

The first step is to evaluate the electron motion in a uniform and static magnetic field orientated along the X axis like $\vec{B}=(B,0,0)$ as shown in Figure (\ref{fig:UniformB_drawing}), this study is useful to show the features of the electron motion in the RF tracker, which lies in the uniform region of the PTOLEMY magnet, and in the case of the uniform field of the R\&D setup called electron trap. It is worth pointing out that the following calculations are performed, for the moment, in the non relativistic case \cite{plasma}.\\

Taking $\vec{v}$ as the velocity of the electron in the laboratory reference frame, the equation of motion can be derived by using the Lorentz force:

\begin{equation}
\begin{cases}
\Vec{F}=q\Vec{v} \times \Vec{B} \\

\Vec{F}=m\frac{d\Vec{v}}{dt}

\label{eq:LorentzF1}
\end{cases}
\end{equation}

where $m$ is the mass and $q=-|q|$ is the charge of the electron.\\
The equation of motion in the transverse plane with respect to $\Vec{B}$ direction can be derived by Equation (\ref{eq:LorentzF1}) 

\begin{equation}
\begin{cases}
	
\dot{v}_y=-(q/m)Bv_z=+(|q|/m)Bv_z\\

\dot{v}_z=(q/m)Bv_y=-(|q|/m)Bv_y

\end{cases}
\end{equation}

that gives

\begin{equation}
\begin{cases}
\ddot{v}_y=-(qB/m)^2v_y\\
\ddot{v}_z=-(qB/m)^2v_z
\end{cases}
\label{equation:uniform_eq_motion}
\end{equation}

which are the equations of motion of a harmonic oscillator.
The magnitude of the velocity in the plane orthogonal to the magnetic field lines is $v_{\perp}$ and is such that $\displaystyle \Vec{v}=\Vec{v}_{\perp}+\Vec{v}_{\parallel}$, obtaining
 
 \begin{equation}
 \begin{cases}
 v_y=v_{\perp}e^{i\omega_0t}\\
 v_z=iv_{\perp}e^{i\omega_0t}
 \end{cases}
 \end{equation}
 
with  

\begin{equation}
\centering
\omega_0=\frac{|q|B}{m}
\label{eq:cyclotron_frequency}
\end{equation}

and 

\begin{equation}
\centering
r_L=v_{\perp}/\omega_0=\frac{m}{|q|B}v_{\perp}
\label{eq:Larmor_Radius}
\end{equation}

 In conclusion, the transverse part of the motion is a circular motion with frequency $\omega_0$ which is called \emph{the cyclotron frequency}, and radius $r_L$ which is the \emph{Larmor radius}. While in the longitudinal direction the motion is simply constant, in fact 

\begin{equation}
	\centering
	m\dot{v}_x=0
\end{equation}

and so the longitudinal velocity $v_{\parallel}$ corresponds to the initial velocity

\begin{equation}
	\centering
	v_{\parallel}=v_{x,0}=v_0\cos\theta
\end{equation}

where $\theta$ is the angle between the electron velocity and the magnetic field lines and is also known as the \emph{pitch angle}.\\
The resultant electron motion is a circular motion in the transverse plane and a uniform motion in the longitudinal direction, across the magnetic field lines, as can be seen in Equation (\ref{eq:xyz_gyromotion}).
 
 \begin{equation}
 	\begin{cases}
 		x(t)=v_{x,0}t+x_0\\
		y(t)=r_L\sin(\omega_0t)+y_0\\
 		z(t)=r_L\cos(\omega_0t)+z_0
	 \end{cases}
	\label{eq:xyz_gyromotion}
\end{equation}

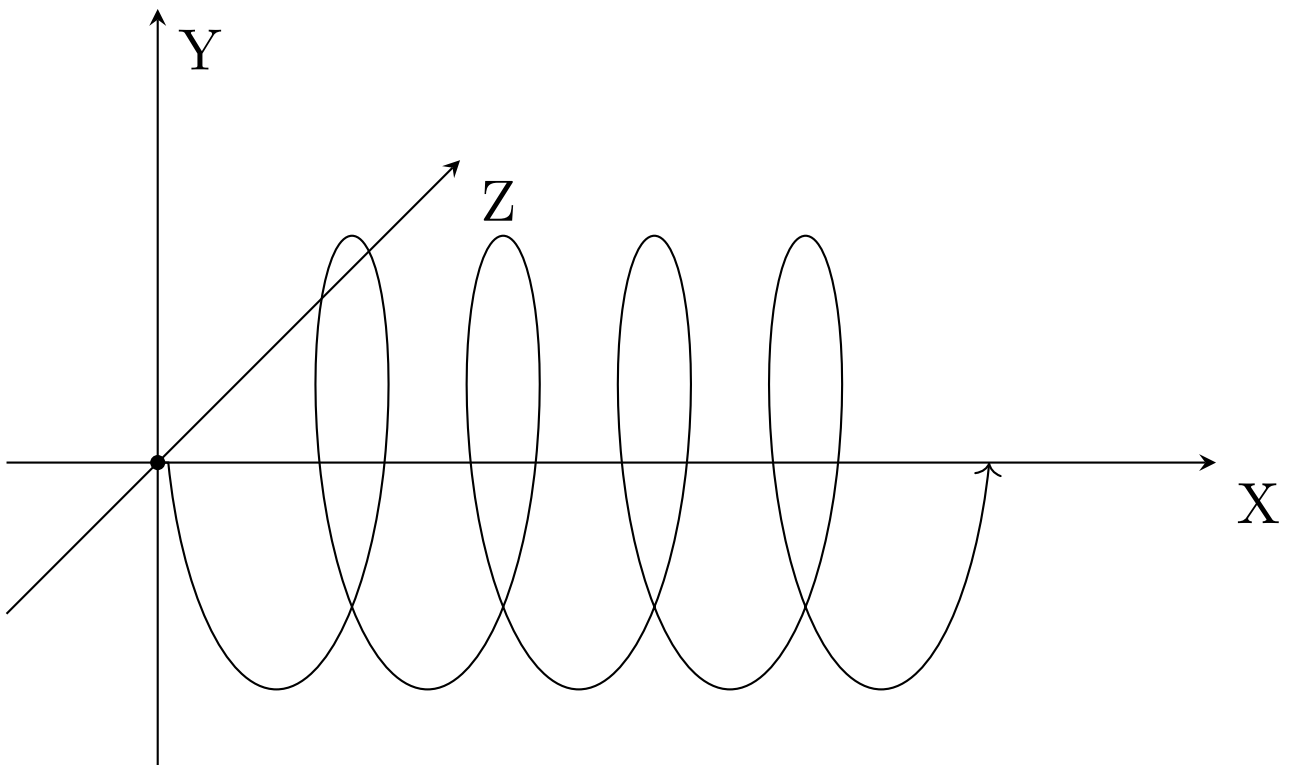
\begin{figure}[h!]
	\centering
	\begin{tikzpicture}
    \coordinate (origo) at (0,0) node[black, anchor = east] {};
    \coordinate (pivot) at (1,5);
    % draw axes
    \fill[black] (origo) circle (0.05);
    \draw[-stealth] (-1,0) -- ++(8,0) node (mary)[black, anchor =  north west] {X};
    \draw[-stealth] (0,-2) -- ++(0,5) node[black,anchor = north west]{Y};
    \draw[-stealth] (-1,-1) -- (2,2) node () [black,anchor = north west]{Z};
    %\draw[thick,->] (origo) -- (2.5,0) node[anchor = north east]{$\Vec{B}$};

    \draw[thin,decoration={aspect=0.31, segment length=10mm,
     amplitude=1.5cm,coil},decorate,arrows = {<[bend]-}] (5.5,0) --(0,0);
    %\draw[thick,->, red] (0,2.5) -- (0,5) node[anchor = west]{$\Vec{v}_\perp$};
    %\draw[thick,->, red] (0,2.5) -- (1,2.5) node[anchor = south east]{$\Vec{v}_\parallel$};
    %\draw[thick,->] (0,2.5) -- (1,5) node[anchor = west]{$\Vec{v}$};
         %\coordinate (electron) at (0,2.5) node[black, anchor = east] {};
         %\fill[black] (electron) circle (0.05);
\end{tikzpicture}
    \caption{Trajectory of a cyclotron motion of an electron in a static and uniform magnetic field. }
	\label{fig:e_gyromotion1}
\end{figure}

This spiraling motion is also called \emph{cyclotron motion} and is depicted in Figure (\ref{fig:e_gyromotion1}).

A convenient way to describe the electron motion in such a combination of fields is to introduce the so called \emph{guiding center system} or GCS reference frame. The GCS is the reference frame with the origin at the average electron position after one turn, indeed the velocity of the origin of the GCS is calculated by averaging the velocity of the electron in the laboratory frame as reported in Equation (\ref{eq:V_GCS_general}).

\begin{equation}
	\begin{cases}
		V_{\parallel}= \langle v_{\parallel}\rangle \simeq v_{\parallel}\\
		V_{\perp}= \langle v_{\perp}\rangle
	\end{cases}
	\label{eq:V_GCS_general}
\end{equation}

In the case of uniform and static $\Vec{B}$ the origin of the GCS moves with longitudinal velocity $V_{\parallel}=v_{\parallel}$ and perpendicular velocity $V_{\perp}=0$, this case is particularly useful, as described in Section (\ref{sec:uniform_montecarlo}).\\
It is possible to obtain a relation between the GCS and the laboratory reference frame, indeed considering that $\Vec{X}$ is the position of the origin of the GCS, while $\Vec{x}^*$ is the position of the electron with respect to the GCS, it is possible to obtain that $\Vec{x}=\Vec{x}^*+\Vec{X}$ is the position of the electron in the laboratory frame.\\
Similarly, $\Vec{V}$ is the velocity of the GCS in the laboratory frame, $\Vec{v}^*$ is the velocity of the electron in the GCS, and  $\displaystyle \Vec{v}=\Vec{v}^*+\Vec{V}$ is the velocity of the electron in the laboratory frame. It is worth to define the kinetic energy related to the velocity of the GCS as $K=K_{\parallel}+K_{\perp}=\frac{1}{2}m(V_{\parallel}^2+V_{\perp}^2)$, where $K_{\parallel}$ ($K_{\perp}$) is the kinetic energy longitudinal (transverse) the magnetic field lines.

%The motion of the electron in the laboratory frame can be approximated to the motion of the origin of the GCS plus a drift orthogonal to $B$ field lines, and in the general case the velocity of the electron in the GCS is

%\begin{equation}
%	\begin{cases}
%		v^*_{\parallel}=0\\
%		v^*_{\perp}=v_{\perp} - V_{\perp}
%	\end{cases}
%\end{equation}

\subsection{Gradient $\Vec{B}$ drift}
%\subsection{Electron motion in non Uniform and Static $\Vec{B}$ field}

 Let's now describe the motion of the electron in non uniform magnetic field, starting with a gradient $B$ oriented across the Z axis, this example is particularly interesting to understand the PTOLEMY filter concept. The presence of a gradient, as explained in detail in \cite{plasma}, introduces a force, leading the GCS to moves with a non zero orthogonal velocity.\\
For the purposes of this work, it is useful to consider a magnetic field with the intensity that decreases while $Z$ coordinates increase thus the gradient is oriented across $-Z$.

\begin{equation*}
\centering
\Vec{B}=(B(z),0,0)
\end{equation*}

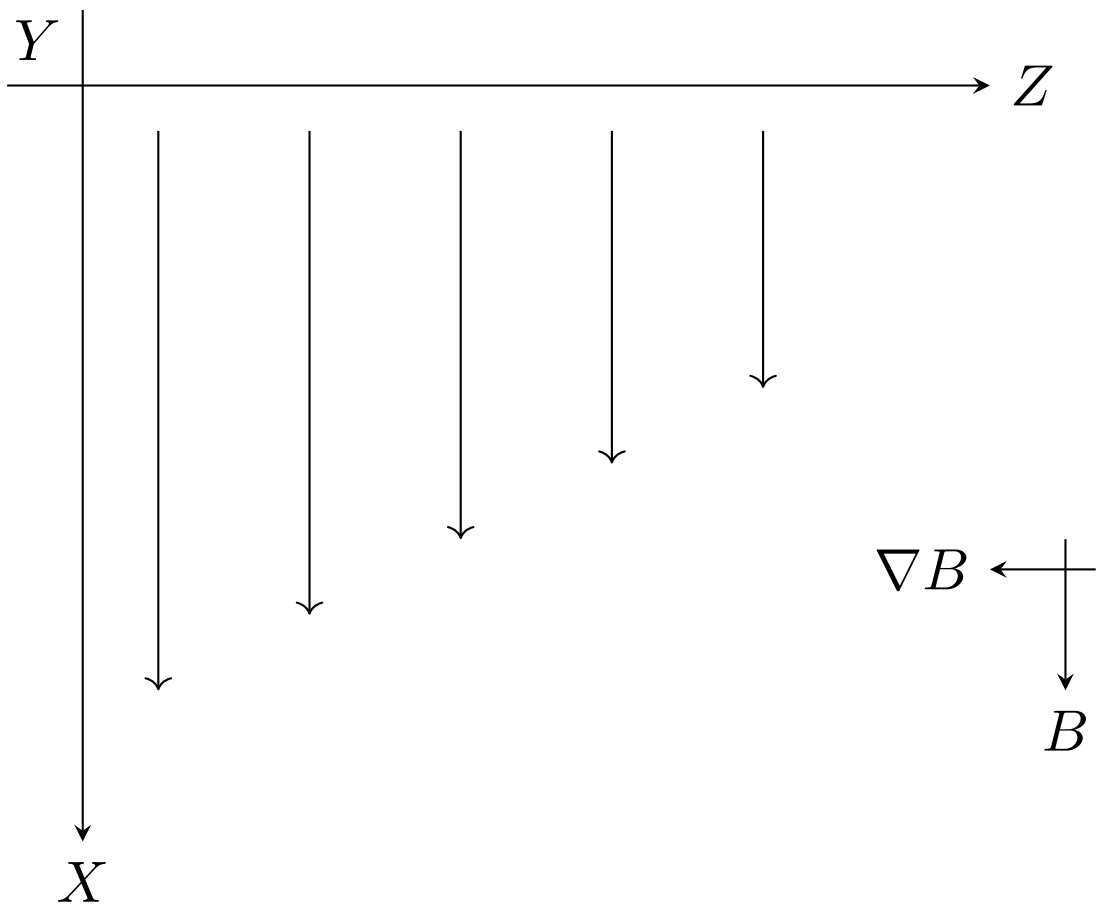
\begin{figure}[h!]
	\centering
	\begin{tikzpicture}
		%axis
		\draw[-stealth](0,0.5)--(0,-5) node[below]{$X$};
		\draw[-stealth](-0.5,0)--(6,0) node[right]{$Z$};
		\node[] at (-0.3,0.3){$Y$};
		
		%fields
		\draw[->](0.5,-0.3)--(0.5,-4);
		\draw[->](1.5,-0.3)--(1.5,-3.5);
		\draw[->](2.5,-0.3)--(2.5,-3);
		\draw[->](3.5,-0.3)--(3.5,-2.5);
		\draw[->](4.5,-0.3)--(4.5,-2);
		
		%legend
		\draw[-stealth](6.5,-3)--(6.5,-4) node[below]{$B$};
		\draw[-stealth](6.7,-3.2)--(6,-3.2) node[left]{$\nabla B$};
		
	\end{tikzpicture}
\caption{Magnetic field lines with gradient field oriented along $-Z$ direction. The $Y$ axis is outgoing the scheme towards the reader.}
\label{fig:B_non_uniform}
\end{figure}

When an electron starts to move in gyromotion around the magnetic field lines like the one depicted in Figure (\ref{fig:B_non_uniform}) it experiences a different magnitude of the Lorentz force depending by the $Z$ coordinate. This can be described by including a generic force $\Vec{F}$ in the equation of motion:

\begin{equation}
\centering
m\frac{d\Vec{v}}{dt}=\Vec{F} +q\Vec{v}\times \Vec{B}
\label{eq:F_gen}
\end{equation}

By averaging in the velocity the first term of Equation (\ref{eq:F_gen}) vanishes, and so the velocity of the electron is $\Vec{V}$ that has been defined in Equation (\ref{eq:V_GCS_general}):

\begin{equation}
\centering
0=\frac{1}{q}\Vec{F}+\Vec{V} \times \Vec{B}
\end{equation}

If we take the $\Vec{B}$ vector product we obtain

\begin{equation}
\centering
\Vec{F} \times \Vec{B} =
-\Vec{B} \times (\Vec{B} \times \Vec{V}) =
\Vec{B} \times (\Vec{V} \times \Vec{B}) =
\Vec{V}B^2-\Vec{B}(\Vec{V} \cdot \Vec{B})
\label{eq:FxB}
\end{equation}

It is possible to obtain the transverse velocity of the CGS due to the force $\Vec{F}$

\begin{equation}
\centering
\Vec{V}_{F, \perp}=\frac{1}{q}\frac{\Vec{F} \times \Vec{B}}{B^2}
\end{equation}

That is obtained by Equation (\ref{eq:FxB}), taking the transverse component the term with the scalar product vanishes. This general description is useful to understand the motion under a generic external force that can, for example, be gravity.\\
In the case of a non uniform magnetic field the formula becomes

\begin{equation}
\centering
\Vec{V}_{\nabla B} = -\frac{1}{|q|} \frac{\Vec{F}_{\nabla B} \times \Vec{B}}{B^2}
\label{eq:V_grad2}
\end{equation}

that is of fundamental importance for PTOLEMY's filter understanding.\\
It is important to point out that the GCS approximation holds when the Larmor radius $r_L$ is small compared to the distance $L$ ($\frac{B}{\nabla B}\gg r_{L}$) which is the scale length of the gradient field \cite{plasma}.\\
%But what is the force made by the gradient field? What we expect is that the motion along the $y$ axis is unchanged\footnote{In the sense that do a circular motion like the B uniform case.} and so the only component of the gradient force is along the $z$ axis.\\ 
%Looking at the magnetic field in Figure (\ref{fig:B_non_uniform}) one can see that the electron will experience a Lorentz force larger for low $z$ position than when the electron is at high $z$ position. This explains the presence of the product $\Vec{F}_{\nabla B} \times \Vec{B}$. Everything holds in the approximation that the Larmor radius is small compared to the distance $L$ which is the scale length of the gradient field \cite{plasma}.\\
It is possible to derive the explicit formula for the force due to the presence of a gradient field, starting with the Lorentz force (\ref{eq:LorentzF1}) and Equation (\ref{eq:xyz_gyromotion}) we obtain that

\begin{equation}
\begin{cases}
F_{\nabla,x}=0\\

F_{\nabla,y}=|q|v_zB_x(y)=i|q|v_{\perp}[cos(\omega_0t)+i~sin(\omega_0t)] \cdot [B_{0,y}+0 \cdot y(t)\frac{\partial B_x}{\partial y}]\\

F_{\nabla,z}=-|q|v_y B_x(z)=-|q|v_{\perp}[cos(\omega_0t)+i~sin(\omega_0t)] \cdot [B_{0,z}-z(t) \frac{\partial B_x}{\partial z}]

\end{cases}
\end{equation}

Where $\Vec{B}$ is expanded at first order

\begin{equation}
\Vec{B}=\Vec{B_0}+(\Vec{r} \cdot \Vec{\nabla})\Vec{B}+...
\label{eq:expansion_B}
\end{equation}

such that $\Vec{B}_0$ is the magnetic field of the uniform case of Figure (\ref{fig:UniformB_drawing}).\\
It is possible to use the GCS approximation to calculate the average gradient force $\bar{F}_{\nabla B}$:

\begin{equation}
\begin{cases}
\bar{F}_{\nabla,x}=0\\

\bar{F}_{\nabla,y}= \langle i|q|v_{\perp}[cos(\omega_0t)+i~sin(\omega_0t)] \cdot [B_{0,y}+0 \cdot r_Lsin(\omega_0t)\frac{\partial B_x}{\partial y}] \rangle =0\\

\bar{F}_{\nabla,z}= \langle -|q|v_{\perp} [cos(\omega_0t)+i~sin(\omega_0t)] \cdot [B_{0,z}-r_Lcos(\omega_0t) \frac{\partial B_x}{\partial z}] \rangle =\\
= \langle +|q|v_{\perp}r_Lcos^2(\omega_0t)\frac{\partial B_x}{\partial z} \rangle =\\
=+\frac{1}{2}|q|v_{\perp}r_L\frac{\partial B_x}{\partial z}
\end{cases}
\end{equation}

where has been used the expression described above such that $\displaystyle r_L\ll L$, under which condition $\partial B_x/\partial z$ is constant.\\
Finally, the only non vanishing component is 

\begin{equation}
\centering
\bar{F}_{\nabla,z}=+\frac{1}{2}|q|v_{\perp}r_L\frac{\partial B_x}{\partial z}
\label{eq:f_y}
\end{equation}

and the resultant velocity is

\begin{equation}
\centering
\Vec{V}_{\nabla B}=-\frac{1}{|q|}\frac{\bar{F}_{\nabla B}\times \Vec{B}}{B^2}=+\frac{1}{2}\frac{v_{\perp}r_L}{B}\frac{\partial B_x}{\partial z} \hat{y}
\end{equation}

that is the velocity of the guiding center system due to the gradient field of Figure (\ref{fig:B_non_uniform}). A first generalization is to take a generic field $\Vec{B}$ and his gradient field, in this case the average force $\bar{F}_{\nabla B}$ is

\begin{equation}
\centering
\bar{F}_{\nabla B}=+\frac{1}{2}|q|v_{\perp}r_L\Vec{\nabla}B
\end{equation}

that gives 

\begin{equation}
\centering
\Vec{V}_{\nabla_{\perp} B}=+\frac{1}{2}v_{\perp}r_L\frac{\Vec{\nabla}_{\perp}B\times \Vec{B}}{B^2}
\label{eq:V_grad}
\end{equation}

%\subsection{Electron motion in Curved and Static $\Vec{B}$ field}
\subsection{Curvature drift}
\begin{figure}[h!]
    \centering

    \begin{tikzpicture}

        \draw[-stealth](0,0)--(0,1.8) node [left]{$R_c$};

        \draw[-stealth](1.47,1.6) arc (45:135:2);
        \draw[-stealth](1.8,2.1) arc (45:135:2.5) node [left]{$\vec{B}$};
        \draw[-stealth](2.2,2.6) arc (45:135:3);

        \draw[-stealth](0,3.7)--(0,4.7) node [right]{$\hat{r}$};
        \draw[-stealth](0,3.7)--(-1,3.7) node [left]{$\hat{\theta}$};              
    \end{tikzpicture}   
    \caption{Curved magnetic field $\Vec{B}$ lines with curvature radius $R_c$. }
    \label{fig:curve_B}
\end{figure}
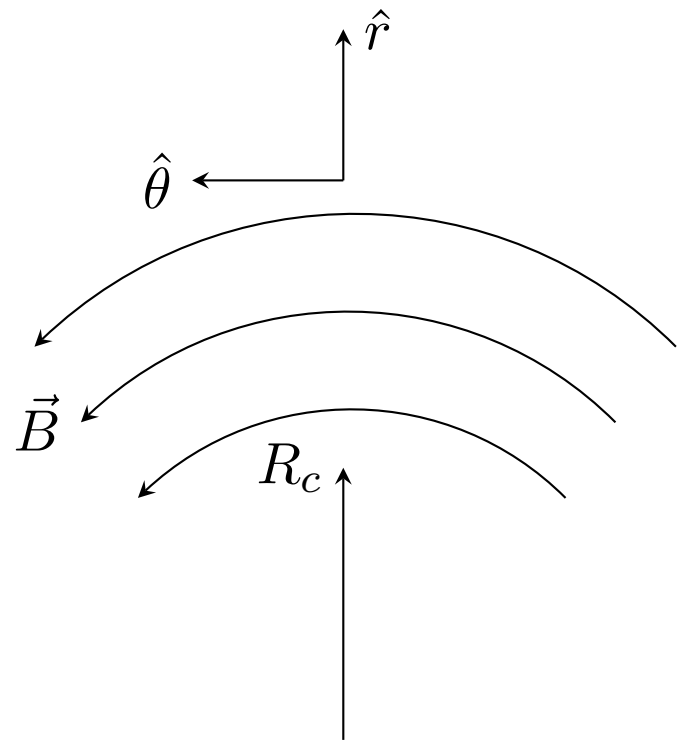

After the introduction of the electron motion in a uniform and non uniform magnetic field, it is useful to introduce the case in which the magnetic field is curved, the equations that are derived in this section are used in the electron motion in the PTOLEMY filter. A curved magnetic field makes two forces on the electron: one is the centrifugal force that GCS experiences in the perpendicular direction of $\Vec{B}$; the other one is the force longitudinal with respect to the field lines which is responsible of the so called \emph{magnetic mirror effect} \cite{plasma}. By imposing $\Vec{\nabla} \times \Vec{B}=0$ and taking $\hat{r}$ and $\hat{\theta}$ as the radial and longitudinal directions, we obtain that the gradient field is perpendicular with respect to the field lines and decreases as $\Vec{|B|}=B\propto 1/R_c$, where $R_c$ is the curvature radius \cite{plasma}. The radial and longitudinal directions in a curved magnetic field are shown in Figure (\ref{fig:curve_B}).\\

This means that the transverse gradient is $\displaystyle \Vec{\nabla}_{\perp}B=R_c\hat{r}\frac{\partial B}{\partial r}\propto -\frac{1}{R^2_c}R_c\hat{r}$ , which can be rewritten as 

\begin{equation}
\centering
\frac{\Vec{\nabla}_{\perp}B}{B}=-\frac{\hat{r}}{R_c}
\end{equation}

The transverse velocity due to the gradient drift force is obtained by Equation (\ref{eq:V_grad}) and corresponds to

\begin{equation}
\centering
\Vec{V}_{\nabla_{\perp}B}=-\frac{1}{2}\frac{mv^2_{\perp}}{qB^2}\frac{\hat{r}\times \Vec{B}}{R_c}
\label{eq:V_grad_B_Curvo}
\end{equation}

Another contribution along the transverse direction is the velocity due to the centrifugal force, which can be calculated in a similar way and the result is described in Equation (\ref{eq:centrifugal_vel}).

\begin{equation}
\centering
\Vec{V}_{CF}=\frac{mv^2_{\parallel}}{qB^2}\frac{\hat{r}\times \Vec{B}}{R_c}
\label{eq:centrifugal_vel}
\end{equation}

in which has been used the formula for the centrifugal force $ \displaystyle \Vec{F}_{CF}=m\frac{v^2_{\parallel}}{R_c}\hat{r}$.\\
The overall formula of transverse velocity in the curve magnetic field is then

\begin{equation}%V o v?
\centering
\Vec{V}_{\nabla_{\perp}B-CF}=\frac{1}{2}\frac{m}{q}\frac{1}{R_cB^2}(v^2_{\perp}+2v^2_{\parallel})\hat{r}\times\Vec{B}=\frac{1}{2}m(v^2_{\perp}+2v^2_{\parallel})\frac{\Vec{B}\times \Vec{\nabla}_{\perp}B}{qB^3}
\label{eq:grad-cf}
\end{equation}

This resultant velocity is used in the PTOLEMY filter to balance the drift motion of the electrons, it is possible to study the consequence of the gradient of the magnetic field along the longitudinal direction $\displaystyle \Vec{\nabla}_{\parallel}B$.\\
In general we find that the divergence of the magnetic field is zero, that in cylindrical coordinates it is given by

\begin{equation}
\centering
\frac{1}{r}\frac{\partial}{\partial r}(rB_r)+\frac{1}{r}\frac{\partial B_{\theta}}{\partial \theta} + \frac{\partial B_x}{\partial x}=0
\label{eq:B_Divergence}
\end{equation}

Assume that our problem has cylindrical symmetry, in this case the term $B_{\theta}$ vanishes. From Equation (\ref{eq:B_Divergence}) we can obtain an expression for $B_r$ in the assumption that the value of $\partial B_x/\partial x$ do not vary too much with respect to the value that has at $r=0$

\begin{equation}
\centering
B_r=-\frac{1}{2}r \left( \frac{\partial B_x}{\partial x} \right) \Bigr\rvert_{r = 0}
\label{eq:B_r}
\end{equation}

This expression is then used to calculate the contributions of the Lorentz force, in particular, it is possible to obtain

\begin{equation}
\begin{cases}
F_r=q(v_{\theta}B_x-v_xB_{\theta})=qv_{\theta}B_x\\
F_{\theta}=q(-v_rB_x+v_xB_r)\\
F_x=q(v_rB_{\theta}-v_{\theta}B_r)=-qv_{\theta}B_r
\end{cases}
\end{equation}

where $F_x$ can be rewritten as

\begin{equation}
\centering 
F_x=\frac{1}{2}qv_{\theta}r\frac{\partial B_x}{\partial x}
\end{equation}

Like in the other cases, the electron moves quickly in gyromotion and so it makes sense to compute the mean force

\begin{equation}
\centering
\bar{F}_x=-\frac{1}{2}|q|v_{\perp}r_L\frac{\partial B_x}{\partial x}=-\frac{1}{2}\frac{mv^2_{\perp}}{B}\frac{\partial B_x}{\partial x}
\end{equation}

where  $v_{\theta}=+v_{\perp}$ and $r=r_L$.\\
This longitudinal force can be rewritten in a useful notation

\begin{equation}
\centering
\Vec{F}_{\parallel}=-\mu \Vec{\nabla}_{\parallel}B
\end{equation}

where $\displaystyle \mu=\frac{mv^{2}_{\perp}}{2B}$ is the magnetic moment of the electron and is a quantity conserved during an adiabatic motion \cite{plasma}.\\

%\subsection{Electron motion in Uniform $\Vec{B}$ and Uniform $\Vec{E}$ fields}
\subsection{$\Vec{E} \times \Vec{B}$ drift}
The remaining case that we need to analyze, is when an electron is immersed in a uniform magnetic and electric \cite{pesce}. Given the electric $\Vec{E} = (0,E,0)$ and uniform magnetic field shown in Figure (\ref{fig:UniformB_drawing}) the Lorentz force is always determined by Equation (\ref{eq:LorentzF1}) and can be rewritten as
 
\begin{equation}\notag
    \begin{cases}
        m\dot{v}_x = 0\\
        m\dot{v}_y = qv_zB + qE\\
        m\dot{v}_z = -qv_yB\\
        \end{cases}
\end{equation}

then applying the derivative over time of both sides, as already done in Equation (\ref{equation:uniform_eq_motion}), it becomes

\begin{equation}\notag
    \begin{cases}
        \Ddot{v}_y = \frac{qB}{m}\dot{v}_z = -\left(\frac{qB}{m}\right)^2v_y\\
        \Ddot{v}_z = -\frac{qB}{m}\dot{v}_y = -\frac{qB}{m}\left(\frac{qB}{m}v_z+\frac{q}{m}E\right)
    \end{cases}
\end{equation}

The solutions of the equation of motion gives a similar result of the uniform magnetic field case, with an additional term that is responsible for an additional drift in the Z direction

\begin{equation}\notag
    z(t) = z_0 + r_L\sin\left(\omega_0 t\right) +\frac{E}{B} t
\end{equation}

In conclusion, the equation of motion of an electron in a region with uniform magnetic and electric field are given by Equation (\ref{eq:EOM_E}), where $ V_{D} = \frac{E}{B}$ .

\begin{equation}
    \begin{cases}
        x(t) =  v_{x,0} t +x_0\\
        y(t) =   r_L\sin(\omega_0t)+ y_0\\
        z(t) =  r_L\cos(\omega_0t) + V_{D}t +z_0\\
    \end{cases}\label{eq:EOM_E}
\end{equation}

In conclusion, the introduction of a static and uniform electric field, rise a linear motion along Z, with the GCS velocity $\vec{V}_{D}$.\\ 
The example just mentioned is useful for understanding the presence of short tracks in the Electron Trap setup. As shown in \cite{plasma} a general results for any electric field is given by the Equation (\ref{eq:Edrift_general}).

\begin{equation}\label{eq:Edrift_general}
    \Vec{V}_{D} = \frac{\Vec{E}\times \Vec{B}}{|B|^2}
\end{equation}

%% file: sections/2PTOLEMY_Project.tex
\chapter{PTOLEMY Project} \label{chap:PTOLEMY_Project}

In Section (\ref{sec:neutrino_mass}) a brief overview of the lagrangian neutrino mass terms was discussed, and the most relevant direct neutrino mass experiments have been presented. In Section (\ref{sec:e_motion}) the electron motion in electric and magnetic fields, which is necessary to understand the working principle of the PTOLEMY detector, has been described in details. The PTOLEMY collaboration is developing a novel detector concept that combines the strengths of each existing neutrino mass experiment.\\
In this chapter, the concept of the PTOLEMY detector is described and relies on relies on the use of Tritium as a target for neutrino interactions. This detector can be used to measure the neutrino mass and, in the future, to directly detect the cosmic neutrino background.\\

\section{PTOLEMY detector concept}

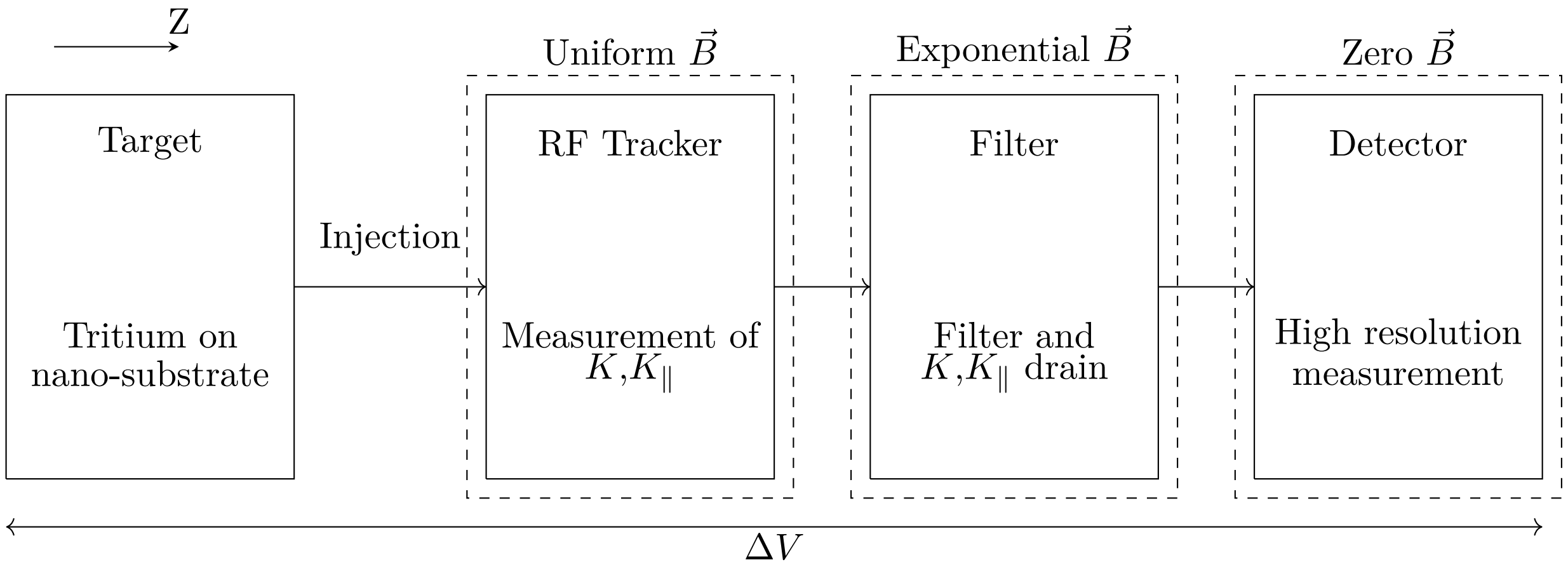
\begin{figure}[h!]
    %\centering
    %\includegraphics[scale=0.35]{imgs/ptolemy_block_diagram.png}
    \hspace{-1.5cm}
    \begin{tikzpicture}

        \draw[-stealth](0.5,2.5)--(1.8,2.5)node[above]{Z};
        \draw[dashed](4.8,2.2)--(8.2,2.2)--(8.2,-2.2)--(4.8,-2.2)--(4.8,2.2);
        \node at (6.5,2.5){Uniform $\vec{B}$};
        \draw[dashed](8.8,2.2)--(12.2,2.2)--(12.2,-2.2)--(8.8,-2.2)--(8.8,2.2);        
        \node at (10.5,2.5){Exponential $\vec{B}$};
        \draw[dashed](12.8,2.2)--(16.2,2.2)--(16.2,-2.2)--(12.8,-2.2)--(12.8,2.2);        
        \node at (14.5,2.5){Zero $\vec{B}$};
    
        \draw(0,-2)--(3,-2)--(3,2)--(0,2)--(0,-2);
        \node at (1.5,1.5){Target};
        \node at (1.5,-0.5){Tritium on};
        \node at (1.5,-0.9){nano-substrate};
        
        \draw(5,-2)--(8,-2)--(8,2)--(5,2)--(5,-2);
        \node at (6.5,1.5){RF Tracker};
        \node at (6.5,-0.5){Measurement of};
        \node at (6.5,-0.9){$K$,$K_{\parallel}$};
        
        \draw(9,-2)--(12,-2)--(12,2)--(9,2)--(9,-2);
        \node at (10.5,1.5){Filter};
        \node at (10.5,-0.5){Filter and};
        \node at (10.5,-0.9){$K$,$K_{\parallel}$ drain};
        
        \draw(13,-2)--(16,-2)--(16,2)--(13,2)--(13,-2);
        \node at (14.5,1.5){Detector};
        \node at(14.5,-0.5){High resolution};
        \node at(14.5,-0.9){measurement};

        \draw[->](3,0)--(5,0);
        \node at(4,0.5){Injection};
        \draw[->](8,0)--(9,0);
        \draw[->](12,0)--(13,0);

        \draw[<->](0,-2.5)--(16,-2.5);
        \node at (8,-2.7){$\Delta V$};

    \end{tikzpicture}
    \caption{Block diagram of the PTOLEMY project. $\Delta V$ is the potential difference between the target and the final detector.}
    \label{fig:ptolemy_block_diagram}
\end{figure}

The concept of the PTOLEMY detector has been under development for more than a decade with the initial goal to create a detector of cosmic neutrino background direct detection, and the design has been updated according to technological advancement over time. In the following, the working principle of the PTOLEMY detector is given. The concept of the PTOLEMY detector can be schematized with a block diagram composed of four sub-sections: the Tritium target, the RF tracker, the novel dynamic filter and the electron analyzer. \\
The detector working principle is described as follows: an electron emitted from the Tritium target is injected into an electrode system and with suited magnetic and electric fields, is driven in the RF tracker, that is located in the central part of the magnet, in which a preliminary spectrum measurement is done with an energy resolution of about $\sigma_{K,RF} \sim 10~eV$. After this, by measuring the longitudinal energy with respect to the B field lines, the potentials at the entrance of the dynamic filter are set according to the measurement done in the RF tracker. The event by event voltage setting serves to optimize the potential electrostatic barrier to force the electron against it and transform the initial kinetic energy in electrostatic potential energy. Finally, the electron is focused with a Einzel lens system in a system capable of measuring the energy of the electrons with $\sigma_{K,det}=50~meV$.\\
The PTOLEMY detector block diagram is shown in Figure (\ref{fig:ptolemy_block_diagram}).

\begin{figure}[h!]
    \centering
    \includegraphics[scale=0.5]{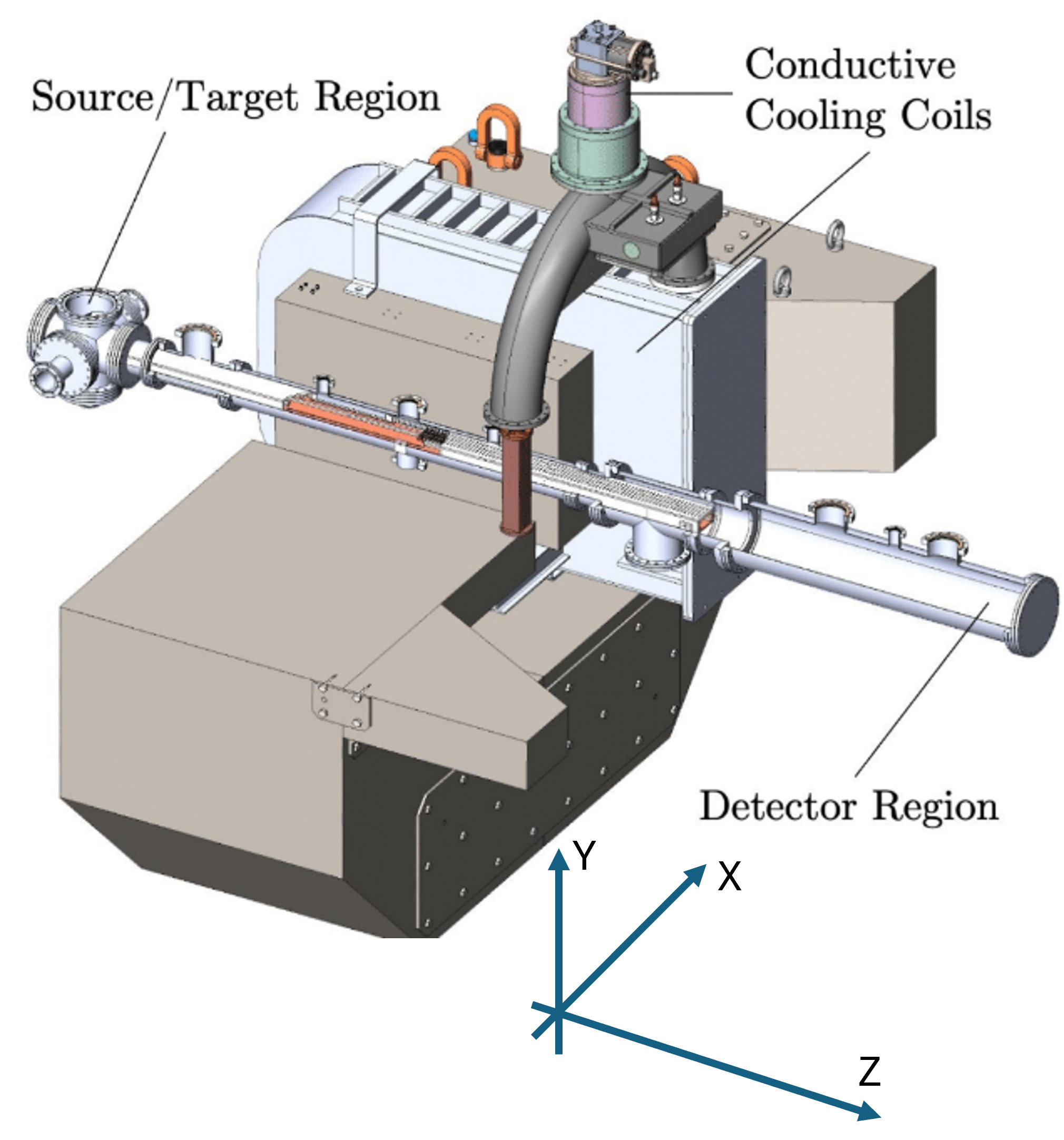}
    \caption{PTOLEMY superferric magnet: two superconductive coils energizes the massive yoke.}
    \label{fig:PTOLEMY_magnet}
\end{figure}

The Tritium energy spectrum (initial kinetic energy $K_i$) can be reconstructed by using Equation (\ref{eq:PTOLEMY_initial_energy_reconstruction})

\begin{equation}
K_i=q \left( V_{target}-V_{det}\right) + E_{RF,corr} +K_{det}
    \label{eq:PTOLEMY_initial_energy_reconstruction}
\end{equation}

where $E_{RF,corr}$ is the correction of the energy loss for cyclotron motion during the electron trajectory and $K_{det}$ is the energy measured at the final stage.\\
The magnet is the pivot device around which the PTOLEMY detector concept turns, in particular it fulfills the requirements imposed by the electron filtering process and cyclotron radiation emission. This will be discussed in more details in Section (\ref{sec:RF_region}, \ref{sec:filter}). The PTOLEMY magnet, that is shown in Figure (\ref{fig:PTOLEMY_magnet}), is a dipole with an iron yoke energized by two superconducting coils. With respect to a standard dipolar magnet, two horns have been added in order to get the desired decay length of the fringe field, that matches the filter requirements.

\newpage

Before describing the PTOLEMY detector, it is useful to introduce the reference system that is going to be used in this work and it is typically used in the PTOLEMY's works. This reference system, that is shown in Figure (\ref{fig:PTOLEMY_magnet}), is based on the orientation of the magnet, in particular the X axis leys on the direction of the dominant field component, the Z direction corresponds to the overall drift of the electrons and the Y axis is the vertical direction across which acts mainly the electric field. In the following pages, each detector part is described and then a comprehensive overview of the experimental methods is given.

\subsection{Tritium Target} \label{sec:tritium_target}

As described in Chapter (\ref{chap:Introduction}), Tritium has the advantage of being the best solution for the C$\nu$B capture by exploiting the NCB reaction:

\begin{equation*}
    \nu_e + ^3_1H \rightarrow ^3_2He^++e^-
\end{equation*}

\begin{figure}[h!]
    \centering
    \includegraphics[scale=0.4]{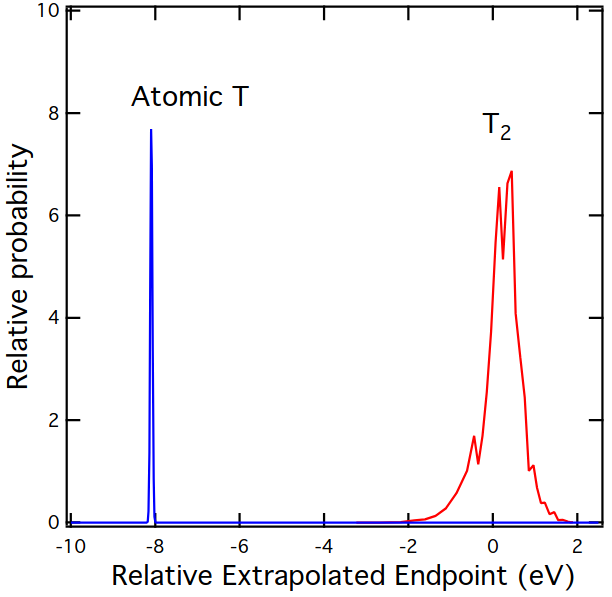}
    \caption{Comparison of energy smearing of bounded molecular Tritium (red) with respect to atomic Tritium (blue) \cite{direct_neutrino_mass}.}
    \label{fig:atomic_vs_molecular_tritium}
\end{figure}

The main background process for a cosmic neutrino background direct detection is the standard beta decay of the Tritium,

\begin{equation*}
         ^3_1H \rightarrow ^3_2He^++e^- +\bar{\nu_e}
\end{equation*}

that can be used, instead, for the neutrino mass measurement by measuring the spectrum distortion at the endpoint of the energy spectrum.\\
The Tritium Kurie plot with the relic neutrino absorption peak is shown in Figure (\ref{fig:tritium_kurie_plot_toy}).
However, in the context of relic neutrino detection, it is necessary to demonstrate the feasibility of instrumenting a large mass of Tritium, much larger than any instrumented mass exploited so far. Then a scalability of the target must be proved to be able to achieve the desired instrumented mass; for example, in Section (\ref{sec:scalabilità}) a proposal for target scalability is described. Tritium $^3_1H$ is typically used in the gaseous state, that comes in the molecular form $T_2$. A molecular Tritium gas target introduce an experimental problem that is the energy smearing of the emitted electron because of the bounded state of the two Tritium atoms; in particular this is due to the activation of possible roto-vibrational degree of freedom of the molecule which turns into a deterioration of the energy prediction of the emitted electrons. The spectra of atomic and molecular Tritium are shown in Figure (\ref{fig:atomic_vs_molecular_tritium}). The solution to this problem was primarily proposed by the PTOLEMY collaboration, suggesting using atoms of Tritium bounded to monoatomic layers of graphene. In this case, Tritium is not really free but is in a covalent bound state. This approach has the advantage of allowing to collect a significant amount of Tritium; in contrast, the KATRIN++ \cite{KATRIN} and PROJECT-8 \cite{p8-atomic} experiment aim at using a really pure atomic Tritium target but in this case there are strong limitations to the amount of instrumented Tritium. In particular in a given volume arbitrarily large quantities of Tritium cannot be stored without increasing the recombination process at the experimental vessel walls, which goes in the opposite direction of having the desired target in atomic form \cite{walls}.\\

\begin{figure}[h!]
    \centering
    \includegraphics[scale=0.5]{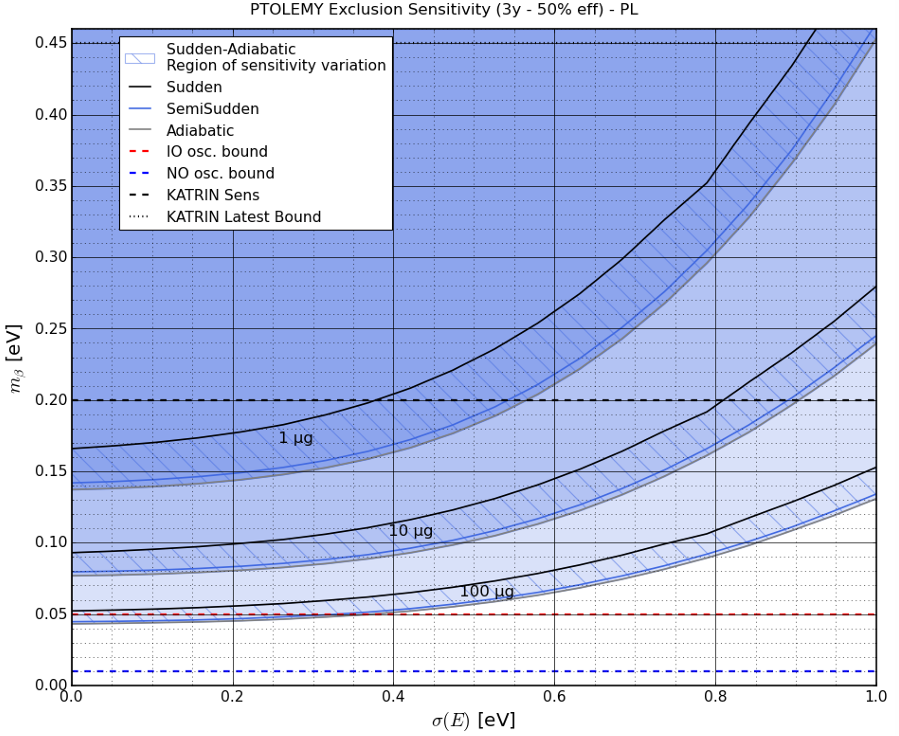}
    \caption{PTOLEMY neutrino mass sensitivity with 1,10 and 100 $\mu g$ of instrumented mass, in which the dashed regions depict the uncertainty of the theory models while the truth is expected to lay in the model of the dashed zone. The sensitivity has been evaluated for three theoretical models \cite{paper_teor2} that need experimental evaluation.}
    \label{fig:PTOLEMY_sensitivity}
\end{figure}

\newpage
Returning to the PTOLEMY Tritium storage option and assuming a $100\%$ of Tritium on graphene loading, which means almost $3\times 10^{15} ~atoms/cm^2$, a $1~\mu g$ of Tritium on single graphene layer corresponds to a rectangular target of dimensions $ \sim7\times7~cm^2$ \footnote{Assuming an efficiency of about $50\%$. For more details see Section (\ref{sec:scalabilità}).}. Another great advantage of using a mono-layer graphene is that the process of stacking tiles can tolerate several graphene layers, given the very low probability of inelastic scattering of the electrons while crossing more than one layer. The quantitative measurement electron inelastic scattering cross section through the graphene layer is under measurements by the Roma3 group of the PTOLEMY collaboration. The bound state of Tritium to the Carbon atoms of graphene layers brings back a problem similar to the one of the activation of roto-vibrational degree of freedoms of the Tritium molecule, and this is due to the bound state which around a minimum of the binding energy is described by an harmonic oscillator. In this case the Tritium state leys in an energy eigenstate where the energy is well defined, while the momentum is defined only as the mean value on an operator on a given quantum state. Thus, when the decay takes place, the free daughter ($^3_2He^+$) and the electron will collapse in a free state where the momentum and energy value will be well defined and the uncertainty on the momentum will be reflected on an uncertainty on energy. This effect is inversely proportional to the mass of the recoiling particle, however it is worth saying that the exact value of this effect is unknown and can only be measured. In the paper \cite{heisenberg} the PTOLEMY collaboration gave a detailed description of the process. The effect just mentioned does not significantly impact the direct neutrino mass measurement. The sensitivity curves, shown in Figure (\ref{fig:PTOLEMY_sensitivity}), for different instrumented target masses, account for different models predicting the decay rate of Tritium atoms when considered in covalent bound with graphene. The dashed area span the theoretical uncertainty on decays.
 
 \begin{comment}
 assuming different theoretical models and several instrumented mass, are shown in Figure (\ref{fig:PTOLEMY_sensitivity}), in which the dashed regions depict the uncertainty of the theory models while the truth is expected to lay in the model of the dashed zone. The lines representing the sensitivity for each instrumented mass reports shown in Figure

 In Section (\ref{sec:systematics}) the systematic effect of theoretical models are discussed.
\end{comment}

\newpage

\subsection{Target arrangement and injection into the magnet} \label{Sec:target_injection}

 As mentioned previously, the implementation of a solid tritium target allows the possibility to store more tritium per unit of volume with respect to the gas source. In the following we show that this imposes constraints on the injection of charged particles from the target region to the high B field; the main issue is the adiabaticity of the motion of an electron, in particular the conservation of the first adiabatic invariant $\mu$,
 
\begin{equation}
\mu=\frac{K_{\perp}}{B}
    \label{eq:first_adiabatic}
\end{equation}

\begin{figure}[h!]
    \centering
    \includegraphics[scale=0.3]{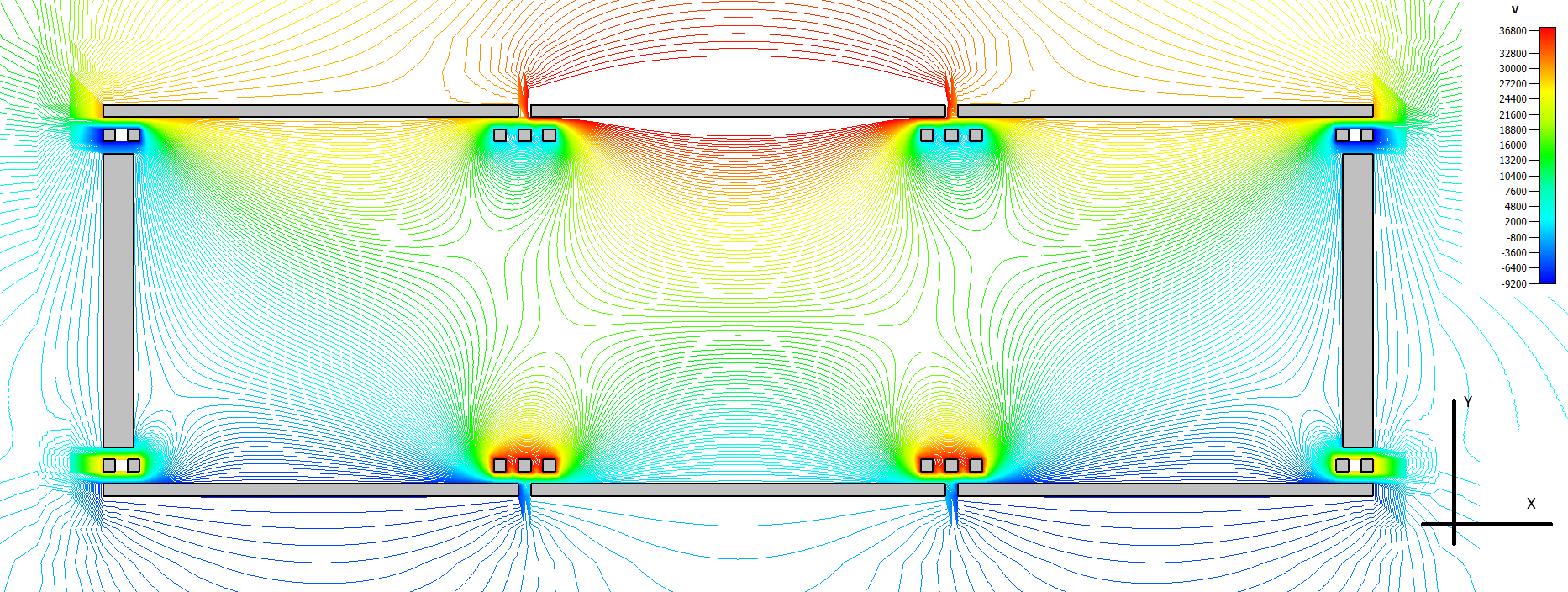}
    \caption{Injection electrodes system with potential isolines.}
    \label{fig:filtro_acceleratore}
\end{figure}

For an adiabatic motion, this value must be conserved during the motion.\\
For example let's consider an electron with initial transverse kinetic energy $K_{\perp,1}=K\sin^2 \theta_1$ emitted in a low magnetic field region $B_1$ and then injected toward a high magnetic field $B_2$, for $\mu$ conservation, it is possible to obtain 

\begin{equation}
    \sin\theta_2=\sin\theta_1\sqrt{\frac{B_2}{B_1}}
\end{equation}

where $\theta_2$ is the pitch angle of the electron in the high field region. Taking $B_1=0.1~T$ and $B_2=1~T$, it follows that in order to obtain a high field pitch angle of $\theta_2^{max}=90^{\circ}$, the maximum initial pitch angle must be $\theta_1^{max}=18.4^{\circ}$. An electron emitted in the low field region with pitch angle greater than $\theta_1^{max}$ then will be reflected back. It is actually possible to avoid the electron reflection and then control the drift toward the high field region by properly accelerating the electron. This injection process can be assisted by emitting the electrons directly in a comparable high field region of about $1~T$, in this way it could be possible to avoid an excessive acceleration of the transverse kinetic energy of the electron. To do so, the PTOLEMY collaboration foresees placing the tritiated graphene on a permanent magnet, with central field between 1 and $1.2~T$, in this way the emitted electron can arrive at the magnet, where the RF tracker is located. The geometry and configuration of this magnetic type target depend on the quantity of Tritium that is desired to store in the target part of the detector. After being emitted, the electrons drift towards the so called "injection electrodes system" that is an electrode system with a potential configuration analogous to that of the transverse dynamic filter, as shown in Figure (\ref{fig:filtro_acceleratore}). \\

\begin{figure}[h!]
    \centering
    \includegraphics[scale=0.5]{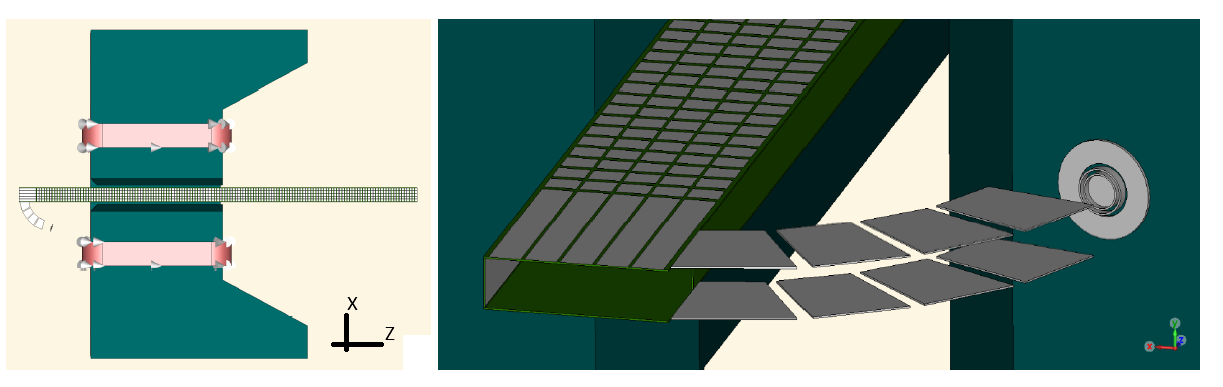}
    \caption{Injection electrodes system with Tritium on graphene target positioned on cylindrical permanent magnet.}
    \label{fig:injection_electrodes_setup}
\end{figure}

An example of target geometry is shown in Figure (\ref{fig:injection_electrodes_setup}), where a cylindrical magnet of $0.5~T$ has been used. This geometry can be used to store a tritium mass capable of reaching sensitivity to neutrino mass, however, a proposal for high tritium mass target arrangement is described in Section (\ref{sec:scalabilità}).\\

\newpage
The main feature of this injection electrodes system is that, by exploiting the bouncing motion of the electrons in the electrodes system shown in Figure (\ref{fig:filtro_acceleratore}), it is also possible to accelerate the electrons by collimating the electron pitch angles from half of the total solid angle at the emission to a smaller phase space ($\theta^{min}<\theta<\theta^{max}$), due to the initial magnetic field. The values of $\theta^{min}$ and $\theta^{max}$ depend on the magnetic field ratio and acceleration.\\

\subsection{The RF Tracker} \label{sec:RF_region}

\begin{figure}[h!]
    \centering
    \includegraphics[scale=0.35]{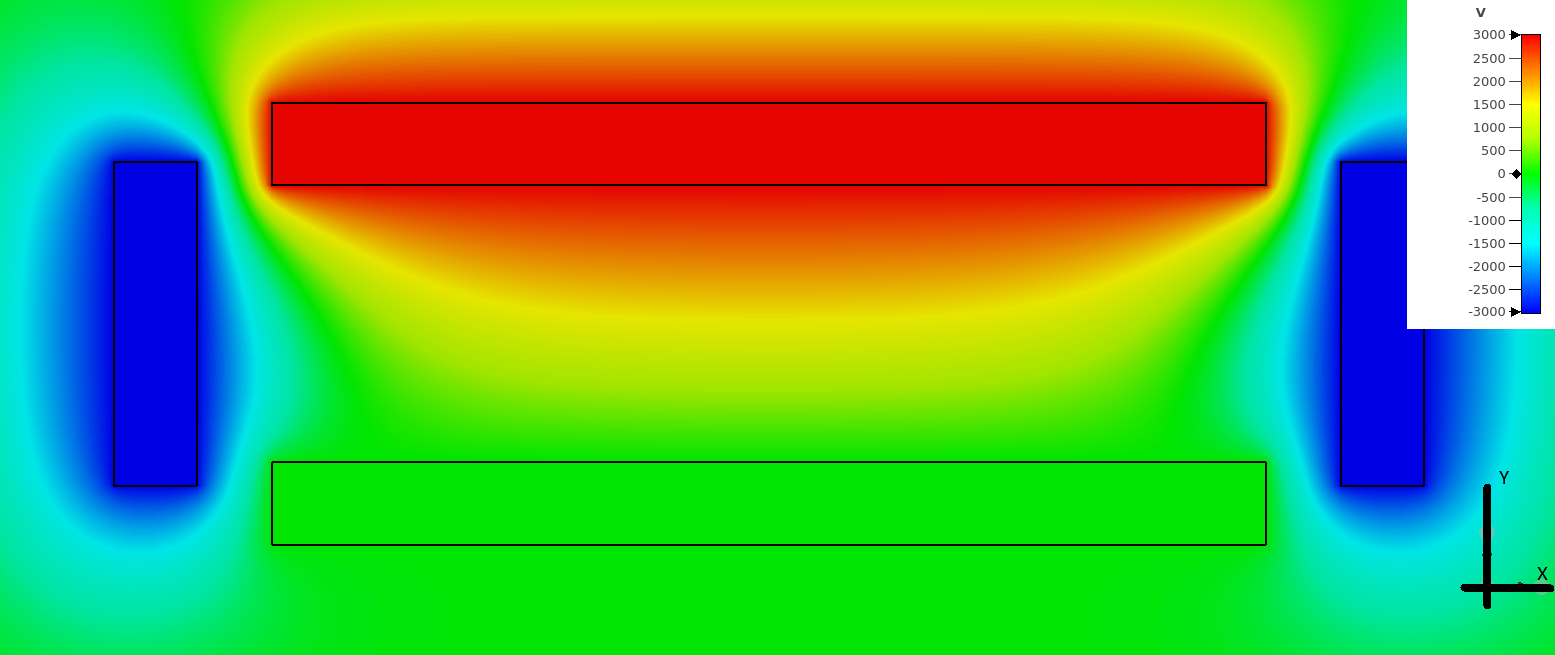}
    \caption{Conceptual design of PTOLEMY RF tracker electrode system with an example of electric potential.}
    \label{fig:rf_region_potential}
\end{figure}

The main goal of the RF tracker sub-detector is to measure the kinetic energy $K$ and the longitudinal kinetic energy $K_{\parallel}$ of the incoming electrons after the injection into the magnet. The utility of this detector part is crucial to trigger and set the first electrodes of the transverse dynamic filter and the main challenge is that these measurements must be performed on an event by event basis for electrons in an ROI at $Q_{\beta}$ within $10~eV$. Starting from the beginning, the electrons that are injected from the target are drifted in a region with bouncing potentials, in which the electron moves in a cyclotron and bouncing-drift motion. The electrode setup consists of the bouncing electrodes that are arranged along the Z axis and are positioned on the side, and two drift electrodes that are positioned on the top and bottom of this RF tracker, the schematic of the RF tracker electrode system is shown in Figure (\ref{fig:rf_region_potential}).

\newpage

\begin{figure}[h!]
    \centering
    \includegraphics[width=0.7\linewidth]{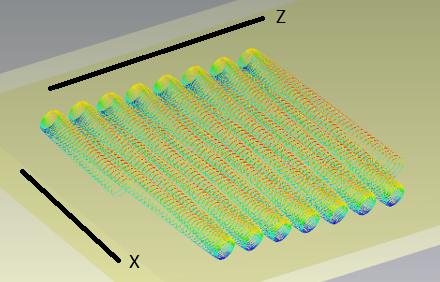}
    \caption{Simulation of the expected electron motion in the RF tracker. The cyclotron motion is due to the presence of a magnetic field, the bouncing motion is given by to the presence of the bouncing electrodes while the Z drift is due to the $\vec{E}\times \vec{B}$ effect.}
    \label{fig:motion_RF}
\end{figure}

The electrodes and potential configuration shown in Figure (\ref{fig:rf_region_potential}), according to the laws that have been described in Section (\ref{sec:e_motion}), cause the electrons to move in cyclotron and drift motion as shown in Figure (\ref{fig:motion_RF}).

By exploiting the feature of this electron motion it is possible to reconstruct the two kinematic quantities of interest, the reconstruction method is described in more detail in Section (\ref{sec:kin_reconstruction}), in particular in the case of the Electron Trap setup.\\
This measurement is realized using the so called \emph{CRES technique}, that has been developed by PROJECT-8 collaboration \cite{p8}. This method consists of measuring the Larmor frequency of the cyclotron radiation emitted by the electron during its motion, in particular as can be seen in Equation (\ref{eq:larmor_frequency})

\begin{equation}
    f=\frac{1}{2\pi}\frac{|q|B}{m}\frac{1}{K/m+1}
    \label{eq:larmor_frequency}
\end{equation}

the frequency depends on the kinetic energy $K$ of the spiraling electron. Then, thanks to the periodicity of the bouncing motion, it is possible to reconstruct the longitudinal kinetic energy $K_{\parallel}$, also this technique is described in Section (\ref{sec:kin_reconstruction}).\\
The radiation emitted by the electron in the mentioned motion is given by the following Formula (\ref{eq:Larmor_Power}).

\begin{equation}
    P=\frac{2}{3}\frac{1}{4\pi \epsilon_0} \frac{q^4B^2}{m^2c^2} \left( \gamma^2 -1 \right) sin^2\theta
    \label{eq:Larmor_Power}
\end{equation}

\begin{figure}[h!]
    \centering
    \includegraphics[scale=0.3]{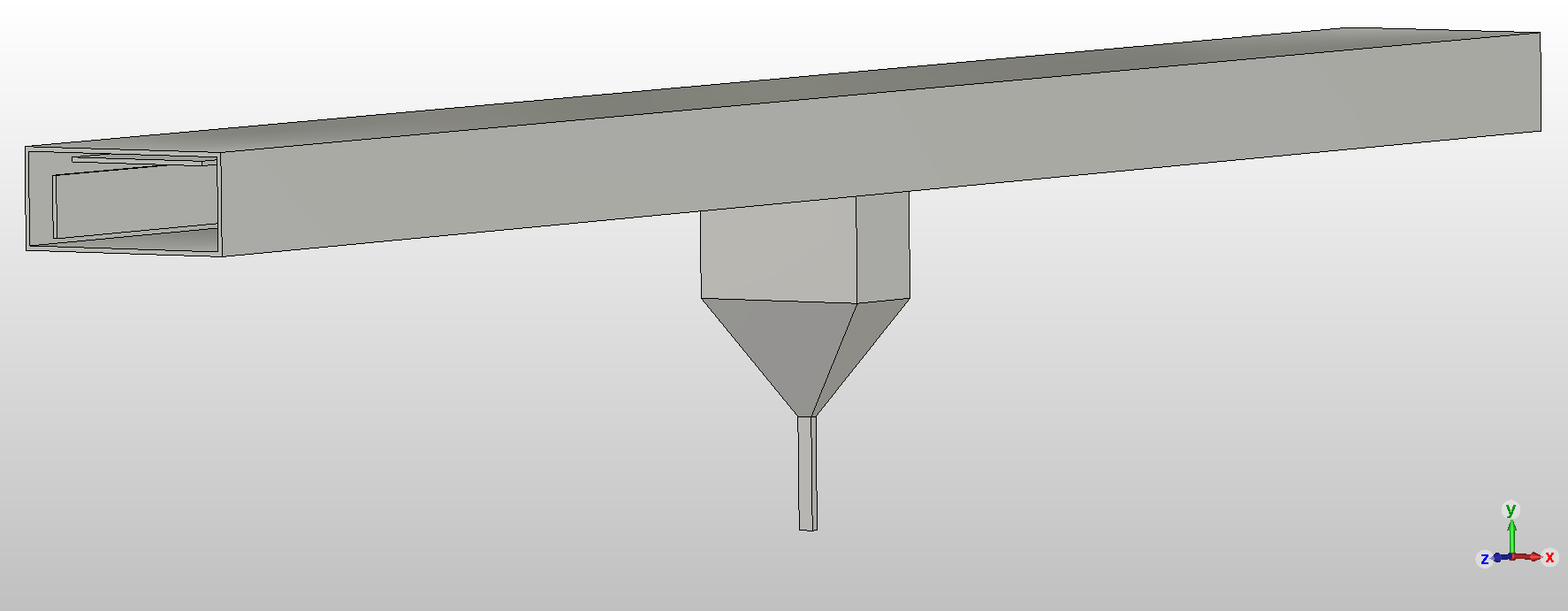}
    \caption{Conceptual design of the PTOLEMY RF tracker. The bouncing and drift motion is done between the electrodes system, the signal is picked up with a transverse rectangular waveguide. The optimization of impedance matching is postponed to future studies.}
    \label{fig:rf_region}
\end{figure}

As a reference, an electron with kinetic energy close to the one of the Tritium endpoint $K=18.6~keV$ in a $1~T$ magnetic field, emits a radiation at a frequency of $27~GHz$ and power of $1~fW$. According to the experimental requirements\footnote{For example like electrode dimensions, low power and high frequency of the signal.}, this radiation can be detected using several approaches, the possibility that is considered in this work is the implementation of a rectangular waveguide. The antenna with rectangular waveguide for the PTOLEMY RF tracker is under development and a conceptual design is shown in Figure (\ref{fig:rf_region}).

Considering the main goal as the measurement of the electron kinetic energy, it is important to note that the electrons are accelerated during the collimation process, and thus, in the RF tracker, they are going to have a higher transverse kinetic energy. Another effect due to the adiabatic collimation, is an electron kinetic energy spread in the RF tracker due to differences of the reference potential lines (which depends on value of the Y position during the drift) at which the electrons are bouncing. For example, in Figure (\ref{fig:rf_region_potential}) are shown the different potential lines by varying the Y coordinate. Another consequence of this Y coordinate drift dependence is that, depending on the Y coordinate, the electrons are going to have a different pitch angle acceptance (the potential well is less deep). \\
With such electrodes setup, the electrons exhibit a variable drift time $t_{D}$ from the RF tracker entrance to the exit, however the potentials are configured such that $t_D \simeq 10 - 100 ~\mu s$.\\
In conclusion, the requirements of the RF tracker are to measure the signal emitted by the electron, with $27~GHz$ and $1~fW$, in a time scale of $50 ~\mu s$ and with an energy resolution of about $10~eV$.

\subsection{Filter} \label{sec:filter}

\begin{figure}[h!]
    \centering
    \includegraphics[scale=0.6]{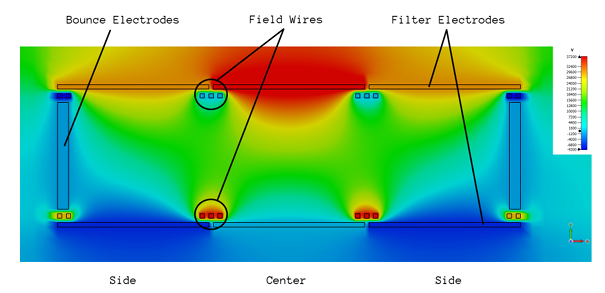}
    \caption{Filter electrodes setup, the geometry is the same as that used for injection, while the fields are optimized for the filtering purpose \cite{filtro2}. }
    \label{fig:filter_electrodes}
\end{figure}

Once electrons have been measured in the RF tracker, they drift toward the PTOLEMY filter. The filter consists of an electrode system, shown in Figure (\ref{fig:filter_electrodes}), with a geometry similar to that of the injection electrode system and the RF tracker (the bouncing and drift geometry allow for a more compact detector geometry). In this geometry, the filter is divided in two electrodes set: the central channel and the side channels.

\begin{figure}[h!]
    \centering
    \includegraphics[scale=0.55]{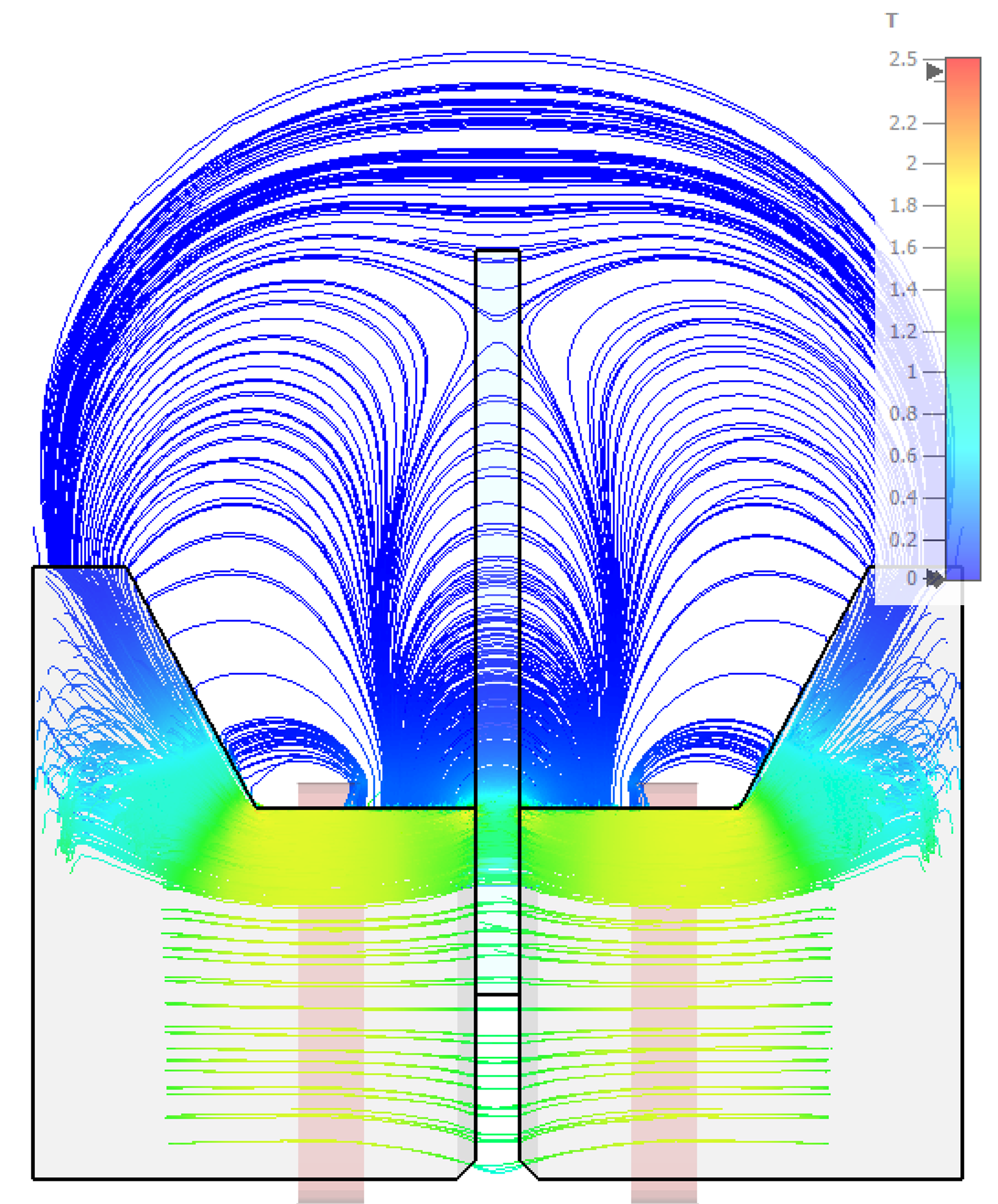}
    \caption{Simulated field map of the PTOLEMY magnet. The horns that are used to tune the field module decay length are also visible.}
    \label{fig:PTOLEMY_Bfield_lines}
\end{figure}

The potentials applied to this electrode system are such that it is capable of filtering the incoming electron and, if it is in the energy ROI, to decelerate the electrons to $O(100~eV)$. This is achieved using the information of the single electron measured in the RF tracker, in particular by setting properly the first electrode of the filter. In this Section the working principle of particle deceleration with the filter is described \cite{filtro1} \cite{filtro2}.\\
The deceleration is done by means of exploiting the controlled motion through a suitable combination of electric and non-uniform magnetic fields. The basic idea is to let the electron lose their kinetic energy at the disadvantage of the potential electric energy by controlling the electron drift. The general expression for GCS electron drift is reported in Formula (\ref{eq:V_gen_perp})

\begin{equation}
	\Vec{V_{\perp}}=\left( q\Vec{E}+\Vec{F} -\mu\Vec{\nabla}B -m\frac{d\Vec{V}}{dt} \right) \times \frac{\Vec{B}}{q|\Vec{B}|^2}
	\label{eq:V_gen_perp}
\end{equation}

in which $\Vec{F}$ is a generic force (like gravity), and $m\frac{d\Vec{V}}{dt}$ is the inertial term, the longitudinal GCS velocity is typically not influenced by the magnetic field.\\
In the case of PTOLEMY, the electric field in the filter is

\begin{equation}
	\begin{cases}
		E_x=0\\
		E_y=E_0\cos\left( y/\lambda\right)e^{-z/\lambda}\\
		E_z=-E_0\sin\left(y/\lambda\right)e^{-z/\lambda}
	\end{cases}
    \label{eq:filter_Efield}
\end{equation}

while the magnetic field is

\begin{equation}
	\begin{cases}
		B_x=B_0\cos(x/\lambda)e^{-z/\lambda}\\
		B_y=0\\
		B_z=-B_0\sin(x/\lambda)e^{-z/\lambda}
	\end{cases}
    \label{eq:filter_Bfield}
\end{equation}

where $\lambda$ is a parameter used to tune the decay of the field and is provided by the magnet already shown that is capable of doing the magnetic field lines shown in Figure (\ref{fig:PTOLEMY_Bfield_lines}).\\

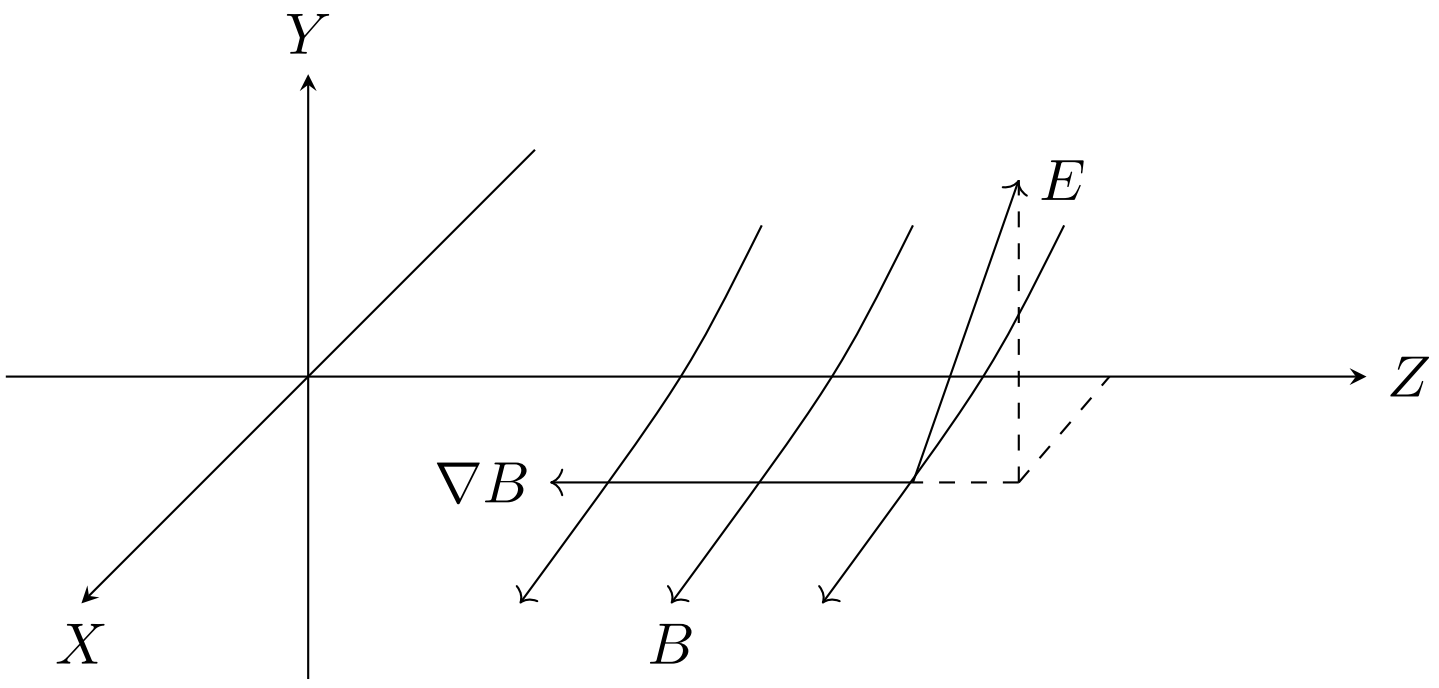
\begin{figure}[h!]
	\centering
	\begin{tikzpicture}
		\draw[-stealth](1.5,1.5)--(-1.5,-1.5) node [below]{$X$};
		\draw[-stealth](0,-2)--(0,2) node [above]{$Y$};
		\draw[-stealth](-2,0)--(7,0) node [right]{$Z$};
		
		\draw[->](3,1)..controls(2.5,0)..(1.4,-1.5);
		\draw[->](4,1)..controls(3.5,0)..(2.4,-1.5)node[below]{$B$};
		\draw[->](5,1)..controls(4.5,0)..(3.4,-1.5);
		
		\draw[->](4,-0.7)--(1.6,-0.7)node[left]{$\nabla B$};
		\draw[->](4,-0.7)--(4.7,1.3) node[right]{$E$};
		
		\draw[dashed](4.7,-0.7)--(4.7,1.3);
		\draw[dashed](4.7,-0.7)--(5.3,0);
		\draw[dashed](4.7,-0.7)--(4,-0.7);
		
	\end{tikzpicture}
	\caption{Schema of field combinations in the PTOLEMY electromagnetic filter.}
	\label{fig:filter_fields}
\end{figure}

Due the shape of the magnet horns, that are shown in Figure (\ref{fig:PTOLEMY_magnet}), the $B_x$ decays exponentially along the Z axis, resulting in a gradient field transverse to the magnetic field lines, as in the example of Section (\ref{sec:e_motion}). The overall field combination of the filter is depicted in Figure (\ref{fig:filter_fields}).

\begin{figure}[h!]
%    \centering
%    \includegraphics[scale=0.06]{imgs/potential_well_filter.jpg}

	\begin{tikzpicture}
		
		\draw[-stealth](-6,0)--(5,0) node[below]{x (a.u.)};
		\draw[-stealth](-5,-1)--(-5,3) node[left]{|V| (a.u.)};
		
		\draw(-4,3)..controls(-4,0)..(-3,0);
		\draw(4,3) ..controls(4,0)..(3,0);

        \draw[red, dashed](-4,3)..controls(-4,0.5)..(-0,0.5);
        \draw[red, dashed](4,3)..controls(4,0.5)..(0,0.5);

        \draw[blue, dashed](-4,3)..controls(-4,1)..(-0,1);
        \draw[blue, dashed](4,3)..controls(4,1)..(0,1);      
		
	\end{tikzpicture}

    \caption{Schematization of the increasing potential well (at different Z coordinate) in the central channel. The highest potential value is achieved in the intersection between the central and the side channel. The black like is the potential well at $z=z_0$, the red dashed line represent the potential after one bounce, which is slightly higher than the previous one. The blue dashed line represent the potential at the second bounce.}
    \label{fig:filter_potential_well}
\end{figure}
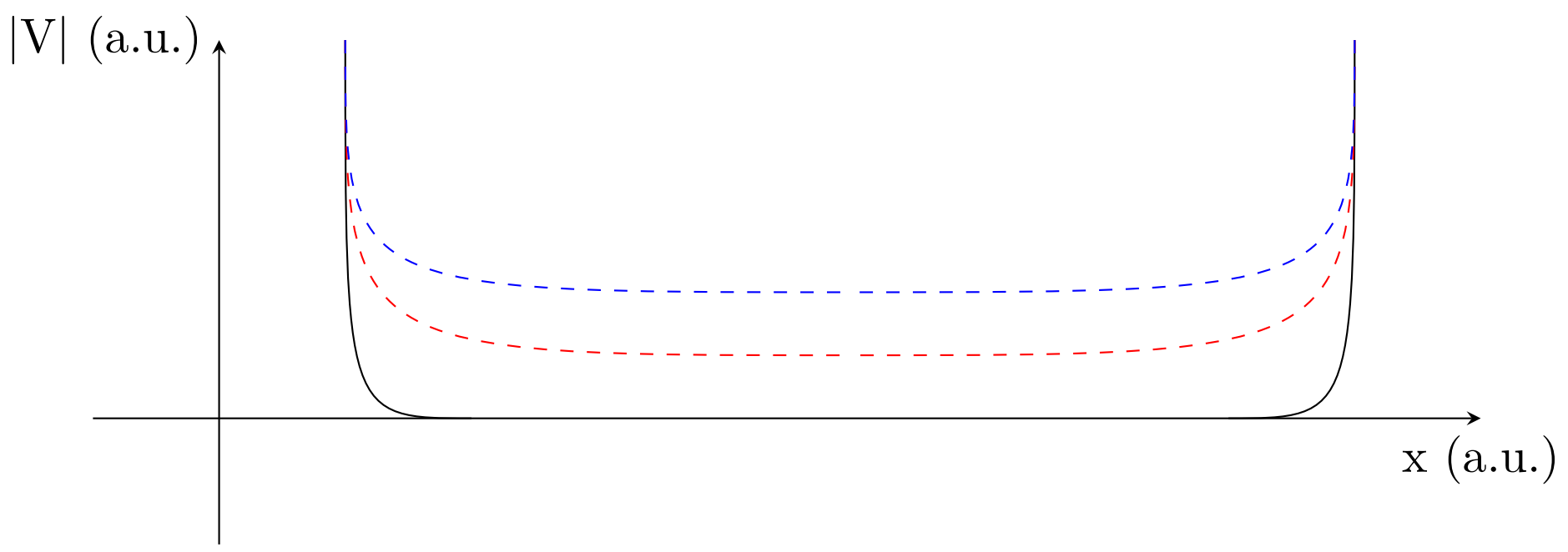

So far, the fields present in the filter and the electron motion in this field configuration have been described. It is now possible to introduce energy draining. The emission at the target is isotropic in pitch angle, but thanks to the adiabatic collimation of the injection electrode system, the pitch angle phase space is squeezed. However, at the RF tracker entrance, each electron has a different pitch angle, in a range that depends on the difference of potential of the bouncing electrodes. This means that, in order to drift the electrons toward a high precision detector after the filter, it must have all the longitudinal kinetic energy drained (no energy along X and $\mathcal{O}(100~eV)$ along Z). A first simple attempt, already discussed in the first chapter, could be to put a potential barrier transverse the spiraling electron, in order to drain the kinetic energy, in a similar way to the KATRIN filter, however the PTOLEMY's electrons are drifting in a bouncing geometry and so there is an additional energy component that needs to be drained, the one along X.\\
In conclusion, the transverse energy draining can be done by increasing the potential along Z while the draining along the longitudinal direction is more complicated.\\
This problem can be solved by using Formula (\ref{eq:grad-cf}), from which we obtain that the combination between the curvature drift when the electron is bouncing and the transverse gradient drift.
This can be rewritten as

\begin{equation}
    \centering
\Vec{V}_{\nabla_{\perp}B-CF}= \left( K_{\perp}+2K_{\parallel} \right)\frac{\Vec{B}\times \Vec{\nabla}_{\perp}B}{qB^3}
\label{eq:parallel_energy_filter}
\end{equation}

where the kinetic energy is considered the instantaneous kinetic energy. However in the PTOLEMY design, the electron is moving in a bouncing geometry and so the squared average of the longitudinal velocity\footnote{In the harmonic oscillator approximation} is like described in Equation (\ref{eq:filter_max_para_vel})

\begin{equation}
    \langle v^2_{\parallel} \rangle=\frac{1}{2}(v^{max}_{\parallel})^2
    \label{eq:filter_max_para_vel}
\end{equation}

where $(v^{max}_{\parallel})^2$ is the velocity in the centre of the filter (bottom of the potential well). This can be exploited in the following way: when the electron is bouncing from the central channel to the side channel, it experiences a potential barrier that temporarily reduces $K_{\parallel}$, so once the electron is in the side channel, $K_{\parallel}\simeq0$. In the side channel the electron feels high $\Vec{E} \times \Vec{B}$ and that pushes the electron along Z. After a bouncing, the electron is drifted back to the central channel, in which it gains back the majority of its longitudinal kinetic energy but not all of it, in fact along the Z direction, the potential of the central channel is increasing. This means that the electron is losing the longitudinal energy over its drift motion along Z. A schematic of the increasing potential of the central channel is shown in Figure (\ref{fig:filter_potential_well}).

\begin{figure}
    \centering
    \includegraphics[scale=0.8]{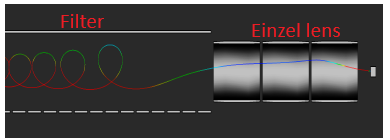}
    \caption{Electron trajectory simulation at the end of the filter, with focusing Einzel lens.}
    \label{fig:filter_trajectory}
\end{figure}

The electric field required for the filter, which is reported in Formula (\ref{eq:filter_Efield}), can be obtained with the following expressions:

\begin{equation}
    \begin{cases}
        \Phi_{c}(z)=K_{\perp}^{RF}e^{-z/\lambda_c}+K_{\parallel}^{RF}e^{-z/\lambda_s}\\
        \Phi_{s}(z)=\Phi_{c}(z)-K_{\parallel}^{RF}e^{-z/\lambda_s}=K_{\perp}^{RF}e^{-z/\lambda_c}
    \end{cases}
    \label{eq:pot_filter}
\end{equation}

where $\lambda_c, \lambda_s$ are the characteristic lengths of the fields in the central and side channels and $\Phi_c$, $\Phi_s$ are the potentials and $K_{\parallel}^{RF}$, $K_{\perp}^{RF}$ are the longitudinal and transverse kinetic energy in the RF tracker. At the end of the filter the electrons have around $100~eV$ and are in a low B field region that is the saddle point of the magnet. In this point the electrons have lost the majority of the energy along X direction and are therefore ready to be focused using an \emph{Einzel lens} toward the final sub-detector: the hemispherical analyzer. In Figure (\ref{fig:filter_trajectory}) the simulation of an electron trajectory outgoing from the filter toward the final detector is shown .

\subsection{Hemispherical electron analyzer}

\begin{figure}[h!]
    \centering
    \includegraphics[scale=0.5]{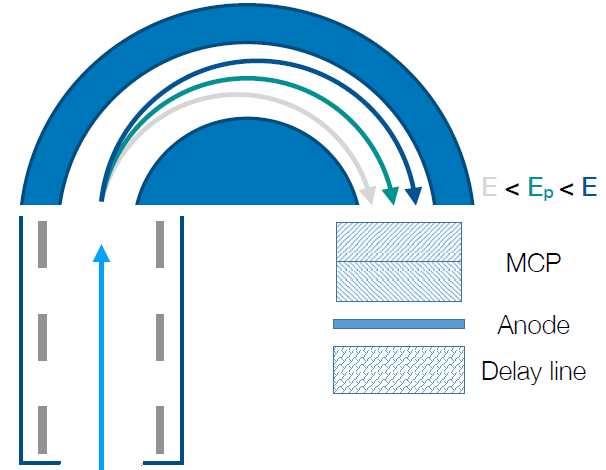}
    \caption{Scheme of the PTOLEMY final detection unit composed by Einzel lens, hemispherical analyzer and micro channel plate detector with delay line.}
    \label{fig:hem_analyzer}
\end{figure}

The electrons after the filter are injected in the final part of the detector, which is the part responsible for the high precision energy measurements. The electrons coming from the filter are expected to have a small angle spread and a kinetic energy of about $\mathcal{O}(100~eV)$. Those electrons are measured with a hemispherical analyzer, that is composed of two hemispheres with an input and an output port, as shown in Figure (\ref{fig:hem_analyzer}). Inside the gap between the two hemispheres, an electric field that deflects the electrons depending on the kinetic energy\footnote{This detector must be used in a zero magnetic field environment like the saddle point of the PTOLEMY magnet.} is applied. Depending by the position of the outgoing electron, it is possible to reconstruct the curvature (energy) of the electron. The energy resolution is expected to be $50~meV$. After measuring the filtered electron and the potential difference between the target and the analyzer, it is possible to reconstruct the initial energy using Equation (\ref{eq:PTOLEMY_initial_energy_reconstruction}).

\begin{equation}
E_i=q \left( V_{target}-V_{det}\right) + E_{RF,corr} +E_{det}
\label{eq:energy_reconstruction_PTOLEMY}
\end{equation}

\newpage
\section{The PTOLEMY experimental method} \label{sec:PTOLEMY_phases}

In the previous pages, each sub-detector of the PTOLEMY project has been described. In this section instead, the feature of the whole detector principle is discussed.

\subsection{The PTOLEMY experimental method}

The main issue of the Tritium endpoint experiments is the background rejection, which in this case means several types of background particles:

\begin{itemize}
    \item Low energy electrons from Tritium beta decay
    \item Electrons emitted from the interaction of cosmic rays
    \item Electrons emitted from particle emitted in the experimental vessel
\end{itemize}

Those types of background are a limiting factor for the aforementioned Tritium neutrino mass experiments. For example in KATRIN, in order to solve the problem of the background electrons with energy less than the Tritium endpoint (filtering electrons from the target), a central electric field in the spectrometer with low magnetic field is used.
Exploiting the time of flight (TOF) with the RF tracker, in PTOLEMY it could be possible to reject the background electrons; in particular the PTOLEMY design allows the rejection of the majority of the background electrons by implementing an offline analysis of the time of flight of the electron between the RF tracker and the final analyzer, this is done by a calibration campaign in order to have the time of flight in the RF tracker and filter. The electrons from cosmic rays interacting with the vessels can be rejected because in the central region of the experiment, a high magnetic field is present and so the electron emitted from the interaction is immediately re-absorbed back by the vessel surface because of the fast cyclotron motion. A similar background are the electrons emitted by impurities in the copper electrodes that are emitted toward the bottom of the RF tracker potential well; however the well is symmetric, and so the electron is absorbed back in the opposite electrode, without possibilities of emitting an RF signal for enough time.

%considerazioni sullo spazio delle fasi?
%considerazioni su invarianza adiabatica?

\subsection{Systematics}\label{sec:systematics}
%possibilità di compensare gli effetti quantistici con molta massa di trizio
There are two main sources of systematics that are expected in PTOLEMY: 

\begin{itemize}
    \item Experimental systematics
    \item Theoretical systematics
\end{itemize}

The main expected experimental systematic is the reference potential instability for the initial energy reconstruction of Equation (\ref{eq:energy_reconstruction_PTOLEMY}), however, in a recent work \cite{HV} has shown that by means of the "REF5010" voltage reference board it is possible to achieve a voltage stability of $\sigma_{HV}=0.9~mV$ at $20~kV$, proving that this systematic error can be controlled in the PTOLEMY detector. Another expected source of systematic, that will be discussed in Section (\ref{sec:resol_systematics}), is the error on the reconstruction of the energy loss during the cyclotron radiation emission phase. There is also a third systematic error that is the reconstruction of the TOF of the electron between the RF tracker and the detector after the filter. This error has a double effect: it decreases the capability of the background rejection and ruins the reconstruction of the drift time for the energy loss reconstruction just mentioned.\\
The theoretical systematic uncertainty is related to the quantum energy spread of the final state electron due to the covalent bound of Tritium to the graphene substrate, as has been described in Section (\ref{sec:tritium_target}) \cite{heisenberg}. The quantification of the theoretical uncertainty comprehensive of condensed matter effect is currently under investigation. This aspect leads to two problems: the smearing of the C$\nu$B capture peak such that the expected capture ratio in one year for $100~g$ of Tritium is suppressed, and the distortion of the Kurie plot.\\

%mettere imamgine spettro teorico aggiornato
\begin{figure}[h!]
    \centering
    \includegraphics[scale=0.4]{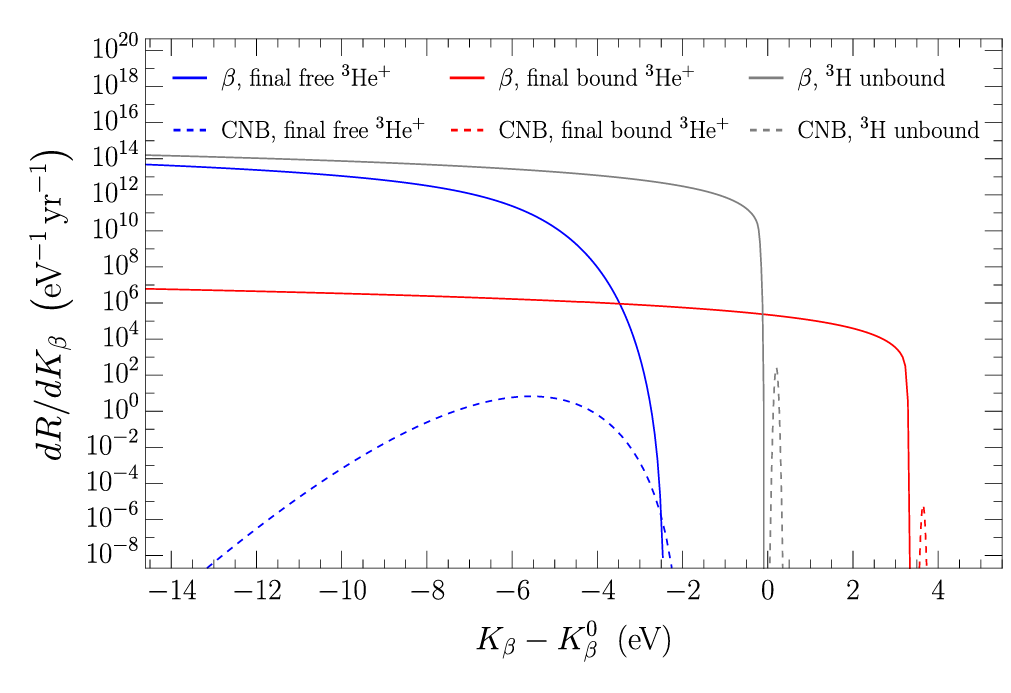}
    \caption{Kurie Plot and the signal from C$\nu$B capture. The gray line corresponds to the ideal case of unbound (free) tritium atoms, where the electron spectrum has a sharp endpoint. The blue line shows a more realistic situation in which the tritium atom is bound (e.g. in molecules or materials). In this case, the initial momentum of the decaying system causes a smearing of the electron energy, spreading the beta-decay spectrum and partially overlapping with the expected C$\nu$B signal. The red line represents rare kinematic configurations in which this smearing effect is effectively canceled, allowing a sharper feature near the endpoint. The size of this effect requires experimental confirmation. At present, the theoretical uncertainty in the models describing this smearing is still large.}
    \begin{comment}
    \caption{Tritium Kurie plot with expected C$\nu$B capture peak in two different scenarios. In the final state free Helium model (blue line), because of the energy smearing, the rate of C$\nu$B capture is hidden below the beta decay spectrum; instead the model in which final state Helium is bound to the ground state with no emission of vibrational modes is represented with a red line. The scale at which the energy smearing takes place requires experimental proof \cite{heisenberg}.}
    \end{comment}
\begin{comment}
        \caption{Tritium Kurie plot with C$\nu$B peak at $2m_{\nu}$ from the tritium endpoint. Because of the energy smearing, the rate of C$\nu$B capture is hide below the beta decay spectrum. The red line represent the case when the kinematics reduce at negligible level the smearing effect. The scale at which the energy smearing takes place requires experimental proof. }
\end{comment}
    
    \label{fig:tritium_kurie_plot}
\end{figure}

\newpage
The distortion of the Kurie plot can influence the sensitivity to neutrino mass, as has been described in \cite{paper_teor2}, however PTOLEMY has an advantage that other Tritium experiments do not have, that could actually solve the problem: the possibility to increase the target mass. In the next Section, a proposal for target arrangement for larger target mass and electron transport is described.\\

\subsection{Target scalability and electron transport}\label{sec:scalabilità}

\begin{figure}[h!]
    \centering
    \includegraphics[width=0.5\linewidth]{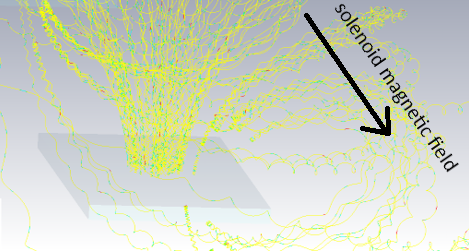}
    \caption{Simulation of electrons outgoing from tritium atoms, sitting on a permanent magnet, and captured by the solenoidal transverse field.}
    \label{fig:magnetino}
\end{figure}

In Section (\ref{Sec:target_injection}) the idea of using a permanent magnet as a magnetic support for adiabatic collimation for electrons has been described. That concept is supposed to work for a neutrino mass measurement with a target scale of about $1~ \mu g$ of Tritium or less, however it is difficult to use that concept for higher masses. A proposal under development is described in the following.\\
Considering a rectangular permanent magnet with a central field of $1-1.2~ T$ and dimensions $a \times b \times c$ with $a=b \sim 12~cm$ and $c\simeq 5~mm$. To this magnet, a Tritiated graphene layer is applied to both faces and at the center of the magnet.\\
Because of this configuration, the electrons that are emitted toward the magnet are going to be absorbed by the magnet. An array of magnets is then created by placing several magnets for a distance $A,B,C$ between them, where $A$ is the displacement along X direction, $B$ is the distance along Y and $C$ along Z. The magnets are then displaced in planes one in front of another. The electrons are collected by implementing a longitudinal solenoidal magnetic field, with an intensity of about $B_{solenoid}=0.1~T$.\\
This concept also allows the possibility of rotating the target in order to minimize the scattering with cosmic rays. The magnetic field generated from the magnet decreases rapidly in the transverse direction and so, when the electrons are emitted at the beginning they feel the target field, while drifting in the direction outgoing from the target face, the magnetic field from the solenoid becomes predominant, changing the direction of the drift. A simulation of electron drift motion is shown in Figure (\ref{fig:magnetino}).

\begin{figure}[h!]
    \centering
    \includegraphics[width=0.8\linewidth]{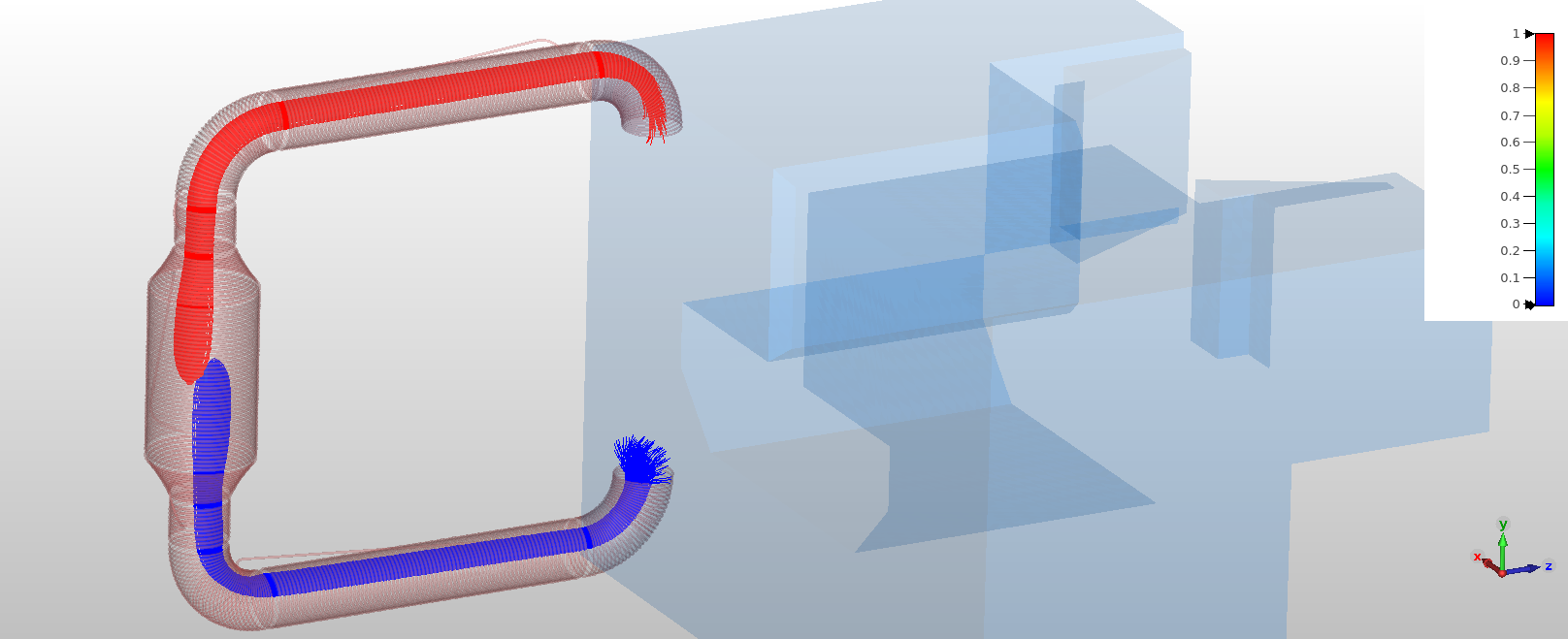}
    \caption{Electrons trajectories from the target to the transportation coils (the color represent the face of the target in which the electrons are emitted).}
    \label{fig:traiettorie_trasporto}
\end{figure}

\newpage
The electrons from each plane are collected with the solenoid that transports the electrons to the injection electrode system. It is important to note that the solenoidal transport magnetic field is at a low magnetic field ($\mathcal{O}(0.1~T)$), and so the Larmor emission is negligible. A simulation of one tile is shown in Figure (\ref{fig:traiettorie_trasporto}) in which are visible the trajectories of several electrons during the transport from the tritiated target.

\begin{figure}[h!]
    \centering
    \includegraphics[width=0.8\linewidth]{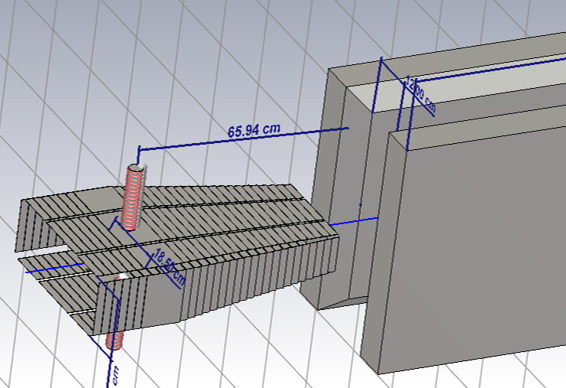}
    \caption{Injection electrode system, the potential concept is similar to the one described in Section (\ref{Sec:target_injection}) with the difference that the electrons are injected from the coils of $\mathcal{O}(0.6~T)$ field. On the right it is possible to see a part of the PTOLEMY magnet that provide the fringe field for trapping the electrons in a bouncing geometry.}
    \label{fig:iniezione_mia}
\end{figure}

The electrons are then injected into the injection electrode system by placing one coil on the top central plate and one on the bottom central plate, as shown in Figure (\ref{fig:iniezione_mia}). 

\begin{figure}[h!]
    \centering
    \includegraphics[width=1\linewidth]{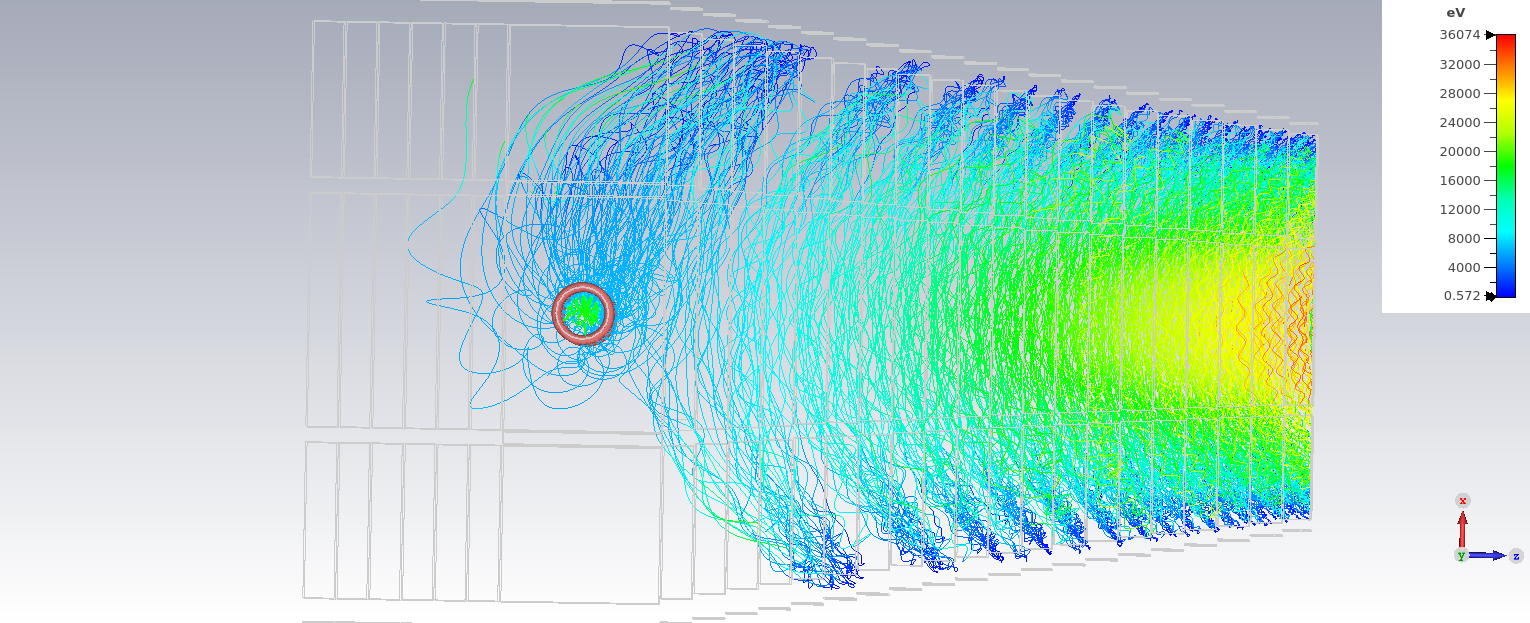}
    \caption{Top view of the simulation of electron trajectories injected from a coil geometry to a bouncing geometry.}
    \label{fig:iniezione_traiettorie}
\end{figure}

\newpage
A simulation of electrons injected from a coil geometry to a bouncing electrode geometry is shown in Figure (\ref{fig:iniezione_traiettorie}), once the electrons are entered into the electrode system, the fringe field of the PTOLEMY magnet becomes predominant with respect to the field of the coils.

%% file: sections/3RF_theory.tex
\chapter{Electron cyclotron radiation} \label{chap:RF_theory}

In the last Chapter the PTOLEMY detector concept has been described, with particular attention to the role of the RF Tracker whose conceptual design is shown in Figure (\ref{fig:rf_region}). It is composed of two main parts: the electrodes that will adjust the motion of the electrons and the rectangular waveguide which is required to collect the RF, emitted by a single electron, towards a dedicated electronic chain. The RF Tracker allows detecting single electron by recording the signal emitted during the gyro-motion, where the power associated with the signal is expected to be on the scale of $1~fW$, a frequency of about $27~GHz$ and a short time scale (the drift time of the electron) of about $50~ \mu s$. The goal is to measure the electrons in a region of interest around the endpoint with a resolution of about $10~eV$.\\
In this Chapter the theory of cyclotron radiation emission will be described together with the signal propagation in waveguides, and finally in Section (\ref{sec:signal_detection}) the experimental limitations of those measurements will be discussed.

\section{Propagation of radio-frequency in waveguide}\label{sec:rf_propagation}

In this Section a general description of a radio-frequency propagation through a rectangular waveguide is given \cite{pozar}. Considering a linear, isotropic, homogeneous medium and \emph{time harmonics} electric and magnetic fields as

\begin{equation}
    \begin{cases}
        \Vec{E}=\Vec{E}(x,y,z)e^{j\omega t}\\
        \Vec{H}=\Vec{H}(x,y,z)e^{j\omega t}
    \end{cases}
\end{equation}

the Maxwell equations, by deriving the exponential over time, are reduced to

\begin{equation}
    \begin{cases}
        \vec{\nabla}\times \vec{E}= -j\omega \mu\vec{H}-\vec{M}\\
        \vec{\nabla}\times \vec{H}= j\omega \epsilon\vec{E}+\vec{J}
    \end{cases}
    \label{eq:max1}
\end{equation}

\begin{equation}
    \begin{cases}
        \vec{\nabla}\cdot \vec{D}=\rho\\
        \vec{\nabla}\cdot \vec{B}=0
    \end{cases}
    \label{eq:max2}
\end{equation}

where $\Vec{M}$ and $\Vec{J}$ are the magnetic and current sources. Instead in the case of source-less region, the Equations (\ref{eq:max1}) are reduced to 

\begin{equation}
    \begin{cases}
        \vec{\nabla}\times \vec{E}= -j\omega \mu\vec{H}\\
        \vec{\nabla}\times \vec{H}= j\omega \epsilon\vec{E}
    \end{cases}
    \label{eq:max_to_hel}
\end{equation}

Considering that $\vec{\nabla}\times\vec{\nabla}\times\vec{A}=\nabla(\vec{\nabla}\cdot\vec{A})-\nabla^2\vec{A}$ and applying the rotor operator to the Equation (\ref{eq:max_to_hel}), we obtain

\begin{equation}
\begin{cases}
    \vec{\nabla}\times\vec{\nabla}\times\vec{E}= -j\omega \mu\vec{\nabla}\times\vec{H}=\omega^2\mu\epsilon\vec{E}\\
    \vec{\nabla}\times\vec{\nabla}\times\vec{H}=j\omega \mu\vec{\nabla}\times\vec{E}=\omega^2\mu\epsilon\vec{H}
\end{cases}
\end{equation}

that are named Helmholtz equations, and with $k^2=\omega\mu\epsilon$ is the \emph{propagation constant}, can be rewritten in the following form

\begin{equation}
\begin{cases}
    \nabla^2\vec{E}+k^2\vec{E}=0\\
    \nabla^2\vec{H}+k^2\vec{H}=0
\end{cases}
\label{eq:helmholtz}
\end{equation}

\begin{figure}[h!]
    \centering
    \includegraphics[width=0.5\linewidth]{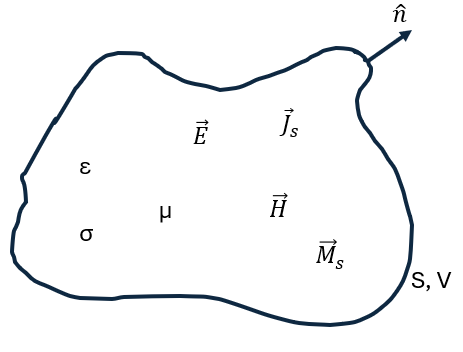}
    \caption{Scheme of a generic medium with surface S, volume V and conductivity $\sigma$.}
    \label{fig:generic_medium}
\end{figure}

The Helmholtz Equations (\ref{eq:helmholtz}) are key to studying the propagation of waves, that in our case will take place in a rectangular waveguide. However, before describing propagation in a rectangular waveguide, it is useful to introduce some general aspects of RF. In a generic medium with external surface $S$, volume $V$ it is possible to define the flow of power. In particular, it is possible to consider an internal source that emits power $P_s$ (like an electron), some of this power will exit from the medium (i.e. a rectangular waveguide) with power $P_o$ and some will be dissipated as $P_l$ (because of parasitic currents through the skin layer of the waveguide). A schematic of the power budget in a generic volume is shown in Figure (\ref{fig:generic_medium}).

The components of the power budget can be given as functions of fields $E$ and $B$ such as the current density $\vec{J}_s$ and magnetization density $\vec{M}_s$ and the Poynting vector $\vec{S}$ in the following formulas \cite{pozar}

\begin{equation}
    \begin{cases}
        P_s=\frac{1}{2}\int_V\left( \vec{E}\cdot\vec{J}^*_s+\vec{H}^*\cdot \vec{M}_s \right)dV\\
        P_o=\oint_S\vec{S}\cdot d\vec{s}\\
        P_l=\frac{\sigma}{2}\int_V|\vec{E}|^2dV+\frac{\omega}{2}\int_V\left( \epsilon^{''}|\vec{E}|^2+\mu^{''}|\vec{H}|^2 \right)dV
    \end{cases}
\end{equation}

$P_l$ depends on conductivity, dielectric, and magnetic losses (where complex permeability $\epsilon=\epsilon^{'}-j\epsilon^{''}$ and $\mu=\mu^{'}-j\mu^{''}$ have been introduced \cite{pozar}). It is possible to obtain \cite{pozar} that the power loss, in the case of rectangular waveguide, can be described by the average power absorbed $P_{l,avg}$ by the conductor

\begin{equation}
    P_{l,avg}=\frac{1}{2}\int_{S_0+S}\left( \vec{E}\times \vec{H}^* \right)\cdot \hat{n}dS=\frac{R_s}{2}\int_{S_0}|\vec{H}|^2dS
\end{equation}

where $R_S$ is the surface resistance of the conductor (typically copper); this means essentially that if the waveguide surface is not smooth enough, there could be attenuation of the signal power. In the case of the Electron Trap, the attenuation of the waveguide has been measured to be almost $2~dB$.

\subsection{Introduction to Radio-frequency}\label{sec:rectangular_waveguide}

\begin{figure}[h!]
    \centering
    \begin{tikzpicture}
        \draw[-stealth](-1,0)--(7,0)node[below]{X};
        \draw[-stealth](0,-1)--(0,4)node[left]{Y};

        \draw[blue](0,0)--(4,0)--(4,2)--(0,2)--(0,0);
        \draw[blue](4,2)--(6,4);
        \draw[dashed,blue](0,0)--(4,4)node[right]{Z};
        \draw[blue](0,2)--(2,4);
        \draw[blue](4,0)--(6,2);

        \node at(4,-0.5){a};
        \node at(-0.5,2){b};

        \node at(2,0.5){$\epsilon$,$\mu$};

    \end{tikzpicture}
    \caption{Schema of a rectangular waveguide with inner dimension of $a\times b$ with inside a generic medium with dielectric constant $\epsilon$ and magnetic permeability $\mu$. }
    \label{fig:rectangular_waveguide}
\end{figure}
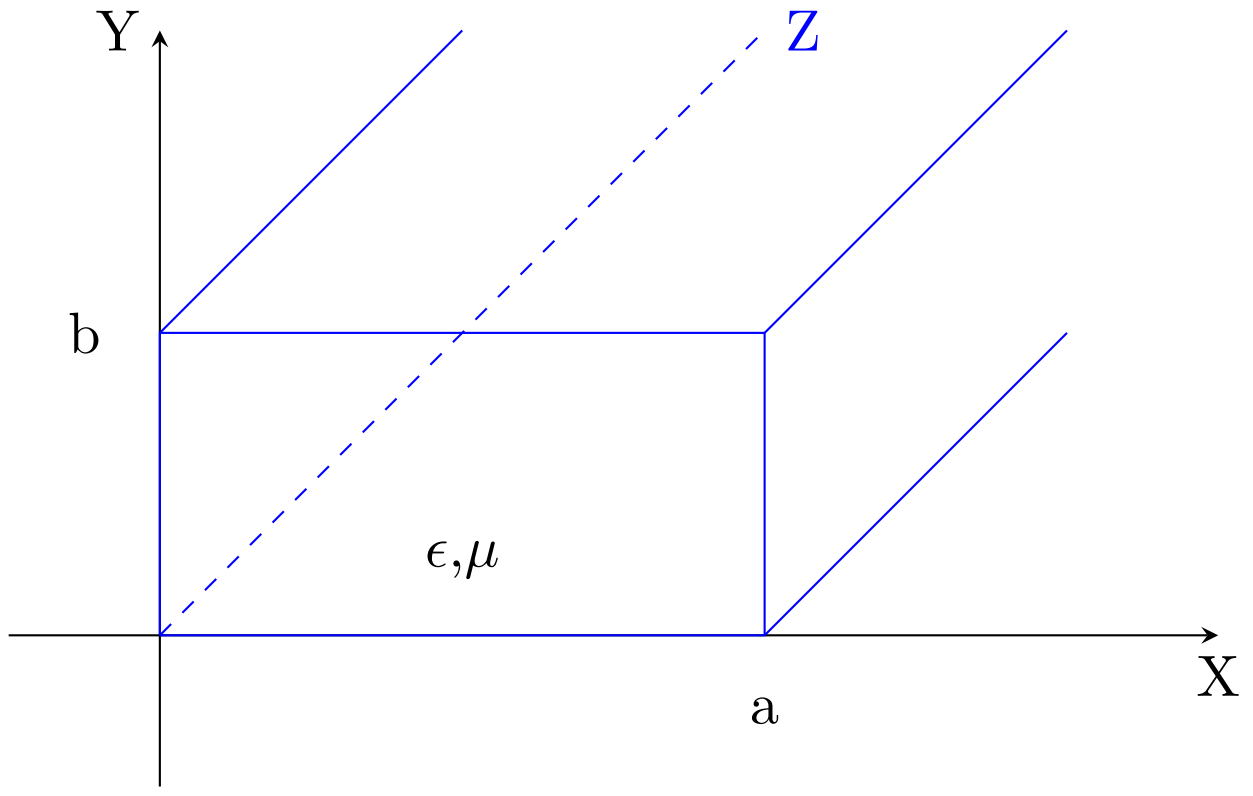

So far, the motion of electrons that are expected to happen in the PTOLEMY detector has been described and has been introduced how this motion implies an emission of cyclotron radiation. The measurement process of this radiation can be divided into different steps: the radiation emission that is described together with the coupling to the waveguide modes, described in Section (\ref{sec:cyclotron_radiation}), the wave propagation in the rectangular waveguide, shown in Section (\ref{sec:rectangular_waveguide}), and then the signal detection electronic chain described in Section (\ref{sec:cyclotron_radiation_detection}).\\
Let's describe the radio-frequency signal propagation trough a rectangular waveguide of dimensions $a$ and $b$ as shown in Figure (\ref{fig:rectangular_waveguide}).

To have the solution of the wave propagation in the waveguide, as a first step we assume time harmonic fields and wave propagation along the Z direction as described in Equation (\ref{eq:time_harm_fields}).

\begin{equation}
    \begin{cases}
        \vec{E}(x,y,z)=\left( \vec{e}(x,y)+\hat{z}e_z(x,y) \right)e^{-j\beta z}\\
        \vec{H}(x,y,z)=\left( \vec{h}(x,y)+\hat{z}h_z(x,y) \right)e^{-j\beta z}
    \end{cases}
    \label{eq:time_harm_fields}
\end{equation}

where $\beta$ is the propagation constant. In general, by using Equation (\ref{eq:max1}) it is possible to obtain

\begin{equation}
    \begin{cases}
        H_x=\frac{j}{\omega\mu}(\partial_yE_z+j\beta E_y)\\
        H_y=\frac{j}{\omega\mu}(-j\beta E_x-\partial_x E_z)
    \end{cases}
\end{equation}

\begin{equation}
    \begin{cases}
        E_x=-\frac{j}{\omega\epsilon}(\partial_yH_z+j\beta H_y)\\
        E_y=-\frac{j}{\omega\epsilon}(-j\beta H_x-\partial_x H_z)
    \end{cases}
\end{equation}

and by merging the last two equations and defining $k_c^2=k^2-\beta^2$ where $\beta$ is the propagation constant, it is possible to obtain the first set of Equations (\ref{eq:eq_wgE1}) (\ref{eq:eq_wgH1}) that describe the wave propagation in the waveguide.

\begin{equation}
    \begin{cases}
        H_x=-j\frac{\beta}{k^2_c}\partial_xH_z\\
        H_y=-j\frac{\beta}{k^2_c}\partial_yH_z
    \end{cases}
    \label{eq:eq_wgH1}
\end{equation}

\begin{equation}
    \begin{cases}
        E_x=-j\frac{\omega\mu}{k^2_c}\partial_yH_z\\
        E_y=-j\frac{\omega\mu}{k^2_c}\partial_xH_z
    \end{cases}
        \label{eq:eq_wgE1}
\end{equation}

The solutions of such equations are divided into \emph{modes}. The most important one is the \emph{TE} (transverse electric, defined by the condition $E_z=0$, $H_z\neq 0$). This is the most relevant mode for signal detection in rectangular waveguides.\\
The solutions are obtained by solving the Helmholtz Equations (\ref{eq:helmholtz}) and searching for solutions as shown in Equation (\ref{eq:time_harm_fields}). For example, it is possible to get that 

 \begin{equation}
     \left(\partial^2_x +\partial^2_y+k_c^2 \right)h_z(x,y)=0
     \label{eq:time_harm1}
 \end{equation}

and by checking if a solution of the form $h_z(x,y)\equiv f_x(x)f_y(y)$ is a solution of Equation (\ref{eq:time_harm1}), it is possible to obtain

\begin{equation}
    f_y\partial^2_xf_x+f_x\partial_y^2f_y+k_c^2f_xf_y=0
\end{equation}

that gives

\begin{equation}
    \begin{cases}
        \partial^2_xf_x+k^2_xf_x=0\\
        \partial^2_yf_y+k^2_yf_y=0
    \end{cases}
\end{equation}

that admit solutions of the form of

\begin{equation}
    h_z(x,y)=\left( A \cos(k_xx)+B\sin(k_xx)\right)\left(C\cos(k_yy)+D\sin(k_yy)\right)
\end{equation}

It is possible to obtain a similar result for the electric field. Then by imposing boundary conditions typical of a waveguide, we have

\begin{equation}
    \begin{cases}
        e_x=0 \hspace{1cm} y=0,b\\
        e_y=0 \hspace{1cm} x=0,a
    \end{cases}
    \label{eq:condiz_contorno}
\end{equation}

in fact by using Equation (\ref{eq:eq_wgE1}) and using the solution of $h_z$ just obtained, we get 

\begin{equation}
    \begin{cases}
        e_x=-j\frac{\omega\mu}{k_c^2}\partial_yh_z=-j\frac{\omega\mu k_y}{k_c^2}(A\cos(k_xx)+B\sin(k_xx))(-C\sin(k_yy)+D\cos(k_yy))\\
        e_y=j\frac{\omega\mu}{k_c^2}\partial_xh_z=j\frac{\omega\mu k_x}{k_c^2}(-A\sin(k_xx)+B\cos(k_xx))(C\cos(k_yy)+D\sin(k_yy))
    \end{cases}
\end{equation}

By applying the boundary conditions and renaming the amplitudes, we obtain the solution of the magnetic field as shown in Equation (\ref{eq:H_z}).

\begin{equation}
    H_z(x,y,z)=A_{mn}\cos\left( \frac{m\pi}{a}x \right)\cos\left( \frac{n\pi}{b}x \right)e^{-j\beta z}
    \label{eq:H_z}
\end{equation}

It is now possible to obtain the overall solutions for the $TE_{m,n}$ mode by using (\ref{eq:H_z}) in (\ref{eq:eq_wgH1}) and in (\ref{eq:eq_wgE1}), obtaining

\begin{equation}
    \begin{cases}
        H_x=j\frac{\beta m\pi}{ak^2_c}A_{mn}\sin\left( \frac{m\pi}{a}y \right)\cos\left( \frac{n\pi}{b}y \right)e^{-j\beta z}\\
        H_y=j\frac{\beta n\pi}{bk^2_c}A_{mn}\cos\left( \frac{m\pi}{a}y \right)\sin\left( \frac{n\pi}{b}y \right)e^{-j\beta z}\\
    \end{cases}
    \label{eq:TEmn1}
\end{equation}

 \begin{equation}
    \begin{cases}
        E_x=j\frac{\omega\mu n\pi}{bk^2_c}A_{mn}\cos\left( \frac{m\pi}{a}y \right)\sin\left( \frac{n\pi}{b}y \right)e^{-j\beta z}\\
        E_y=-j\frac{\omega\mu m\pi}{ak^2_c}A_{mn}\sin\left( \frac{m\pi}{a}y \right)\cos\left( \frac{n\pi}{b}y \right)e^{-j\beta z}\\
    \end{cases}
    \label{eq:TEmn2}
\end{equation}

where $\beta=\sqrt{k^2-k^2_c}=\sqrt{k^2-(\frac{m\pi}{a})^2-(\frac{n\pi}{b})^2}$. Note that considering just the definition of $\beta$, it is possible for wave propagation if and only if $k>k_c$. This fact is typically used in the frequency domain by defining the cut frequency, which is shown in Equation (\ref{eq:cut_frequency}) and is such that there is wave propagation if $f>f_{cut}$.

\begin{equation}
    f_{cut\hspace{1mm}m,n}=\frac{\sqrt{(\frac{m\pi}{a})^2+(\frac{n\pi}{b})^2}}{2\pi\sqrt{\mu\epsilon}}
    \label{eq:cut_frequency}
\end{equation}

In a rectangular waveguide, the most commonly used mode for signal propagation is $TE_{10}$, as shown in Equations (\ref{eq:TE10_E}) and (\ref{eq:TE10_H}).\\

\begin{equation}
    \begin{cases}
        H_x=j\frac{\beta a}{\pi}A_{10}\sin(\frac{\pi}{a}x)e^{-j\beta z}\\
        H_y=0\\
        H_z=A_{10}\cos(\frac{\pi}{a}x)e^{-j\beta z}
    \end{cases}
    \label{eq:TE10_H}
\end{equation}

\begin{equation}
    \begin{cases}
        E_x=0\\
        E_y=-j\frac{\omega \mu a}{\pi}A_{10}\sin(\frac{\pi}{a}x)e^{-j\beta z}\\
        E_z=0
    \end{cases}
    \label{eq:TE10_E}
\end{equation}

\begin{figure}[h!]
    \centering
    \includegraphics[width=0.9\linewidth]{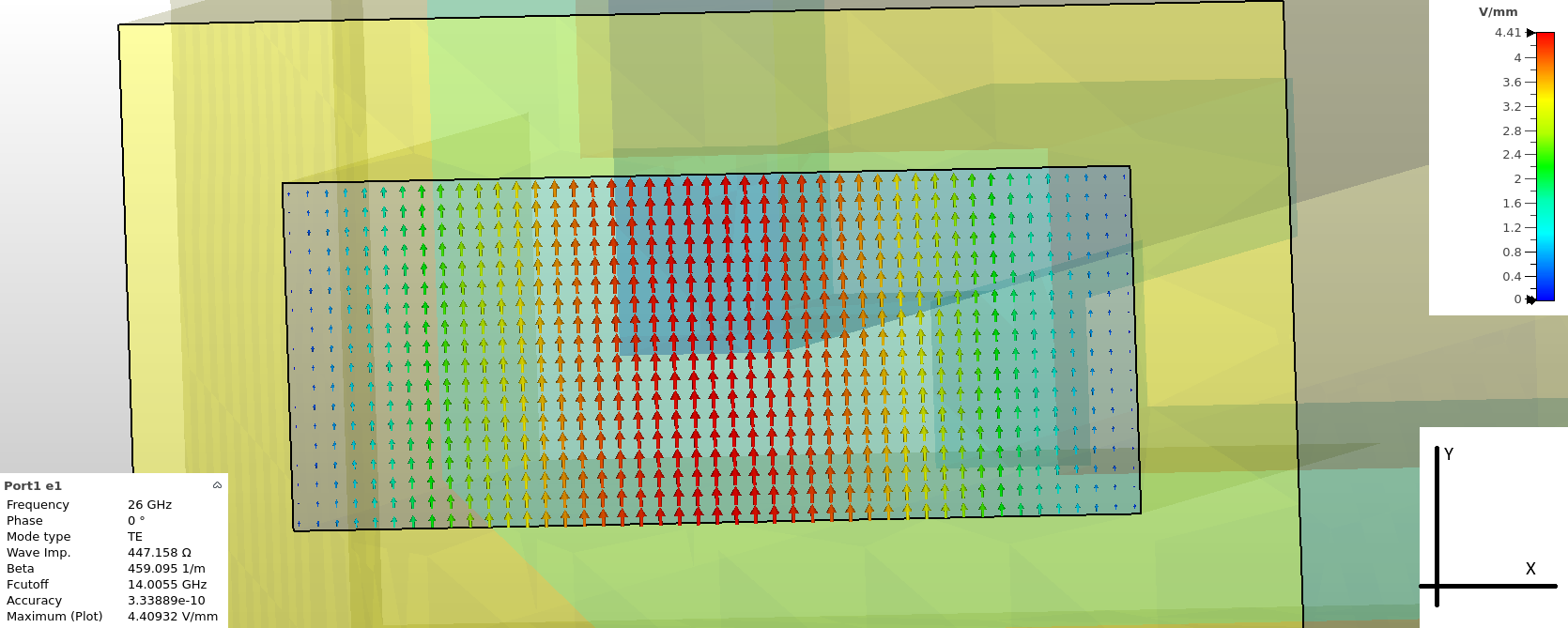}
    \caption{Mode $TE_{10}$ in the electron trap, the color represent the intensity of the electric field.}
    \label{fig:TE10}
\end{figure}

The resultant wave is shown in Figure (\ref{fig:TE10}), where the intensity along the Y direction is oscillating over time and phase.\\

As mentioned before, the main purposes of the RF tracker is the detection of the cyclotron radiation, that has a frequency of about $27~GHz$, however it is important to note that the spatial scale of this physics phenomenon is the Larmor radius, which must fit the waveguide dimensions, and the frequency radiation must be larger that the cut frequency of the $TE_{10}$, which is given by the Formula (\ref{eq:cut_frequency_TE10}).

\begin{equation}
    f_{cut\hspace{1mm}1,0}=\frac{(\frac{\pi}{a})}{2\pi\sqrt{\mu\epsilon}}
    \label{eq:cut_frequency_TE10}
\end{equation}

Another useful quantity is the characteristic impedance, reported in Equation (\ref{eq:impedance}).

\begin{equation}
    Z_{TE}=\frac{k\eta}{\beta}
    \label{eq:impedance}
\end{equation}

\newpage
\subsubsection{Scattering parameters}\label{sec:scattering_parameters}

\begin{figure}[h!]
    \centering
    \begin{tikzpicture}
        \draw[-stealth](-4,0)--(4,0) node[below]{z};
        \draw[-stealth](0,-1)--(0,4) node[left]{x};

        \node[] at(-3,2){$\epsilon_0$,$\mu_0$};
        \node[] at(3,2){$\epsilon$,$\mu$};

        \node[waves,scale=1] at (-2,3){};
        \begin{rotate}{180}
        \node[waves,scale=0.4] at (2,-1){};
        \end{rotate}

        \node[waves,scale=0.8] at (1,3){};

        \filldraw[fill=black!5](0.1,-1)rectangle(4.1,4)[fill opacity=0.3];
        
    \end{tikzpicture}
    \caption{Scheme of transmission and reflection of an incident wave trough a generic medium.}
    \label{fig:scattering1}
\end{figure}
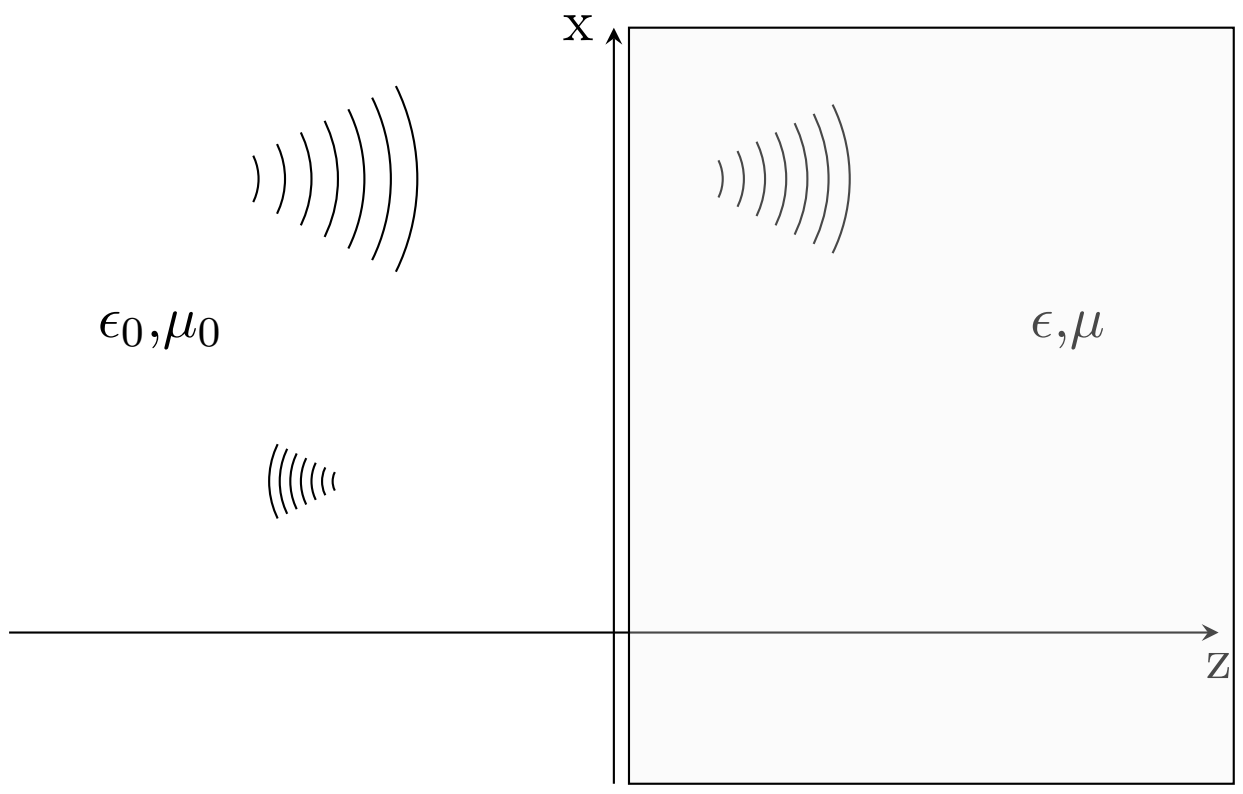

When a signal is propagating through several devices in cascade, it is important to take into account possible reflections of the signal that could cause a power loss of the low power electron signal \cite{pozar}. It is possible to study the wave reflections by starting from a simple example, where at the beginning the wave is propagating in vacuum and it pass through a generic medium at $z=0$, as shown in Figure (\ref{fig:scattering1}).

Then let's define the \emph{reflection coefficient} $\Gamma$ and the \emph{transmission coefficient} $T$.\\
In this situation, at $z<0$ two waves are present, an incident one

\begin{equation}
    \begin{cases}
        \vec{E}_i=\hat{x}E_0e^{-jk_0z}\\
        \vec{H}_i=\hat{y}\frac{E_0}{\eta_0}e^{-jk_0z}
    \end{cases}
\end{equation}
 and the reflected radiation
\begin{equation}
    \begin{cases}
        \vec{E}_r=\hat{x}\Gamma E_0e^{jk_0z}\\
        \vec{H}_r=-\hat{y}\Gamma \frac{E_0}{\eta_0}e^{jk_0z}
    \end{cases}
\end{equation}

while at $z>0$ there is only the transmitted one

\begin{equation}
    \begin{cases}
        \vec{E}_T=\hat{x}T E_0e^{-\gamma z}\\
        \vec{H}_T=-\hat{y}T \frac{E_0}{\eta_0}e^{-\gamma z}
    \end{cases}
\end{equation}

where $\gamma=\alpha+j\beta$ is the generalized propagation constant, where $\alpha$ is the attenuation and $\beta$ is the phase constant. Taking into account the continuity of the fields when the wave is passing at $z=0$ \cite{pozar}, it is possible to define the coefficients as 

\begin{equation}
    \begin{cases}
        \Gamma=\frac{\eta-\eta_0}{\eta+\eta_0}\\
        T=\frac{2\eta}{\eta+\eta_0}
    \end{cases}
\end{equation}
Then passing to the logarithmic of those quantity we can define the \emph{return loss} $RL$ and the \emph{insertion loss} reported in dB. An ideal waveguide gives $RL=-\infty$ and in the worst case, total reflection, $|\Gamma|=1$ and $RL=0~dB$.\\
While the \emph{insertion loss} $IL$, a number expressing how much power is lost or absorbed by parasitic currents, is also reported in dB. 
\begin{equation}
    RL=20\log_{10}|\Gamma|;~~~~IL=-20\log_{10}|T| 
\end{equation}

\begin{figure}[h!]
    \centering
    \begin{tikzpicture}
        \draw(-1,-1)--(1,-1)--(1,1)--(-1,1)--(-1,-1);

        \draw[->](-2,0.5)--(-1.1,0.5);
        \draw[->](-1.1,-0.5)--(-2,-0.5);

        \draw[-stealth](-1.5,0.4) arc (90:-90:0.4);
        \node at (-1.5,0)[]{$S_{11}$};

        \draw[-stealth](1.5,-0.4) arc (270:90:0.4);
        \node at (1.5,0)[]{$S_{22}$};

        \draw[-stealth](-0.8,0.5)--(0.8,0.5);
        \node at (0,0.7)[]{$S_{12}$};

        \draw[-stealth](0.8,-0.5)--(-0.8,-0.5);
        \node at (0,-0.7)[]{$S_{21}$};
        
        \draw[->](2,-0.5)--(1.1,-0.5);
        \draw[->](1.1,0.5)--(2,0.5);
        
    \end{tikzpicture}
    \caption{Scheme of the scattering parameters.}
    \label{fig:scattering_scheme}
\end{figure}
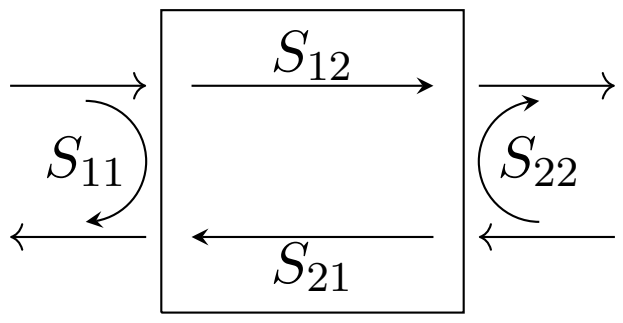

Those parameters may be rewritten in the \emph{scattering matrix} $S$, to describe the behavior of circuital devices. Taking a general two port device as shown in Figure (\ref{fig:scattering_scheme})

\begin{equation}
S=\left(
    \begin{matrix}
        S_{11}&S_{12}\\
        S_{21}&S_{22}
    \end{matrix}
    \right)
\end{equation}

where $IL=-20\log_{10}|S_{21}|$, $RL=-20\log_{10}|S_{11}|$. $G=S_{12}$ is the gain and $S_{22}$ is the reflection at the output port.

\section{Electron cyclotron radiation detection}\label{sec:cyclotron_radiation_detection}

In Section (\ref{sec:rectangular_waveguide}) it has been described how a signal can propagate through a rectangular waveguide. It is now possible to apply such theory to the signal emitted by an electron in cyclotron motion.

\subsection{Electron cyclotron radiation}\label{sec:cyclotron_radiation}

The radio-frequency emitted by an electron moving in cyclotron motion can be described, as shown in \cite{jackson}, by looking at the frequency $f_c$ and the power of the signal $P$. Those two quantities depend on two main factors: the electron kinematics and electromagnetic effects.\\
In this Section a description of the frequency and power of an electron spiraling as in Section (\ref{sec:e_motion}) is given, however the motion that we want to analyze is slightly complicated as described in Section (\ref{sec:RF_region}), this case will be discussed in Chapter (\ref{chap:Electron_Trap}) which is likely similar to the case of the expected motion in the RF tracker. The frequency of the signal emitted by an electron in cyclotron motion depends on the circular motion of the electron, in fact in Equation (\ref{eq:cyclotron_frequency}) the non relativistic frequency is described and if we divide by the Lorentz factor

\begin{equation}
    \omega_c=\frac{\omega_0}{\gamma}
\end{equation}

the relativistic formula $\omega_c$ is obtained. In Equation (\ref{eq:larmor_frequency2}) one can see the explicit dependence of the frequency by the kinetic energy.

\begin{equation}
    f_{cy}=\frac{1}{2\pi}\frac{|q|B}{m}\frac{1}{K/m+1}
    \label{eq:larmor_frequency2}
\end{equation}

In Equation (\ref{eq:larmor_frequency2}) the lower case "cy" has been used because this is called \emph{Cyclotron frequency} and corresponds to the electron instantaneous frequency.
In the next Chapter other features of the frequency spectrum will be presented and discussed in particular the \emph{Carrier frequency} $f_c$, that is the main contribution of the signal, and the \emph{Sidebands frequencies} which allow to an estimate of the parallel momentum of the particle. In the case of our interest, $B=1~T$ and for an electron at Tritium endpoint energy ($K=18.6~keV$), the Larmor revolution time is $t_{larmor} \sim 37~ps$ corresponding to a frequency of about $f_c\sim 27~GHz$, and considering  $\theta=90^{\circ}$ the Larmor radius is $r_L\sim0.6~mm$.\\
The power of the cyclotron signal can be obtained, as described in general in \cite{jackson}, in a relativistic description with the solid angle dependence $\frac{dP}{d\Omega}$. This description is particularly useful when this spectroscopy technique is used for very high energy particles, such as in accelerators. However, in this case the Larmor radius is very large ($r_L\gg\mathcal{O}(10~m)$) compared to the one of our scale, in conclusion for the PTOLEMY purposes it is convenient to use an angle averaged quantity\footnote{It is important to point out that the pitch angle dependence is crucial also for our case.} because $t_{larmor}=1/f_c \ll t_{D}=\mathcal{O}(10~\mu s)$ where the time of the electron drift ($t_D$) is also the time scale of the signal detection\footnote{This is related to the resolution bandwidth parameter.}.\\
The power emitted by an electron in acceleration can be derived starting from the Poynting vector, which gives the general formula for the emitted instantaneous relativistic power

\begin{equation}
    P=-\frac{2}{3}\frac{1}{4\pi\epsilon_0}\frac{q^2}{m^2c^3}\left( \frac{dp_{\mu}}{d\tau}\frac{dp^{\mu}}{d\tau} \right)
    \label{eq:power_1}
\end{equation}

where $p^{\mu}$ is the four-momentum $p^{\mu}=(E/c,p_x,p_y,p_z)$ and $\tau$ is the proper time. In Equation (\ref{eq:power_1}) the four-vector scalar product can be written as

\begin{equation}
    -\frac{dp_{\mu}}{d\tau}\frac{dp^{\mu}}{d\tau}=\left( \frac{d\vec{P}}{d\tau} \right)^2-\frac{1}{c^2}\left(\frac{dE}{d\tau}\right)^2
\end{equation}

However, the energy loss per revolution $dE/d\tau$ is negligible because, considering $1W=6.2\times10^{18}~eV/s$, it is possible to see that $1fW=6.2\times10^{-9}~eV/ps$; then, by using that $|d\vec{p}/d\tau|\equiv\gamma\omega_0|\vec{p}_{\perp}|$ from the circular motion equations, the Equation (\ref{eq:power_1}) becomes

\begin{equation}
\begin{split}
    P&=\frac{2}{3}\frac{1}{4\pi\epsilon_0}\frac{q^2}{m^2c^3}\gamma^2(\frac{q^2B^2}{m^2})(\gamma^2m^2|\vec{v}_{\perp}|^2)=\\
    &=\frac{2}{3}\frac{1}{4\pi\epsilon_0}\frac{q^4B^2}{m^2c}\gamma^2\beta^2\sin^2\theta=\\
    &=\frac{2}{3}\frac{1}{4\pi\epsilon_0}\frac{q^4B^2}{m^2c}(\gamma^2-1)\sin^2\theta
\end{split}
\label{eq:larmor_power}
\end{equation}

that for a reference an electron with $K=18.6~keV$ and $\theta=90^{\circ}$ gives $P\sim 1~fW$.\\
Now that the signal emitted by the electron has been described, it is possible to see how much of this radiation couples to the $TE_{10}$ mode of a rectangular waveguide. It is possible to obtain from \cite{jackson} and \cite{P8_ampiezze} that the amplitude $A$ of the coupled signal is

\begin{equation}
    A \propto \int\vec{v}\cdot\vec{E}e^{j\omega t}e^{-jkz}dt
\end{equation}

where $\omega$ is the frequency, $\vec{v}$ is the velocity of the electron in the laboratory frame. In the case of the $TE_{10}$ mode, the electric field is shown in Equation (\ref{eq:TE10_E}) and because $E_z=0$ only the transverse velocity (kinetic energy) matter.\\
It is possible to obtain \cite{P8_ampiezze} that the power of the signal, that propagates along the waveguide after being coupled with the electron signal is

\begin{equation}
    P_{TE10}=\frac{q^2Z_{10}}{4ab}v^2_{\perp}\cos^2 \left(\frac{\pi}{a}x \right)
\end{equation}

where $Z_{10}$ is the impedance of the first TE mode obtainable from Equation (\ref{eq:impedance}). It is useful to consider the average power because the electron is moving in bouncing motion along X direction, in this case considering that the average of the cosine squared at maximum bouncing over one period is 1/2, the average power carried from the wave is

\begin{equation}
    \langle P_{TE10} \rangle =\frac{q^2Z_{10}}{4ab}v^2_{\perp}\frac{1}{2}
\end{equation}

where during the bouncing the $v_{\perp}$ does not vary (only the $v_{\parallel}$ is reduced in the potential well), and so the signal-waveguide coupling depends essentially on the range of the electron bouncing along the X direction. This is particularly interesting because this distance depends on the pitch angle $\theta$ of the electron and also depends on the bouncing period $T_b$. In conclusion, the average power coupled with the $TE_{10}$ mode and carried by the waveguide, now named $P_{wg}=\langle P_{TE10} \rangle$, is supposed to be in the range 

\begin{equation}
    \frac{P_{e^-}}{2}<P_{wg}<P_{e^-}
\end{equation}

where $P_{e^-}$ is the power emitted by the electron.

\subsection{Electron cyclotron Signal detection} \label{sec:signal_detection}

In this Section, an introduction about the noise in high frequency systems is given, this is particularly interesting in order to understand the capability of the detection of the cyclotron radiation of the electron, in particular this requires the estimate of the signal to noise ratio $SNR$. Firstly, it is useful to introduce the power unit that is typically used in RF systems, that is the $dBm$:

\begin{equation}
    P[dBm]=10\log_{10}\left(\frac{P[W]}{1mW}\right)
    \label{eq:dBm}
\end{equation}

\subsubsection{Thermal Noise}

\begin{figure}[h!]
    \centering
    \includegraphics[width=0.9\linewidth]{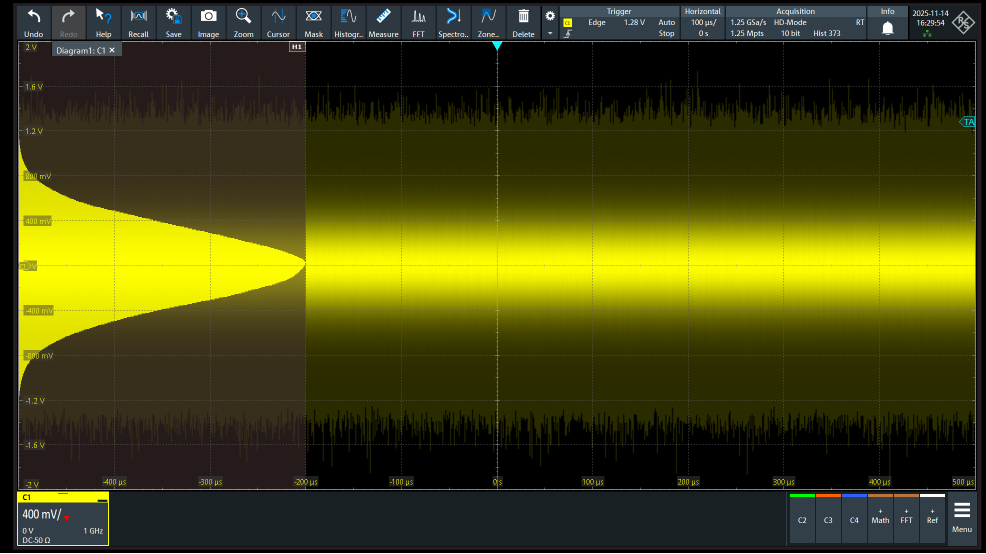}
    \caption{White noise on the oscilloscope. It is possible to see the occupancy histogram (graphical entries per voltage bin) on the left part of the picture, showing the gaussian distribution around the \emph{average noise}.}
    \label{fig:white_noise}
\end{figure}

A high frequency system is typically dominated by the white \emph{thermal noise} (Johnson-Nyquist noise). The power of a white noise can be schematized by a gaussian with central value the average noise, as can be seen in Figure (\ref{fig:white_noise}) and it is a random fluctuation with a standard deviation. 

and its average power is defined as

\begin{equation}
    N_t[W]=k_BT\cdot RBW
    \label{eq:thermal_noise}
\end{equation}

that depends on the temperature $T$ and by the \emph{resolution bandwidth} $RBW$, which is the inverse of the observation time.\\
This last parameter is particularly important because it must be directly compared with the electron drift time $t_D$. The thermal noise is dominant at the typical power and frequency range of operation of PTOLEMY. However, if an antenna for the electron signal is connected directly to DAQ like a \emph{spectrum analyzer}, it is possible to measure a noise power higher than the expected from Equation (\ref{eq:thermal_noise}). This means that there is additional noise is present into the system. To understand that, we first define the additional noise of a generic two port device with gain $G$. The additional noise is defined with the \emph{noise factor} $F$ that is given by the ratio between the input and output $SNR$

\begin{equation}
    F=\frac{SNR_i}{SNR_o}
    \label{eq:F}
\end{equation}

and in log scale gives the so called \emph{noise figure} $NF$

\begin{equation}
    NF=10\log_{10}(F)
    \label{eq:NF}
\end{equation}

The additional noise aforementioned is the noise from the DAQ device, a typical value of the noise figure of a spectrum analyzer can be $NF=40~dB$, this means that if we consider a temperature of $T=30~K$ and a resolution bandwidth of about $\Delta f=20~kHz$ (those are the parameters typically used in the electron trap setup) we should expect a noise power of about $N=\mathcal{O}(-140~dBm)$ which considering a signal power of about $P=1fW=-120~dBm$ would correspond to an $SNR=20~dB$.\\
However, if in this example we measure directly this $1~fW$ signal, we introduce the noise figure of the DAQ, that means that the noise power becomes $N=N_t+NF$ and so the noise power becomes larger than the signal. Another problem for the detection of the electron signal is that the frequency is too high to be sampled.\\
To solve these problems in the RF systems it is typically used the \emph{Friis law} for the noise reduction and a \emph{down-conversion} is then used in order to convert a $27~GHz$ signal down to  an intermediate frequency of about $f_{IF}=\mathcal{O}(500~MHz)$.\\
The Friis law, that is reported in Equation (\ref{eq:friis}), says that given $N$ two port devices in cascade, the overall noise factor $F_{tot}$ is

\begin{equation}
    F_{tot}=F_1+\sum_{k=2}^N\frac{F_k-1}{\prod_{i=1}^{k-1}G_i}
    \label{eq:friis}
\end{equation}

where $G_i$ is the gain expressed in linear units. \\

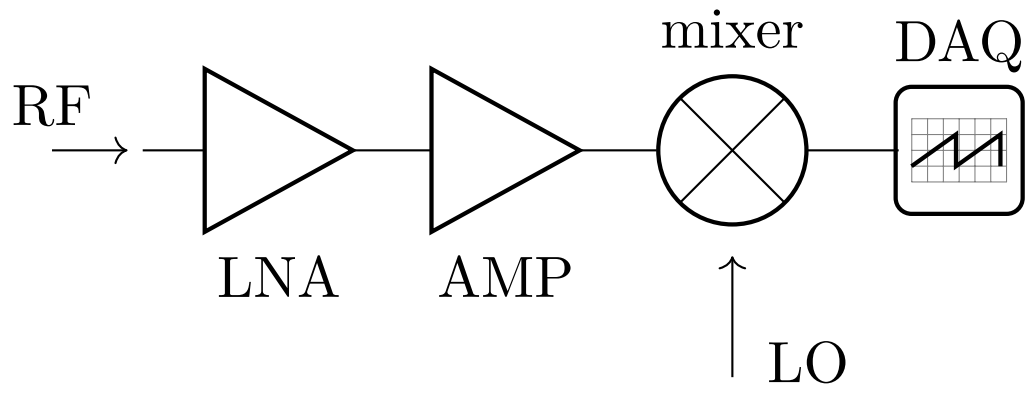
\begin{figure}[h!]
    \centering
    \begin{circuitikz}
        \draw[->](-5,0)--(-4.5,0);
        \node[] at(-5,0.3){RF};
        \draw(-4.4,0)--(-4,0);

        \draw
		(-4,0) to [amp,l_=LNA](-3,0)
		to [amp,l_=AMP](-1,0);

        \node[mixer] at (-0.5,0){};
		\node[] at(-0.5,0.8){mixer};
        \draw[->](-0.5,-1.5)--(-0.5,-0.7);
        \node[] at(0,-1.4){LO};

        \draw
        (0,0) to [short,-](0.6,0);
        \node[oscopeshape] at (1,0){DAQ};
        
    \end{circuitikz}
    \caption{Scheme of a generic electronic chain needed for the detection of low power and high frequency signals.}
    \label{fig:chain}
\end{figure}

\newpage
It is possible to note that, for example, also in the case of a three component electronic chain composed by two amplifiers and a high noise receiver, the total additional noise depends essentially on the first amplifier. In this way, by using a \emph{cryogenic low noise amplifier} (LNA) as the first amplification stage, it is possible to limit the additional noise even to $NF<1~dB$, opening up the possibility of measuring the cyclotron signal. The problem of high frequency sampling can be solved using a \emph{mixer}, that is, a device capable of translating the input frequency $f_{RF}$ using an internal reference frequency $f_{LO}$, giving an intermediate frequency as $f_{IF}=|f_{RF}-f_{LO}|$. A prototype of the electronic chain required for the electron signal detection is shown in Figure (\ref{fig:chain}), in the next Chapter the electronic chain implemented in the Electron Trap setup will be described.\\

In conclusion the detection of the cyclotron signal is possible if a proper electronic chain is used. An important parameter for the detection is also the resolution bandwidth that must be set according to the experimental drift time of the electrons.

%% file: sections/4Electron_Trap.tex
\chapter{The Electron Trap} \label{chap:Electron_Trap}

In the previous Chapters the role of the RF tracker in the PTOLEMY detector has been described and the expected electron motion and RF signal have been shown, then in Section (\ref{sec:cyclotron_radiation_detection}) an introduction to low power and high frequency signal detection has been given. At the beginning of 2022, an R\&D programme started at \emph{Laboratori Nazionali del Gran Sasso} with the goal of studying electron detection by means of Radio Frequency (RF) detection with a dedicated test setup: the Electron Trap. The goal of this test is to gain experience with the electron cyclotron radiation spectroscopy technique and to learn how to measure the electron kinematic variables by the RF pattern, as anticipated in Section (\ref{sec:RF_region}). In this Chapter the design and simulation of the Electron Trap setup is shown, and in Section (\ref{sec:kin_reconstruction}) the kinematics variable reconstruction methods are described, too.

\section{The Electron Trap setup}\label{sec:etrap}

The key elements for the development of an electron cyclotron radiation spectroscopy test setup are: 

\begin{itemize}
    \item Electron source ($^{83m}Kr$)
    \item Magnetic field (Permanent Hallbach Magnet)
    \item System for electron track containment (Electrodes system)
    \item A cryogenic system for cavity and LNA cooling
    \item High frequency readout electronic chain
    \item Trigger
\end{itemize}

In the following the Electron Trap setup will be described by looking in detail at the item listed above.

\begin{figure}[h!]
    \centering
    \includegraphics[width=0.7\linewidth]{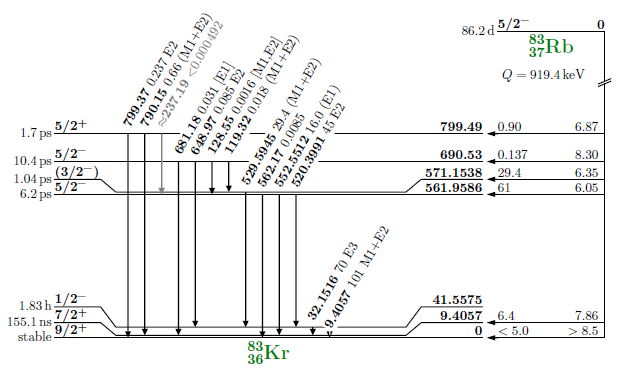}
    \caption{Rubidium source nuclear transitions.}
    \label{fig:linee_Rb}
\end{figure}

\newpage
In the PTOLEMY experiment, the electrons are produced by means of beta decay of Tritium embedded on a solid substrate; however, for Electron Trap studies, a solid source of Rubidium $^{83}Rb$ has been used. This source, as shown in Figure (\ref{fig:linee_Rb}), produces $^{83m}Kr$ in gaseous form by electron capture.

\begin{figure}[h!]
    \centering
    \includegraphics[width=0.5\linewidth]{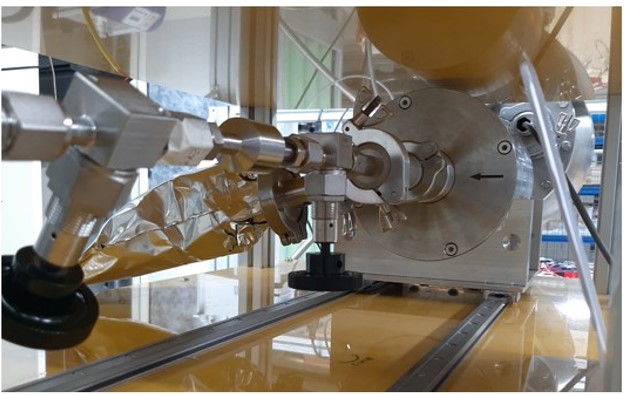}
    \caption{Rubidium source with gauge valve for Krypton gas inlet towards the experimental apparatus.}
    \label{fig:kr_source}
\end{figure}

Rubidium is sealed in a cylindrical container with a gauge valve that allows the emission of the krypton gas to enter the experimental apparatus as shown in Figure (\ref{fig:kr_source}).

The Krypton meta-stable atoms decay to $^{83}Kr$ through several channels, emitting two electrons in cascade by EC transitions of energy $32.2~keV$ and $9.4~keV$, respectively, where the lifetime of the first transition is $\tau_{Kr}=158.1~min$ while the second transition is a prompt decay. In Table (\ref{tab:Kr_emission_lines}) the Krypton emission lines (only the $32.2~keV$ gamma transition has been considered in the Electron Trap setup studies and data taking) are shown together with the branching ratio.

\begin{table}[h!]
    \centering
    \begin{tabular}{|c|c|c|}
    \hline
        Line&K[eV]&BR[\%]\\
        \hline
    K & 17824.2(5) & 24.8(5) \\
    \hline
    L1 & 30226.8(9) & 1.56(2) \\
    \hline
    L2 & 30419.5(5) & 24.3(3) \\
    
    L3 & 30472.2(5) & 37.8(5) \\
    \hline
    M1 & 31858.7(6) & 0.249(4) \\

    M2 & 31929.3(5) & 4.02(6) \\

    M3 & 31936.9(5) & 6.24(9) \\

    M4 & 32056.4(5) & 0.0628(9) \\

    M5 & 32057.6(5) & 0.0884(12)\\

    N1 & 32123.9(5) & 0.0255(4) \\

    N2 & 32136.7(5) & 0.300(4)\\

    N3 & 32137.4(5) & 0.457(6) \\
    \hline
    \end{tabular}
    \caption{$^{83m}Kr$ electron emission lines together with the kinetic energy and the corresponding branching ratio. Only the L3 has been studied in this work \cite{kripton}. }
    \label{tab:Kr_emission_lines}
\end{table}

%attività sorgente?

\newpage
\subsection{Magnet}

\begin{figure}[h!]
    \centering
    \includegraphics[width=0.6\linewidth]{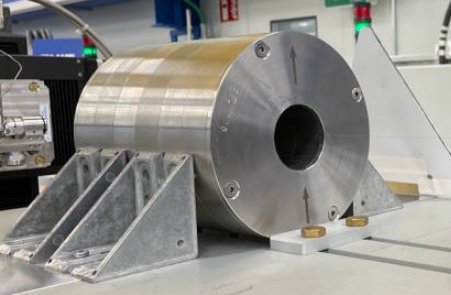}
    \caption{Cylindrical permanent magnet ($L=193~mm$ and $\diameter=  55~mm$) used in the electron trap setup. The X axis is oriented across the arrows visible in the photograph.}
    \label{fig:magnet_trap}
\end{figure}

The characteristics of the permanent magnet used in the electron trap setup are a key element for the design of the whole setup, in particular, the magnet geometry and intensity have a crucial role on the experimental design.\\
The magnet, which is shown in Figure (\ref{fig:magnet_trap}), have been purchased in 2021 by \emph{China Magnets Source Material}; it is built with permanent magnetic components in a Halbach configuration in cylindrical geometry. This magnet is $193~mm$ long and has an outer radius of $65~mm$ and a bore hole of about $\diameter = 55~mm$. The field generated in the bore hole is a dipolar field with an amplitude of about $1~T$ in the central region. 

\begin{figure}[h!]
    \centering
    \includegraphics[width=0.8\linewidth]{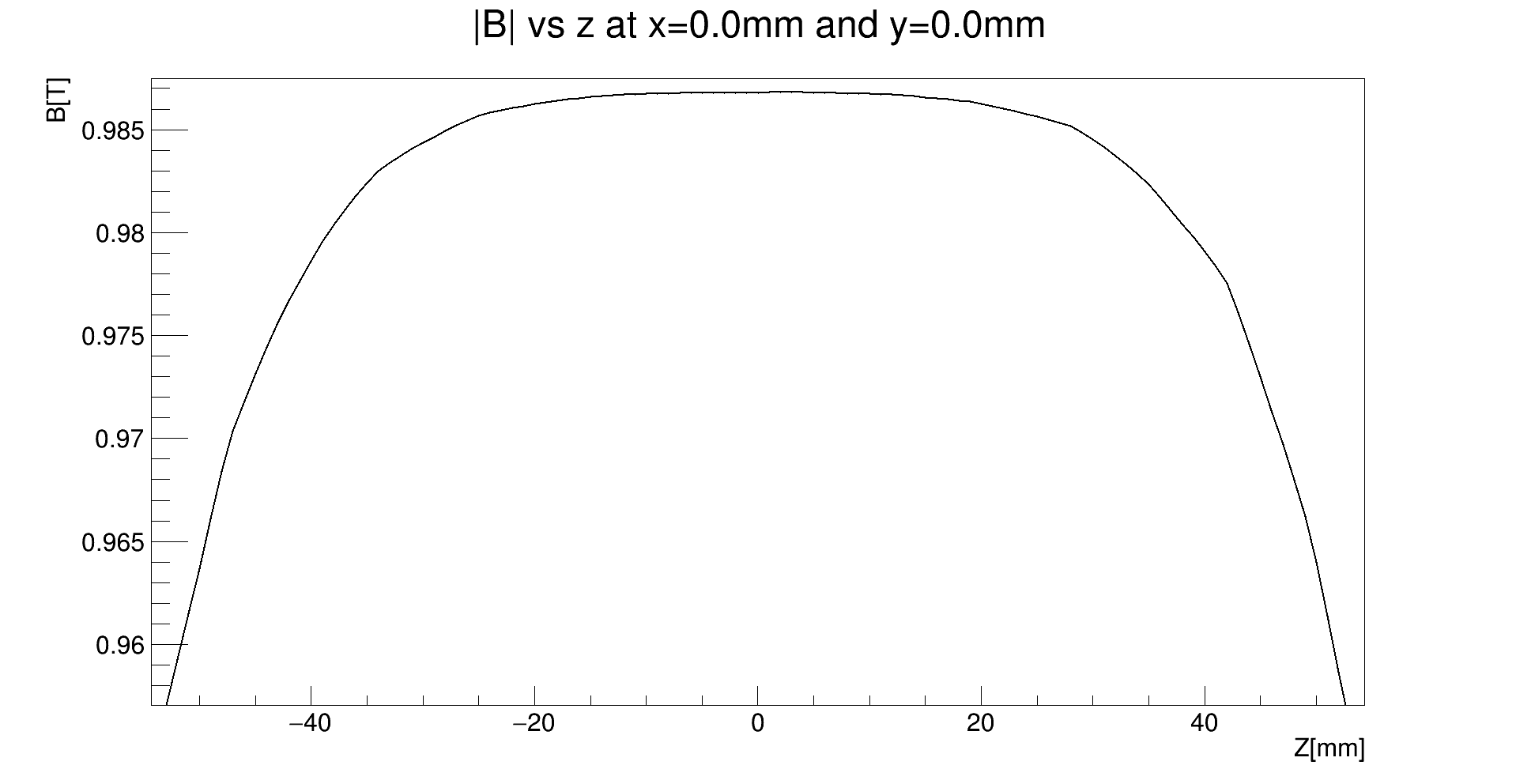}
    \caption{Simulated magnetic field along the central line as function of Z. It is possible to note that the most uniform region of the magnet is in the region between $-20~mm$ and $+20~mm$.}
    \label{fig:B_z}
\end{figure}

\newpage
It is important to recall the standard reference system used in the PTOLEMY's works, in which the main dipolar magnetic field is aligned along the X axis and the magnet is placed across the Z direction. The correct knowledge of the magnetic field is necessary to predict the correct Larmor frequency and to study the electron trajectories in the Electron Trap. For this reason, simulation of the whole magnetic field map has been done. In Figure (\ref{fig:B_z}) the simulated value of the magnetic field modulus ($|B|\sim B_x$) is shown at the central part of the magnet. It is visible a region, of almost $4~cm$ across Z axis,
where the field has an acceptable constant value.\\

\begin{figure}[h!]
    \centering
    \includegraphics[width=1\linewidth]{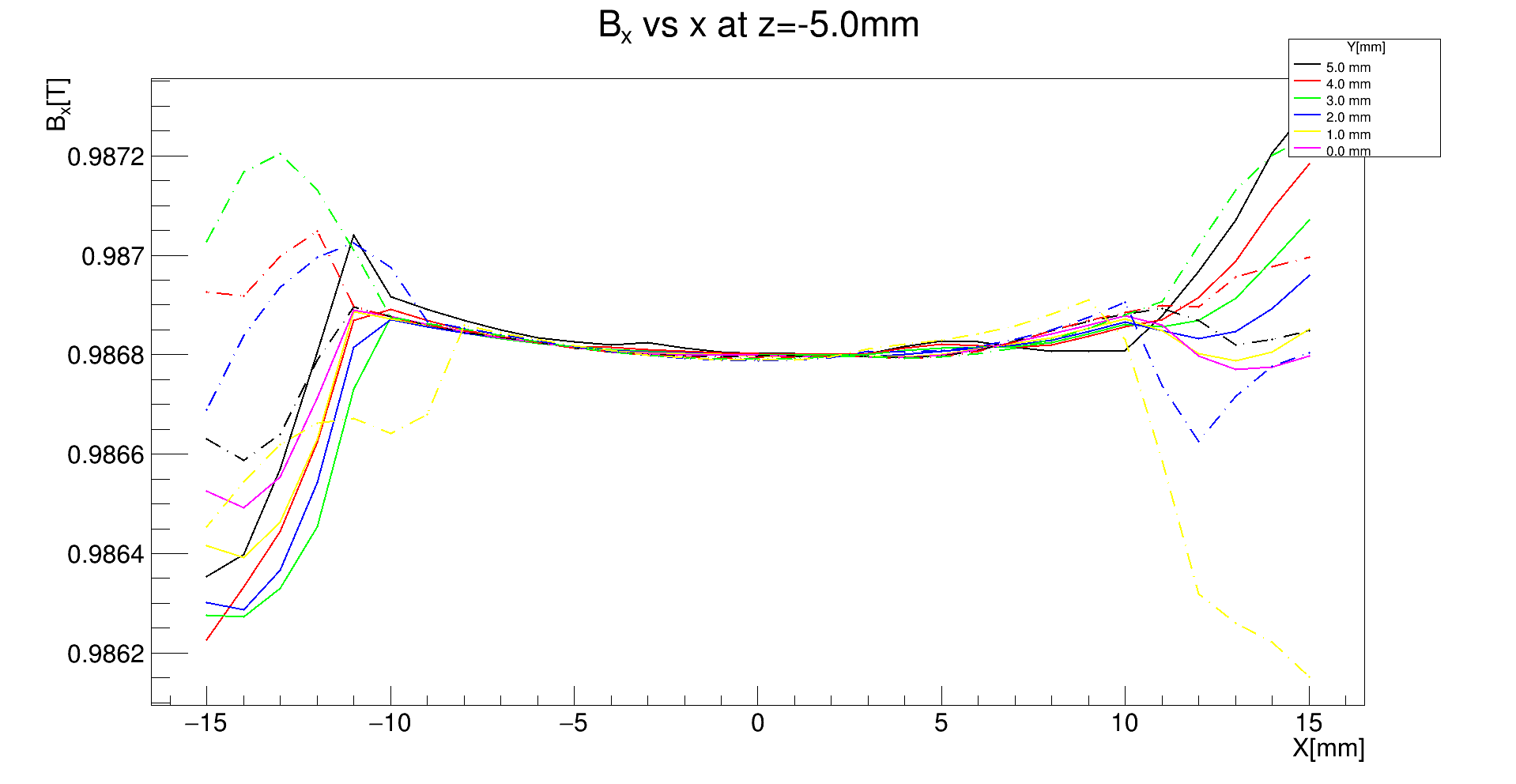}
    \caption{$B_x$ vs X (the electron trap electrodes are placed at $\pm15~mm$ over X) at $z=-5~mm$. The color represent different the value at different Y and the dotted line represent negative value of Y.}
    \label{fig:B_x}
\end{figure}

From the magnetic field map it is possible to obtain the dependence of the $B_x \sim |B|$ component vs X coordinate and at different Y, as shown in  Figure (\ref{fig:B_x}). From the field map in Figure (\ref{fig:B_x}) it is visible that when the electron is bouncing across the X direction with a span of $2~cm$, the variation of the magnetic field is small, which implies that no frequency modulation, because of magnetic field variation, should be expected.

\newpage
\subsection{Electrodes and waveguide system}

\begin{figure}[h!]
    \centering
    \includegraphics[width=0.5\linewidth]{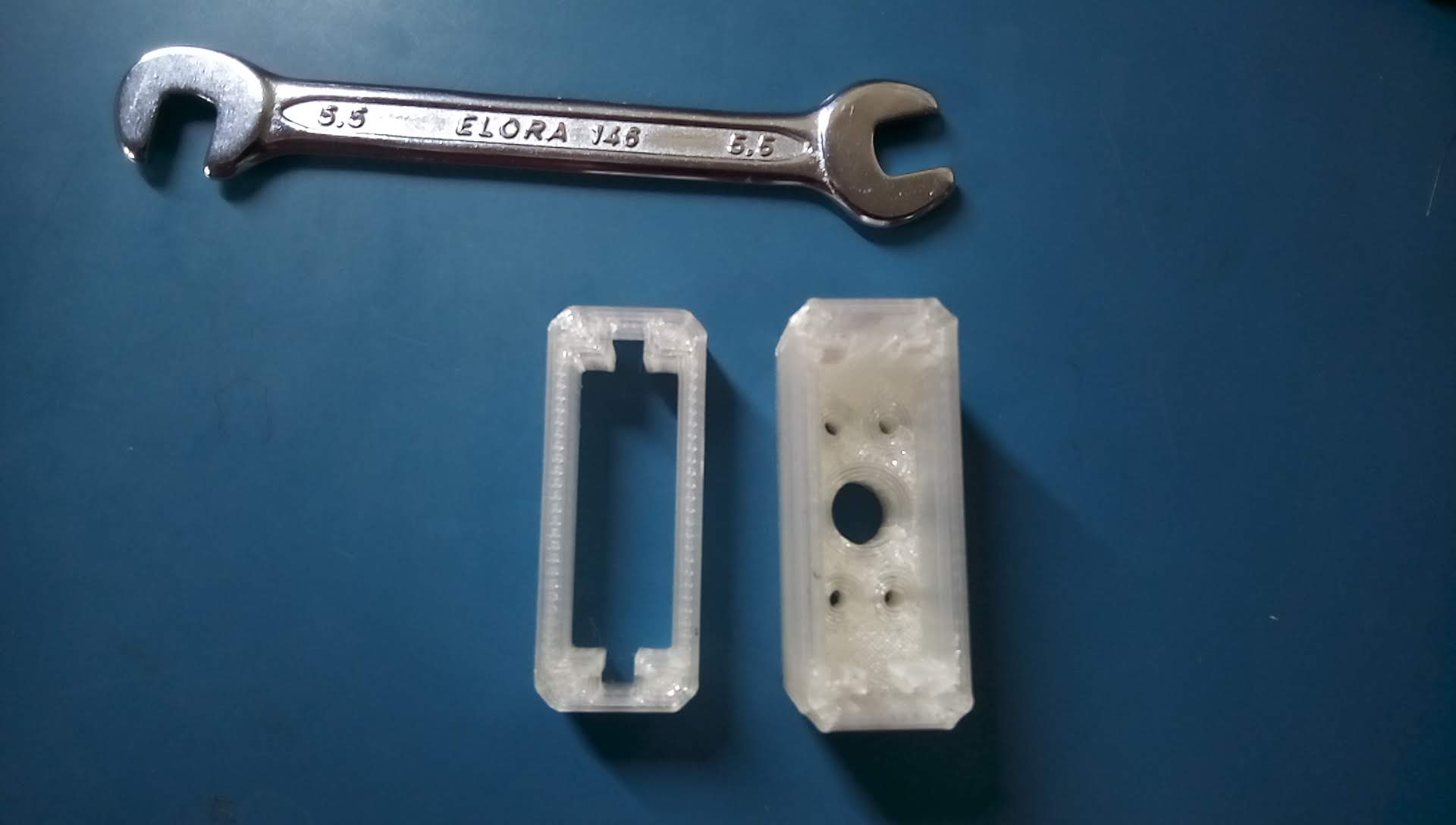}
    \caption{Plastic supports used inside the Electron Trap. The support on the left is used on the waveguide extremity in which the RF is detected while the support on the right is used to sustain electrodes and Krypton pipe trough the big circular hole at the center. The four small holes are required for electrodes biasing.}
    \label{fig:plastic_supports}
\end{figure}

\begin{figure}[h!]
    \centering
    \includegraphics[width=0.5\linewidth]{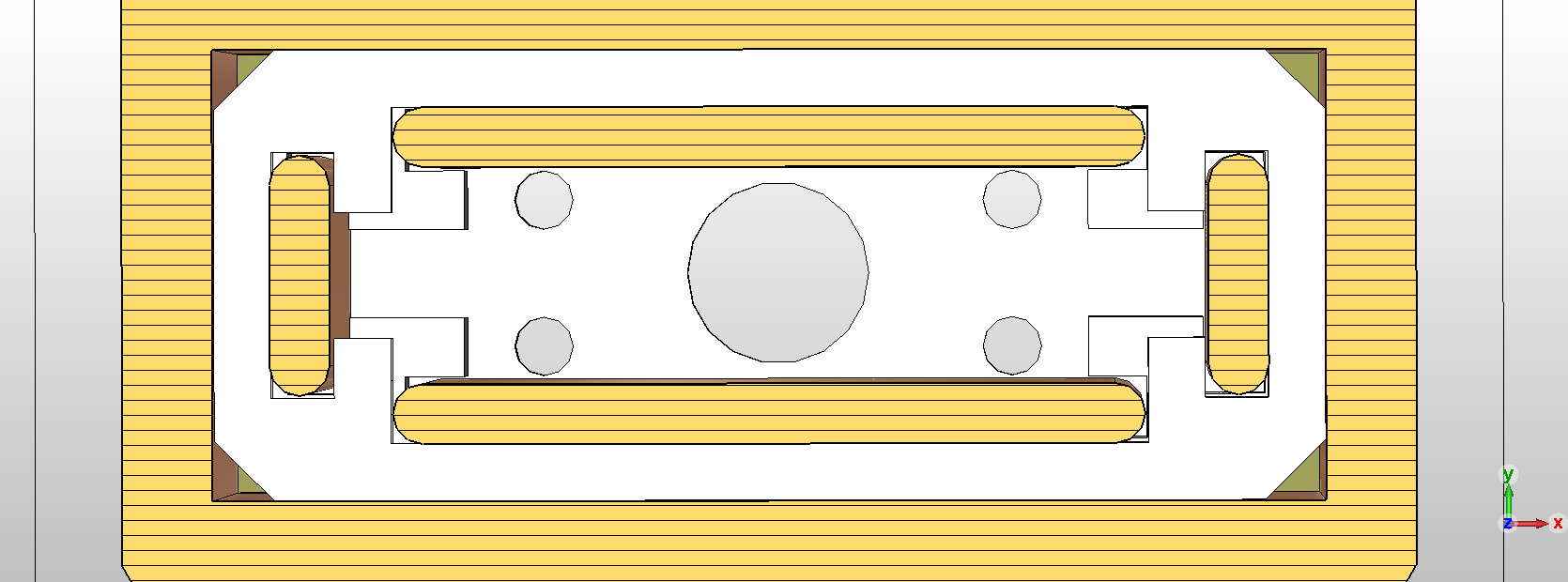}
    \caption{Inner view of the Electron Trap setup in the four electrodes configuration. The electrode support are capable to host four electrodes, however only bouncing electrodes has been used in this work.}
    \label{fig:e_trap_cut_view}
\end{figure}

In this Section the Electron Trap geometry is presented. The Electron Trap has been developed to study detection of high frequency cyclotron radiation emitted by electrons in bouncing motion, in a simplified geometry with respect to the real experiment. The setup is composed of a rectangular waveguide of $L=198~mm$ long made of pure copper of outer dimensions $42\times20~mm^2$ and inner dimensions $37\times 15~mm^2$. Inside this volume, electrodes for bouncing motion and for Z drift can be inserted, as shown in Figure (\ref{fig:e_trap_cut_view}), in particular the bouncing electrodes have a short side of $10~mm$ and are displaced along the X direction of $\pm15.5~mm$ (the inner face of the electrodes is at $x=\pm15~mm$) while top and bottom electrodes, which are not used in the simulations and in the data taking described in Chapter (\ref{chap:Electron_Trap_Measurements}), are at $y=\pm11~mm$. Because the main goal of this setup is the detection of single electrons, as a first step it is convenient to have the lowest electron $V_z$ (and so $E_y\times B_x$ drift) possible to facilitate the detection; thus, to reduce the drift across Z, the top and bottom drift electrodes have not been used\footnote{This has been done also for an impedance improvements.}, resulting in a fiducial volume (F.V.) of about 

\begin{equation}
    F.V.=30\times10\times40~mm^2
    \label{eq:fiducial_volume}
\end{equation}

where the limitations over Z are due to the decrease of the magnetic field as shown in Figure (\ref{fig:B_z}). The electrodes are supported by two 3D printed supports in dielectric material, which are shown in Figure (\ref{fig:plastic_supports}) and are located at the two extremities of the waveguide.

\newpage

\begin{figure}[h!]
    \centering
    \includegraphics[width=0.5\linewidth]{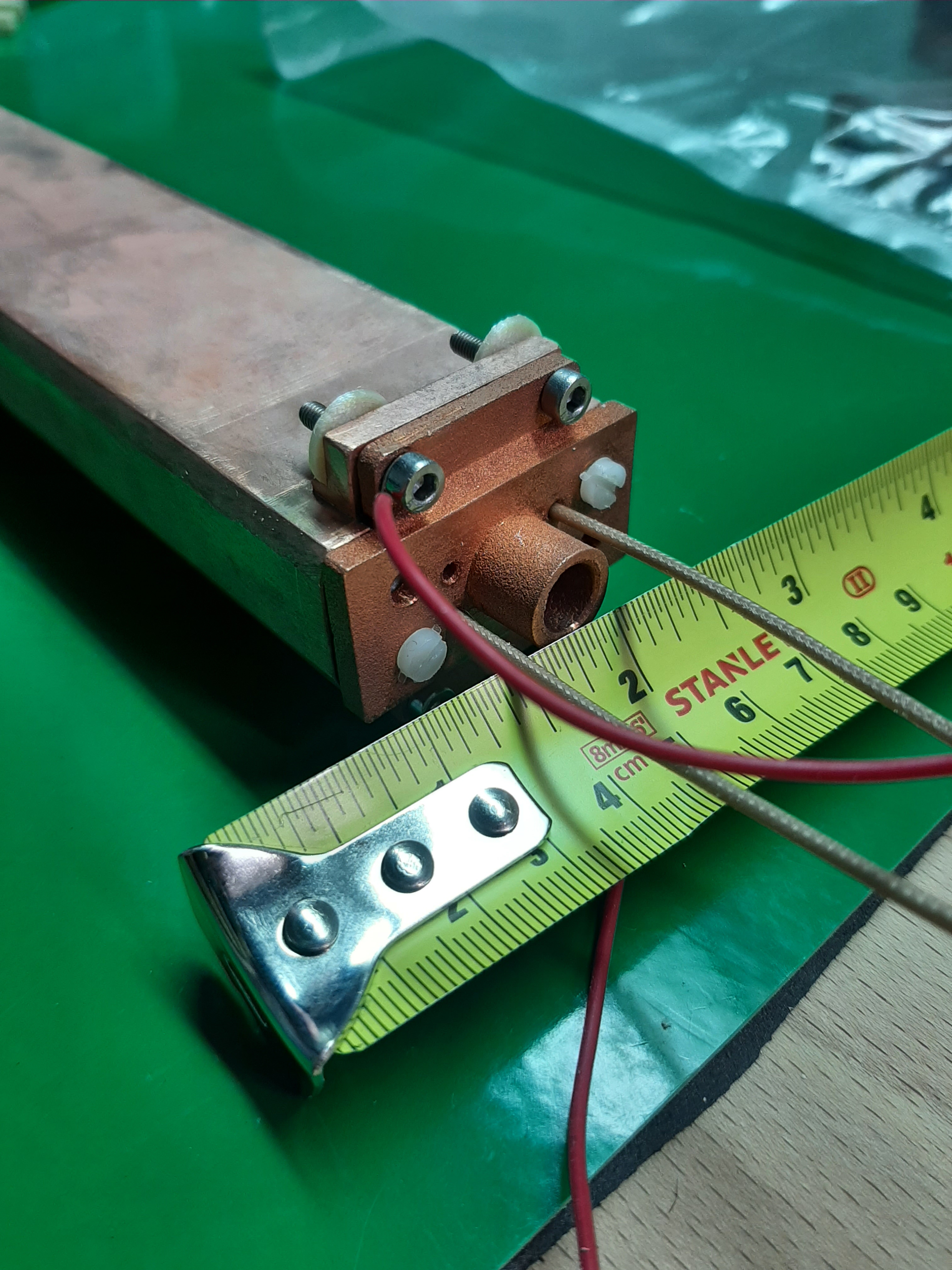}
    \caption{Photograph of the Krypton port. The Krypton gas is injected by silicone tube in the cylindrical hole. It is also possible to see the two HV cables used for bouncing electrodes bias. The red cable is the ground, that is in common with the HV.}
    \label{fig:Kr_port}
\end{figure}

\begin{figure}[h!]
    \centering
    \includegraphics[width=0.5\linewidth]{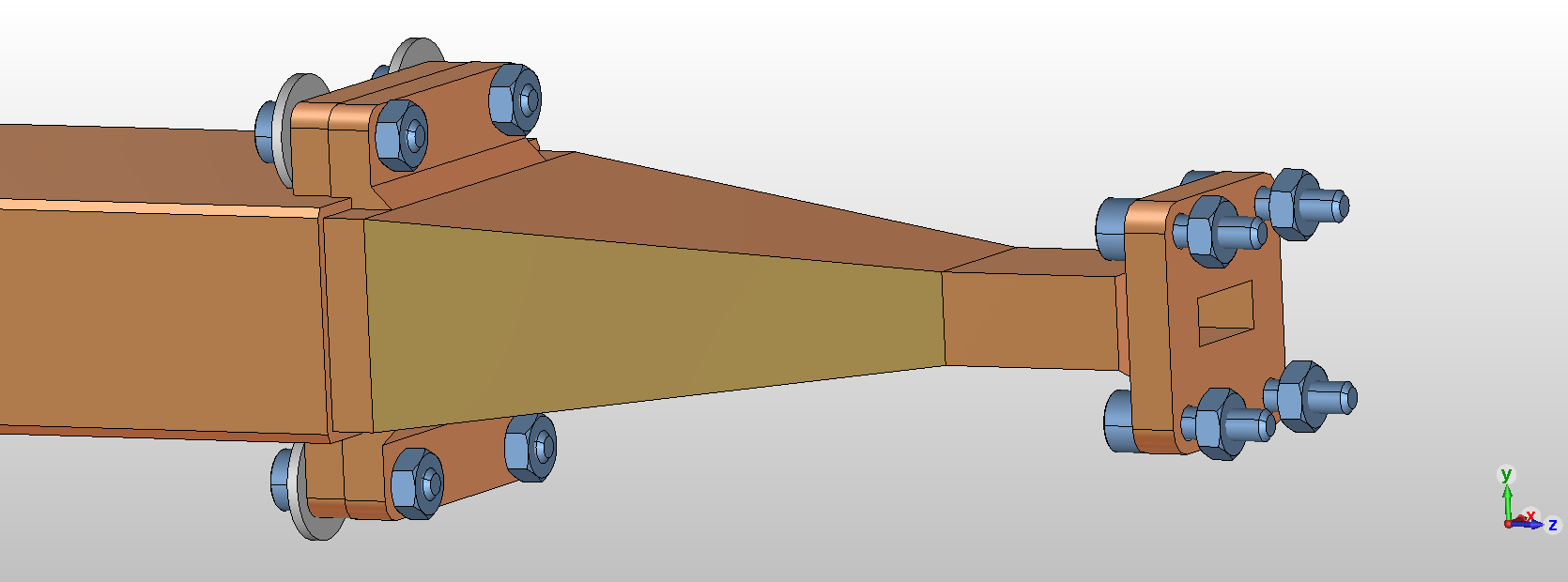}
    \caption{Horn used at the RF port for wave propagation from the Electron Trap to the standard WR42 waveguide.}
    \label{fig:horn}
\end{figure}

Those two extremities, also called \emph{ports}, have two important roles: one port, which is shown in Figure (\ref{fig:Kr_port}), is used for Krypton inlet and is also used to support the electrode setup (Krypton port) while the other one is used for the RF readout by means of a 3D copper printed horn (RF port) that converts the inner section of the trap to a standard WR42 waveguide ($10.7\times 4.32~mm^2$), as shown in Figure (\ref{fig:horn}).

\newpage

\begin{figure}[h!]
    \centering
    \includegraphics[width=0.5\linewidth]{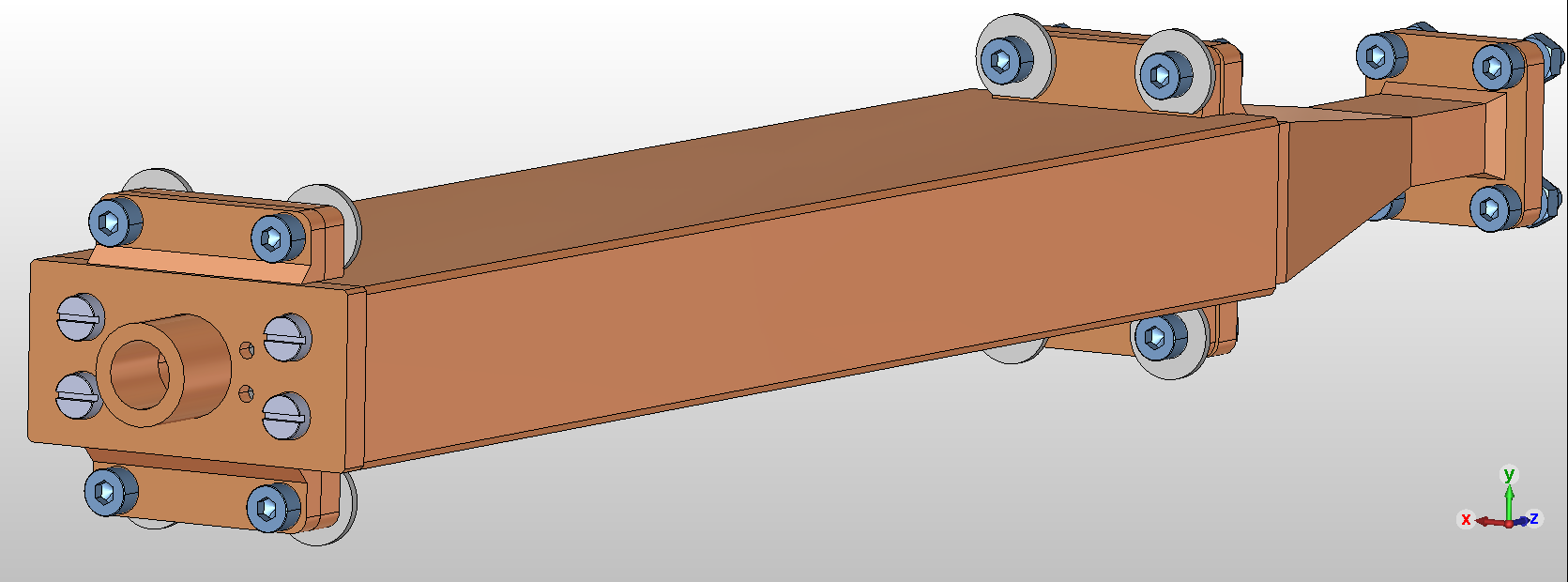}
    \caption{The Electron Trap setup. During the data taking the whole setup is in high vacuum, cooled down and wrapped with Mylar super-insulation layers. It is possible to note the PEEK washer used to hold the Electron Trap to prevent thermal contact with the external vessel. }
    \label{fig:e_trap3D}
\end{figure}

The inner view of the electron trap electrodes is shown in Figure (\ref{fig:e_trap_cut_view}) while the Electron Trap 3D model is shown in Figure (\ref{fig:e_trap3D}).

\subsection{Cryogenic setup}

\begin{figure}[h!]
    \centering
    \includegraphics[width=0.5\linewidth]{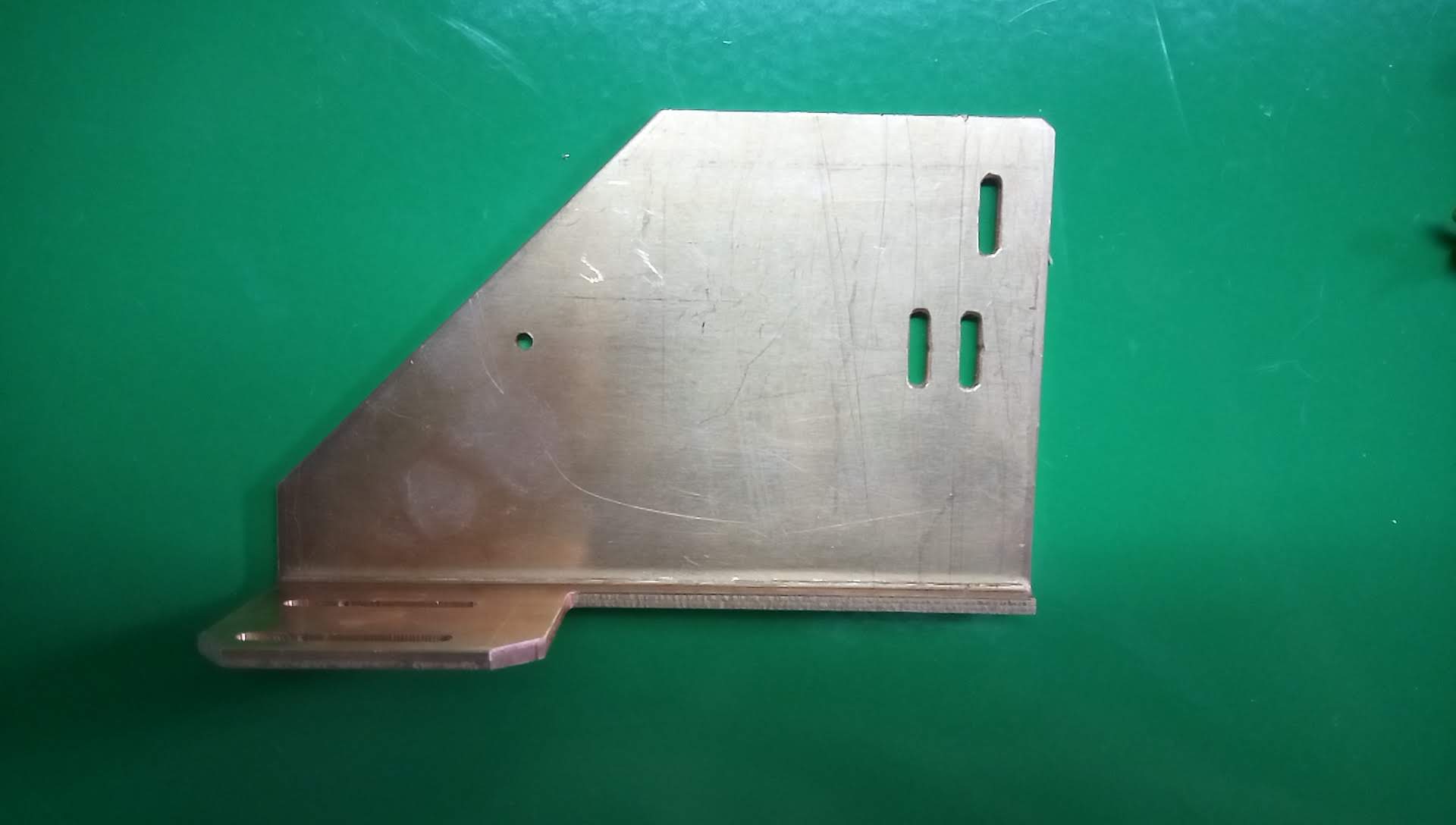}
    \caption{Copper support for the LNA. The LNA is placed on the top right while the thermal sensor is on the circular hole on the left. The scale length is $10~cm$.}
    \label{fig:LNA_support}
\end{figure}

The detection of cyclotron radiation emitted by an electron in cyclotron motion requires an electron drift time of at least $\mathcal{O}(1~\mu s)$ without unwanted scattering with gas. Because of this requirement, it is important to put the Electron Trap in high vacuum, then, as is shown in Equation (\ref{eq:thermal_noise}), the system must also be cooled down to reduce the thermal noise power. The Electron Trap is then inserted into an aluminium vacuum tube with an outer diameter $50~mm$ that is inserted into the cavity of the permanent magnet. One side of the tube is connected with a Tee vacuum tube that is connected with the Rubidium source, while the other part is connected with the larger part of the vacuum vessel, which will be discussed in the following. The main vacuum volume is connected to the cold source, that is a \emph{Gifford-McMahon} (GM) cold head, capable to cool down to $11.7~K$ in the best conditions\footnote{Without thermal load.} and is connected with a copper bar that is extended towards the central part of the main vacuum volume. The copper bar host a $50~\Omega$ resistance for thermal stabilization. On top of this bar, a copper support for the LNA is placed with good thermal connection. This support, that is shown in Figure (\ref{fig:LNA_support}), is used to cool down the LNA and host a diode to measure the temperature.

\begin{figure}[h!]
    \centering
    \includegraphics[width=0.6\linewidth]{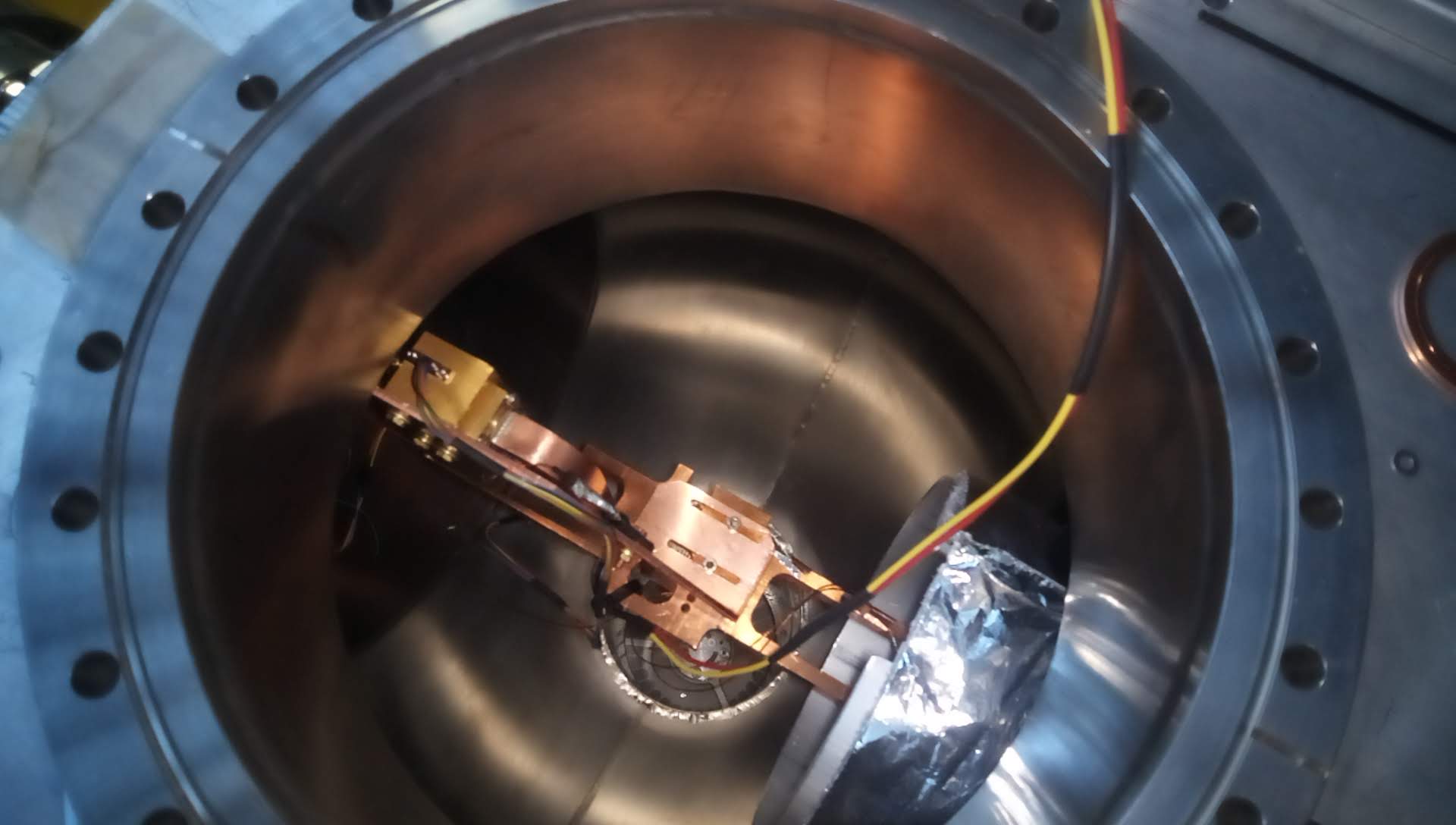}
    \caption{Top view of the inner part of the main vacuum vessel. One can see the LNA on the left and the connection between the LNA support and the copper bar on the right.}
    \label{fig:cryo_inner}
\end{figure}

The top view of the inner part of the cryogenic system is shown in Figure (\ref{fig:cryo_inner}) in which the LNA connected to the copper bar is shown and temperature sensor, too.

\begin{figure}[h!]
    \centering
    \includegraphics[width=0.8\linewidth]{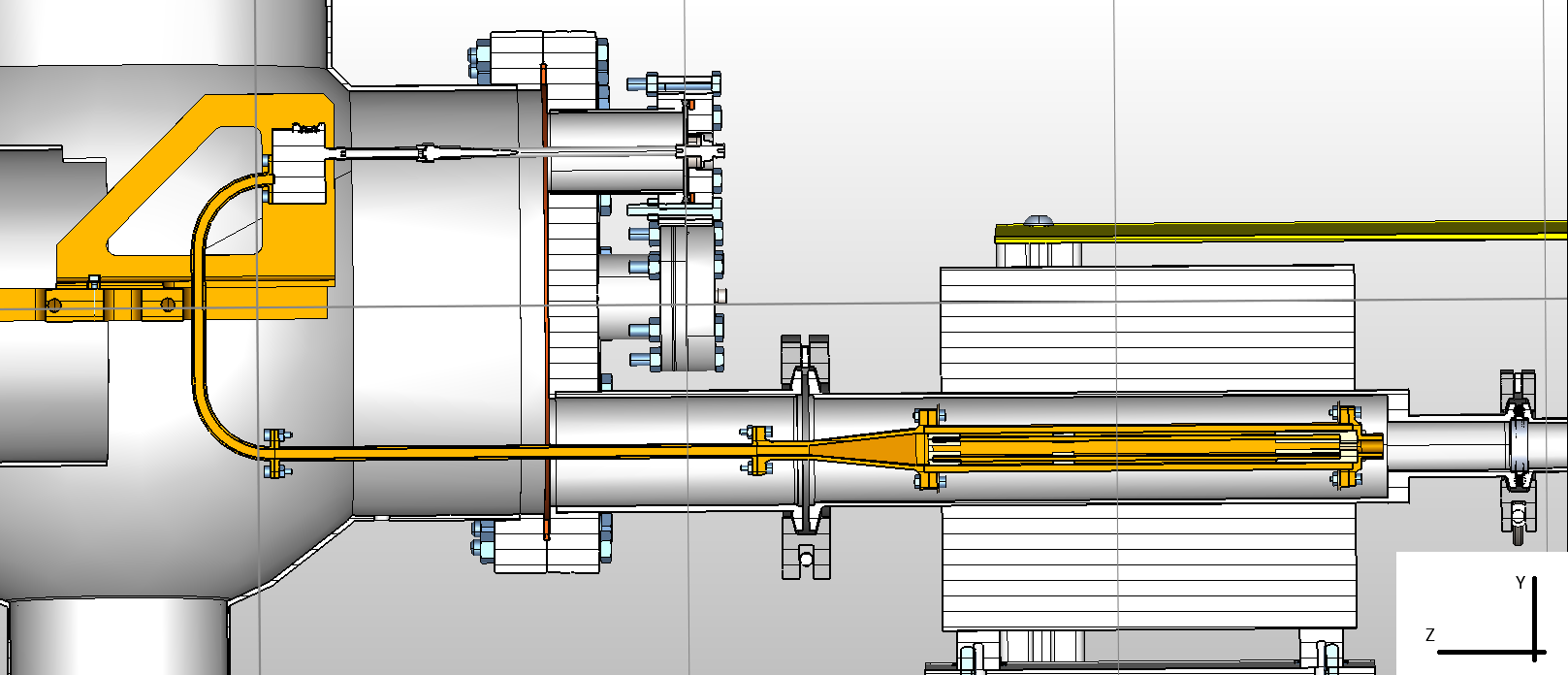}
    \caption{Cross section view of the Electron Trap setup. From right to left: KF flange for Rubidium source, Electron Trap inserted in the cylindrical magnet, horn for RF readout, linear WR42, bended WR42 , LNA and finally coaxial cable. It is possible to note on the left the support for the LNA placed on top of the copper bar connected to the cold head.}
    \label{fig:section_vessel}
\end{figure}

The Electron Trap horn and the input port of the LNA are connected by two WR42 waveguides that guarantee low signal losses even at low temperature. Indeed, the signal outgoing from the LNA in connected with a semi-rigid coaxial cable with 2.92 (K) connector to a high frequency RF feed-trough flange. An additional flange for LNA bias has been used. A picture of the inner section of the Electron Trap vacuum vessel is shown in Figure (\ref{fig:section_vessel}).

\begin{figure}[h!]
    \centering
    \includegraphics[width=0.8\linewidth]{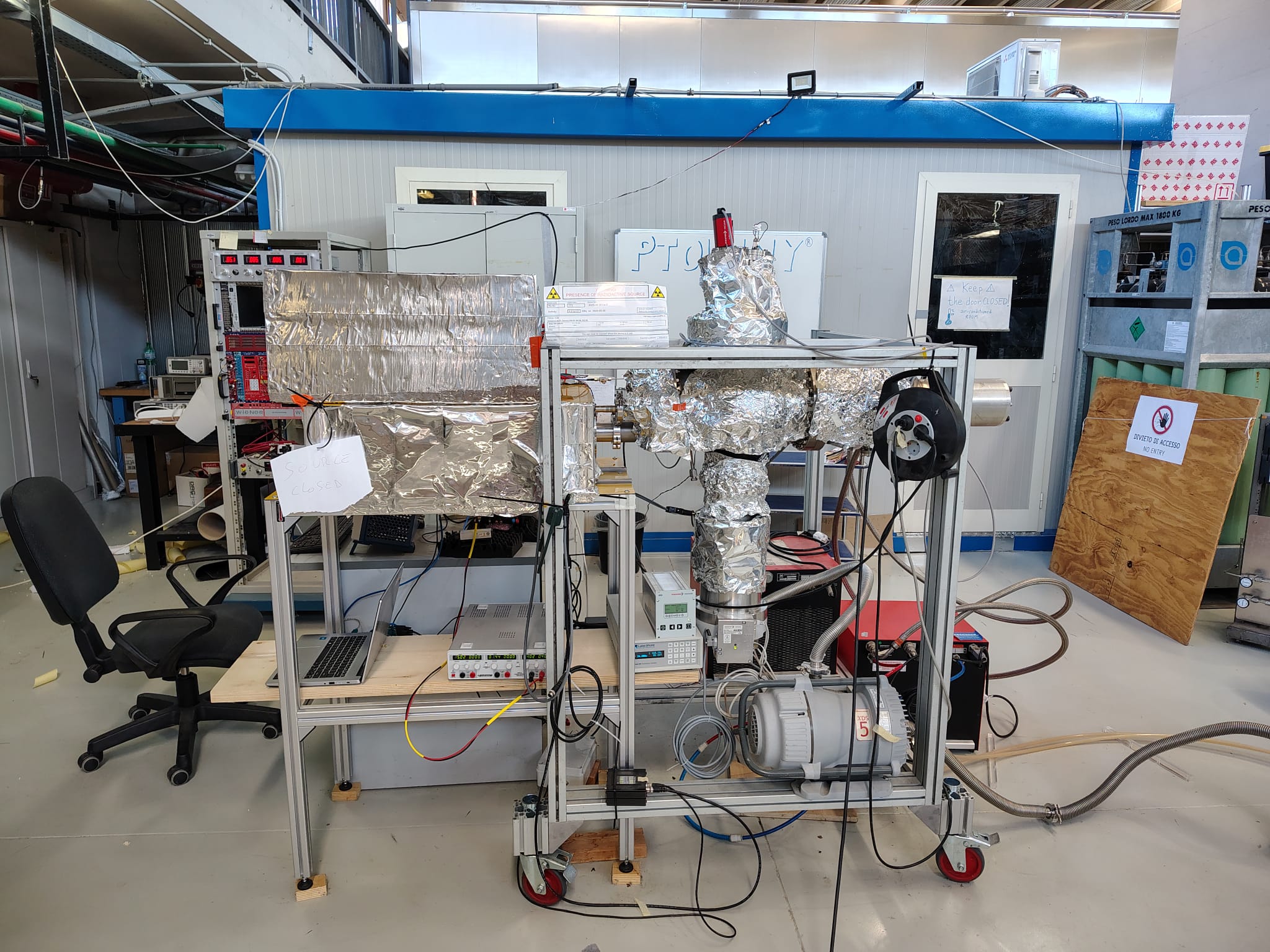}
    \caption{The Electron Trap setup. It is possible to note the primary and turbo pumps together with the compressor for the cold head (right). The system is wrapped in aluminum foils for electromagnetic shielding.}
    \label{fig:etrap_frontale}
\end{figure}

On the top of the vacuum system the vacuum gauge sensor is installed for vacuum pressure recording. All surfaces are wrapped with mylar super-insulation layers in order to avoid heating by black body radiation of the external vessel. A picture of the Electron Trap facility is shown in Figure (\ref{fig:etrap_frontale})

\newpage
\subsection{Electronic Chain}

In Section (\ref{sec:cyclotron_radiation_detection}) the main concepts of electron cyclotron radiation detection have been described, let's now describe the electronic apparatus of the Electron Trap setup. The signal emitted by the electron in cyclotron motion with power $P_{e^-}$, as mentioned in Section (\ref{sec:cyclotron_radiation_detection}), is coupled to the $TE_{10}$ mode of the electron trap with a resulting signal power of $P_{wg}$. This signal is then attenuated by the presence of the electrodes and dielectric material inside the waveguide, that changes the impedance of the waveguide such that an effective attenuation factor of $ATT_{trap}=4.8~dB$ is present. This attenuation has to be added to take into account for a non ideal behavior of the WR42 waveguides, that are present between the Electron Trap and the LNA. The attenuation factor is about $ATT_{wg}=2.3~dB$. As shown in Figure (\ref{fig:section_vessel}), after the "C" waveguide the first amplifier of the electronic chain is present, that is a \emph{LNF-LNC16-28WB}, with an amplification of about $G_{LNA} \sim 36~dB$ and a noise figure $NF_{LNA} \sim 0.16~dB$ at $5K$. The signal is then propagated outside the vacuum vessel by means of a coaxial cable (for thermal decoupling) and RF feed-trough with 2.92 (k) connectors. Outside the vacuum vessel, a custom down-converter developed by the Nikhef group of the PTOLEMY collaboration, is used to do a second amplification of about $G_{d}=60~dB$ and a down-conversion. The intermediate frequency IF then becomes 

\begin{equation}
    f_{IF}[GHz]=f_{LO}[GHz]-f_{RF}[GHz]+0.45[GHz]
    \label{eq:IF_downconverter}
\end{equation}

\begin{figure}[h!]
    \centering
    \includegraphics[width=0.5\linewidth]{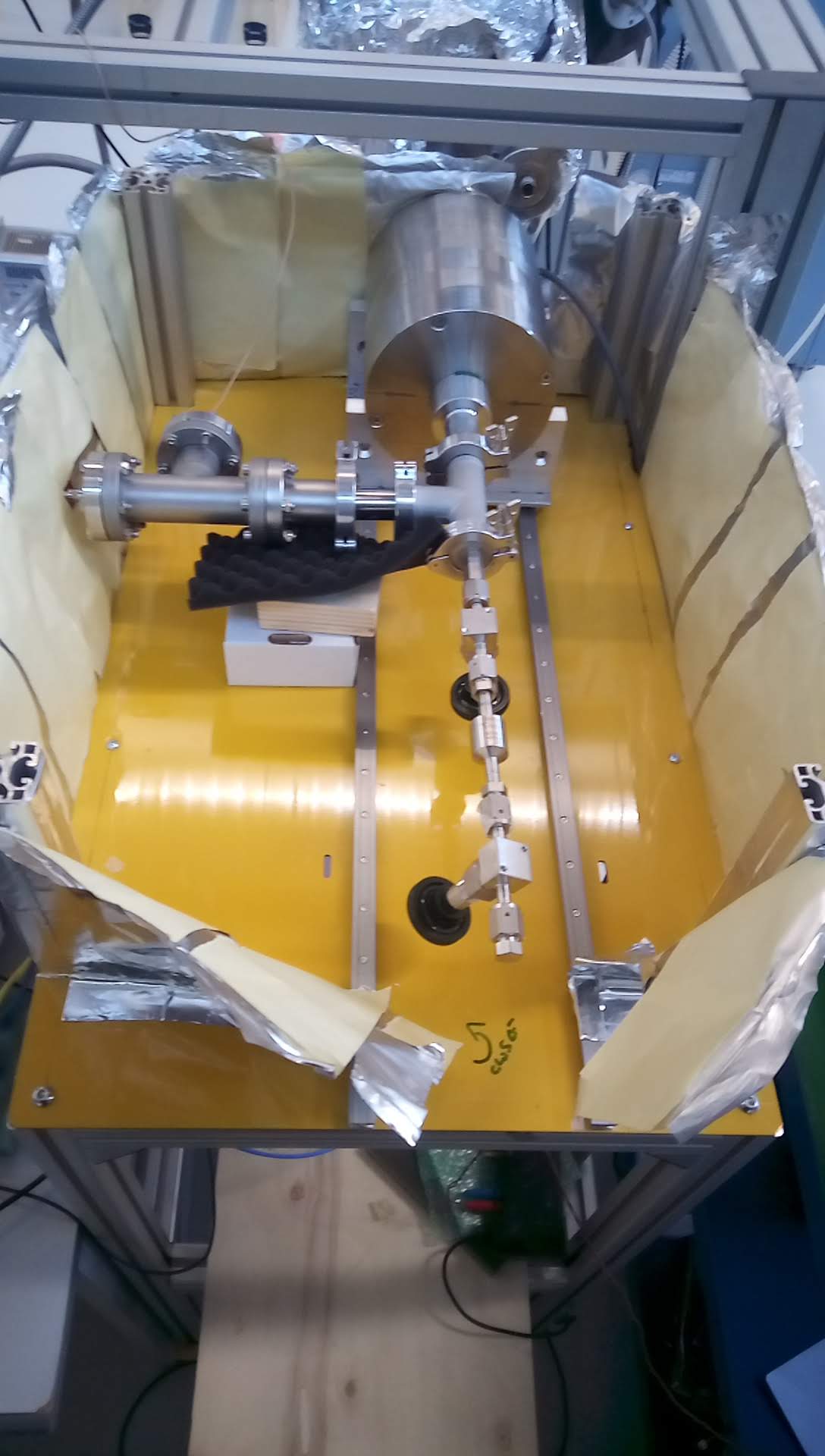}
    \caption{Electron Trap calibration chain. It is possible to see the magnet and the Rubidium source. The calibration chain is hosted in the transverse vacuum vessel together with the HV cable feed-trough. The frequency synthesizer, frequency doubler and external attenuator are hided behind the aluminium foil on the left.}
    \label{fig:calibration_top}
\end{figure}

Between the down-converter and the RF feed-trough flange, a diplexer \emph{DPX-9516} is installed (with the low frequency port closed at $50 ~\Omega$) as high pass filter and to prevent reflections (because of the good scattering parameters of the diplexer) between the two amplification stages. After the down-conversion the signal is propagated to an RTO-6 oscilloscope that performs a trigger and saves the raw data. The dimensions of the Electron Trap waveguide are such that the cut frequency of the waveguide is $f_{cut}=4.0~GHz$, so only signals with higher frequency can be collected and measured. To emulate the signal of the electron, a calibration chain for the electron trap has been implemented. The calibration chain is composed by a frequency synthesizer \emph{ADF-5356SD1Z-UG-1087} capable to provide a single output signal of about $P_{synt}\sim0~dBm$ and a maximum frequency of $13.6~GHz$, this signal must be propagated through the electron trap with a frequency of about $26~GHz$ and a power of about $2~fW$. Firstly the frequency of the signal is doubled to the desired frequency by means of a frequency doubler \emph{MMD-1030H} then the signal is attenuated with an attenuator of variable intensity $ATT_{att}$ to the desired power. After this attenuation stage, the signal is injected in the cryogenic system by means of coaxial cables with an overall coaxial cable propagation attenuation of about $\sim3~dB$. The calibration signal is propagated with a coaxial cable to the Tee flange located between the electron trap and the Rubidium source. The vacuum tube containing the calibration coaxial cable, together with the HV cables of the Electron Trap electrodes is shown in Figure (\ref{fig:calibration_top}).

%calibrazione segnale entra dentro gli elettrodi

\newpage

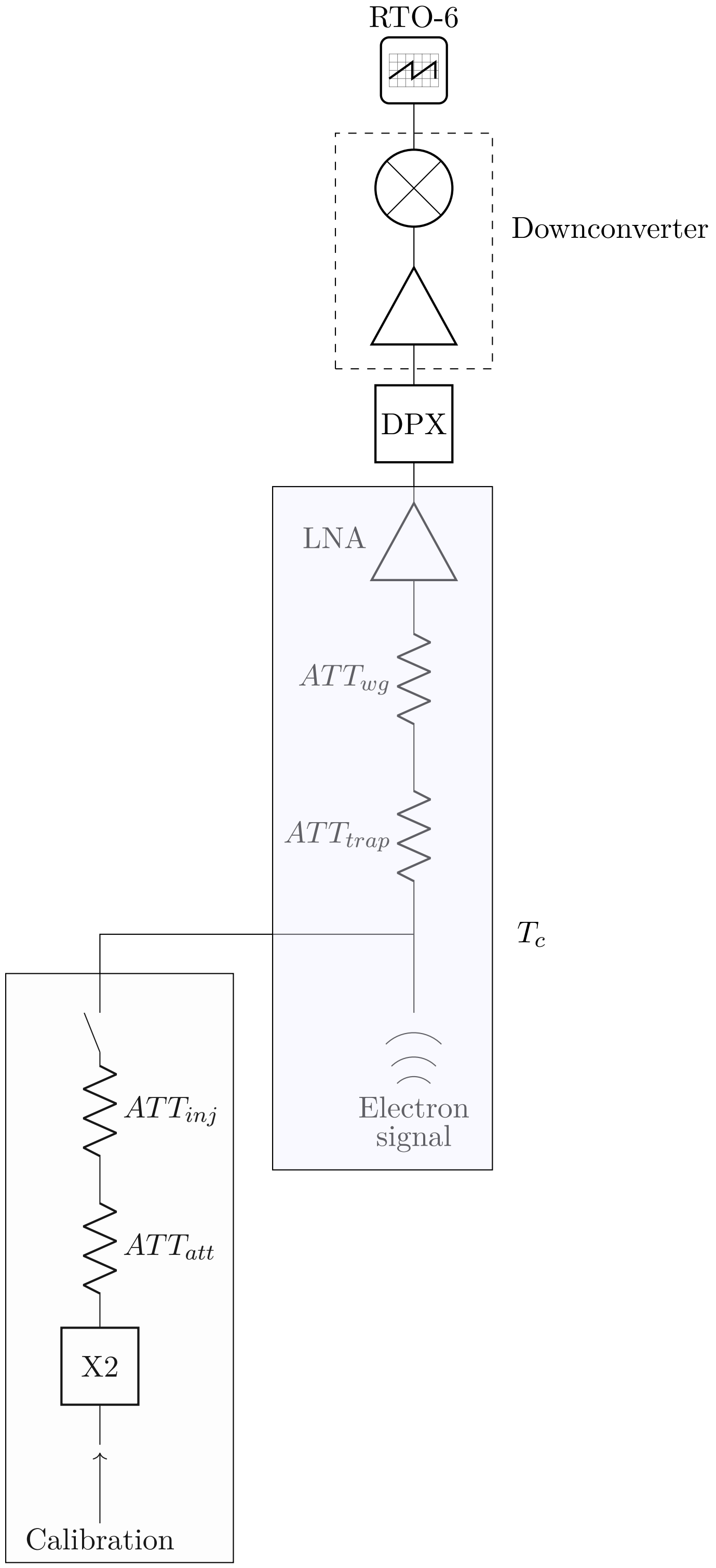
\begin{figure}[h!]
    \centering
    \begin{circuitikz}

    \node[] (origin) at (1,0){};
    \draw(1,-1) to [short,-](1,0);
    \draw(1,0) to [short,-](-3,0)--(-3,-1);
    \draw(-3,-1.5) to (-3.2,-1);
    \draw(-3,-1.5) to [resistor={$ATT_{inj}$}](-3,-3) 
                to [resistor={$ATT_{att}$}](-3,-5);
    \node[twoportshape,t=X2] at(-3,-5.5){};
    \draw(-3,-6)to(-3,-6.5);
    \draw[->](-3,-7.5)--(-3,-6.6);
    \node[] at (-3,-7.7){Calibration};
    \filldraw[fill=black!5](-4.2,-8)rectangle(-1.3,-0.5)[fill opacity=0.1];

    \draw[](1.35,-1.4) arc (45:135:0.5);
    \draw[](1.28,-1.68) arc (45:135:0.4);
    \draw[](1.21,-1.9) arc (45:135:0.3);
    \node[] at (1,-2.2){Electron};
    \node[] at (1,-2.6){signal};
    
    \draw(1,0) to [resistor={$ATT_{trap}$}](1,2.5) 
                to [resistor={$ATT_{wg}$}](1,4)
                to [amp={LNA}](1,6);
    \filldraw[fill=blue!5](-0.8,-3)rectangle(2,5.7)[fill opacity=0.4];
    \node[] at (2.5,0){$T_c$};

    \node[twoportshape,t=DPX] at(1,6.5){};
    
    \draw[](1,7) to [amp={}](1,9);
    \node[mixer] at (1,9.5){};
    \draw(1,10) to (1,10.59);
    \draw[dashed](0,7.2)rectangle(2,10.2);
    \node[] at(3.5,9){Downconverter};
    
    \node[oscopeshape] at (1,11){RTO-6};

    \end{circuitikz}
    \caption{Scheme of the Electron Trap electronic chain. On the bottom left the calibration chain is shown. The electron signal is emitted in the cold temperature stage in which the Electron Trap and the cryogenic amplifier are schematized with symbols. After that the diplexer and the down-converter are also depicted.}
    \label{fig:electronic_chain}
\end{figure}

The signal is then injected inside the Electron Trap by propagation in vacuum, giving an additional effective attenuation $ATT_{inj}$ of the signal; in fact, only a portion of the signal is capable of entering the electron trap and is then measured with the readout electronics. The value of the injection attenuation depends on the environmental conditions. It is possible to estimate it by injecting a signal with known power and by measuring it again with the LNA, with this method the attenuation is measured to be $ATT_{inj} \sim 35~dB$. A scheme of the overall electronic chain is shown in Figure (\ref{fig:electronic_chain}).

\newpage

\newpage

\begin{comment}
It is important to note that in principle also the noise is attenuated, however this part of the detector is before the first amplification stage, so 

\begin{equation}
    P=P_{wg}-Att_{trap}-Att_{wg} +G_{tot}
\end{equation}
\end{comment}

\begin{figure}[h!]
    \centering
    \includegraphics[width=1\linewidth]{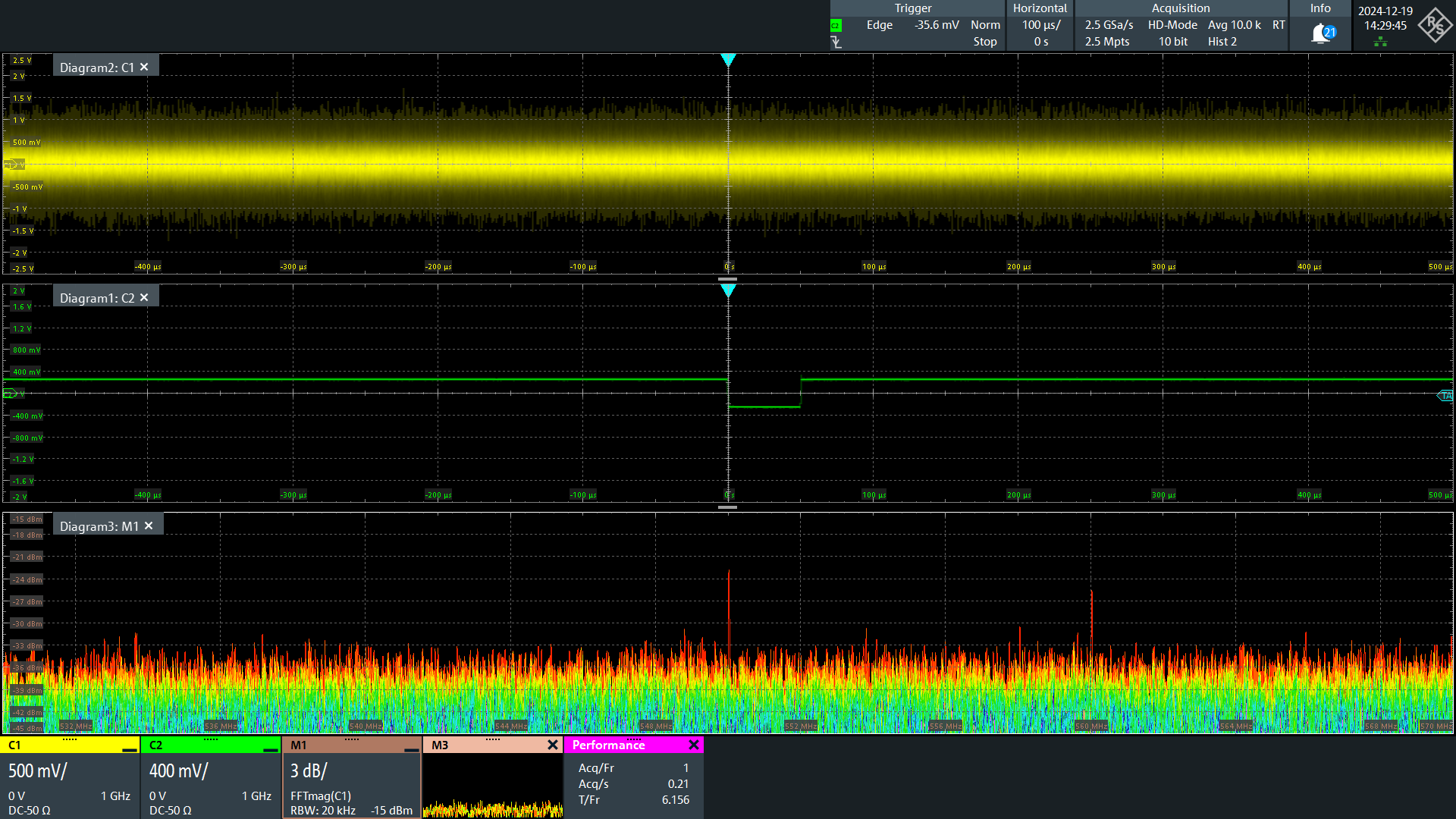}
    \caption{Calibration signal of $SNR=13.4~dB$ at $RBW=25~kHz$. From top to bottom: time series, time pulse of the calibration signal and FFT.}
    \label{fig:calibration_FFT}
\end{figure}

As has been mentioned before, implementing the desired power in the calibration signal is non trivial, however, it is possible obtaining it with an indirect method. This method consists of measuring the SNR in the trap by measuring the physical temperature of the electron trap. In fact, by measuring the temperature in the electron trap and by knowing the total noise figure of the electronic chain, it is possible to estimate the expected SNR at the end of the readout electronic chain. The thermal noise power (in dBm) is

\begin{equation}
    N_t[dBm]=10\log_{10}\left( \frac{k_BT\Delta f}{1~mW} \right)
\end{equation}

starting from this is possible to get the formula of the average noise power of readout electronics chain

\begin{equation}
    N[dBm]=N_t(T,\Delta f)+G_{tot}(f,T)+NF_{tot}(f,T)
\end{equation}

where the dependence of the NF are negligible and the temperature dependence of the total gain depends\footnote{For a maximum 0.2~dB excursion.} by the down-converter temperature.\\
The signal from calibration chain is propagated in the cavity of the Electron Trap with power $P_{wg}$, as shown in Equation (\ref{eq:P_wg_cal}).

\begin{equation}
    P_{wg}[dBm]=P_{synt}-ATT_{doubler}-ATT_{att}-ATT_{inj}
    \label{eq:P_wg_cal}
\end{equation}

where the noise power is $N_t$. At this stage the signal has the power of an electron cyclotron signal, which has another power loss due to the losses in the Electron Trap and the waveguide given by $ATT_{trap}+ATT_{wg}$. At this point, it is possible to measure the SNR of the signal and see if it matches the expected noise power. 

\begin{equation}
\begin{split}
    SNR=P_{wg}-Loss+G_{tot}-N_t-NF_{tot}-G_{tot}=\\
    (P_{wg}-N_t)-Loss-NF_{tot}
    \end{split}
\end{equation}
This measurement can be performed with the electronic chain described in Figure (\ref{fig:chain}) and shown in Figure (\ref{fig:electronic_chain}). An example of calibration signal is shown in Figure (\ref{fig:calibration_FFT}) in which the average of the \emph{Fast Fourier Transform} (FFT) is shown, and the SNR is compatible with the expectations. 

\newpage

The inclusion of electrodes in the electron Trap is considered to be the main source of SNR loss of the electron cyclotron signal. However, there is also another possible SNR loss, because the waveguide, which hosts the electrode system, is matched only on the LNA port, while the Kr port is not completely closed, this could have effects on the reflected signal in the Kr port towards the LNA as described in \cite{P8_ampiezze}.\\

\subsection{Trigger setup}
\label{sec:trigger}
%descrivere trigger RTO
It is now possible to describe the Trigger and DAQ used in the Electron Trap setup. During the electron trap data taking, mainly has been used one instrument, the RTO-6 oscilloscope, that is capable to trigger and save the signal. However, also another trigger option has been identified.

\subsubsection*{Trigger Option A}
\begin{figure}[h!]
    \centering
    \includegraphics[width=0.85\linewidth]{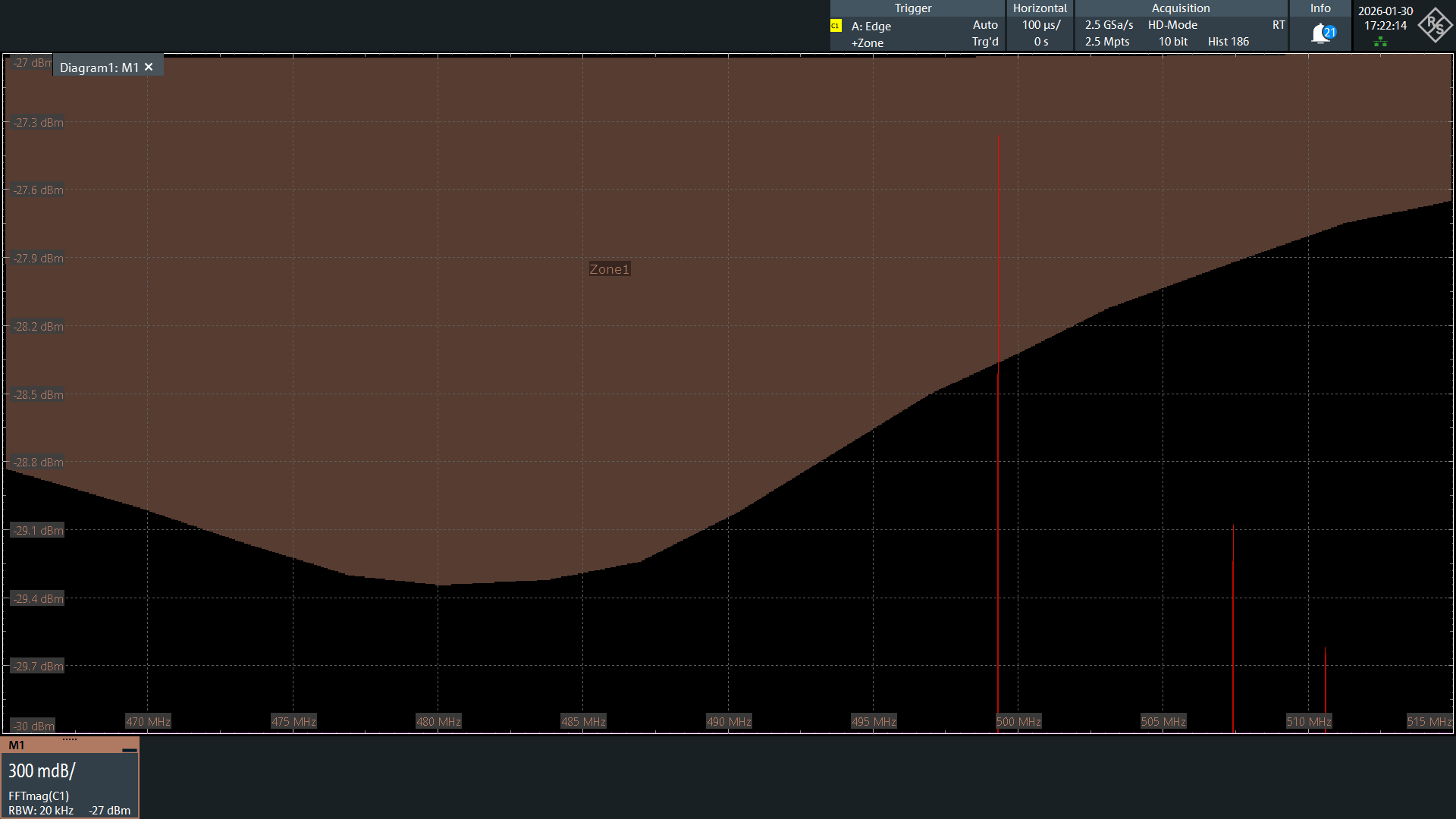}
    \caption{Signal triggered in the frequency domain with the RTO-6 oscilloscope. This is the event 4820 from the RUN94. It is possible to see that the only signal above the trigger limit is the one at about $f_{IF}\sim500~MHz$.}
    \label{fig:triggered_signal}
\end{figure}

The down-converted signal is propagated in the CH1 of the oscilloscope, in which a FFT is performed. This operation requires a certain time, that stops the acquisition of the oscilloscope, generating dead time. This dead time depends by the sampling frequency, the length of the buffer (and so how many FFT it has to do) and the FFT Resolution Band Width (RBW). With this oscilloscope it is then possible to trigger the signal, that is, a peak in the frequency domain, by drawing a \emph{trigger zone} as depicted in Figure (\ref{fig:triggered_signal}), if the signal enters the trigger zone, the signal is triggered and the time-series is saved.

\newpage

\begin{figure}[h!]
    \centering
    \includegraphics[width=0.75\linewidth]{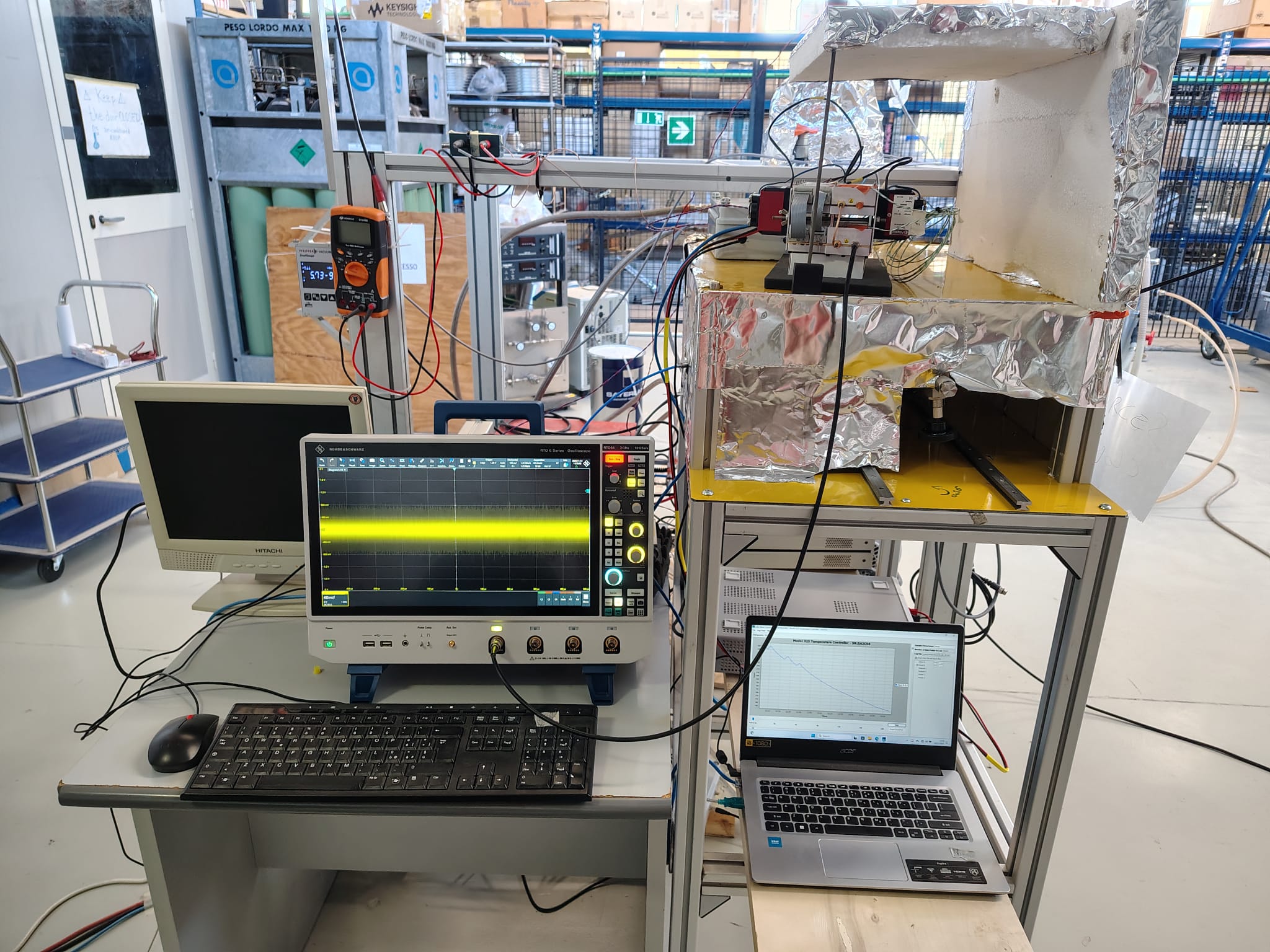}
    \caption{DAQ system of the Electron Trap setup. From left to right: RTO-6 connected to the down-converter (with temperature stabilization system).}
    \label{fig:etrap_daq}
\end{figure}

%parametri tipici oscilloscopio?
The typical parameters of the RTO-6 used during the data taking are at sampling frequency of about $2.5~GSample/s$ used in HD mode (for a total of  12 bit ADC) with a bandwidth of 600~MHz and an FFT of $RBW=20~kHz$. The down-converted signal is expected to be in a range between $[350,600]~MHz$.\\
The DAQ setup of the Electron Trap is shown in Figure (\ref{fig:etrap_daq}).

%metodo alternativo con anritsu
\newpage
\subsubsection*{Trigger Option B}
An alternative method for triggering the signal has been implemented using the \emph{Real time spectrum analyzer} (RTSA) \emph{Anritsu MS2070}, capable of triggering at $100\%$ of \emph{probability of intercept} (POI) signals of at least $30~\mu s$. After the down-conversion, the signal is split in two, a first part goes in the RTSA for triggering, while the other one goes to the RTO that is used to save the CH1 time series. After a trigger out signal is sent to the trigger-in port of the RTO and the data are saved with a total time window of $60~ms$.\\
This trigger option has the characteristic of saturating quickly when the trigger level is set close to the expected electron signal power. In conclusion, two trigger possibilities have been identified; however, only the RTO-6 solution has been used during the data taking because of the reliability of the time window of the saved data and of the trigger capabilities even at the cost of trigger efficiency due to dead time.\\

\newpage
\section{Electron Trap Simulation}\label{sec:montecarlo}

\begin{figure}[h!]
    \centering
    \includegraphics[width=0.7\linewidth]{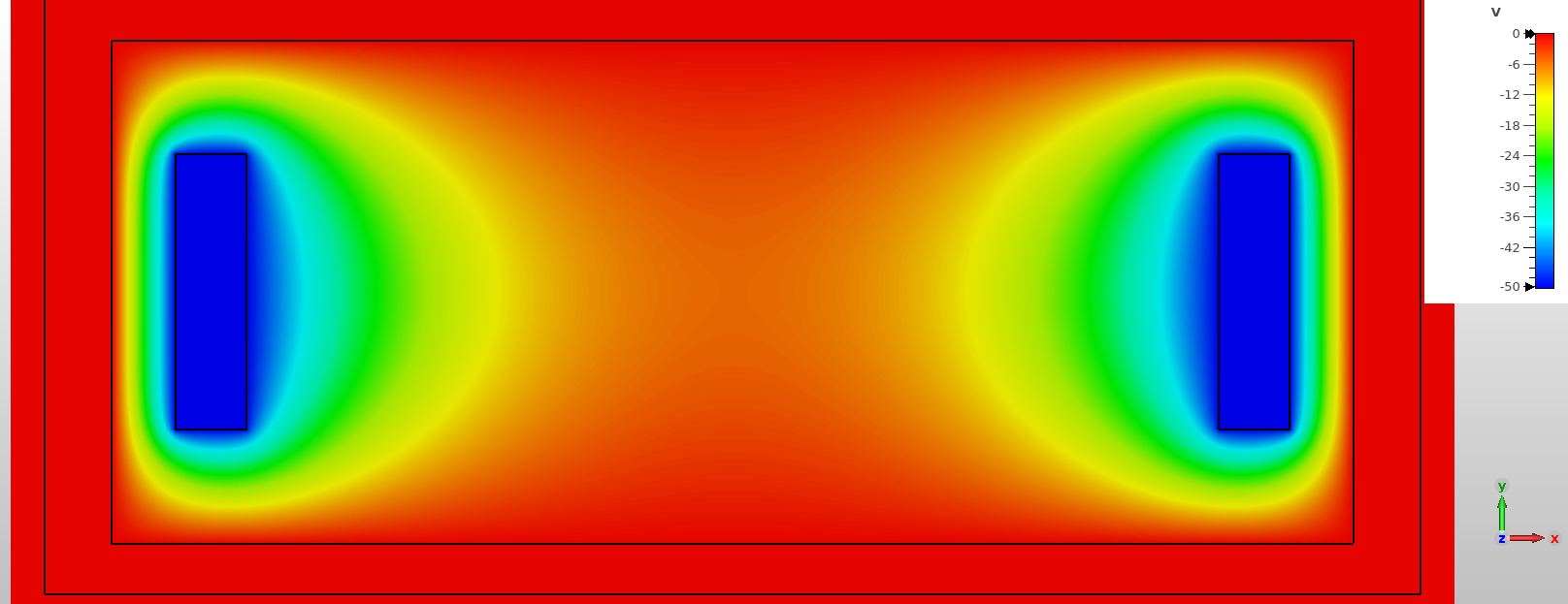}
    \caption{Potential map of the Electron Trap with $V_b=50~V$. The magnetic field is oriented transverse the electrodes such that the electrons are trapped in a cyclotron bouncing motion.}
    \label{fig:V_map50}
\end{figure}

In the previous Section the Electron Trap setup has been described in technical details, this setup has been designed for testing RF spectroscopy measurement for the PTOLEMY RF tracker. In this Section the test that are possible to do with the Electron Trap are described, with the support of a simulation tool. The simulations have been done in two main cases: the case of ideal Uniform magnetic field of and in the case of realistic magnetic field map obtained in Section (\ref{sec:etrap}) by means of simulation. The goal of this comparison is to understand, from the Uniform magnetic field case, which are the measurements that are possible to do in the Electron Trap and what we could expect in the PTOLEMY RF tracker (that is going to be likely similar to the Uniform magnetic field case), while from the realistic magnetic field map simulations it is possible to understand what we could expect from the Electron Trap data and the effect of the magnetic field systematics. The simulation studies are done with the \emph{LORENTZ4} C++ simulation package, which has been developed by the PTOLEMY collaboration with the purposes of simulating electron trajectory and RF emission in magnetic and electric fields. The simulation studies are carried out mainly with the parameter of $V_b=-50~V$, which is the value used in the typical data taking of the Electron Trap. From now on, in this work the potentials are expressed in absolute value for simplicity. A simulation of the Potential map with $V_b=50~V$ is shown in Figure (\ref{fig:V_map50}).

The main measurement under study in this work is the reconstruction of the electron kinematics quantities $K$ and $K_{\parallel}$ from the RF pattern, this procedure is described in detail in Section (\ref{sec:kin_reconstruction}). As first step, the \emph{Potential well effect} that affects the goodness of the energy estimator is described, and the results of the simulations are given.

\newpage
\subsubsection*{The Potential well effect}
%effetto buca

\begin{figure}[h!]
    \centering
    \hspace*{-1cm}
    \includegraphics[width=1.2\linewidth]{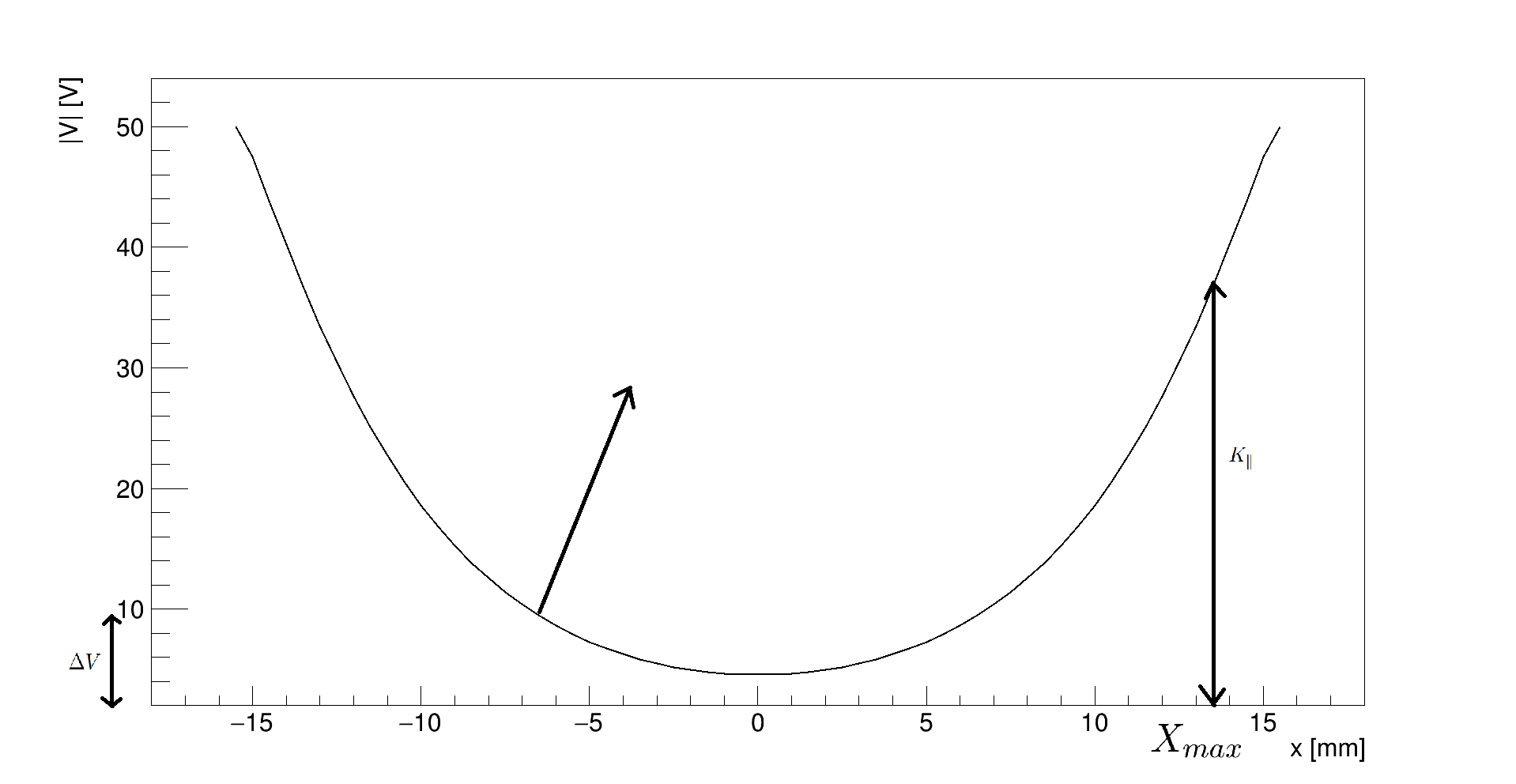}
    \caption{Potential well in the Electron Trap at $y=0$.}
    \label{fig:potential_well}
\end{figure}

Taking as an example the potential map of Figure (\ref{fig:V_map50}), an electron emitted in the center with an initial $K_{\parallel}^0$ will move in bouncing motion along the X axis up to a maximum (minimum) $x=\pm|X_{max}|$. This value depends on the pitch angle and on the shape of the potential well at which the electron is bouncing. In the Electron Trap setup, the electrons are randomly emitted in the volume from a gaseous source, then the electrons are emitted at different starting potential $V^i(x,y)$ that depends on the initial position, as can be seen in Figure (\ref{fig:potential_well}).\\
%questa è la buca di potenziale:1 la buca effettiva è più bassa.

\begin{figure}[h!]
    \centering
    \includegraphics[width=0.8\linewidth]{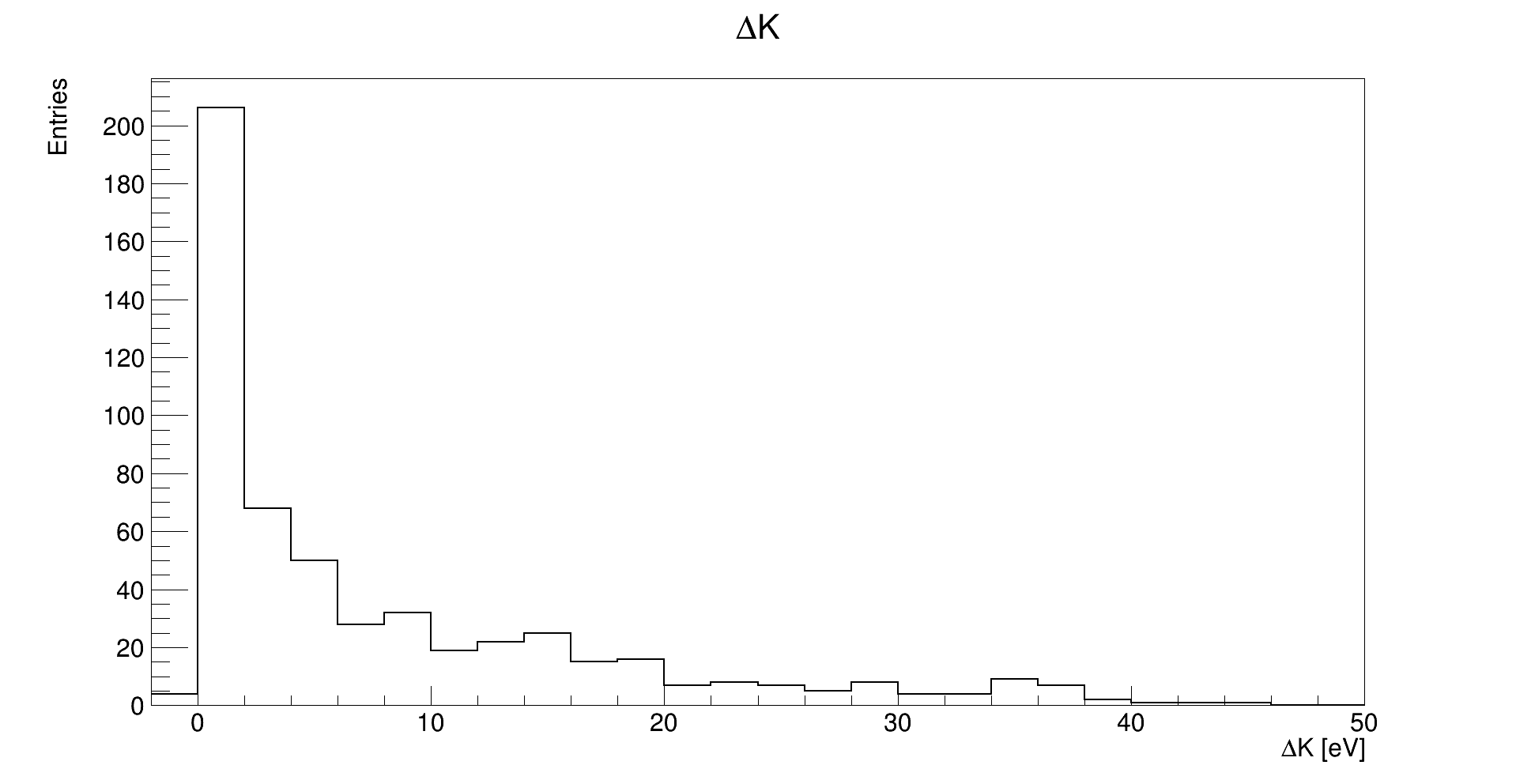}
    \caption{Distribution of the additional kinetic energy gained by the electrons because of the potential well effect ($V_b=50~V$).}
    \label{fig:DeltaK}
\end{figure}

The initial potential $V^i$ must be compared to the lowest reference potential $V_0$ during motion, in fact each electron will have a different additional kinetic energy $\Delta V=|V^i-V_0|$ which is different event by event.\\
The potential well effect is the distortion of the electron kinetic energy spectrum, which is mono-chromatic in the case of Krypton. Considering that the expected carrier frequency of the RF pattern of an electron in bouncing motion depends on the average kinetic energy during the motion, the effective energy measurable from the RF pattern is going to be

\begin{equation}
    K \equiv\langle K(t) \rangle=\langle K_0+U(t) \rangle
\end{equation}

where $K_0=30472.2~eV$ and $U(t)$ is the potential energy. This average could be calculated in the case of the harmonic oscillator; however, the potential lines are not necessarily that of the ideal case. The distribution of the additional kinetic energy $\Delta K$, which is defined as

\begin{equation*}
    \Delta K=\langle K \rangle-K_0
\end{equation*}

is shown in Figure (\ref{fig:DeltaK}) in the case of $V_b=50~V$ and uniform magnetic field.

The distribution shown in Figure (\ref{fig:DeltaK}) has a statistically significant peak at zero, because most of the electrons are likely to be emitted in an almost zero potential region\footnote{In particular, in Figure (\ref{fig:V_map50}) the low potential part of the map is the majority of the fiducial volume.} while the emission of electrons in the high potential region is unlucky and the electrons emitted at that potential also have a reduced angular acceptance for bouncing. The potential well effect is not meant to be present in the PTOLEMY experiment because all electrons will  be emitted at the same reference potential (graphene), while the Electron Trap is affected by this problem.

\subsection{Ideal $\Vec{B}$ field map}\label{sec:uniform_montecarlo}

\begin{figure}[h!]
    \centering
    \includegraphics[width=1\linewidth,trim={9cm 0 0 0},clip]{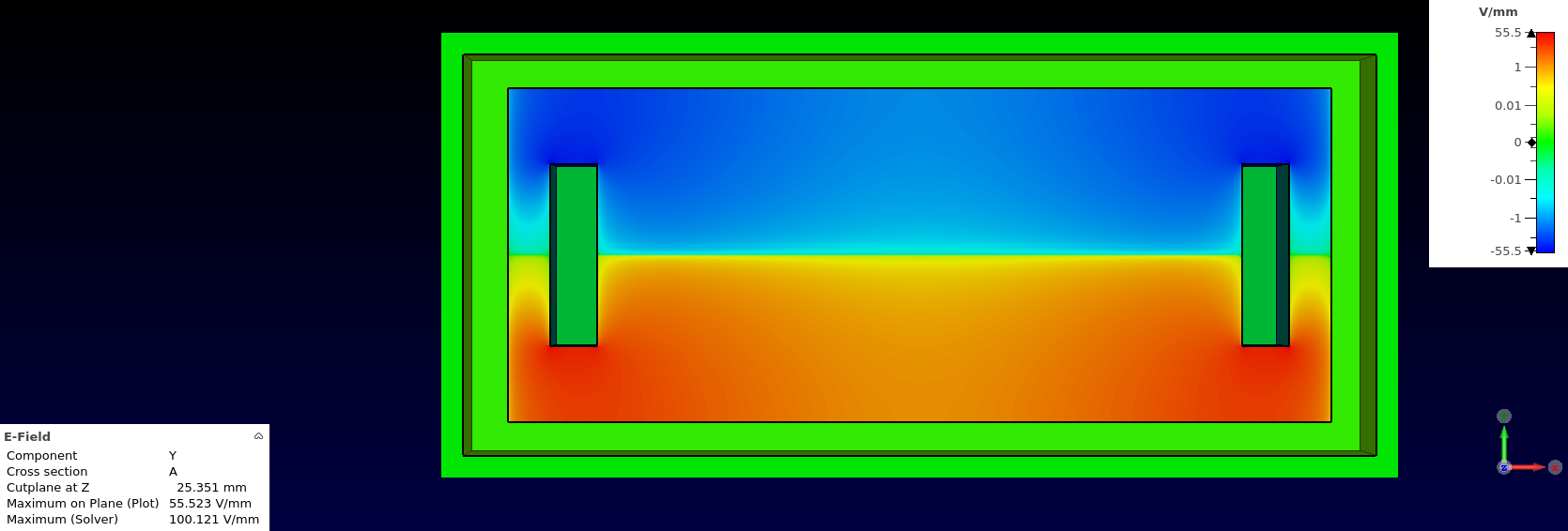}
    \caption{$E_y$ component of the electric field in the Electron Trap with $V_b=50~V$. The electrons emitted on the top part, are drifted toward the screen while the electrons emitted in the bottom part are drifting in the opposite direction. }
    \label{fig:E_field50V}
\end{figure}

In this Section the results of the simulation analysis in the Uniform magnetic field case are reported; as a first step, the basic properties of the electron motion are summarized. Firstly, the electron is moving in cyclotron motion, with a speed component that depends on the longitudinal momentum component. Because the magnetic field is uniform, no Y drift is expected. Regarding the electron drift across Z axis, even though no drift electrodes have been inserted in the system, it is still possible to have a small Z drift. In particular, as shown in Figure (\ref{fig:E_field50V}), the curvature of the potential lines, introduced by the bouncing electrodes, generates a $E_y(x,y)$ component of the electric field that, as shown in Equation (\ref{eq:Edrift_general}), induces a Z drift; this Z drift depends on the $y_0$ of the electron at the emission, and can be forward or backward with respect to the Electron Trap horn. Since there are two opposite directions for the possible electron Z drift, it is possible to study the Doppler shift as is described in Section (\ref{sec:rf_pattern}).

\begin{figure}[h!]
    \centering
    \hspace*{-2cm}% shift slightly to the left
    \includegraphics[width=1.3\linewidth]{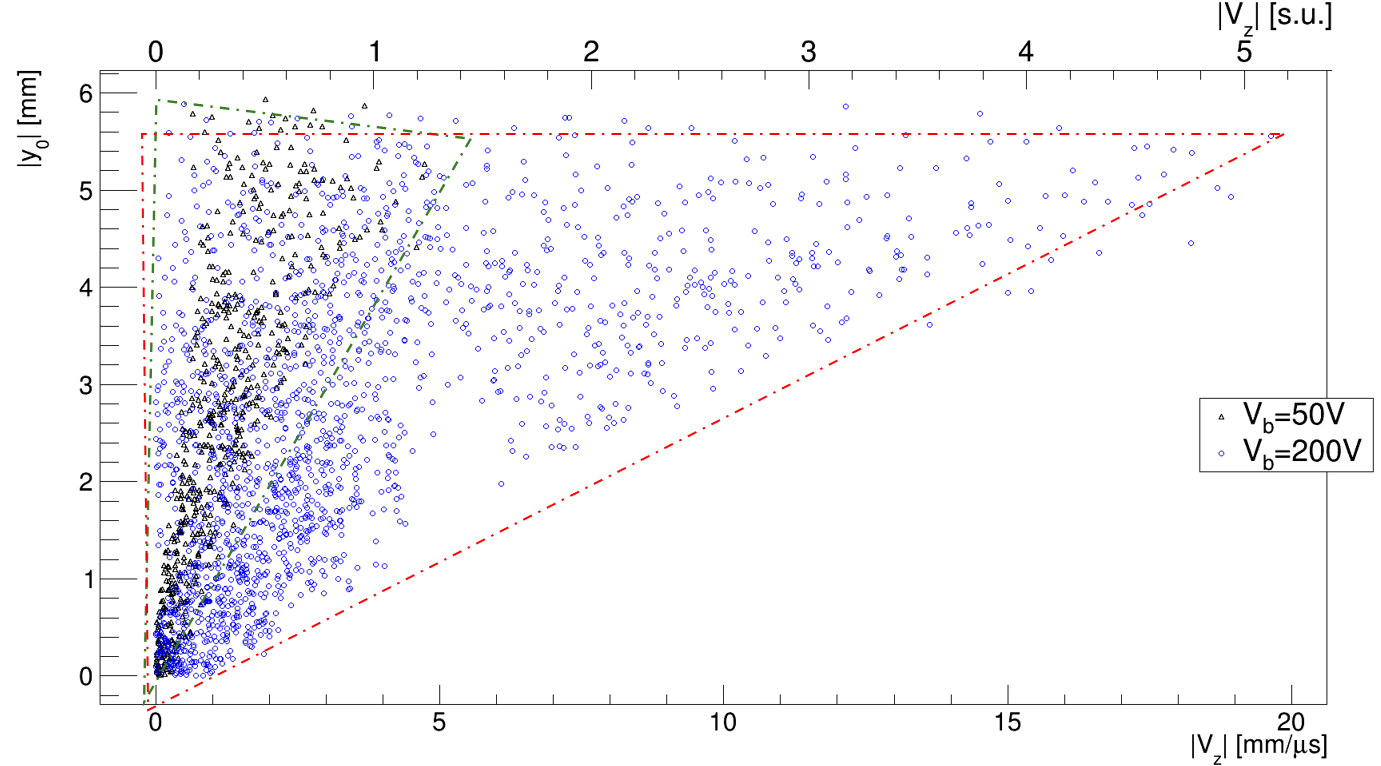}
    \caption{Average Z component of electron GCS velocity $V_z$ vs electron emission $y_0$ position. In blue circle the simulations at $V_b=200~V$ is depicted, while the black triangles are referred to $V_b=50~V$. The slowest electrons are those one drifting at the central Y coordinate and with low $V_b$. It is possible to see that the electrons simulated at $V_b=50~V$ are grouped inside the green shape, while the electrons simulated at $V_b=200~V$ are confined inside the larger red shape. }
    \label{fig:Vz}
\end{figure}

 The intensity of $V_z$ depends on the $y_0$ and the value of $V_b$, if it is higher, a larger Z drift is expected, as shown in Figure (\ref{fig:Vz}) in which the velocity $V_z$ is reported at different $y_0$ with $V_b=50~V$ and $V_b=200~V$. In Figure (\ref{fig:Vz}) a new velocity unit has been used, which is the \emph{Spectroscopy unit} or (s.u.) defined as 

\begin{equation}
    1[s.u.]=\frac{20~cm}{50~\mu s}
    \label{eq:su_def}
\end{equation}

and represent the standard scale units of electron velocity in the Electron Trap and in prospective in the PTOLEMY RF Tracker. In particular, an electron with velocity $V_z<1$ s.u. will give RF tracks that will be longer than the reference time of $50~\mu s$, and so the electrons will lose more energy by RF emission; while if $V_z>1$ s.u. averaging effects along the bouncing motion are present in the measured frequencies.

\newpage

\begin{figure}[h!]
    \centering
    \includegraphics[width=1\linewidth]{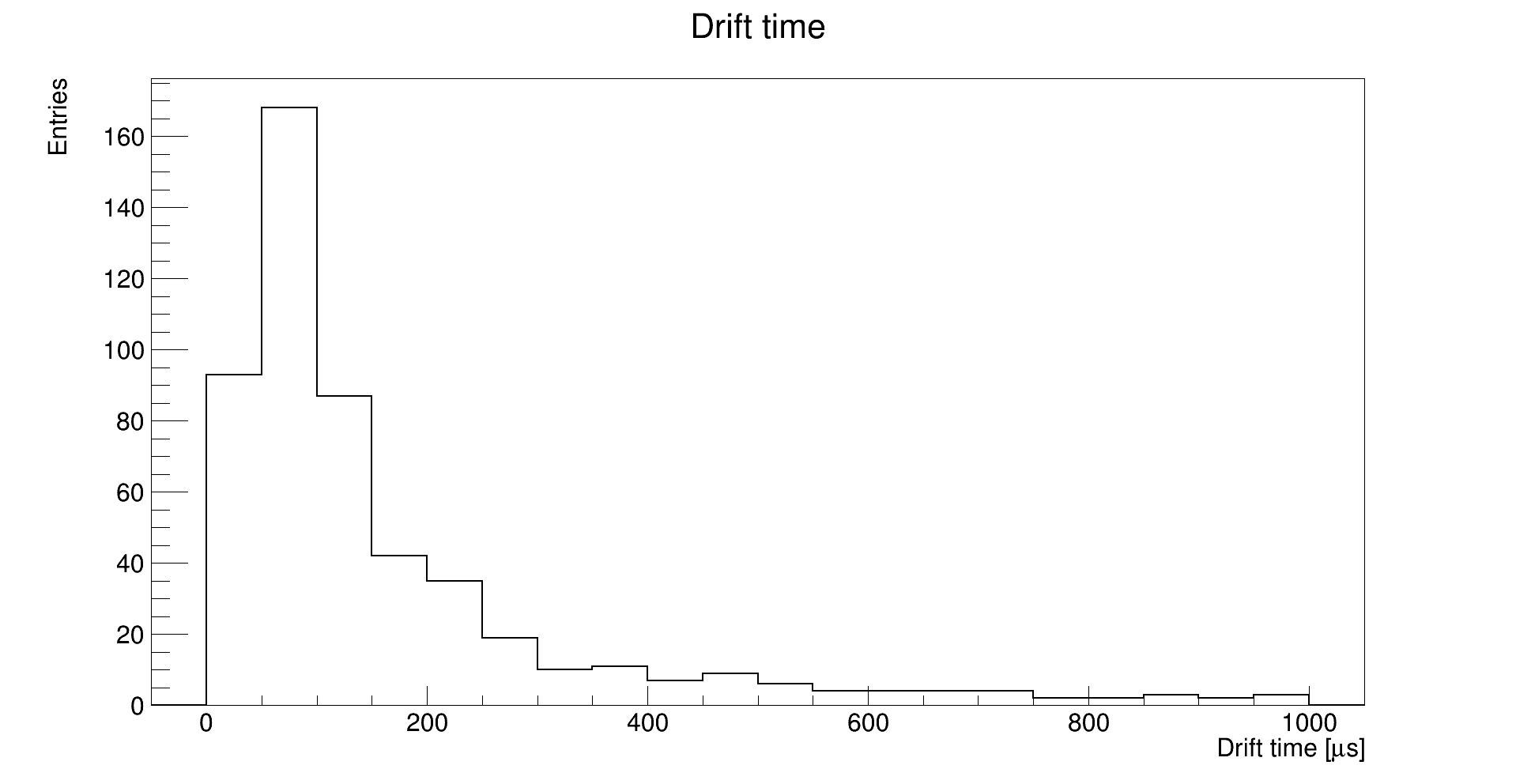}
    \caption{Simulated Electrons drift time in the Electron Trap in the uniform magnetic field case with $V_b=50~V$. It is possible to note that milliseconds tracks are possible.}
    \label{fig:recon_drift_time}
\end{figure}

The plot shown in Figure (\ref{fig:Vz}) is interesting not only for studying Doppler effects on the RF pattern, but also for understanding an important quantity for the RF measurement: the drift time $t_D$. The drift time of the electron, in the uniform magnetic field case, depends on the $V_z$ and the $z_0$ position\footnote{The electrons will experience a small drift time if they are emitted close to the boundaries of the fiducial volume.}. The simulations show that only the $14.6\%$ of the simulated electrons have a drift time smaller than $30~\mu s$, which is the lowest drift time recorded for the Trigger option B of the Electron Trap Setup. The distribution of the drift times of the electrons with a drift time greater than $30~\mu s$ is shown in Figure (\ref{fig:recon_drift_time}).

\newpage

\begin{figure}[h!]
    \centering
    \includegraphics[width=0.9\linewidth]{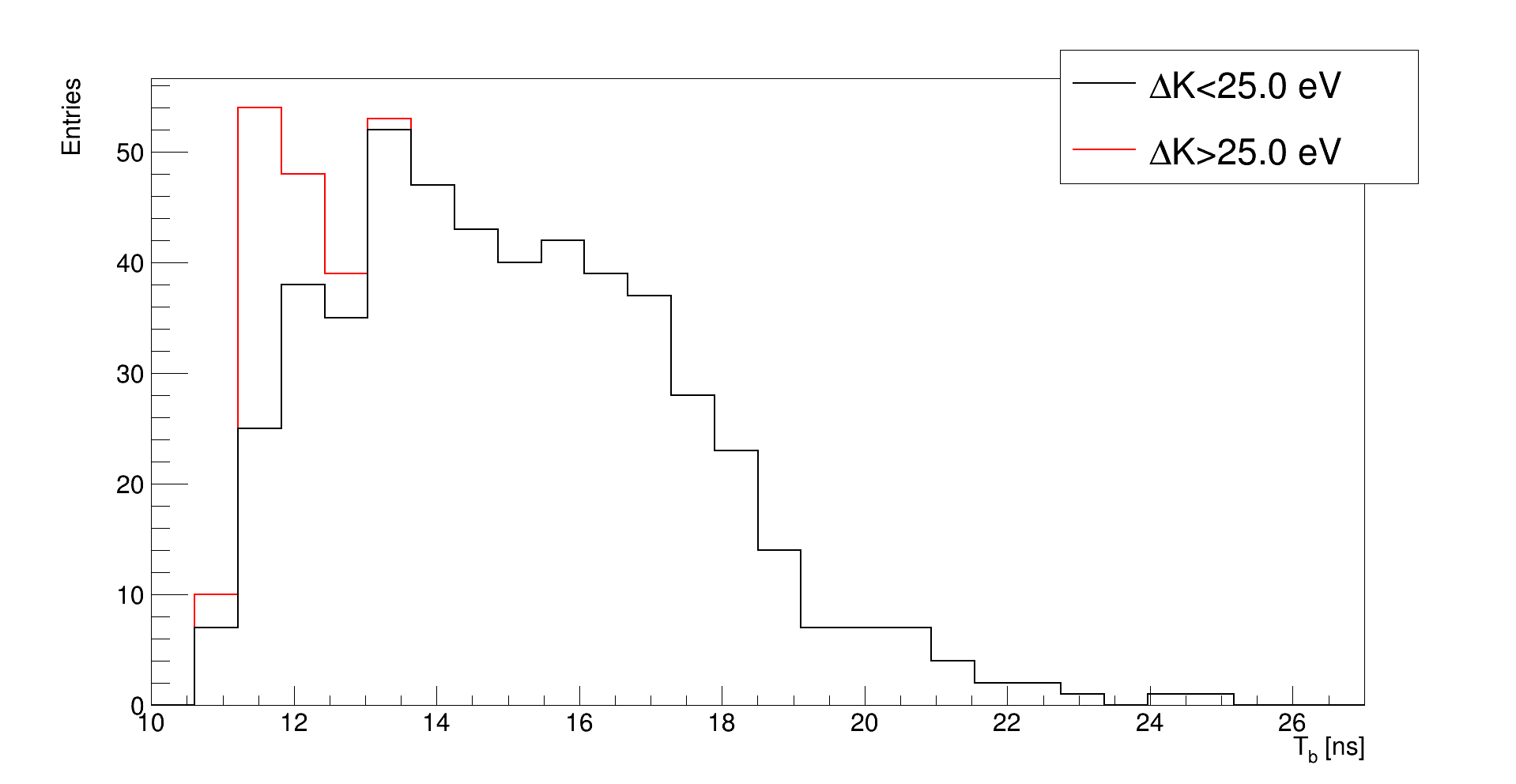}
    \caption{Bouncing period distribution for $V_b=50~V$. Events with high potential well effect are expected in the lowest part of the distribution.}
    \label{fig:Tb}
\end{figure}

\begin{figure}[h!]
    \centering
    \includegraphics[width=0.9\linewidth]{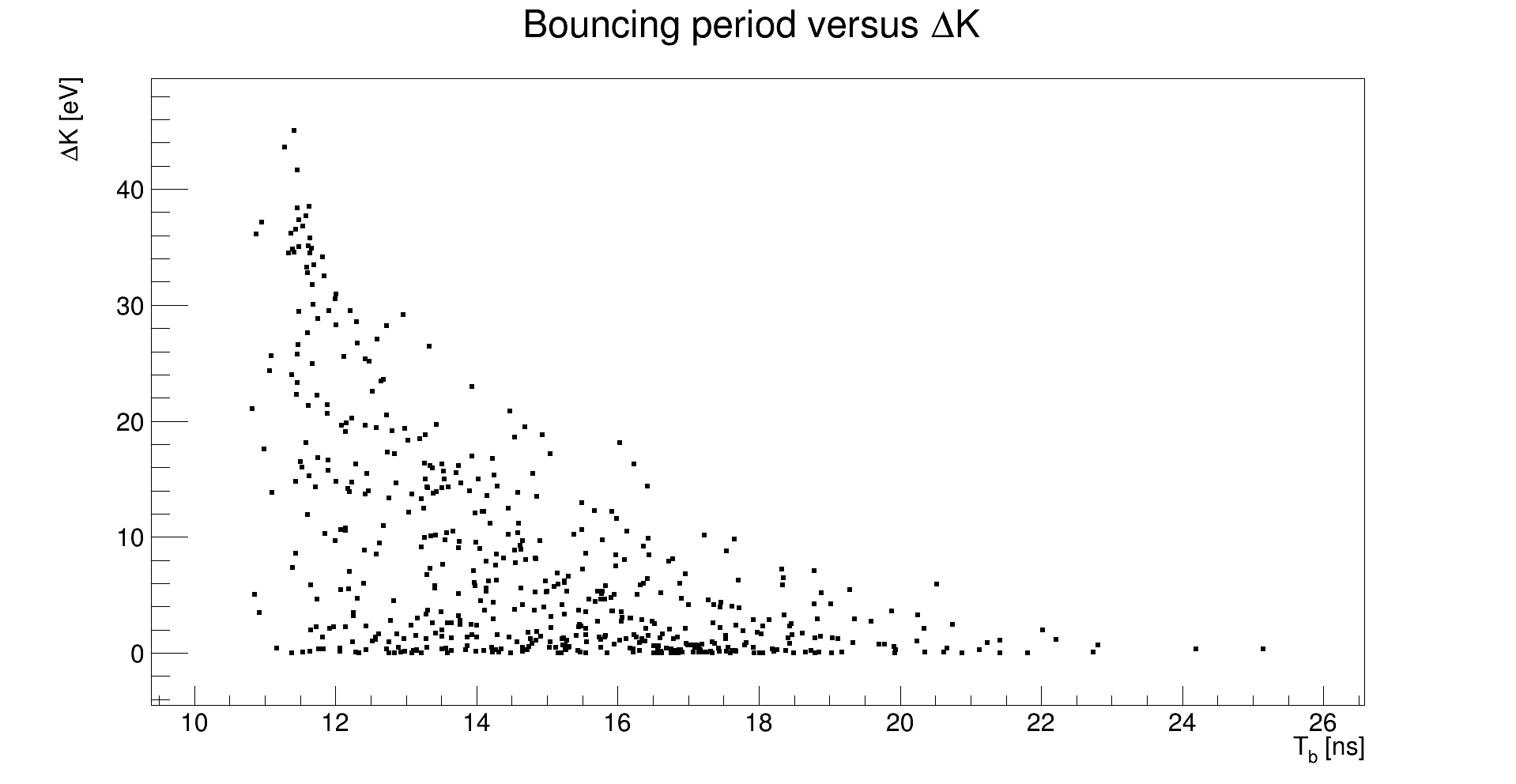}
    \caption{Potential well effect $\Delta K$ vs bouncing period. The electrons with low $T_b$ are at spread along large range of potential well effect. The electrons with large bouncing period show low potential well effect. Therefore, if a large $T_b$ is measured, the electron is expected to have the energy close to $E_0$.}
    \label{fig:Tb_DeltaK}
\end{figure}

The description of the detailed motion, for the RF pattern study, is reported in Section (\ref{sec:rf_pattern}). From the trajectory simulation it is possible to see the distribution, which is shown in Figure (\ref{fig:Tb}), of the bouncing period $T_b$, that is relevant for the reconstruction of the $K_{\parallel}$. The bouncing period distribution can be divided into two parts: the low $T_b$ are typically for electrons with a large $K_{\parallel}$ and so with possible potential well effect, while the high $T_b$ is typically due to low $\Delta \theta$ and $\Delta K$ electrons. Depending on the potential at which the electrons are emitted, $\Delta \theta$ is going to have a different angular acceptance; this means that low $\Delta \theta$ are more likely to be affected by the potential well effect. The electron-waveguide coupling could be studied in principle with electrons with high $\Delta \theta$ and low $\Delta K$ or with low $\Delta \theta$ and high $\Delta K$. In Figure (\ref{fig:Tb_DeltaK}) the correlation between the bouncing period and the potential well effect is shown for each simulated electron. Events with low $\Delta K$ are distributed throughout the range, while electrons with a high potential well effect are restricted in the lower $T_b$ range.

\newpage

\begin{figure}[h!]
    \centering
    \includegraphics[width=1\linewidth]{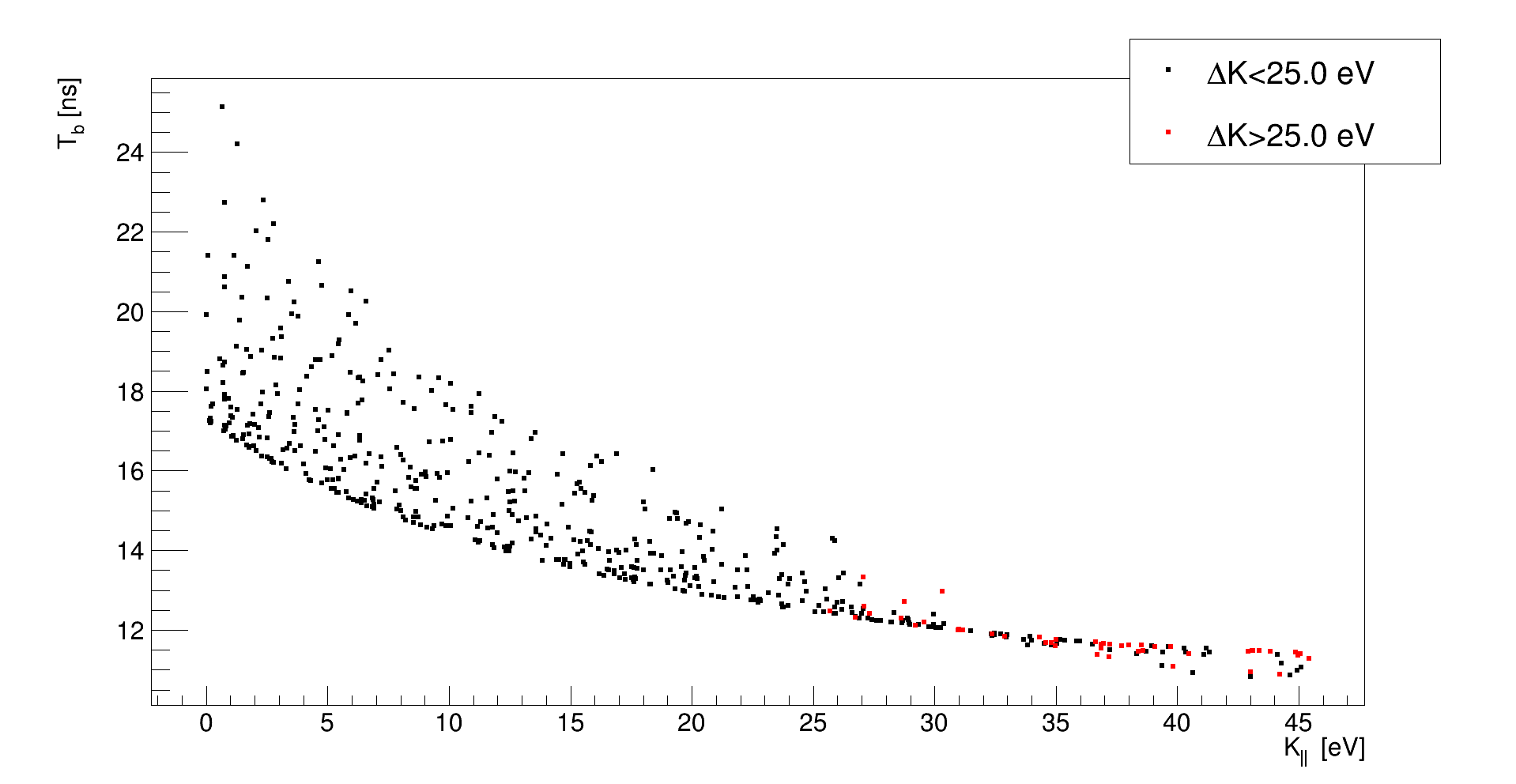}
    \caption{Bouncing period versus maximum longitudinal kinetic energy. The electrons with high potential well effect are depicted in red.}
    \label{fig:Tb_KL_montecarlo}
\end{figure}

It is now possible to study the correlation between the bouncing period and the longitudinal kinetic energy as shown in Figure (\ref{fig:Tb_KL_montecarlo}). This correlation can be deduced from the plot shown in Figure (\ref{fig:Tb_DeltaK}) and is particularly important for the measurement needed in the RF Tracker, as described in detail in Section (\ref{sec:kin_reconstruction}).

\newpage

As can be seen in Figure (\ref{fig:Tb_KL_montecarlo}), electrons with a low potential well effect ($\Delta K<V_b/2$) are along the whole distribution, while electrons with a high potential well effect are forced to have higher $K_{\parallel}$, because they are faster and are located in the lower right part of the plot. In principle a measurement of an high $T_b$ electron is likely be a measurement with small $K_{\parallel}$ and reduced potential well and averaging effects.

%\begin{figure}[h!]
 %   \centering
  %  \includegraphics[width=0.8\linewidth]{plot/fc_vs_Tb.png}
   % \caption{Caption}
    %\label{fig:placeholder}
%\end{figure}

%fare plot a Vb alto di delta theta v s x a due tagli di delta K
%Tb vs Delta K?

\subsection{Simulated $\Vec{B}$ field map}\label{sec:non_uniform_montecarlo}

\begin{figure}[h!]
    \centering
    \includegraphics[width=0.85\linewidth]{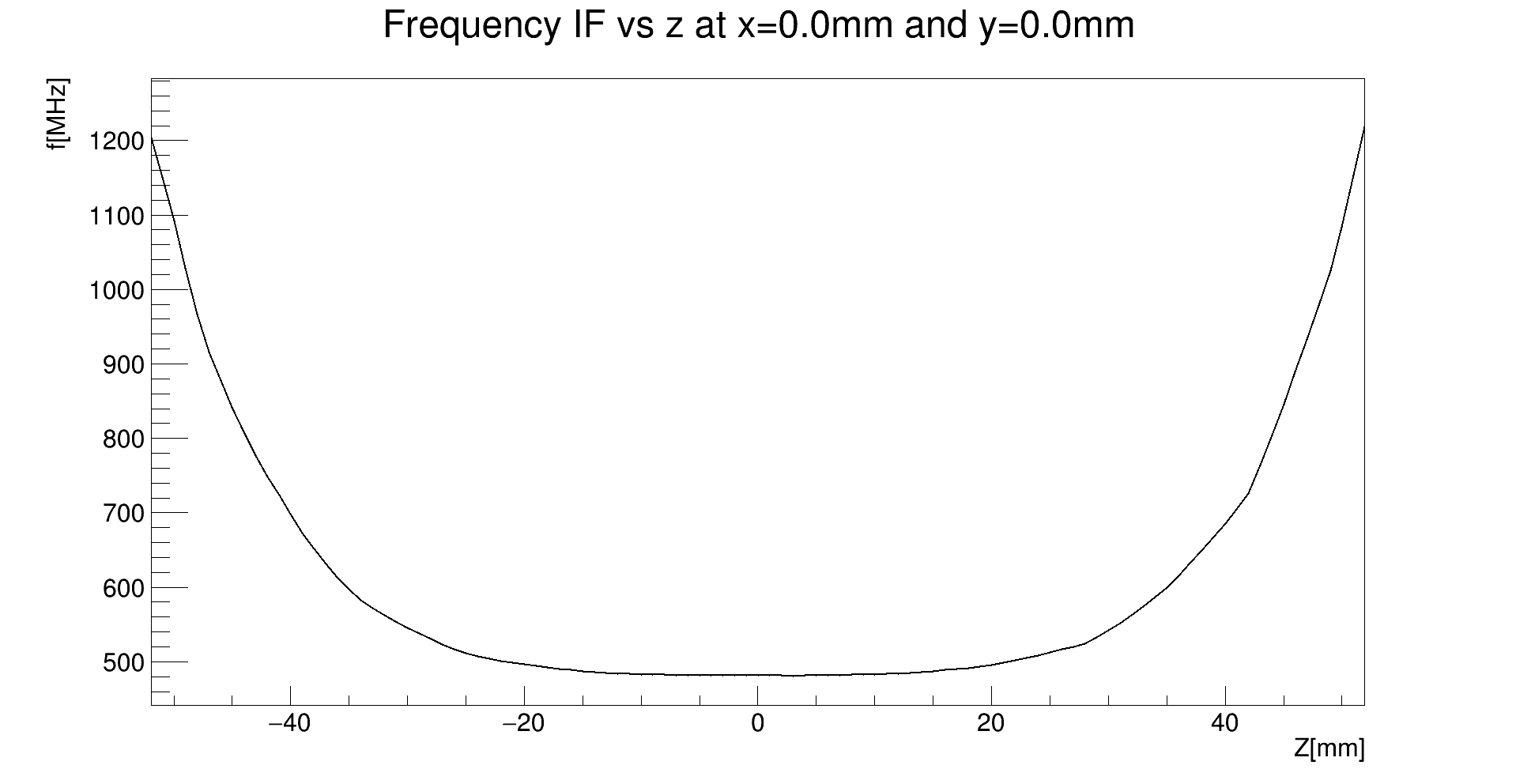}
    \caption{Intermediate frequency of an electron with $K=E_0$ and with $f_{LO}=26.1~GHz$. It is important to note that this frequency is calculated with constant electron energy, so it represent only the variation due to magnetic field.}
    \label{fig:IF_z}
\end{figure}

The analysis of simulated electrons gives very different results when using the simulated magnetic field map, this study is important to guide the Electron Trap data taking. The main difference is the non uniformity of the magnetic field. In particular, as already shown in Figure (\ref{fig:B_z}), this non uniformity of the magnetic field reduces the fiducial volume. For example, the Intermediate frequency IF, which is obtained by Formula (\ref{eq:IF_downconverter}), is strongly affected even by a small variation of the magnetic field. In Figure (\ref{fig:IF_z}) is shown the IF for $f_{LO}=26.1~GHz$ versus the central longitudinal axis of the magnet. It is visible that the region of uniform field is in the range $(-20,+20)~mm$.

\newpage

\begin{figure}[h!]
    \centering
    \includegraphics[width=0.85\linewidth]{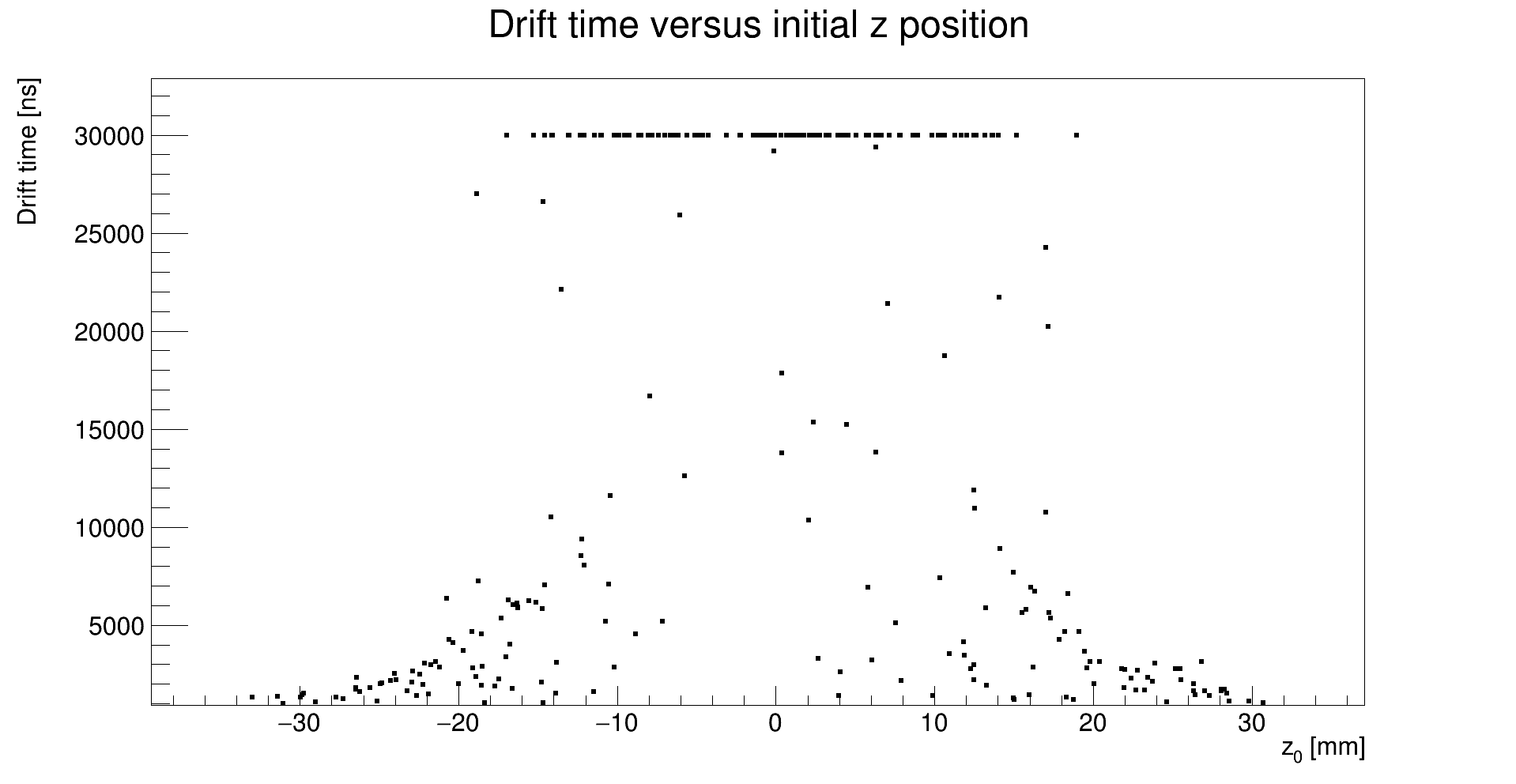}
    \caption{Electron Drift time versus initial Z emission.
    The drift time distribution shows a saturation feature at $30~\mu s$ which is due to the events whose have lasted longer. It is possible to note that outside the fiducial volume (\ref{eq:fiducial_volume}), the electrons are drifting for less time.}
    \label{fig:tD_z0}
\end{figure}

In principle it is possible to measure the electron signal even outside the fiducial volume, or by setting the DAQ at higher frequencies or by changing the down-converter $f_{LO}$, however in those parts of the magnet the magnetic field has a gradient along Z not negligible and so the electrons start to drift across Y as described in Equation (\ref{eq:grad-cf}). This concept is described in Figure (\ref{fig:tD_z0}) where the correlation plot between the electron drift time and the $z_0$ of emission is shown, this plot shows that the electrons outside the fiducial volume are drifting for less time than the electrons in the central part of the magnet.

\begin{figure}[h!]
    \centering
    \includegraphics[width=1\linewidth]{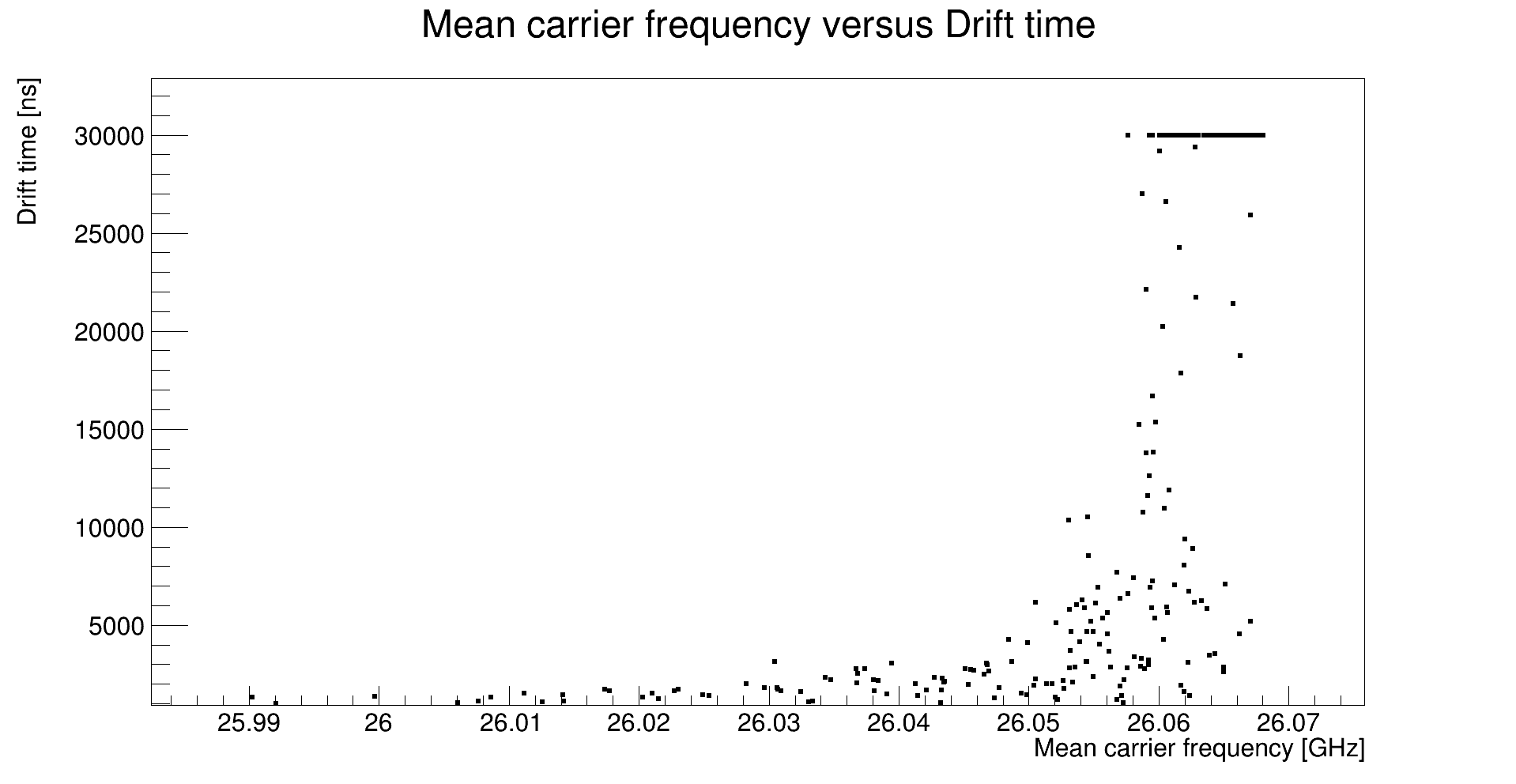}
    \caption{Electron drift time versus mean carrier frequency. The electrons with longest drift time have $f_{RF}=(26.05,26.07)~GHz$. }
    \label{fig:tD_fc}
\end{figure}

\newpage
This information can be used to predict the expected electron frequency, in particular in Figure (\ref{fig:tD_fc}) is shown that electrons with longer drift time, and so with higher trigger probability, are expected to be in a restricted $f_{RF}$ range $f_{RF}=(26.05,26.07)~GHz$.

\begin{figure}[h!]
    \centering
    \includegraphics[width=1\linewidth]{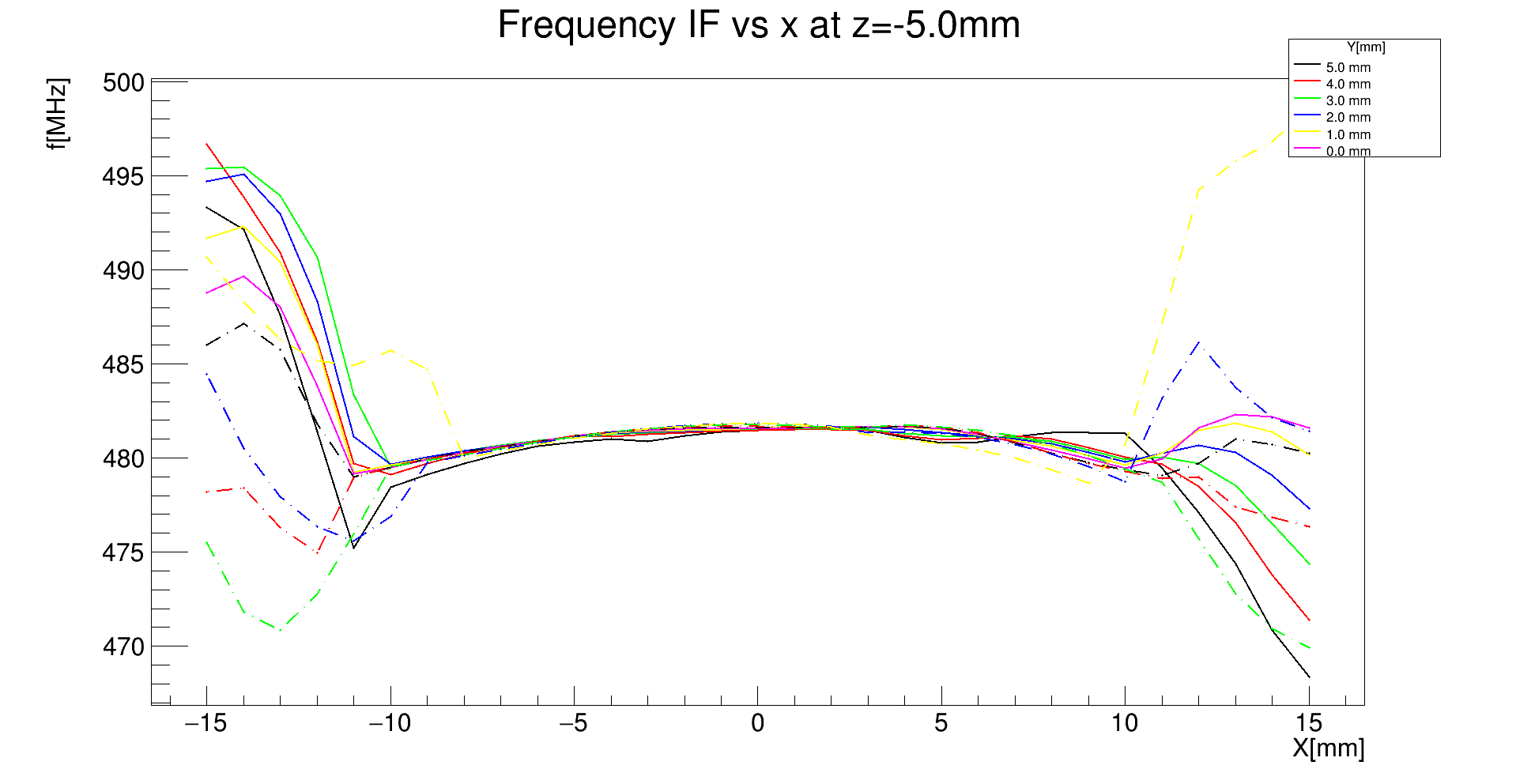}
    \caption{IF frequency versus X at different Y of bouncing. The down-conversion is done with $f_{LO}=26.1~GHz$.}
    \label{fig:IF_x}
\end{figure}

The RF frequency reported in Figure (\ref{fig:tD_fc}) is the average carrier frequency along the electron motion; it is now useful to study in detail the expected variation of the frequency during one bounce. This can be understood by studying the variation of the magnetic field along X direction, as reported in Figure (\ref{fig:B_x}). In Figure (\ref{fig:IF_x}) the intermediate frequency, in the x-y plane, in the central region at the indicative $z=-5~mm$ is shown. As can be noted, electrons that bounce across the region of uniform field $\Vec{B}$ ($\pm 20~mm$) are expected to experience a small frequency variation, while when the electron is bouncing close to the electrodes ($\pm 15~mm$) it experiences a larger frequency modulation, which depends on the starting Y coordinate of the bouncing motion. As in the case of Figure (\ref{fig:IF_z}), this frequency is calculated considering only the variation of the magnetic field, and so at $K=const$.

\newpage
The last physics quantity that is interesting for the measurement of the electron RF signal is the power of the radiation. By exploiting the field simulations, it has been obtained that the mean cyclotron power, which can be calculated by using Equation (\ref{eq:larmor_power}), depending by the kinetic energy and by the pitch angle, is $P=(1.9\pm7.3\times 10^{-4})~fW$. It is now possible to study the electron RF pattern characteristics and how to reconstruct the electron kinematics variables by the RF emission.

\subsection{Electron RF pattern}\label{sec:rf_pattern}

%simulazione RF elettrone

In this work the motion of electrons in magnetic and electric fields has been described, in particular in Section (\ref{sec:e_motion}) we have seen that the motion depends on two factors: the electron kinematic variables, the kinetic energy $K$ and pitch angle $\theta$, and on the combination of Electric and Magnetic fields. In the case of the Electron Trap, and also in the future case of the PTOLEMY RF tracker, the fields are going to be assumed by electric field simulations and magnetic field by the UnivGe and CERN division of magnetic fields mapping. With the assumption that the fields are known, it is possible to measure and reconstruct the kinematic variables of interest from the electron RF pattern, in this Section a detailed description of the electron RF pattern is given, while in Section (\ref{sec:kin_reconstruction}) an overview of the $K$ and $K_{\parallel}$ reconstruction methods is presented. The description of the electron RF pattern is given in the ideal and uniform magnetic field case.\\
As already discussed in this work, the electron is moving in a cyclotron motion with an additional bouncing motion along the X direction. The cyclotron radiation is then a signal (which is going to be called $S$ with a generic amplitude $A$) with an instantaneous frequency that depends on $K(t)=K_{\perp}+K_{\parallel}(t)$ by the Formula (\ref{eq:larmor_frequency}).

\begin{equation}
    S=A \sin(f(t)\cdot t)
\end{equation}

The Fourier spectrum of the signal $S$ is the electron RF pattern.
The modulation of the frequency $f$ depends on the potential well and on $K_{\parallel}$. The RF pattern is characterized by the carrier frequency $f_c$, which is already mentioned in the Potential well Section, which corresponds to the averaged kinetic energy. The frequency $f(t)$ is higher during the bounce, when the electron is at the top of the potential well, this produces another Fourier peak, called \emph{a secondary peak}, which depends on $K_{\parallel}$ and with frequency $f_2$

\begin{equation}
    f_2=\frac{1}{2\pi}\frac{|q|B}{m}\frac{1}{(\langle K\rangle-K_{\parallel})/m+1}
    \label{eq:f2}
\end{equation}

\begin{figure}[h!]
    \centering
    \includegraphics[width=1\linewidth]{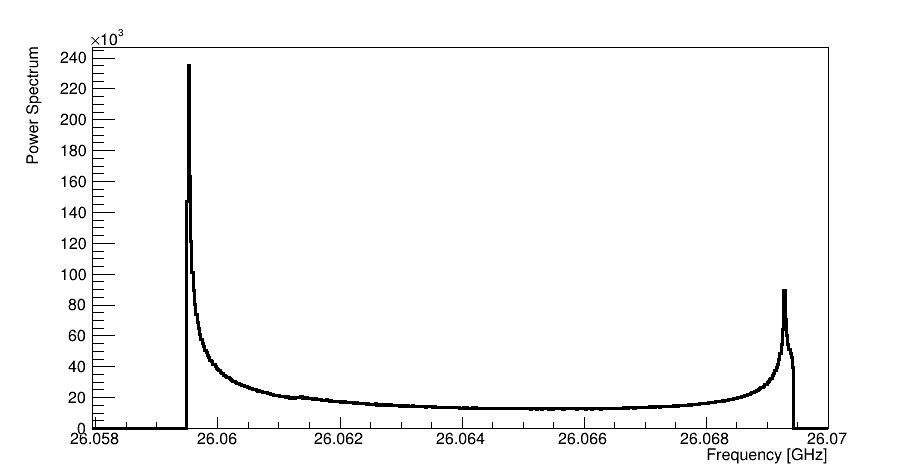}
    \caption{Kinematic spectrum obtained by LORENTZ4 simulation of an electron with $E_0=30472.2~eV$ and $K_{\parallel}=20.8~eV$.}
    \label{fig:kin_spectrum}
\end{figure}

The carrier and secondary peaks of the RF pattern are strictly related to electron kinematics, in fact, from the kinematic spectrum of a simulated electron with frequency $f(t)$, which is shown in Figure (\ref{fig:kin_spectrum}), it is possible to see the presence of the carrier frequency and a smaller secondary peak at higher frequency. In Figure (\ref{fig:kin_spectrum}) the secondary peak is expected to have 
a smaller amplitude because the electrons drift for a shorter time compared to the case where the bouncing motion takes place in the lower part of the potential well. Furthermore, the secondary peak is expected to have a smaller coupling with the $TE_{10}$ mode of the waveguide because the signal is emitted at the sides of the waveguide\footnote{ However, this is not represented in the kinematic spectrum, that is actually just a histogram of the frequency entries.}.

\begin{figure}[h!]
    \centering
    \includegraphics[width=1\linewidth]{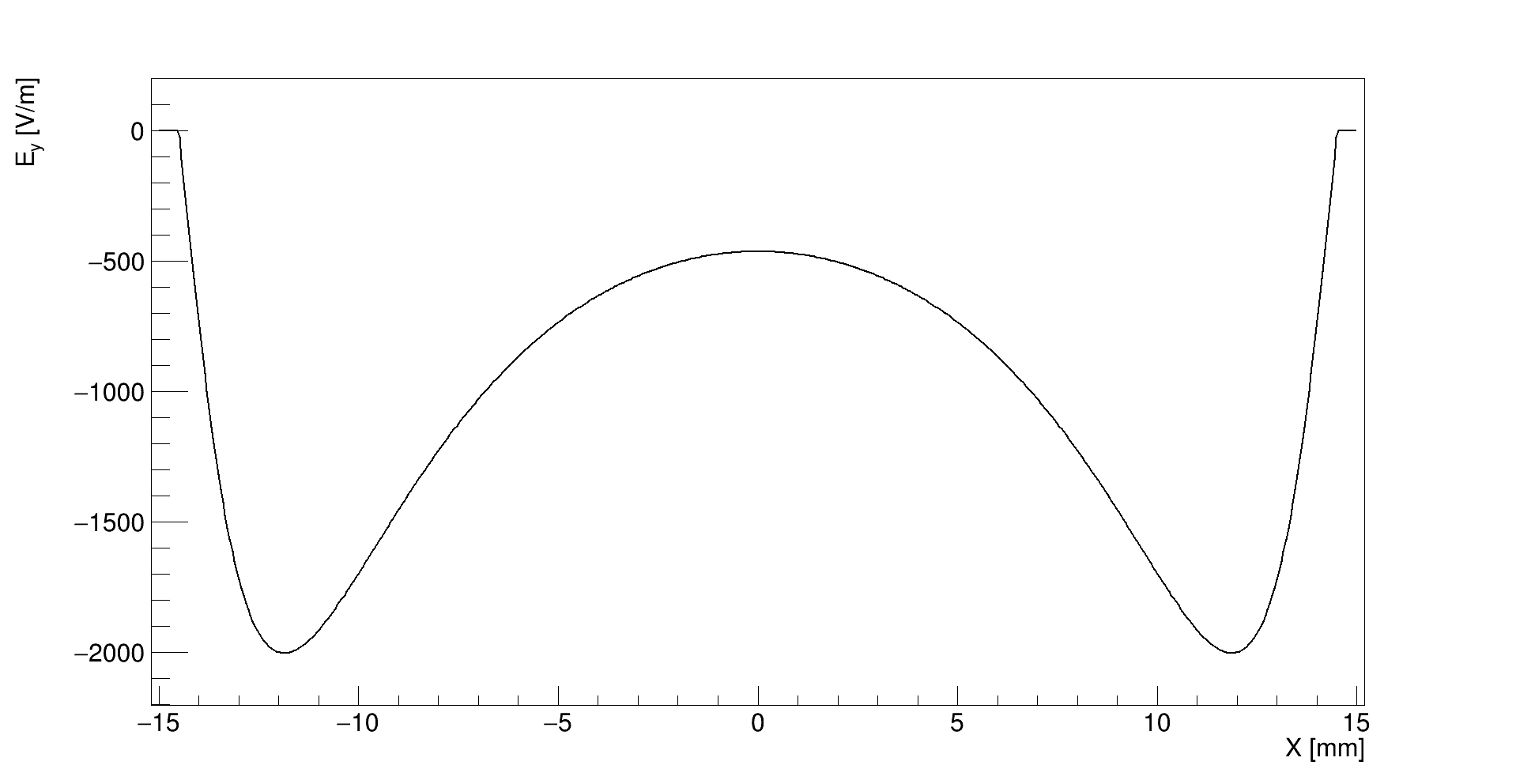}
    \caption{Electric field $E_y$ as function of X. This plot has been obtained with $V_b=50~V$ and at $y=2~mm$. It is possible to see that the $E_y$ is larger close to the electrodes.}
    \label{fig:Ey_x}
\end{figure}

The fact that the frequency is varying with a stable periodicity introduces an electromagnetic effect on the signal: a frequency modulation (FM). This introduces the presence of sidebands on the left (L) and right (R) of the carrier frequency separated by a distance $\Delta f=1/T_b$ that is called \emph{frequency deviation}.\\
It is important to note that the carrier and sidebands are related to electron movement along the X direction, that is, transverse to the readout direction (Z axis). Possible Doppler effect is expected to be present in the secondary peak, that is the one emitted when the electron is also moving along Z.
As described in Figure (\ref{fig:Vz}) the drift velocity of the electron GCS along Z depends on the $V_b$ and Y in which the electron is moving. Another contribution of electron Z drift is the $X_{max}$ of the oscillation, which depends on the $K_{\parallel}$, in fact, the $E_y$ depends on X as shown in Figure (\ref{fig:Ey_x}). Therefore only the electrons that have enough $K_{\parallel}$ to enter in the high field region are going to have a Doppler effect on the secondary peak\footnote{Assuming $B=B_x$ as in Section (\ref{sec:e_motion}).}.

\begin{figure}[h!]
    \centering
    \includegraphics[width=0.8\linewidth]{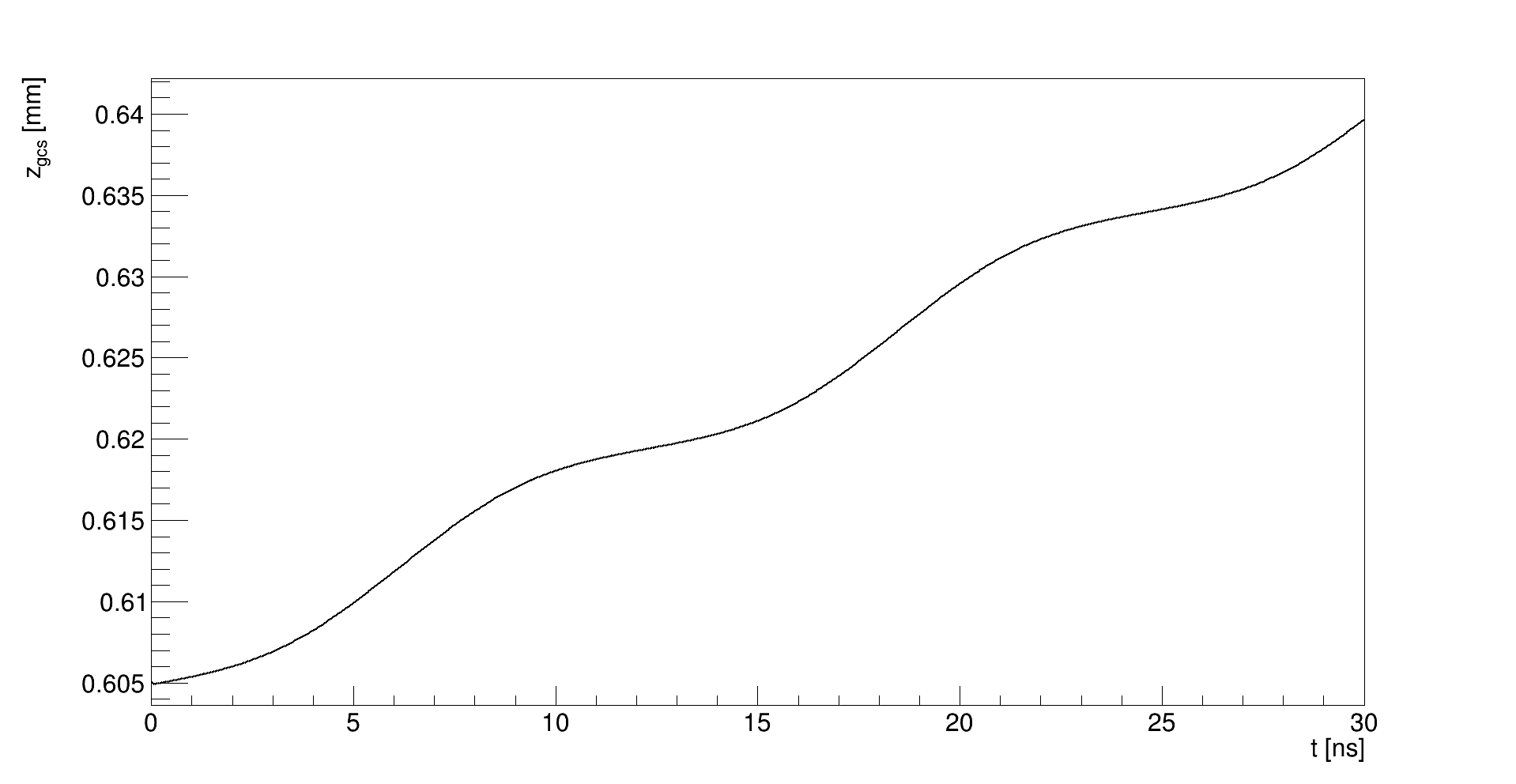}
    \vspace{1cm} 
    \includegraphics[width=0.8\linewidth]{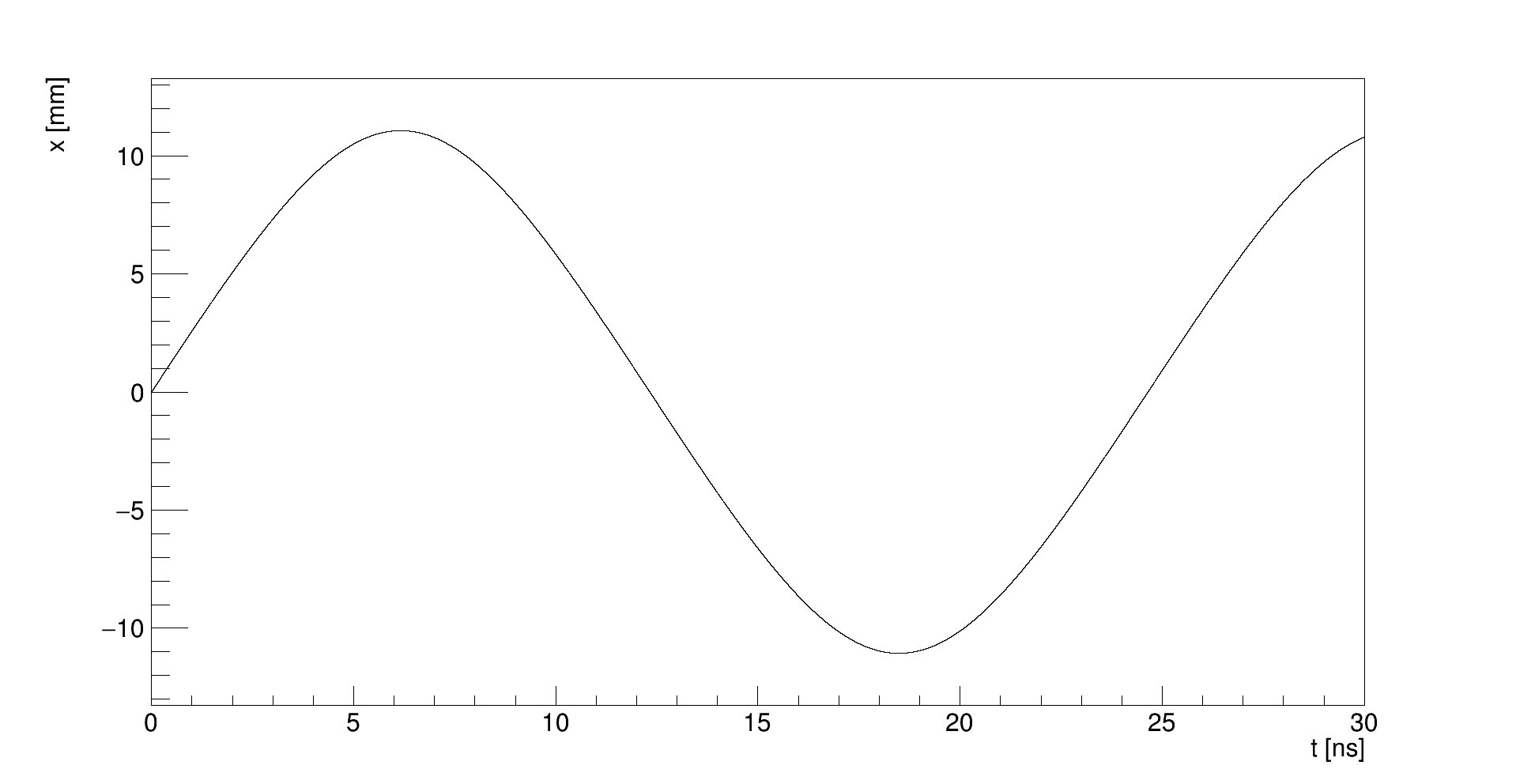}
    \caption{Electron Z (top figure) and X (bottom figure) coordinate computed in GCS as function of the time. It is possible to note that the higher z drift correspond to electrons bouncing close by electrodes.  Initial conditions: $E_0=30472.2~eV$, $\theta =88.5^{\circ}$, $V_b=50~V$ and $Y_0=2~mm$.}
    \label{fig:z_gcs}
\end{figure}

\newpage
The shape of the electric fields, and in particular the Y component, is responsible for the electron Z drift, in particular in Figure (\ref{fig:z_gcs}) the time dependence of the Z component of an electron GCS trajectory together and the X as function of time, is shown, respectively.

\begin{figure}[h!]
    \centering
    \includegraphics[width=1\linewidth]{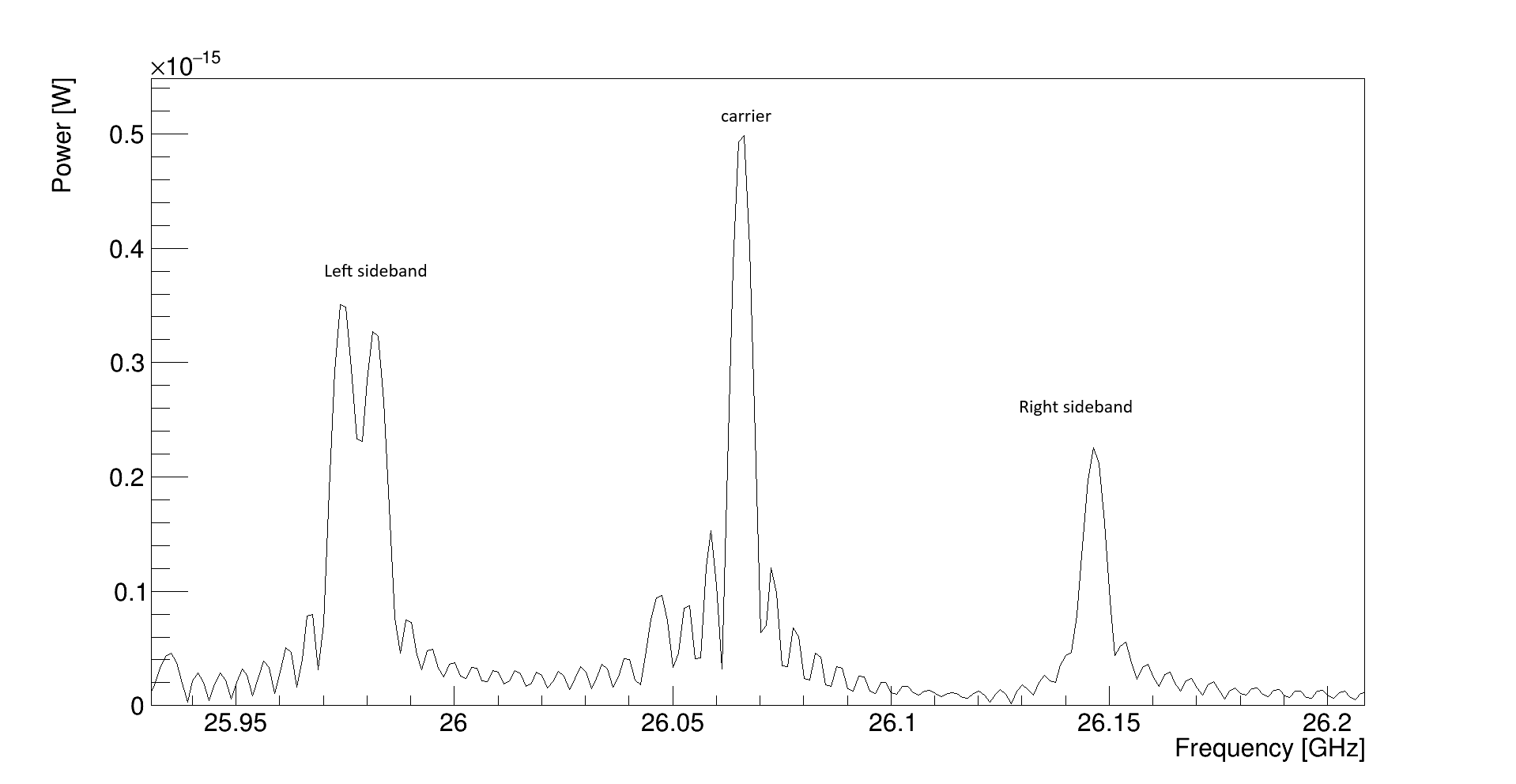}
    \caption{CST simulation of RF pattern of single electron in cyclotron and bouncing motion. It is possible to observe the carrier peak at $26.068~GHz$ and the two sidebands with frequency deviation of $\Delta f \simeq 80~MHz$. The plot is done with a frequency step of $RBW=1.25~MHz$ and so the secondary peak and Doppler shifts are not visible. The power is lower than $1~fW$ because a simulation limit.}
    \label{fig:CST_RF}
\end{figure}

 For example, from the simulated events in Figure (\ref{fig:z_gcs}), the electron is drifting at about $9\times 10^{-3}~mm$ in $5~ns$ when the electron is bouncing, that correspond to $V_z=1.8\times 10^{-3}~mm/ns=0.45~s.u.$ Then the bouncing period is $T_b\simeq12.5~ns$. Considering this Z velocity, it is possible to compute the Doppler shift by using Formula (\ref{eq:Doppler})

\begin{equation}
    f_2'=f_2\sqrt{\frac{1+\beta_z}{1-\beta_z}}
    \label{eq:Doppler}
\end{equation}

where $f_2$ is the secondary frequency obtained by Equation (\ref{eq:f2}). In this case, the Doppler correction is $\pm156~kHz$. In conclusion a Doppler shift is expected only in the secondary peak because it is the only signal related to a transverse drift, longitudinal movements are expected to have a negligible Doppler\footnote{In principle it could be possible to have a transverse Doppler shift (transverse with respect to the waveguide), however the $V_{\parallel}$, of the electron is non relativistic.}.
A complete simulation of the single electron RF pattern has been performed in CST by using the \emph{particle in cell} (PIC) solver; the result is shown in Figure (\ref{fig:CST_RF}), and this simulation is particularly useful for understanding the single electron RF pattern because it is comprehensive of kinematic and electromagnetic effects.

%la potenza? bisognerebbe scrivere che la P di carrier e secondario è 2fW e che dipende da K e Kperp. poitenza delle sideband invece da funzioni di bessel?
\newpage
Up to now, the RF pattern of single electron has been discussed and can it be summarized in four main peaks in the Fourier spectrum:

\begin{itemize}
    \item Carrier peak: It is the main peak, which frequency depends on the average kinetic energy $\langle K \rangle$.
    \item Secondary peak: Has a frequency that is larger than the carrier frequency and depends by the longitudinal kinetic energy $K_{\parallel}$, has Doppler shift.
    \item Sidebands peaks: Peaks that are shifted on the left and right of the carrier peak by a quantity $\Delta f$ that depends by the bouncing period, that depends by the longitudinal kinetic energy. The sidebands are produced by FM of the carrier frequency.
\end{itemize}

\begin{figure}[h!]
    \centering
    \includegraphics[width=0.8\linewidth]{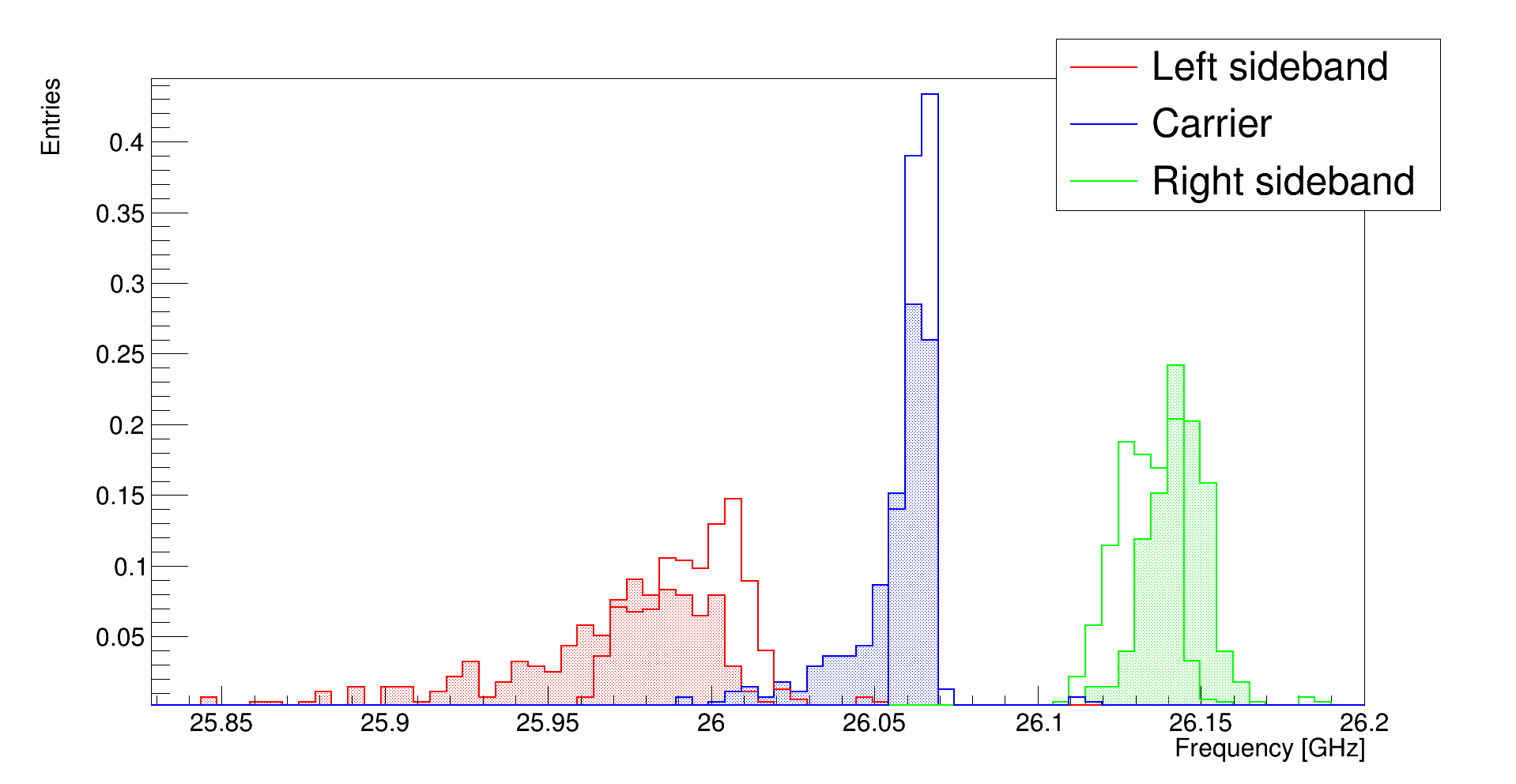}
    \caption{Comparison of the simplified RF pattern in the ideal and simulated field map case (dashed). The left and right distribution correspond to the sidebands at high and low frequency, respectively, while the central distribution correspond to the carrier frequency.}
    \label{fig:RF_pattern}
\end{figure}

It is now possible to study the distribution of the RF pattern of several simulated electrons to have a distribution for signal search for the data taking. Even though a complete simulation of the RF pattern of single electron is possible in CST, a long simulation for multiple electrons is not possible because of the long simulation time (two weeks for one electron with an high performance workstation). By assuming the RF pattern for single electron, it is possible to simulate the \emph{simplified RF pattern} of a single electron, that is obtained by the average frequency of the LORENTZ4 track simulations and including on the left and right the sidebands at $\pm 1/T_b$ from the carrier frequency. In Figure (\ref{fig:RF_pattern}) the simplified RF pattern for the ideal magnetic field map and the simulated magnetic field map (dashed) are shown for electrons emitted with random initial conditions in the fiducial volume with initial kinetic energy of $E_0=30472.2~eV$.

\newpage
\section{Reconstruction of Electron kinematics variables} \label{sec:kin_reconstruction}

In Section (\ref{sec:etrap}) the Electron Trap setup, a test setup for the PTOLEMY RF tracker, has been described and in Section (\ref{sec:montecarlo}) the simulations on the physical quantities of interest have been reported with particular attention for the RF pattern. The RF pattern has been presented in two cases: the ideal and the simulated field map cases; this comparison is primarily useful to understand the effect of the non uniformity systematics and to study the distributions that we could expect in the PTOLEMY magnet which is going to be close to the ideal case. It is now possible to study the methods to reconstruct the electron kinematics variables ($K$,$K_{\parallel}$) from the RF pattern of a single electron. In Section (\ref{sec:filter}) is described the key information needed to set potentials of the filter as reported in Formula (\ref{eq:pot_filter}). Measurement of mean kinetic energy is done by measuring the carrier frequency that depends directly on the average kinetic energy of the electron in the potential, as shown in Equation (\ref{eq:fc}).

\begin{equation}
    f_c=\frac{1}{2\pi}\frac{|q|B}{m}\frac{1}{\langle K\rangle /m+1}
    \label{eq:fc}
\end{equation}

The $K_{\parallel}$ reconstruction requires the previous $K$ measurement and it is more complicated; however, several methods have been identified.\\

\subsubsection{Carrier power method}
A first method is to measure the power of the carrier peak, in fact the instantaneous Larmor power, depends by transverse kinetic energy as shown in Formula (\ref{eq:P}).

\begin{equation}
    P=\frac{2}{3}\frac{1}{4\pi\epsilon_0}\frac{q^4B^2}{m^3c}\left(\frac{K}{m}+2\right)K_{\perp}
    \label{eq:P}
\end{equation}

This Formula has been obtained by rewriting Equation (\ref{eq:larmor_power}).\\
After the measurement of $\langle K \rangle$ and $K_{\perp}$ it is possible to obtain $K_{\parallel}$. Even though this method could appear to be very simple and straightforward it is affected by several systematic errors that are difficult to estimate, like the RF-waveguide coupling, difficult to predict a priori\footnote{In principle this is possible on statistical basis by knowing the electron phase space after the injection.} and, the electronic gain instability by temperature variations.
%scrivere quanto verrebbe incertezza?
\newpage
\subsubsection{Secondary peak method}

A second method deals with the reconstruction of the longitudinal kinetic energy by measuring the frequency deviation between the secondary peak and the carrier peak. This frequency deviation, as shown in Formula (\ref{eq:f2}), depends on $K_{\parallel}$. The problem of this method is primarily that the power of the second peak is smaller than that of the carrier and it could be difficult to measure it. Furthermore, this peak is affected by the Doppler shift that has to be subtracted and is not known a priori in the Electron Trap. However, in the PTOLEMY RF tracker the velocity of the electrons could be predicted because the electrons are emitted at the same reference potential and are collimated to a specific range of Y, in which the electrons are expected to have a $V_z$ in a restricted range. This method could be improved with a previous calibration campaign with electrons from an electron gun.
%un altro metodo è ricotruire la KL media dal carrier e misurare la distanza tra carrier e secondario. tuttavia il secondario ha effetto doppler quindi è necessario conoscere quanto sia la Vz dell'elettrone per correggere lo shift. in ptolemy la direzione degli elettroni sarà univoca però nella trappola l'elettrone si può avvicinare o allontanare. questo metodo è utilizzabile se la velocità lungo z è nota e quindi se gli elettorni dal'iniezione sono collimati in un unico Y noto.

\subsubsection{Sidebands method}

\begin{figure}[h!]
    \centering
    \includegraphics[width=1\linewidth]{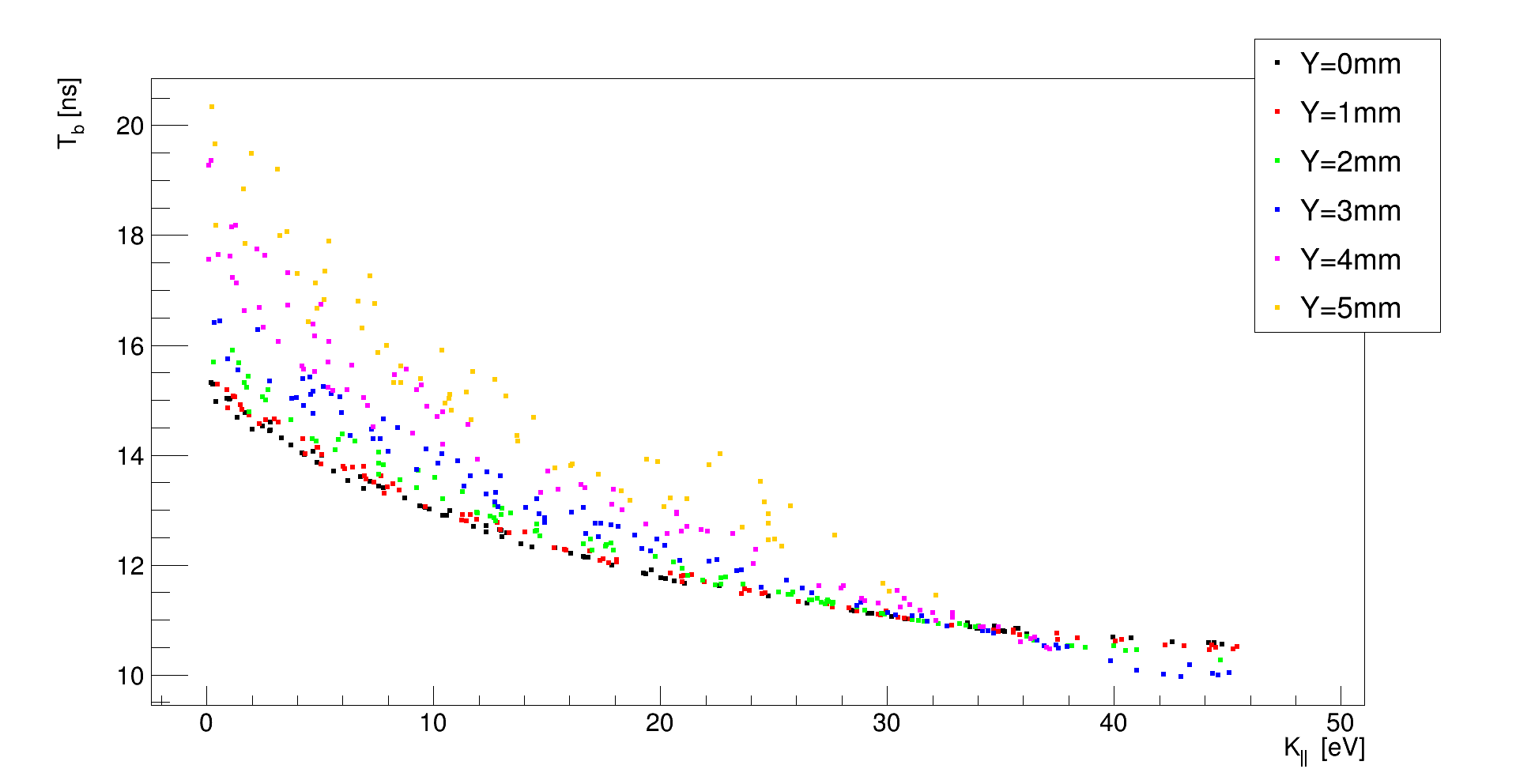}
    \caption{Bouncing period versus longitudinal kinetic energy at different Y in the Electron Trap. It is possible to note that in the central $\Delta y=4~mm$ are spread in a small range.}
    \label{fig:Tb_KL_Y}
\end{figure}

The last reconstruction method is based on the frequency deviation of the sidebands with respect to the carrier frequency. As has been already discussed in Section (\ref{sec:rf_pattern}) the sideband deviation is $\Delta f=1/T_b$ and the bouncing period is strongly dependent on the longitudinal kinetic energy (see Figure (\ref{fig:Tb_KL_montecarlo})). The distribution of the possible $K_{\parallel}$ at fixed $T_b$ depends on the Y coordinate at which the electrons drift. This effect is due to the variation of the potential well shape along Y. The distribution of the bouncing period versus the longitudinal kinetic energy, at different Y of bouncing, is shown in Figure (\ref{fig:Tb_KL_Y}).

After measuring the bouncing period, it is possible to see in Figure (\ref{fig:Tb_KL_Y}) what is the possible interval of available $K_{\parallel}$ for that measured $T_b$. It is possible to note that there are optimal ranges for this reconstruction method, for example, a favorable working point at $T_b\simeq11~ns$, $K_{\parallel}\simeq35~eV$. To summarize if the Y coordinate of the motion is not known a priori, with this method it is possible to set lower and upper limits to the longitudinal kinetic energy. However, if the Y coordinate of the motion is known, the relation between $T_b$ and $K_{\parallel}$ becomes univocal, improving the uncertainty of the reconstruction. This improvement could be obtained in two ways: by collimating the electrons from the target in a narrow range of Y after the injection phase or by measuring the Y of the Doppler shift of the secondary peak. This can be done with the following steps:

\begin{enumerate}
    \item Measure $K_{\parallel}$ from the sidebands with a range, without any assumption on Y
    \item Assume the most probable value of $K_{\parallel}$. This value could be, depending by the cases, the mean value of the range or the most probable value taken from a previous calibration campaign
    \item Use the measured $K_{\parallel}$ to estimate the Doppler shift on the secondary peak
    \item Estimate the Y of motion from the Doppler shift
\end{enumerate}

%L'ultimo metodo, e quello che sarà utilizzato è ricotruire K media dal carrier e KL dalla distanza delle sidebands.
%questo permette di sfruttare le buone risoluzioni strumentali ottenibili con la misura di frequenza.
%come annunciato prima, deltaf dipende da 1/Tb e quindi si misura Tb. Tb è una uantità che dipende dal KL. si può ricostruire KL dal grafico in Figura (\ref{fig:Tb_KL_Y}) dove una volta misurato Tb si traccia una linea e si misura KL. in linea di principio i possibili KL corrispondenti ad un solo Tb  hanno uno span che dipende dalla Y del moto, che a priori non è nota. quindi uno misura KL e mette i limiti. 
%Un ulteriore improvwment della misura si può fare conoscendo la Y dell'elettorne che o puiò essere nota a priori con la collimazione nella fase di iniezione, oppure in linea di prioncipio è possibilie misurare la Y dal doppler sul picco secondario prendendo come riferimento il KL medio.
To conclude, the sidebands method has been chosen as a baseline option for the electron kinematics variables reconstruction method for the PTOLEMY experiment. The main source of $K_{\parallel}$ uncertainty is the different potential shape, and possible improvements can be made by 

\begin{itemize}
    \item Collimating electrons at a restricted range of Y.
    \item Estimate Y by measuring the Doppler shift on the secondary peak.
    \item Shaping the potential lines such that the Y dependence is reduced.
\end{itemize}

\begin{figure}[h!]
    \centering
    \includegraphics[width=0.55\linewidth]{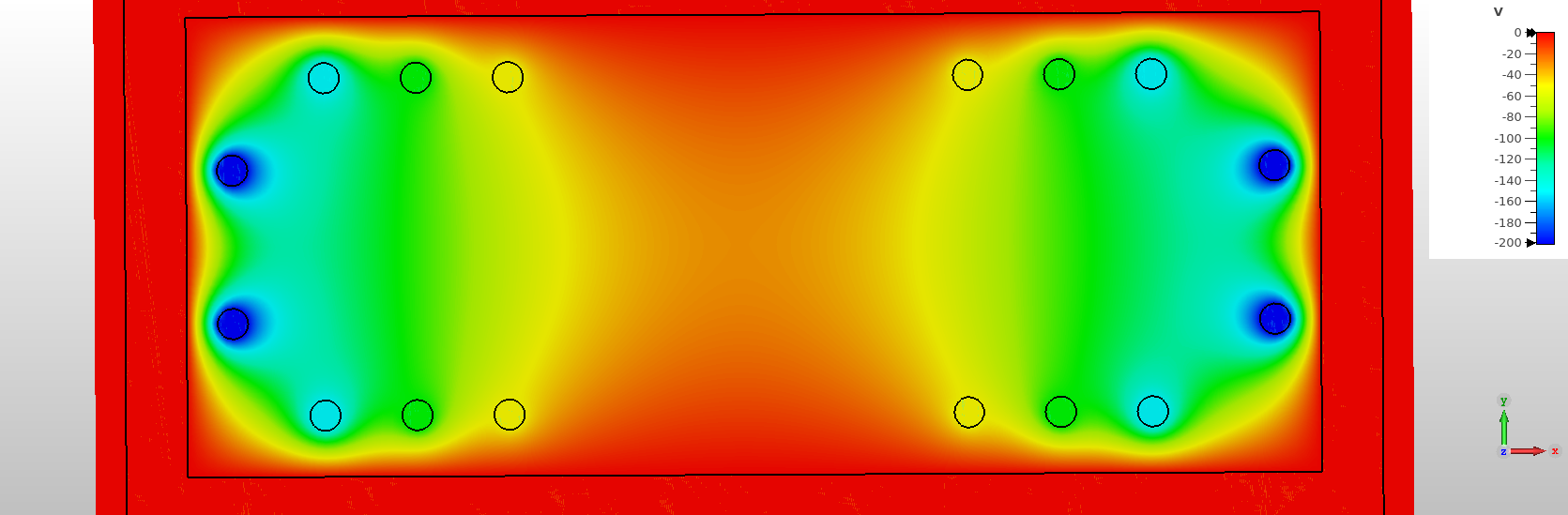}
    \caption{Alternative electrode setup by using wires. It is possible to note that the potential well is constant along Y and the presence of undesired $E_y(x,y)$ is reduced. An additional $V_z$ can be obtained by including wires on the central top part of the setup.}
    \label{fig:wire_trap}
\end{figure}

An example of an alternative electrode setup is shown in Figure (\ref{fig:wire_trap}), where wires are used instead of full plane electrodes. This setup has the advantage of reducing the amount of conductor material in the waveguide and is capable of providing a better electric field shaping for the reconstruction of $K_{\parallel}$.
%(\ref{fig:wire_trap}).

As can be seen in Figure (\ref{fig:wire_trap}), the potential lines are more constant across Y direction, and so the Doppler shift are uniformed and there is a reduced Y dependence for $K_{\parallel}$ reconstruction.
 %metodo df/dt?
%ricostruzione K da <K> e KL e Tb
So far, the method for reconstructing the average kinetic energy $\langle K\rangle$ and the maximum longitudinal kinetic energy $K_{\parallel}$ has been shown and can be summarized as follows:

\begin{enumerate}
    \item Measurement of $\langle K \rangle$ from the carrier frequency $f_c$
    \item Measurement of $K_{\parallel}$ from the bouncing period $T_b$ by using the sideband method
\end{enumerate}

However, as has been discussed in Equation (\ref{eq:pot_filter}), the proper filter setting requires the previous measurement of the kinetic energy at the bottom of the potential well $K$, where

\begin{equation}
    K=K_{\perp}^{RF}+K_{\parallel}^{RF}
\end{equation}

It is now possible to describe the final step for the reconstruction of $K$ using the previous measurement of $\langle K \rangle$ and $K_{\parallel}$.\\
Firstly, it is possible to note that during the cyclotron and bouncing motion, the perpendicular component of the kinetic energy remains constant\footnote{The only energy loss is due to RF emission, which for the moment is not taken into account.} and so the average kinetic energy can be rewritten as

\begin{equation}
    \langle K \rangle =K_{\perp}+ \langle K_{\parallel} \rangle
    \label{eq:K_avg}
\end{equation}

where $ \langle K_{\parallel} \rangle $  is the average longitudinal kinetic energy in the bounce. This quantity can be calculated, in the non relativistic case and by assuming energy conservation, with the following expression

\begin{equation}
    \langle K_{\parallel}\rangle=\frac{1}{T_b}\sqrt{\frac{m}{2}}\int_{-X_b}^{X_b}\sqrt{K_{\parallel}-V(x)}dx
    \label{eq:avg_K_par}
\end{equation}

where has been used that $K_{\parallel}(x)=K_{\parallel}-V(x)$ and $X_b$ (previously called $X_{max}$) are the positions where the electrons are bouncing, that satisfy $V(X_b)=K_{\parallel}$. In PTOLEMY, and also in the Electron Trap, the potential $V(x)$ is assumed because the potentials are fixed and known a priori, this means that the integral can be solved numerically event by event. After the computation of $\langle K_{\parallel} \rangle$ it is possible to get the transverse kinetic energy

\begin{equation}
    K_{\perp}=\langle K \rangle-\langle K_{\parallel} \rangle
    \label{eq:K_perp_recon}
\end{equation}

and so it is possible to reconstruct the desired quantity of the kinetic energy at the bottom of the potential well

\begin{equation}
    K=K_{\perp}+K_{\parallel}
    \label{eq:K_recon}
\end{equation}

At this stage the kinematics variables required for the proper use of the PTOLEMY filter are measured on an event by event basis.

%% file: sections/5Electron_Trap_Results.tex
\chapter{Electron Trap Measurements} \label{chap:Electron_Trap_Measurements}

In the previous Chapter the Electron Trap setup has been presented in Section (\ref{sec:etrap}), with particular emphasis on the electron trapping system, the radio-frequency electronics and the Magnet. Furthermore, simulation studies on electron kinematics have been discussed in Section (\ref{sec:montecarlo}) focusing on the electron drift time and bouncing period, while the Electron RF pattern has been presented in Section (\ref{sec:rf_pattern}). The simulation studies are meant to show the features of the Fourier spectrum of the cyclotron radiation (RF) emitted in physics events. In the present Chapter, the experimental procedure and the results obtained are described and the energy resolution and systematics are analyzed in Section (\ref{sec:resol_systematics}). The Electron Trap has been designed to study the RF emission of single electron in cyclotron motion to properly design the PTOLEMY RF tracker. As a consequence of this goal, the Electron Trap setup has been designed for the detection of low rate events because we were interested in the detection of few events for understanding the reconstruction of the electron's kinematic variables, so the drift (top and bottom) electrodes has been removed to have the longest possible drift time. Also, a low bouncing potential has been set to have a limited pitch angle acceptance for studying electron signals from a limited electron phase space, at the cost of a lower event rate.

\section{Electron Trap experimental procedure}
\label{sec:exp_procedure}

In the following the experimental operation for the Electron Trap is described. The first step is creating a high vacuum in the experimental vessel so that the pressure will be $\mathcal{O}(10^{-8}~ mbar)$, after that the outlet gauge\footnote{It is the gauge between the room and the outlet of the primary pump.} is closed and the Krypton source gauge is opened. This procedure ensures that the residual gas in the Rubidium source vessel is pumped out and it reaches the high vacuum regime. Once the pressure will be at equilibrium, the source gauge is closed and after 6 hours the outlet gauge is opened. At this point the vacuum vessel is in high vacuum state and all undesired residual gas have been removed. The electronics are now powered on with the LNA at the low temperature settings. At this stage the first important parameter for the data taking, that is the local oscillator frequency of the down-converter $f_{LO}$, is set. This is crucial for reconstructing $f_{RF}$ as shown in Equation (\ref{eq:IF_downconverter}). It is now possible to activate the temperature stabilization system for the down-converter and set the trigger that, as described in Section (\ref{sec:trigger}), is generated by means of the RTO oscilloscope as described in Section (\ref{sec:trigger}). The Electron Trap is now cooled and is thermally stabilized at the desired temperature. The first step of trigger setting is to take an average FFT with $f=(0,f_{max})$, where $f_{max}$ is the Nyquist limit or $600~ MHz$, in the case of the activated HD mode (explained in Chapter (\ref{chap:Electron_Trap})). The averaged FFT is meant to monitor the variation of the total gain of the electronic chain as function of the frequency and time, as can be seen in Figure (\ref{fig:avg_noise_time}). We can now set up the trigger zone by following the average noise and shifting it by $+XdB$, where $X$ is the SNR below the signal level we aim at triggering. In Section (\ref{sec:non_uniform_montecarlo}) it was described that the expected electron power is $P=1.9~fW\sim -117~dBm$, considering all gains and losses in the electronic chain and assuming a temperature in the Electron Trap of $52~K$, we do not have a direct measurement of the temperature in the Electron Trap cavity, and $RBW=20~kHz$, the expected electron signal SNR is 

\begin{equation}
    SNR= (P_{wg}-N_t)-Loss-NF_{tot}\simeq13.9~dB
    \label{eq:SNR52K}
\end{equation}

As previously described, due to the losses in the Electron Trap the SNR is lower than the expected value of $SNR\sim 20~dB$. An example of a triggered event is shown in Figure (\ref{fig:triggered_signal}), if a signal enters the trigger zone, the oscilloscope saves the time series. At this point it is possible to start the data taking: the experimental procedure includes a first run without Krypton source in which fluctuations of the thermal noise are recorded, then a separated run with Krypton source opened is done.

%RUN sorgente
%salita discesa o piatto

%RUN BKG

\section{Electron Trap Data analysis}
\label{sec:Data_analysis}

In this Section the preliminary results obtained in the data taking are presented, and the analysis methodology is described, too. Firstly, evidence of an RF signal from an electron has been probed by measuring the Krypton lifetime from the trigger rate in Section (\ref{sec:lifetime}), while in Section (\ref{sec:sub_data_analysis}) the analysis procedure and the results of RUN 93-94 are described.

\subsection{Krypton lifetime measurement}
\label{sec:lifetime}

\begin{figure}[h!]
    \centering
    \includegraphics[width=0.8\linewidth]{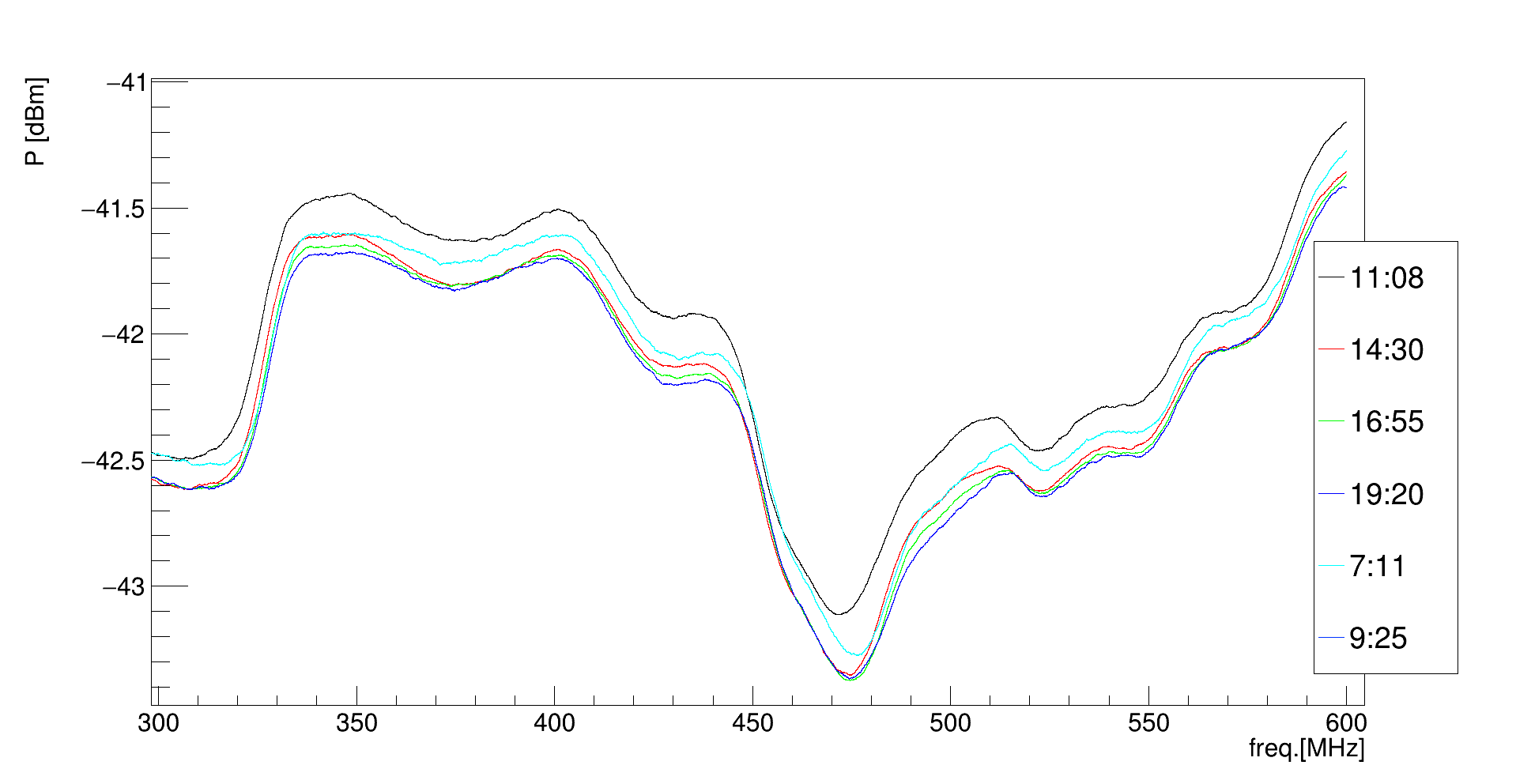}
    \caption{Average noise level measured at different time of the day. It is possible to note that during the day the gain is decreasing or constant while during the night it increases.}
    \label{fig:avg_noise_time}
\end{figure}

Even though the electron trap has been designed to detect a low rate of electrons, it is expected to have sensitivity to the change of the trigger rate due to the increase (decrease) of the activity in the Electron Trap. Before studying the change in the trigger rate in the presence of the Krypton source, it is important to study the behavior of the background. If the thermal stabilization system of the down-converter is activated, the trigger rate remains constant over time because there is no significant variation in the gain of the electronic chain, indeed if the temperature stabilization is not activated, a modulation dependent on the temperature of the down-converter is observed. The temperature of the down-converter increases during the day and decreases during the night, while the gain $G_d$ has an inverse correlation with the temperature, as can be seen in Figure (\ref{fig:avg_noise_time}).

\begin{figure}[h!]
    \centering
    \includegraphics[width=0.8\linewidth]{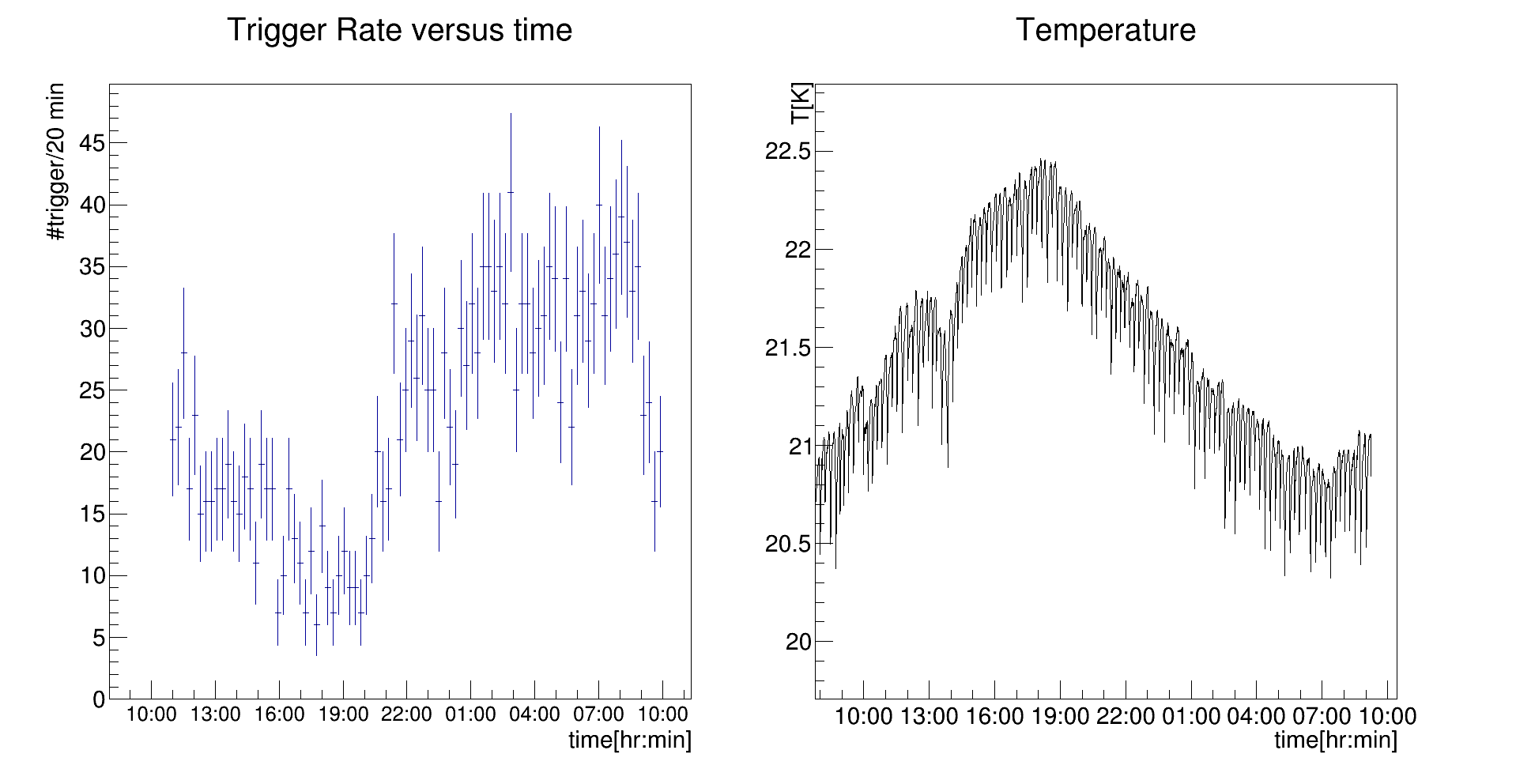}
    \caption{Trigger rate (left) and temperature (right) over time. As can be seen the temperature is increasing during day time and in that period of time the trigger rate decrease. }
    \label{fig:trigger_rate_temperature}
\end{figure}

This variation of the gain has an effect on the measured trigger rate (if the trigger zone is kept fixed), in fact, as can be seen in Figure (\ref{fig:trigger_rate_temperature})\footnote{This is actually the temperature measured inside the vessel at the LNA stage, however, the variation of this temperature is proportional to the variation of the room temperature because the cold head has a cooling power that decreases when the temperature increases.} the trigger rate decreases during the day while it increases when the temperature starts to decrease.  

%sistematica G_d, mettere qui o spostare sopra a cap elettronica?

\begin{figure}[h!]
    \centering
    \includegraphics[width=1\linewidth]{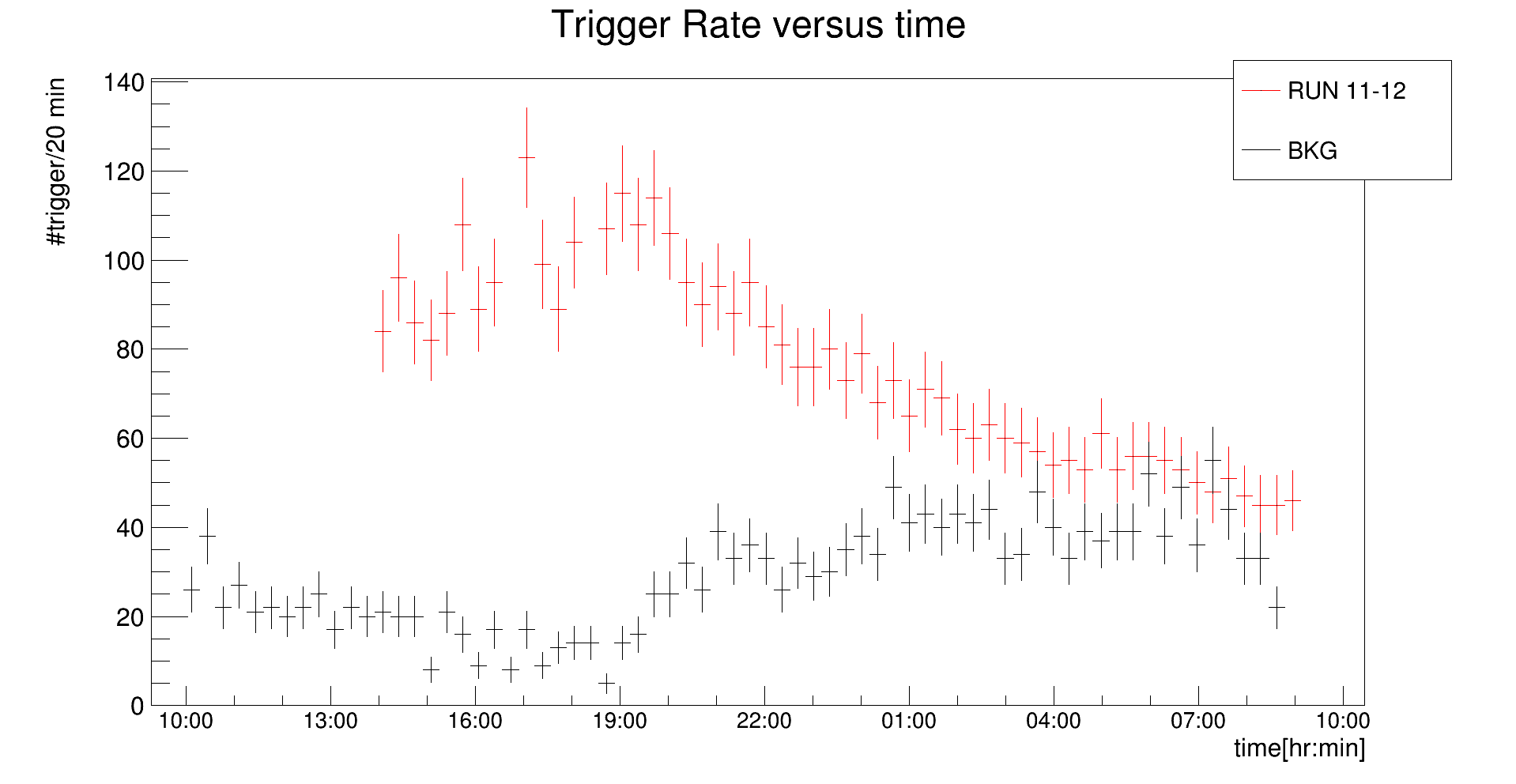}
    \caption{Trigger rate of RUN 11-12 (red) and trigger rate of a reference background RUN (black). It is possible to note that in the RUN 11 the trigger rate is increasing while the background rate is expected to decrease and vice-versa when the source is closed (RUN 12). It is possible to see the two distributions show an opposite trend.}
    \label{fig:trigger_rate_RUN11-12}
\end{figure}

It is important to note that the absolute scale of the trigger rate is not indicative because the absolute value of the gain has several contributions, in Figure (\ref{fig:trigger_rate_temperature}) is shown the gain variation as function of the temperature. It is now possible to study the variation of the trigger rate in the presence of the Krypton source, in fact the following results were taken before the implementation of the temperature stabilization system. The trigger rate over time is shown in Figure (\ref{fig:trigger_rate_RUN11-12}), where the trigger rate of RUN 11, in which the source was opened, and the trigger rate of RUN 12, in which the source was closed, are shown. The source was opened at 11:39 while the RUN 11 started at 13:40\footnote{It started later the source opening for technical reasons.} and the source has been closed at 16:42, when RUN 12 starts. The RUN 11-12 trigger rate is shown in red, while a background RUN has been reported in black. It is worth pointing out that the RUN 11-12 has been taken during summer, while the BKG run has been taken during the winter season, so there was a different time at which the room temperature was at the maximum and so the trigger rate of the BKG starts to increase.
The experimental hall miss a good temperature stabilization system, which will be mandatory in an optimal experiment site.

\begin{figure}[h!]
    \centering
    \includegraphics[width=1\linewidth]{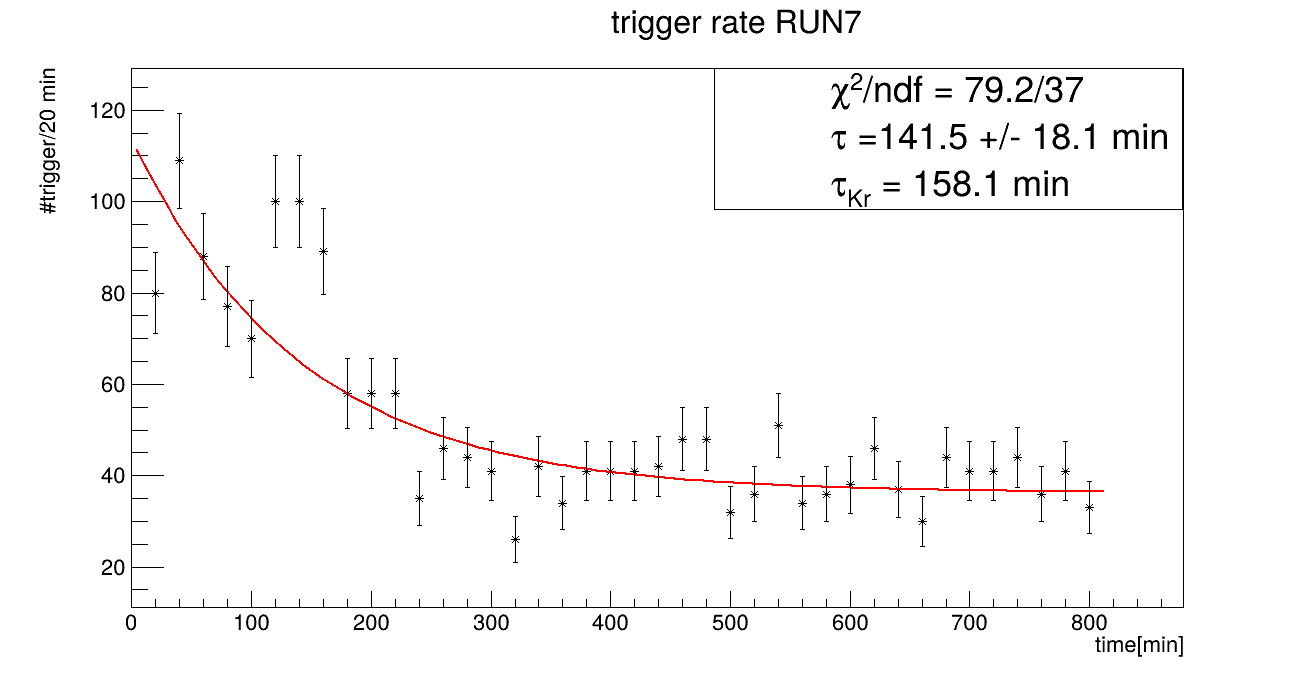}
    \caption{Trigger rate of RUN 7 in which the source has been closed at time 0. It is possible to see the curve obtained by exponential fit.}
    \label{fig:trigger_rate_RUN7}
\end{figure}

Because the two distributions have been taken on different days, it is not possible to subtract the background distribution because the time of maximum temperature is unknown; however the different behavior can be seen. It is possible to fit the decay of the trigger rate to get the Krypton lifetime; this has been done with a dedicated run, the RUN 7, the trigger rate of which is shown in Figure (\ref{fig:trigger_rate_RUN7}). In this RUN the source was opened in the Electron Trap for hours and has been closed at the time 0\footnote{That was at around 18:30.}. The observed decay of the trigger rate has been fitted with an exponential function and as a result gives

\begin{equation}
    \tau_{exp}=141.5 \pm 18.1~ min
\end{equation}

that is compatible with the expected value of $\tau_{Kr}=158.1 ~min$.

%I run riportati sopra sono stati eseguiti con un trigger con forma rettangolare e a SNR di circa 13dB. 

Now that an indication of detection of single electron RF emission, in the Electron Trap, has been obtained, it is possible to study the RF  spectrum.

\newpage
\subsection{Data analysis}
\label{sec:sub_data_analysis}

\begin{figure}[h!]
    \centering
    \includegraphics[width=1\linewidth]{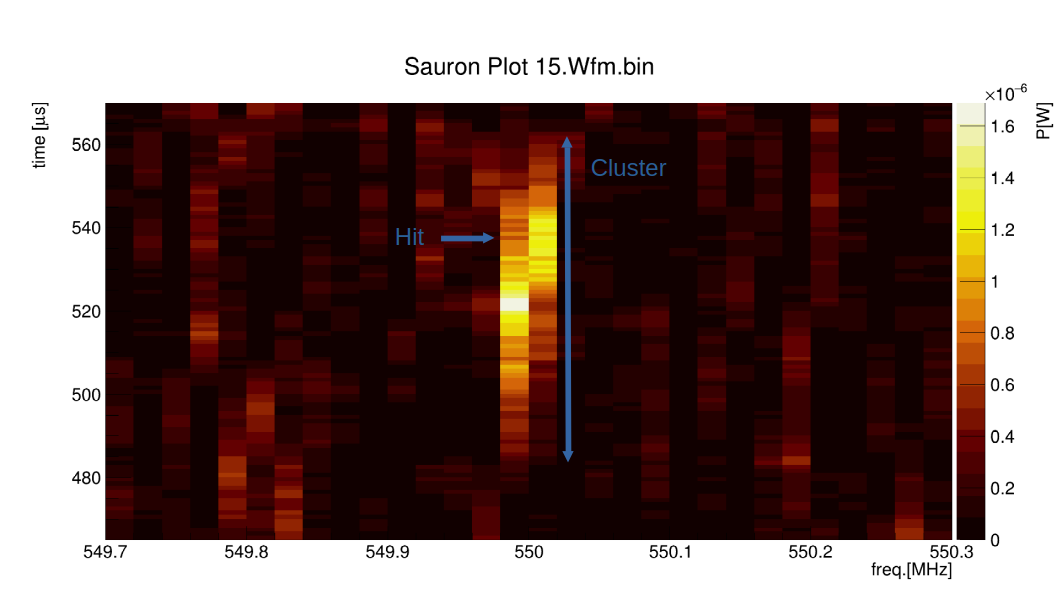}
    \caption{Sauron Plot of a calibration signal. It is possible to see an example of hit and two adjacent clusters.}
    \label{fig:Sauron_Plot_example}
\end{figure}

In the first part of this Section the basic concepts for the analysis of RF signals are described by using a calibration signal, then the results for the RUN 93-98 are given. The expected electron signal is composed of a carrier frequency and two sidebands, and to measure this signal it is useful to introduce the spectrogram of the time series saved by the oscilloscope, which in this work is called \emph{Sauron Plot}.\\
The Sauron Plot is a spectrogram obtained by performing a \emph{Sliding Fourier} on the time series saved once the trigger took place. The FFT is done by choosing the RBW and the \emph{sliding step}, which corresponds to the samples are skipped when shifting the FFT window, which, on the time axis, is set to $1~\mu s$. In Figure (\ref{fig:Sauron_Plot_example}) a Sauron Plot of a reference signal is shown. In this plot, the frequency of the signal (with resolution of $RBW$) is shown in the abscissa while the time of the time series (with resolution of sliding step) is reported on ordinates. The color represents the power in linear scale. The analysis of the Sauron Plot is performed by analyzing each 2D bin, which is also called \emph{Hit}, and by looking if there is a group of neighbor hits (also called \emph{Cluster}) with the same frequency with power greater than a given threshold. An example of Cluster and Hit is shown in Figure (\ref{fig:Sauron_Plot_example}).

%sauron plot

\begin{figure}[h!]
    \centering
    \includegraphics[width=0.9\linewidth]{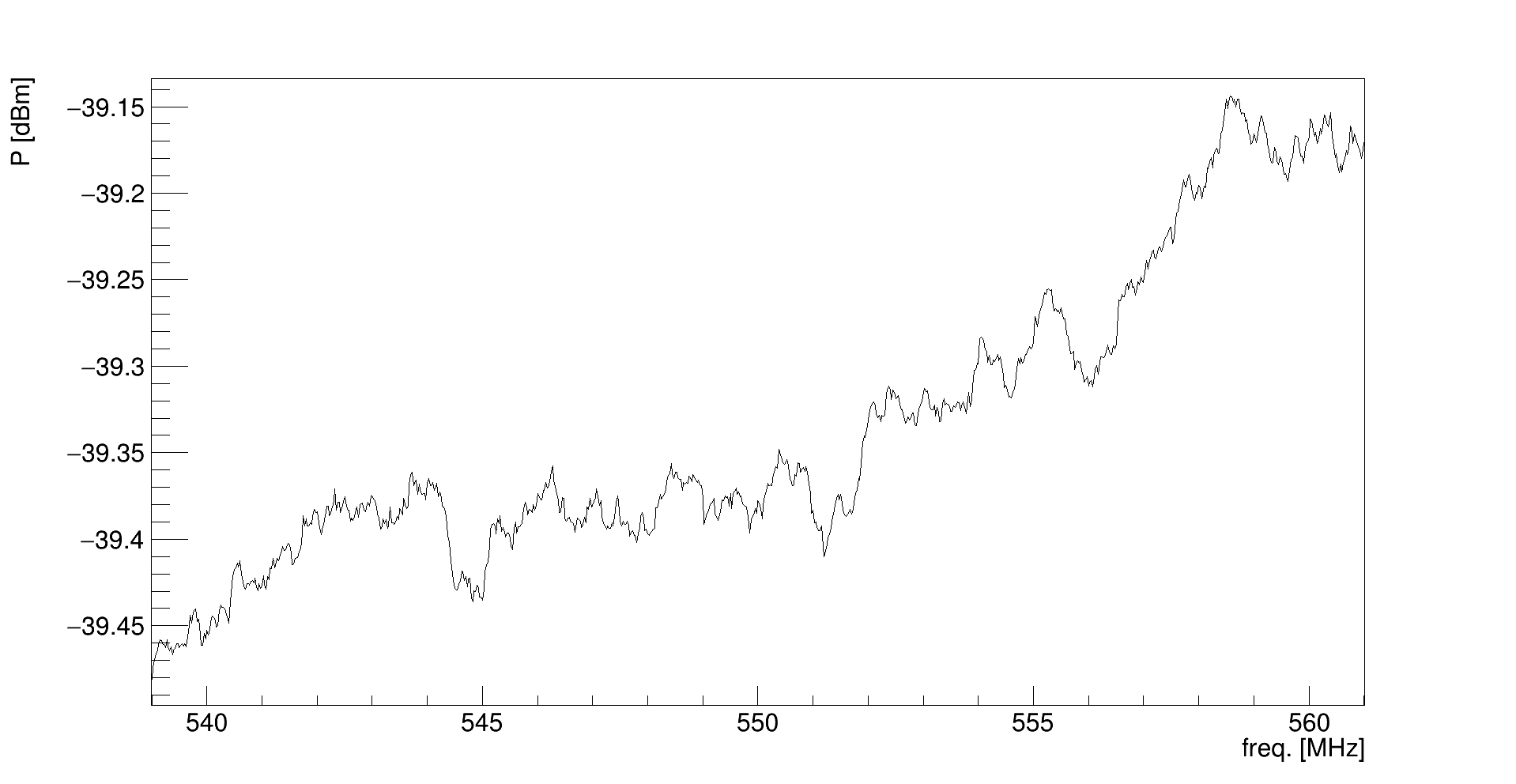}
    \caption{Average noise level of the calibration event. This average noise level is obtained by averaging over $1~ms$.}
    \label{fig:avg_noise_calibration}
\end{figure}

Even though it is possible to see the absolute value of the power of each hit, it is misleading because the gain depends by the frequency. For example in Figure (\ref{fig:avg_noise_calibration}) the average noise level of a calibration event between 540 and 560 MHz is shown; it is possible to note a total 0.3 dB excursion in the gain, this means that if we take as an example two signals with same SNR at 540 and 560 MHz, the signal at 560 MHz will be measured to have a power 1.07 times larger than the signal at 540 MHz. So, it is necessary to analyze the clusters in an SNR basis and not on an absolute power basis.

\begin{figure}[h!]
    \centering
    \includegraphics[width=1\linewidth]{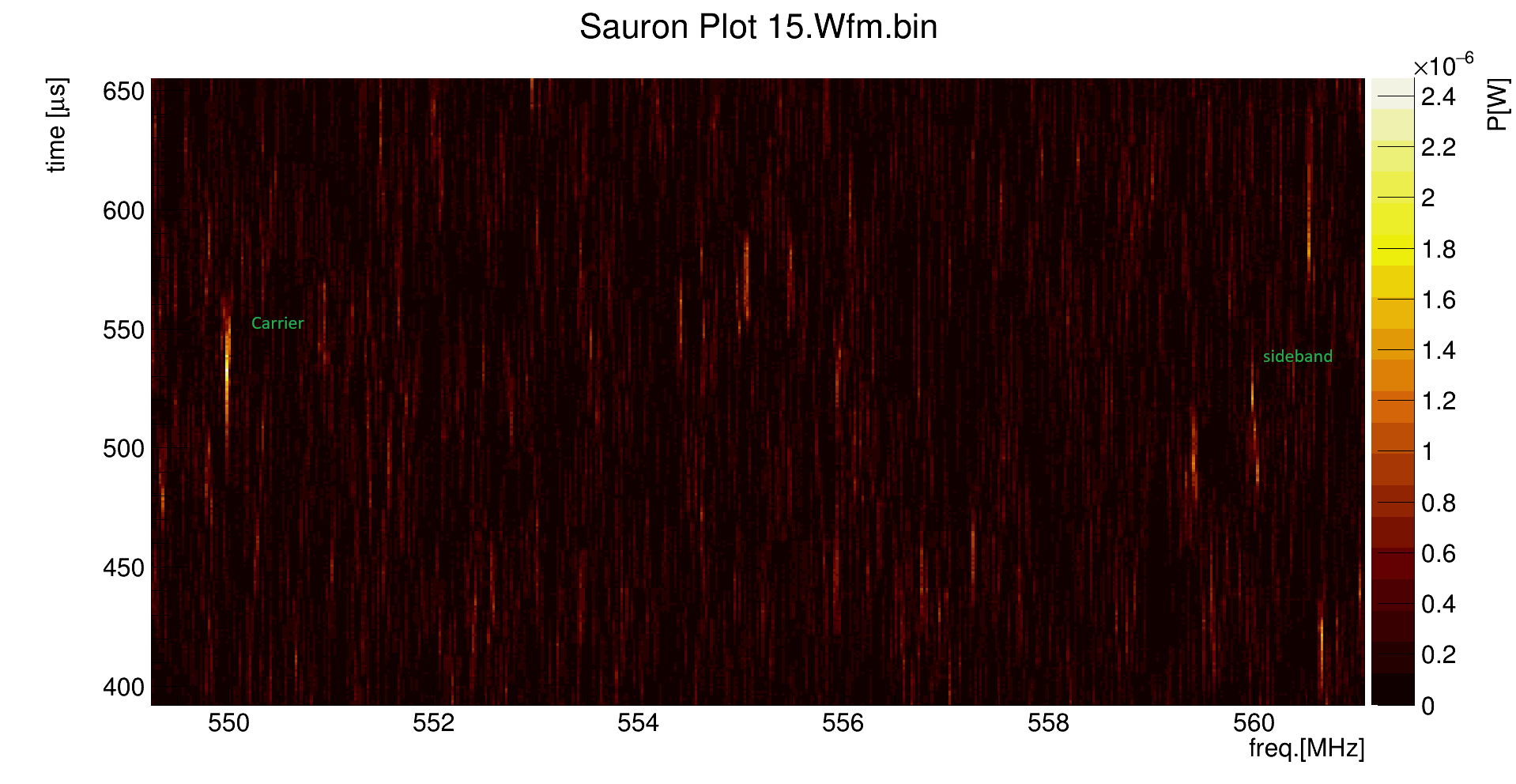}
    \caption{Sauron Plot of a calibration event. It is possible to note the carrier frequency at $550~MHz$ and a sideband track at $560~MHz$ that exist at the same time of the carrier.}
    \label{fig:calibration_fc_sr} 
\end{figure}

We can now introduce the analysis procedure of a Sauron Plot. As described in Section (\ref{sec:exp_procedure}), the first step is to take the average FFT; this is useful not only to properly set the trigger, but also for performing a SNR analysis of the Clusters and Hits. Considering that electrons are expected to produce a track, i.e. a sequence of clusters, in the Sauron Plot, the analysis is mainly centered around the cluster analysis. In the following the quantities that can be obtained in the analysis of Sauron plot, by using a calibration signal, are described.\\
The signal is generated with an \emph{R\&S SMB 100-B} generator, with a carrier frequency of $f_{RF}=26.0~ GHz$, a frequency modulation with deviation $\Delta f=10~ MHz$ for a total time of $t_D=50~\mu s$. The down-converter has been set with $f_{LO}=26.1~GHz$. The carrier frequency and right sideband are shown in Figure (\ref{fig:calibration_fc_sr}) in which it is possible to note the carrier frequency at intermediate frequency of $550~MHz$ and the right sideband at $560~MHz$. We also note that the tracks are enveloped in $\sim50 ~\mu s$ starting from the center of the time series (the total length of the time series is $1~ms$).

\newpage

\begin{figure}[h!]
    \centering
    \includegraphics[width=1\linewidth]{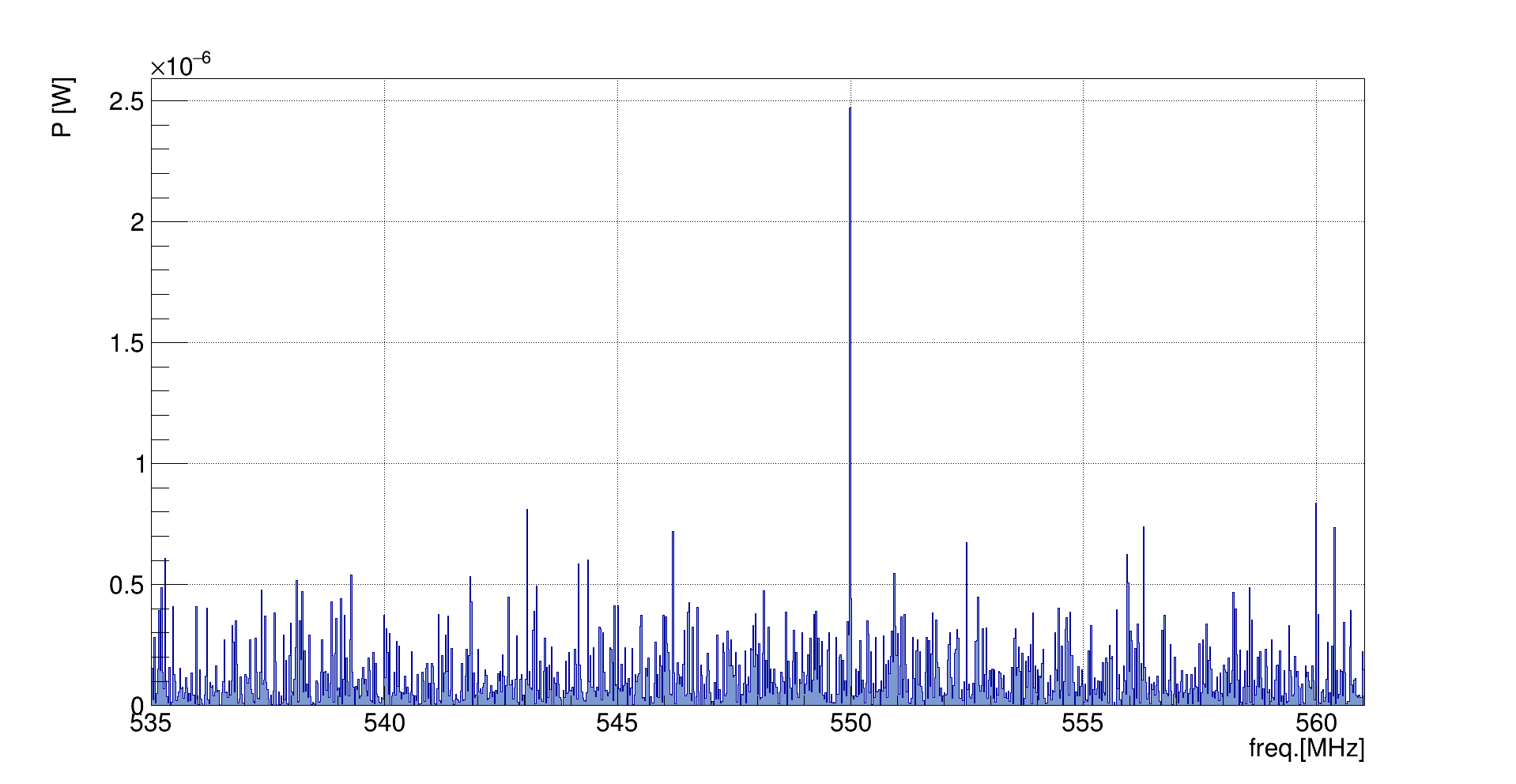}
    \caption{Power versus frequency profile (Fourier transform) of the calibration event at fixed time (time of maximum carrier power).}
    \label{fig:calib_P_f}
\end{figure}

The first information that we get from this Sauron Plot is the FFT at fixed time, as shown in Figure (\ref{fig:calib_P_f}). With this plot it is possible to identify the peaks with an uncertainty that is the RBW. 

\begin{figure}[h!]
    \centering
    \includegraphics[width=1\linewidth]{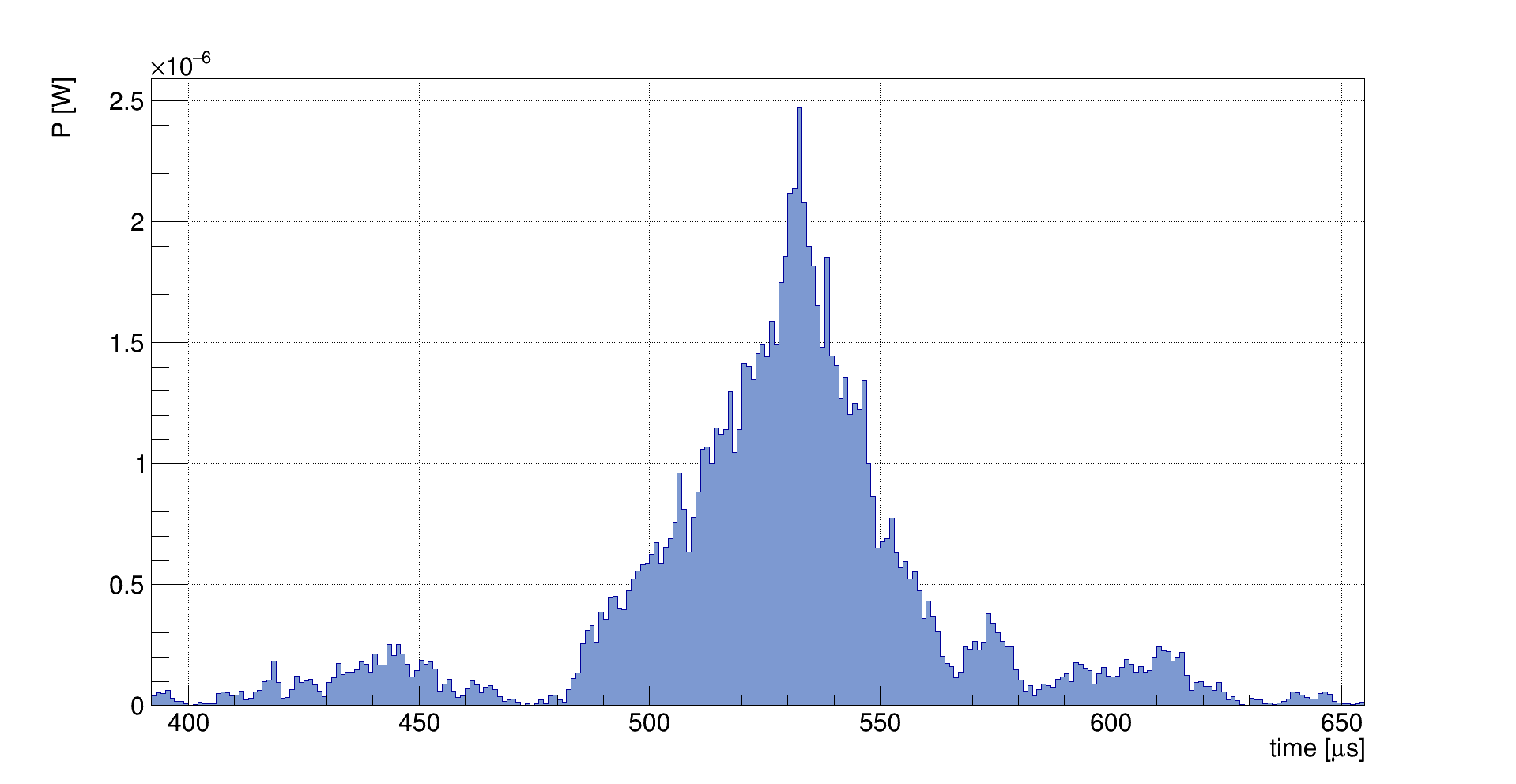}
    \caption{Power versus time profile of the carrier frequency cluster. It is possible to note that the signal has a duration of about $t_D \sim50~\mu s$.}
    \label{fig:calib_P_t}
\end{figure}

Instead, if the frequency of the signal is identified, we can see the profile of the power over time, that is shown in Figure (\ref{fig:calib_P_t}).

\newpage

In Figure (\ref{fig:calib_P_t}) the signal of $t_D=50~\mu s$ is shown with a triangular shape. The feature is an artifact of the Sliding Fourier. It happens that at the beginning (end) of the signal some low-power background samples are integrated\footnote{Like the histogram entries before $470~ \mu s$.}, while at the centre of the signal the FFT integrates the whole signal samples. This effect depends on the time scale of the FFT that is the inverse of the resolution bandwidth; in fact, if the time scale is smaller than the drift time, the signal must look like a trapezoid.\\
From the Cluster analysis it is possible to obtain useful information, like the average SNR, that is useful to select clusters with compatible SNR to the electron expectations, the cluster's Hit with maximum SNR ($SNR^{max}_{hit}$), to estimate the center of the cluster. From the plot of Figure (\ref{fig:calib_P_t}) it is possible to measure the drift time of the track $t_D$, which is obtained by measuring the time difference between the first ($t_{s}$) and the last ($t_{end}$)  hit of the cluster. Another useful quantity is the integrated signal-to-noise ratio, $SNR^{int}$, which is given by the Formula (\ref{eq:SNR_int})

\begin{equation}
    SNR^{int}=\sum_{t=t_{s}}^{t_{end}}SNR_t= \sum_{t=t_{s}}^{t_{end}}P_t -N\cdot t_D
    \label{eq:SNR_int}
\end{equation}

where $P_t$ is the power of the Hit at time $t$ and $N$ is the average noise power at the cluster's frequency. It is important to note that the calculation of Equation (\ref{eq:SNR_int}) must be done in Decibel units. This value can be re-normalized to the power $P^*_t=\mathcal{O}(1~fW)$ by using the value of the gain $G_{tot}$ at that frequency bin.

\newpage
After the normalisation it is possible to obtain the integrated power, which is an estimation of the energy emitted by Larmor emission

\begin{equation}
    P^{int}=\sum_{t=t_{s}}^{t_{end}}P^*_t \cdot \Delta t
    \label{eq:int_power}
\end{equation}

where $\Delta t=1~\mu s$ is the sliding step.

%se avg snr è alto ma segnale ha tD corto, ho segnale pizzutello, vicevrsa ho una traccia lunga ma spompata

%mettere qui risulati analisi da alfredo

\subsubsection{RUN93-94}

\begin{figure}[h!]
    \centering
    \includegraphics[width=1\linewidth]{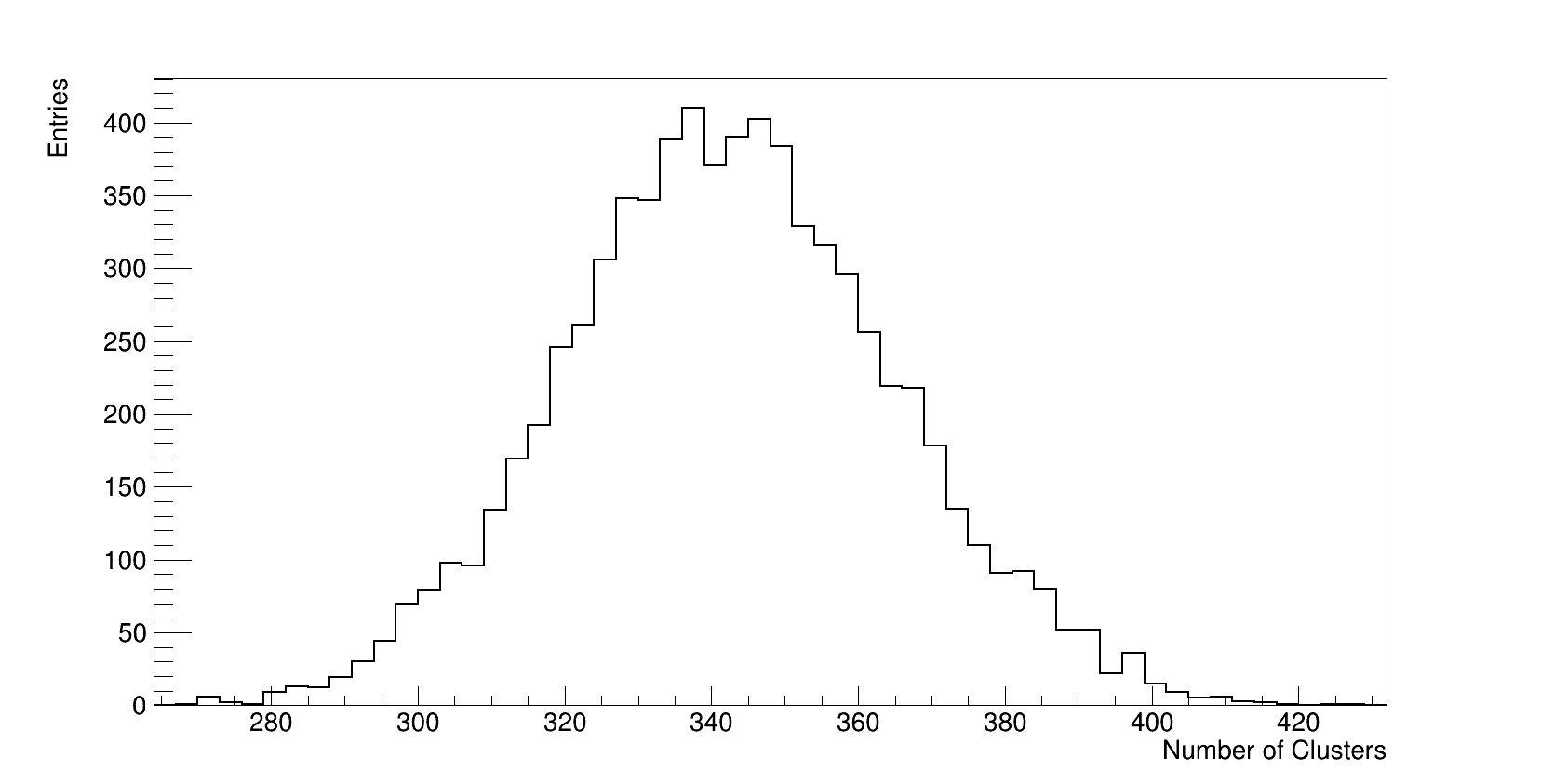}
    \caption{Number of clusters per event without cuts. The large majority are clusters belonging to background signals of thermal noise at low power. The expected number of clusters from the RF pattern should be three: one carrier and two sidebands. }
    \label{fig:NClusters_nocuts}
\end{figure}

Let's now describe the measurements that have been performed for the RF pattern study: the RUNs 93 and 94. In Figure (\ref{fig:RF_pattern}) it is shown that the expected carrier frequency is about $f_{RF}=26.06~GHz$, which can be converted using Equation (\ref{eq:IF_downconverter}) to an intermediate frequency $f_{IF}$. By using $f_{LO}=26.1~GHz$ the carrier signal is expected to be at $f_{IF}=490~MHz$. For these measurements, it was decided to mitigate the condensation of Krypton gas on the Electron Trap surfaces by stabilizing the internal temperature to $T=52~K$. With this temperature, and considering the $G_{tot}$, the power losses and $NF_{tot}$, it is possible to obtain the expected SNR of the electron signal that has shown in Equation (\ref{eq:SNR52K}) is $SNR=13.9~dB$ for a $RBW=20~kHz$. For these reasons, the trigger has been set to $f=[465,515]~MHz$ with a threshold of $SNR=13~dB$ with respect to the average noise. For those RUNs the temperature stabilization of the down-converter has been activated and the data have been taken with the same settings for both RUN with the difference that RUN 93 has been taken without source, while in RUN 94 the source has been opened at the start of the RUN.\\
After the the data taking has been completed, we can analyze the data with the procedure described in the following. As described in Section (\ref{sec:trigger}), for each triggered signal the time series of the oscilloscope is saved and for each time series a spectrogram is performed and clusters are searched. The cluster search works in the following way: the spectrogram is scanned and if a hit with SNR larger than a certain threshold $SNR_{th}^h=11~dB$ is found, the analysis code flags the cluster and carry on the search for next adjacent\footnote{It is important to remember that the cluster is defined with unique frequency, then the hit search is done over time (Y axis of the Sauron Plot).} hits with signal to noise ratio larger than a given threshold $SNR<SNR_{th}^l=9~dB$. The cluster information is saved if the cluster has at least a drift time of about $5~\mu s$. With this procedure clusters with low SNR compatible with fluctuations of the thermal noise are also included. In Figure (\ref{fig:NClusters_nocuts}) the distribution of the number of clusters of RUN 94 for each triggered signal event is shown.

\newpage

By imposing data cuts on drift time $t_D$, integrated SNR ($SNR^{int}$), maximum SNR ($SNR^{max}_{hit}$) and time coincidence between clusters it is possible to select the data. The time coincidence is obtained by measuring the starting and end time of the two clusters, with a time tolerance of $t_{shift}$.
The cuts on main variables are reported in Table (\ref{tab:cuts}).

\begin{table}[h!]
    \centering
    \begin{tabular}{|c|c|c|c|c|}
    \hline
       $SNR^{int} [ dB\cdot\mu s]$ &$ t_D [\mu s]$ & $t_{shift} [\mu s]$  & $SNR^{max}_{hit} [dB]$ & time coincidence \\
       \hline
        600 & 30 & 10 & 13 & yes \\
        \hline
    \end{tabular}
    \caption{Data cuts used in the analysis of RUN 93 and RUN 94.}
    \label{tab:cuts}
\end{table}

\begin{figure}[h!]
    \centering
    \includegraphics[width=0.75\linewidth]{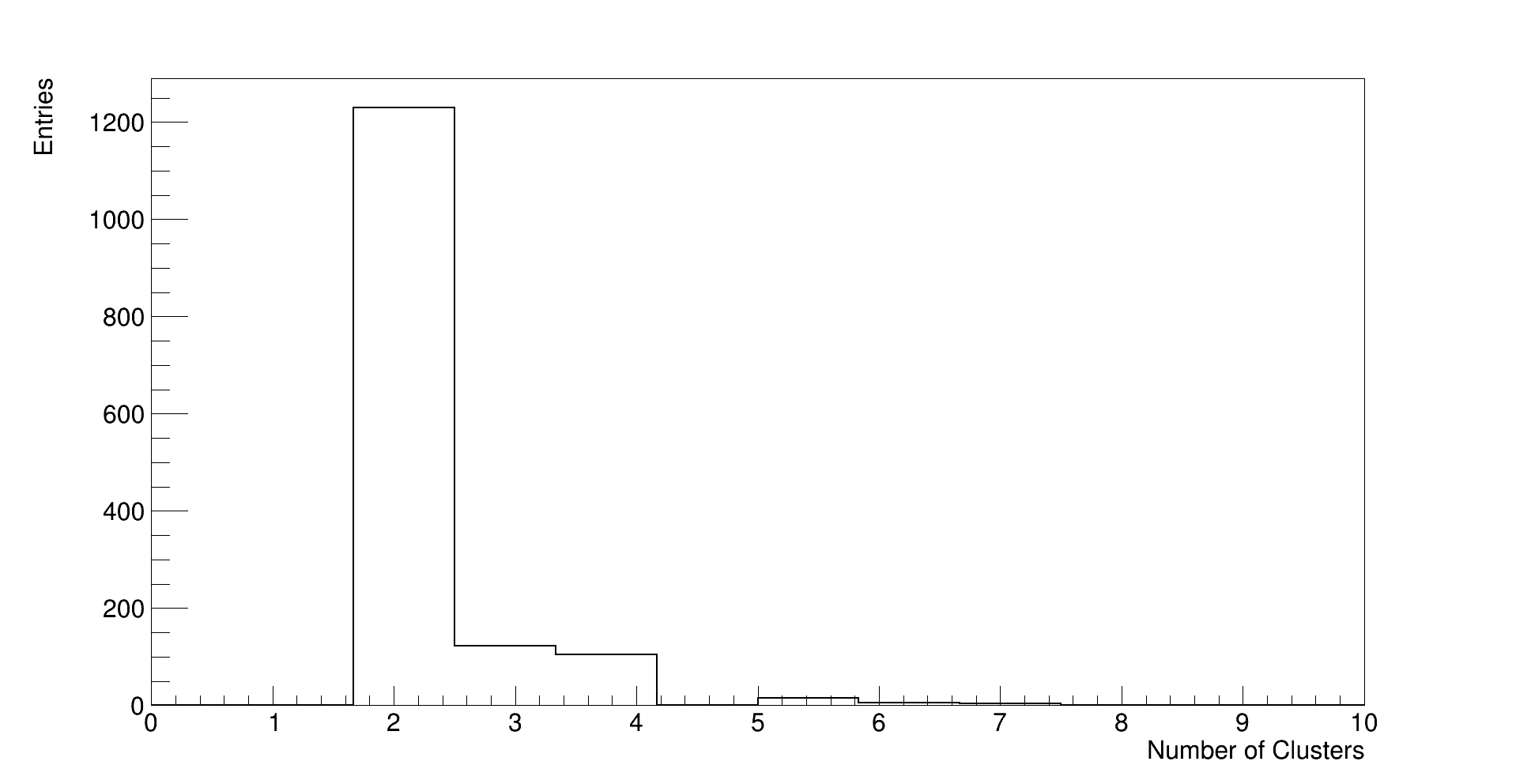}
    \caption{Number of selected clusters per event of RUN 94 (after cuts). The majority of events have two clusters selected per triggered event.}
    \label{fig:NClusters_cuts}
\end{figure}

The cuts applied are interrelated, for example, a cluster with high integrated SNR is likely to have a long drift time, while clusters with high $SNR^{max}_{hit}$ are likely to have high integrated SNR. The time shift tolerance for the time coincidence has been introduced to take into account possible non correct detection of the initial part of the signal. By applying those cuts the distribution of number of clusters per event is shown in Figure (\ref{fig:NClusters_cuts}) in which we note that the majority of the events have two selected clusters.

\newpage

\begin{figure}[h!]
    \centering
    \includegraphics[width=0.9\linewidth]{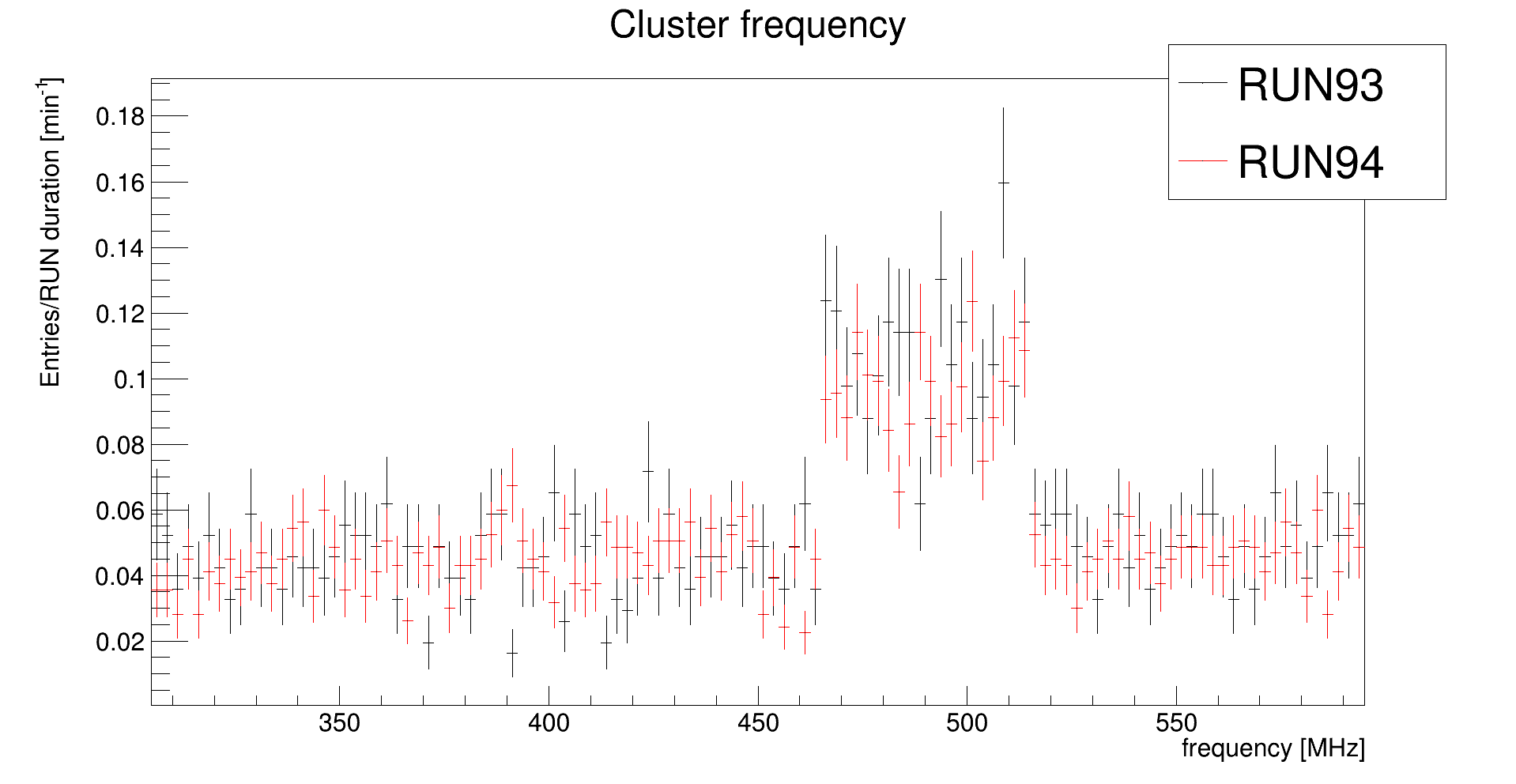}
    \caption{Cluster frequency distribution of RUN 93 (black) without source and RUN 94 (red) with Krypton source opened at the beginning of the RUN.}
    \label{fig:f93-94}
\end{figure}

With the aforementioned cuts, let's now see the distributions and correlation plots of the main variables. The most relevant distribution for an electron spectroscopy experiment with cyclotron radiation is the frequency (energy) distribution of the selected signals. The distributions of the cluster frequency of RUN 93 and RUN 94 are shown in Figure (\ref{fig:f93-94}) where the cluster frequency entries are normalized for the duration of each run (duration of RUN 93 = 310 min and duration of RUN 94 = 534 min).

\newpage

\begin{figure}[h!]
    \centering
    \includegraphics[width=0.7\linewidth]{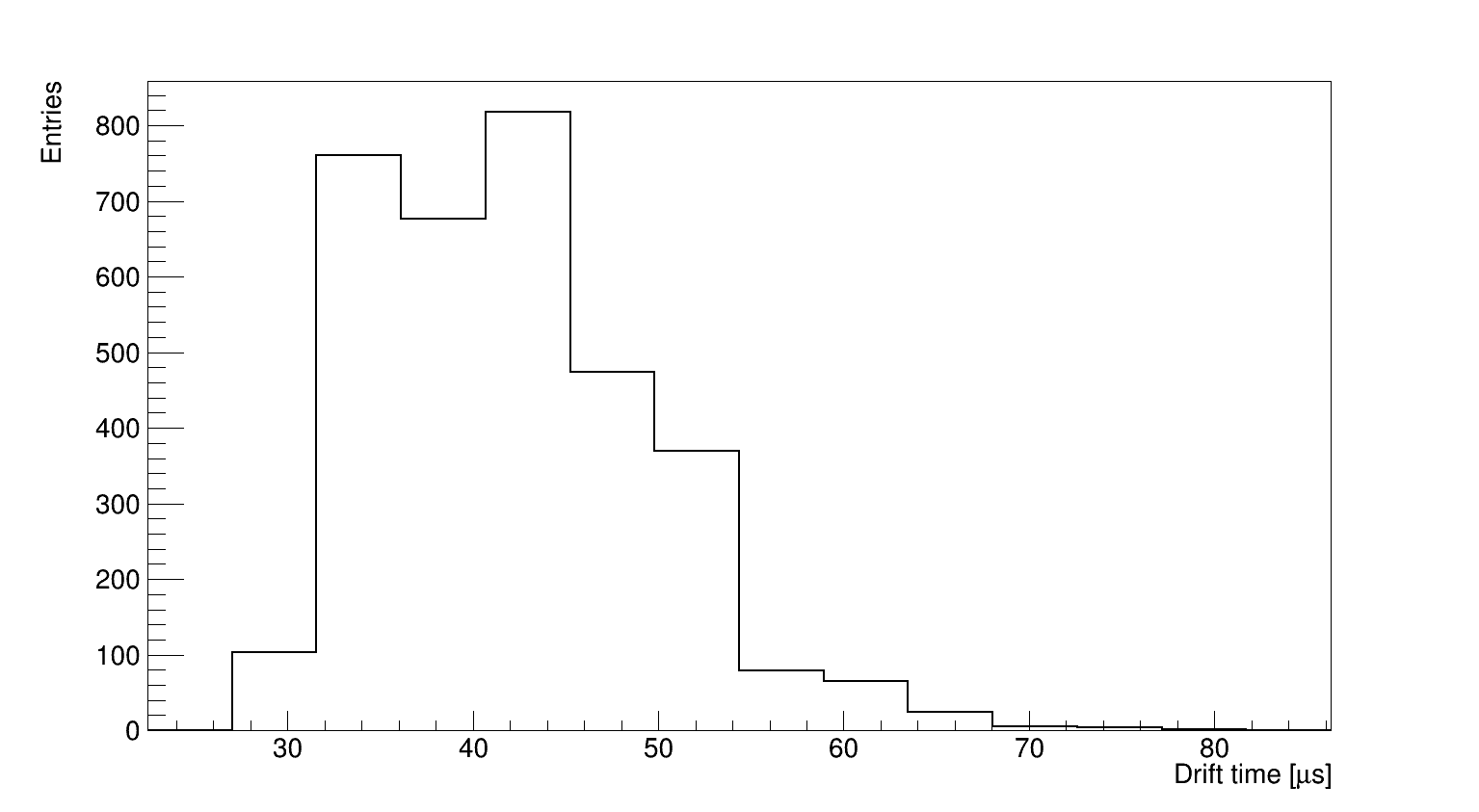}
    \caption{Cluster drift time of RUN 94 selected clusters. Long tracks are not visible.}
    \label{fig:drift_time94}
\end{figure}

In Figure (\ref{fig:f93-94}) one sees that the excess of clusters in the 465-515 frequency window is due to the trigger bias. The rate is similar between the two RUNs and therefore it is not found a significant excess for sideband signals, in fact the two distributions are similar. A possible explanation for why the RF events are scattered is reported in Section (\ref{sec:resol_systematics}) and is the non the expected uniformity of the magnetic field. Let's now see the distributions of the other quantities: the drift time, the average SNR and the maximum SNR. The distribution of the drift time, which is obtained by the time difference between the first hit above the SNR threshold and the last hit of the same cluster, is shown in Figure (\ref{fig:drift_time94}).

\begin{figure}[h!]
    \centering
    \includegraphics[width=0.7\linewidth]{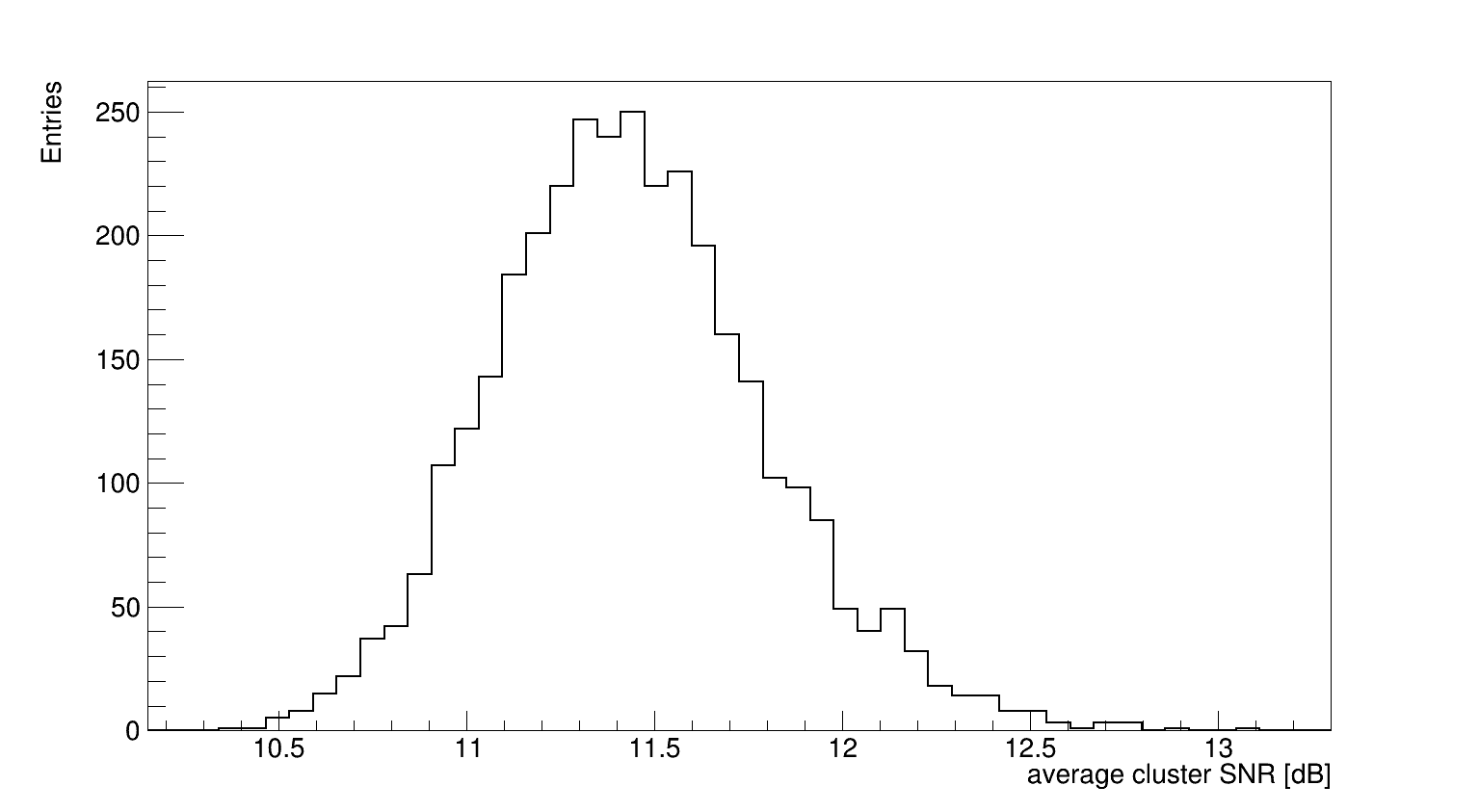}
    \caption{Average SNR of selected clusters of RUN 94. Even though the trigger threshold is set at $13~dB$, the most part of the cluster has lower average SNR because of the sliding Fourier.}
    \label{fig:avg_SNR94}
\end{figure}

As can be seen, all selected clusters have a drift time lower than $100~\mu s$, so milliseconds long tracks are not present, which would be likely identified as electron track. An explanation of why long tracks are not present in the simulations, as it is shown in Figure (\ref{fig:recon_drift_time}) is discussed in Section (\ref{sec:resol_systematics}). The average SNR of the selected clusters of RUN 94 is shown in Figure (\ref{fig:avg_SNR94}).

\newpage

\begin{figure}[h!]
    \centering
    \includegraphics[width=0.8\linewidth]{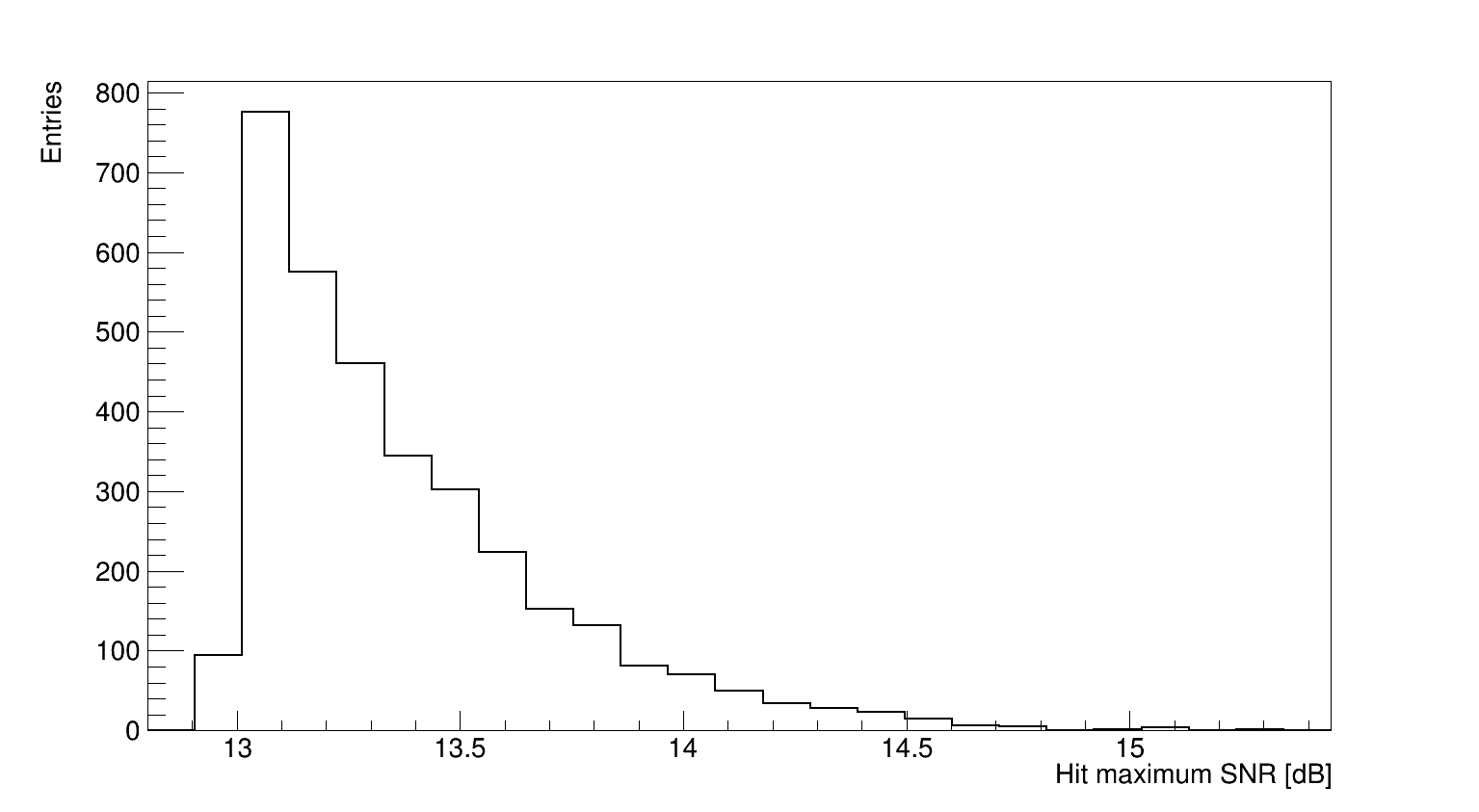}
    \caption{$SNR^{max}_{hit}$ of selected clusters of RUN 94. The trigger threshold is $13~dB$.}
    \label{fig:max_SNR94}
\end{figure}

It is useful to see the distribution of $SNR^{max}_{hit}$ because it is the SNR of the signal triggered by the oscilloscope and because it may happen that on an electron track there could be power fluctuations or low efficiencies in signal detection, therefore the maximum SNR (power) is a useful cut because it is unlucky to have large power thermal noise fluctuations. The distribution of $SNR^{max}_{hit}$ is shown in Figure (\ref{fig:max_SNR94}).

\begin{figure}[h!]
    \centering
    \includegraphics[width=0.7\linewidth]{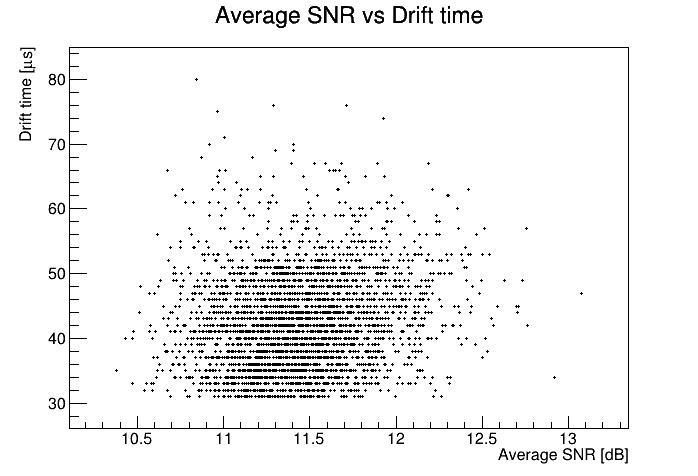}
    \caption{Drift time vs average SNR. The events with high drift time and low SNR are events with low power, while events with high average SNR are more compatible with electron events, however if $t_D<1/RBW=50~\mu s$ the event it is likely to be a background event.}
    \label{fig:drift_time_avg_SNR}
\end{figure}

The quantities shown previously can be investigated with the respective scatter plots to study correlations, for example the scatter plot between drift time and the average SNR is shown in Figure (\ref{fig:drift_time_avg_SNR}).

\begin{figure}[h!]
    \centering
    \includegraphics[width=0.7\linewidth]{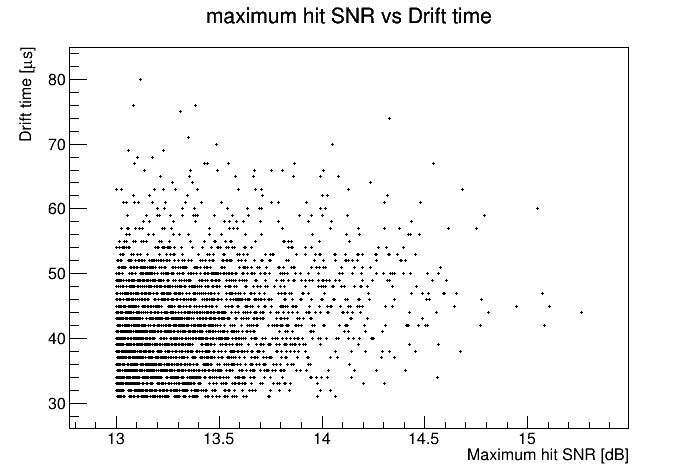}
    \caption{Scatter plot of drift time and $SNR^{max}_{hit}$.}
    \label{fig:max_SNR_drift_time94}
\end{figure}

The plot above is useful to discriminate events with average SNR medium-high (with respect the rejected events) but high drift time that could be identified as electron signature.
\begin{comment}
\begin{figure}[h!]
    \centering
    \includegraphics[width=1\linewidth]{plot/avg_SNR_max_hit_SNR94.png}
    \caption{Caption}
    \label{fig:placeholder}
\end{figure}
\end{comment}
In a similar way, it is possible to see if there are events with high $SNR^{max}_{hit}$ and high drift time in the plot shown in Figure (\ref{fig:max_SNR_drift_time94})

\begin{figure}[h!]
    \centering
    \includegraphics[width=0.8\linewidth]{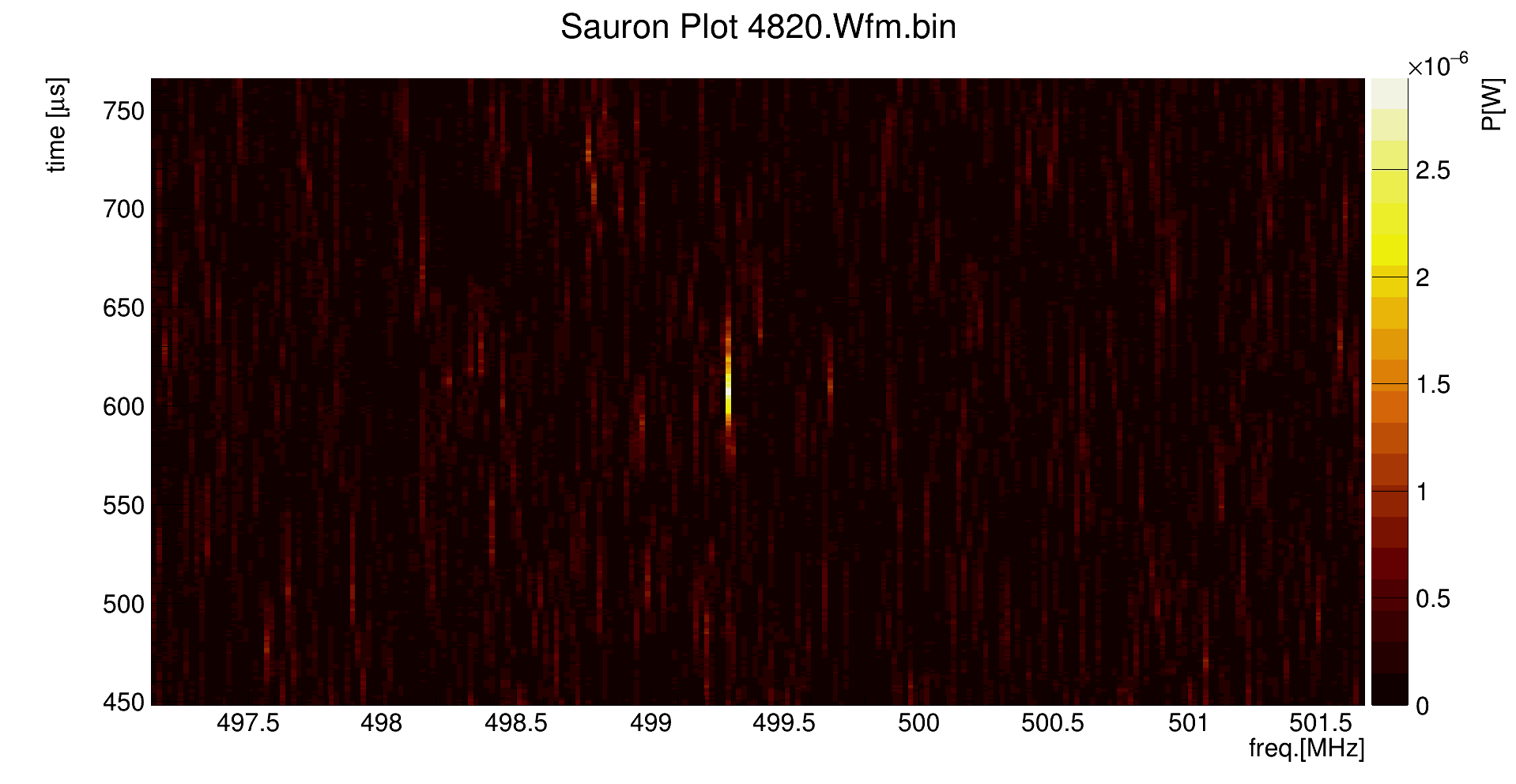}
    \caption{Sauron Plot of event 4820 of RUN 94. This event has been found with the data cuts and present an overall high SNR.}
    \label{fig:E4820c}
\end{figure}

\begin{figure}[h!]
    \centering
    \includegraphics[width=0.9\linewidth]{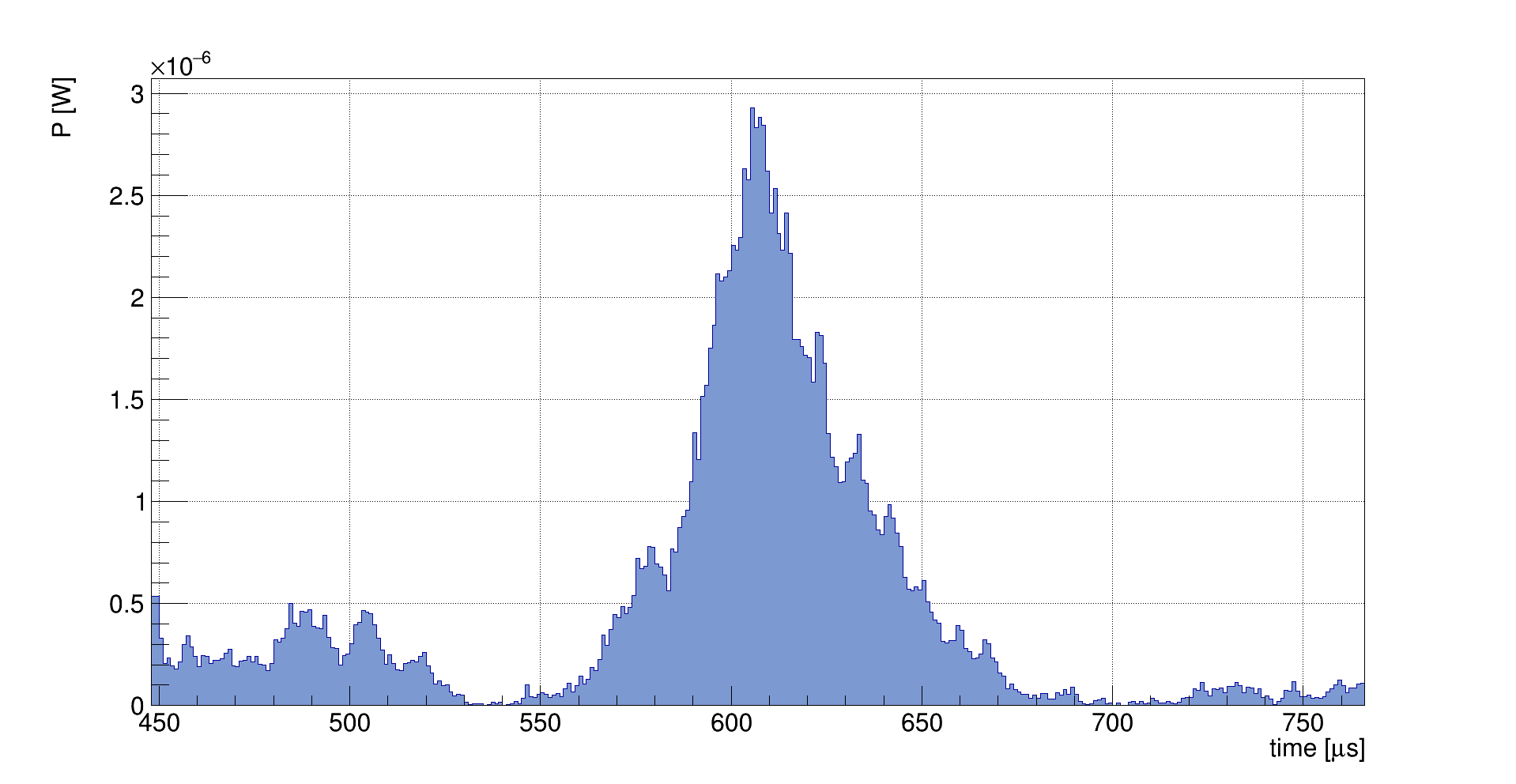}
    \caption{Power profile of event 4820 of RUN 94. The profile is triangular shape as expected from the calibration signal.}
    \label{fig:E4820c_Pt}
\end{figure}

in which events with high drift time and high SNR are likely to be tagged as electron tracks. It is now possible to see a typical Sauron Plot of RUN 94. The Sauron Plot of the candidate signal for the event 4820 is shown in Figure (\ref{fig:E4820c}) in which it is possible to see the high power of this particular event. The power vs time plot (Y profile of the Sauron Plot) is shown in Figure (\ref{fig:E4820c_Pt}) in which the profile is similar to the expectations shown in the calibration signal in Figure (\ref{fig:calib_P_t}).

\newpage

\begin{figure}[h!]
    \centering
    \begin{subfigure}[t]{0.49\linewidth}
        \centering
        \includegraphics[width=\linewidth]{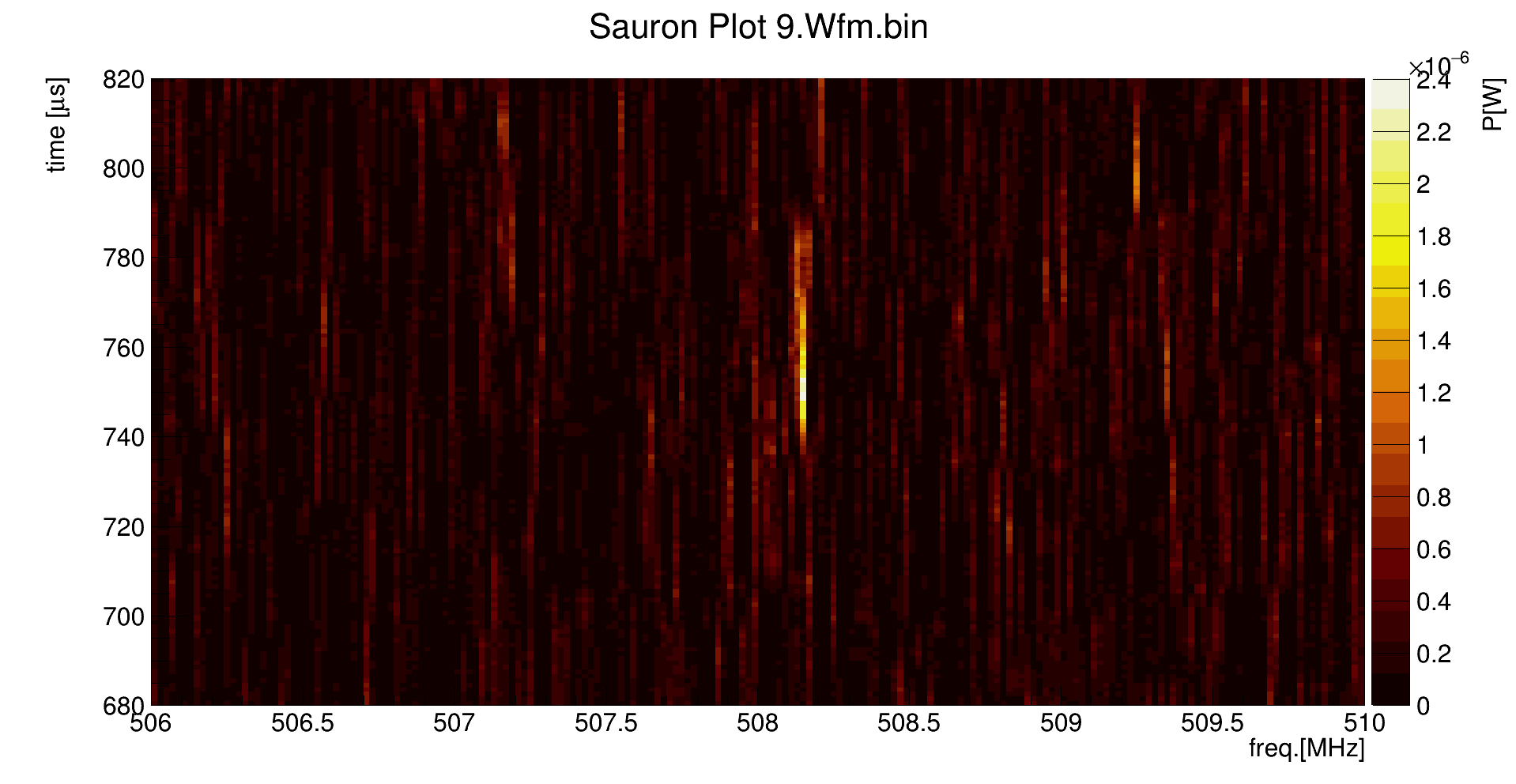}
        \caption{Carrier.}
        \label{fig:E9_sauron_carrier}
    \end{subfigure}\hfill
    \begin{subfigure}[t]{0.49\linewidth}
        \centering
        \includegraphics[width=\linewidth]{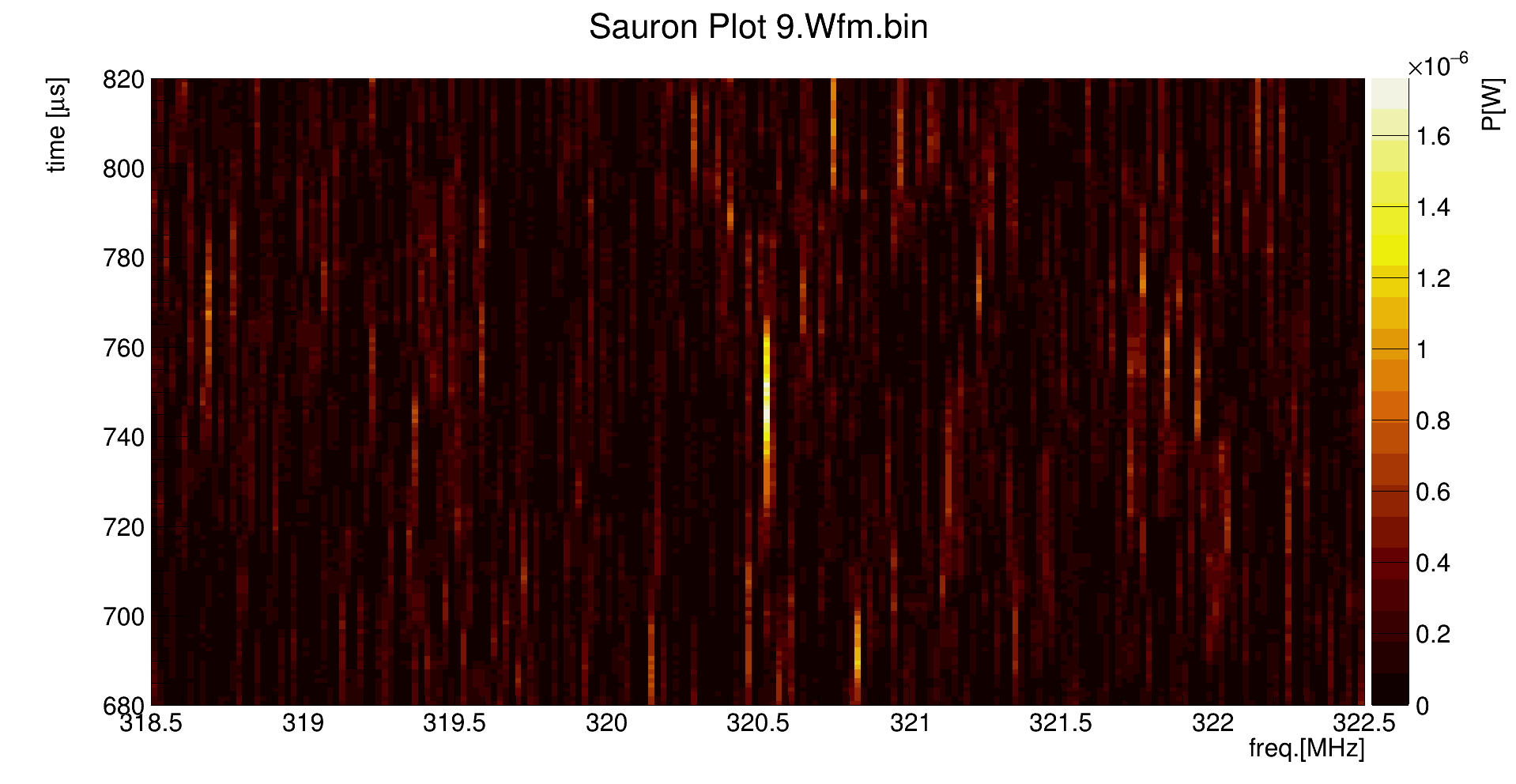}
        \caption{Right sideband.}
        \label{fig:E9_sauron_sideband}
    \end{subfigure}
    \caption{Sauron plots of candidate carrier and sideband signals of event 9. The carrier is in the expected frequency zone, that correspond to $f_{RF}=26.04~GHz$, the sideband present a frequency deviation of $\Delta f=187.7~ MHz$ that would correspond to a bouncing period of $T_b=5.3~ ns$.}
    \label{fig:E9_sauron_plots}
\end{figure}

%mettere come subfigure
Another example of a candidate event is the event 9 in which a candidate carrier and a candidate sideband (only one has been found in this event) are shown in Figure (\ref{fig:E9_sauron_plots}); as it is expected, those two signals are taking place at the same time.

In this Section the main variables for electron tagging have been discussed. Another useful method for the background rejection is the study of the electron energy loss due to RF emission, which corresponds to a constant frequency variation over time; this effect has been observed in PROJECT-8 electron tracks. In the following some considerations about this effect applied to the Electron Trap are discussed.
\newpage
Considering that the electron RF power is $P=1.9~fW$, this corresponds to an energy loss rate of $11~meV/\mu s$ that in a magnetic field of $B=0.9868~T$ corresponds to $df/dt=0.5~kHz/\mu s$. This means that if the electron has a $t_D=50~\mu s$ the frequency change must be approximately $25~kHz$. This value must be compared with the experimental frequency/energy resolution, that is, $RBW=20~kHz$, which means that a small frequency drift is expected; however, there are other parameters to take into account. The main problematic parameter is the knowledge of the magnetic field; in fact, as will be discussed in Section (\ref{sec:resol_systematics}), the magnetic field is not properly uniform in the experimental setup and so the RF signal is subjected to a relatively large averaging effect. Another consideration is that in the Electron Trap case the electron is in a potential well and so the longitudinal energy is continuously changing with a maximum excursion of $V_b$, that is the bouncing potential, which is much larger than the energy loss by RF emission in a short drift time. As discussed in Equation (\ref{eq:fc}), the frequency is obtained by this energy averaging effect. Because the energy loss on the longitudinal component of the electron cyclotron  motion is negligible, the bouncing period measurement should not be affected by the energy loss; however, the kinetic energy is averaged in a smaller range and so a frequency shift effect on the carrier frequency should be visible for high drift time.\\
In conclusion in this Section an hint of the detection of RF signal from electron in cyclotron motion has been found by measuring the source lifetime by trigger rate measurement. Then in a different RUN the RF pattern was searched. Some difficulties of electron signal tagging have been found, and so no evidence of electron signal was claimed, because no significant excess has been found on the analysis variables distributions.

\newpage
\section{Systematics and energy resolution}
\label{sec:resol_systematics}

In this Section the main sources of systematics in the Electron Trap are described and some considerations on the possible energy resolution achievable in an experiment like this are given.

\subsection{Systematics}

\begin{figure}[h!]
    \centering
    \includegraphics[width=1\linewidth]{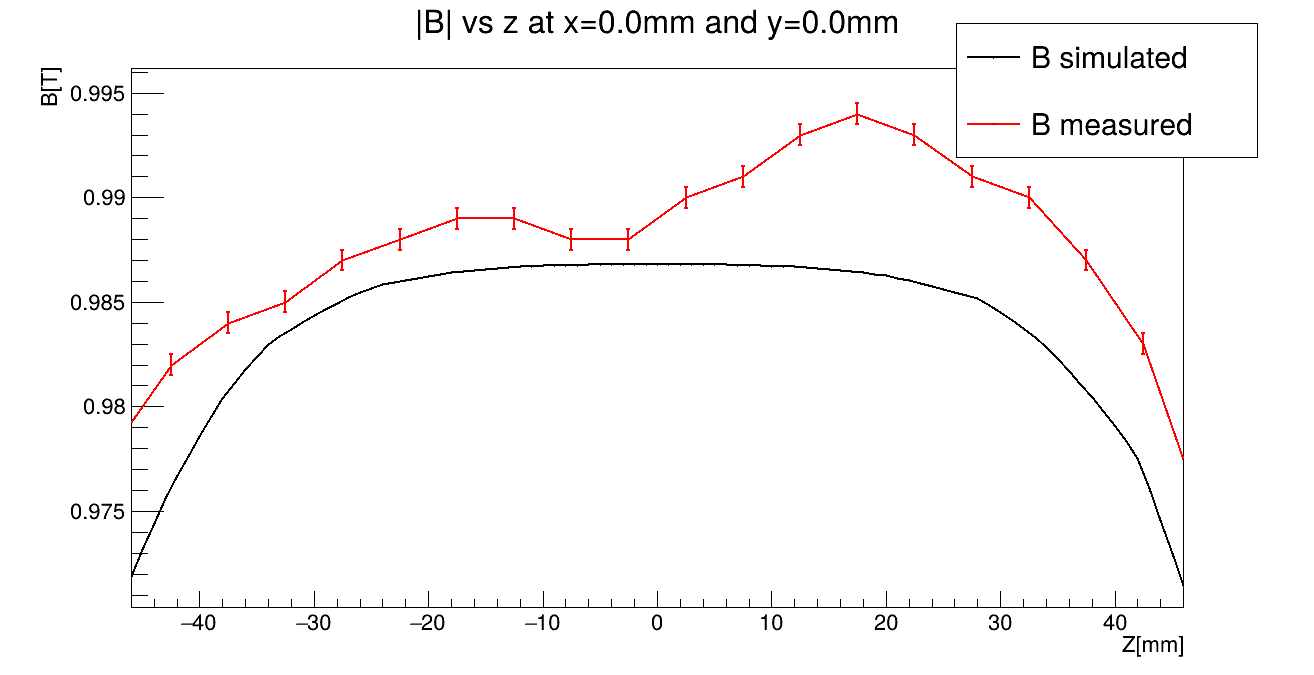}
    \caption{Comparison between the simulated magnetic field on the central magnet line (black) and the magnetic field measured with the Hall probe (red).}
    \label{fig:B_sim_meas}
\end{figure}

As previously described, some systematics are preventing the correct electron tagging. The main source of systematics is the correct knowledge and non uniformity of the magnetic field, in fact the module of the magnetic field has been measured with a transverse hall probe in the central axis of the magnet and the result is shown in Figure (\ref{fig:B_sim_meas}), in which we note that a non negligible gradient on the Z direction is present.

\begin{figure}[h!]
    \centering
    \begin{subfigure}[t]{0.48\linewidth}
        \centering
        \includegraphics[width=\linewidth]{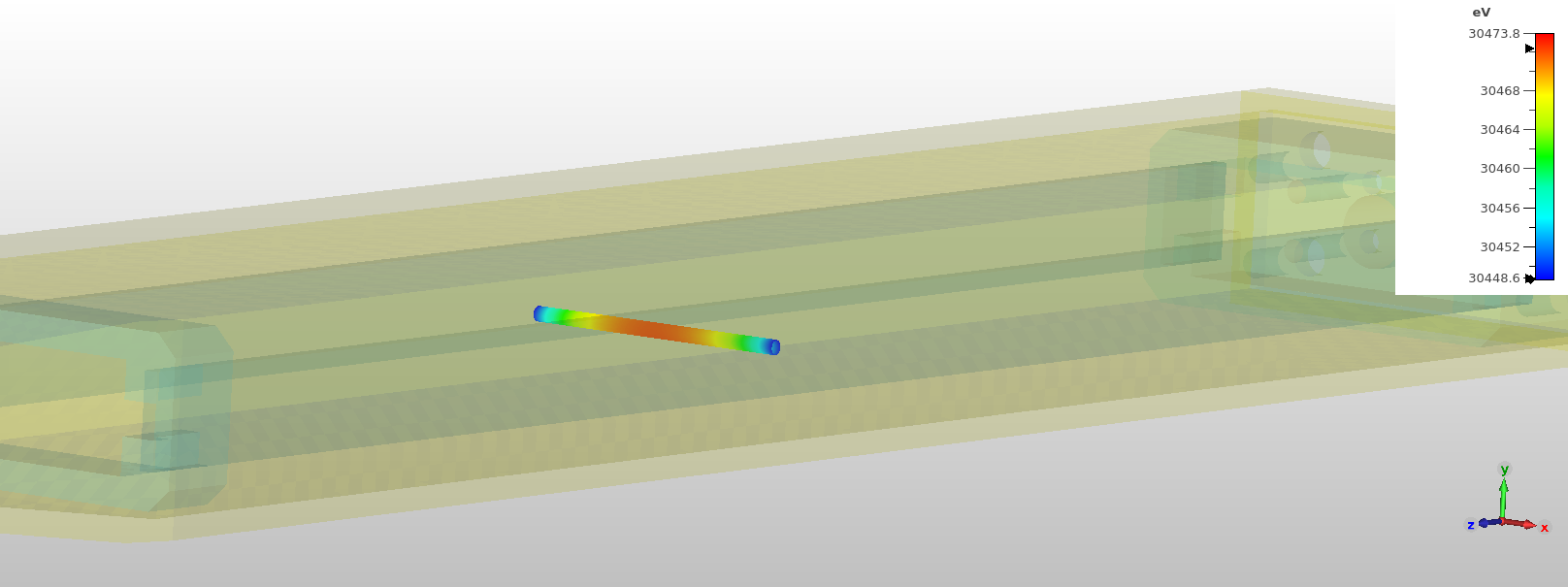}
        \caption{Trajectory in uniform magnetic field. The electron is spiraling and is doing the bouncing motion with a small drift along the Z axis, as predicted in Section (\ref{sec:e_motion}).}
        %\label{fig:trajectories-map}
    \end{subfigure}\hfill
    \begin{subfigure}[t]{0.48\linewidth}
        \centering
        \includegraphics[width=\linewidth]{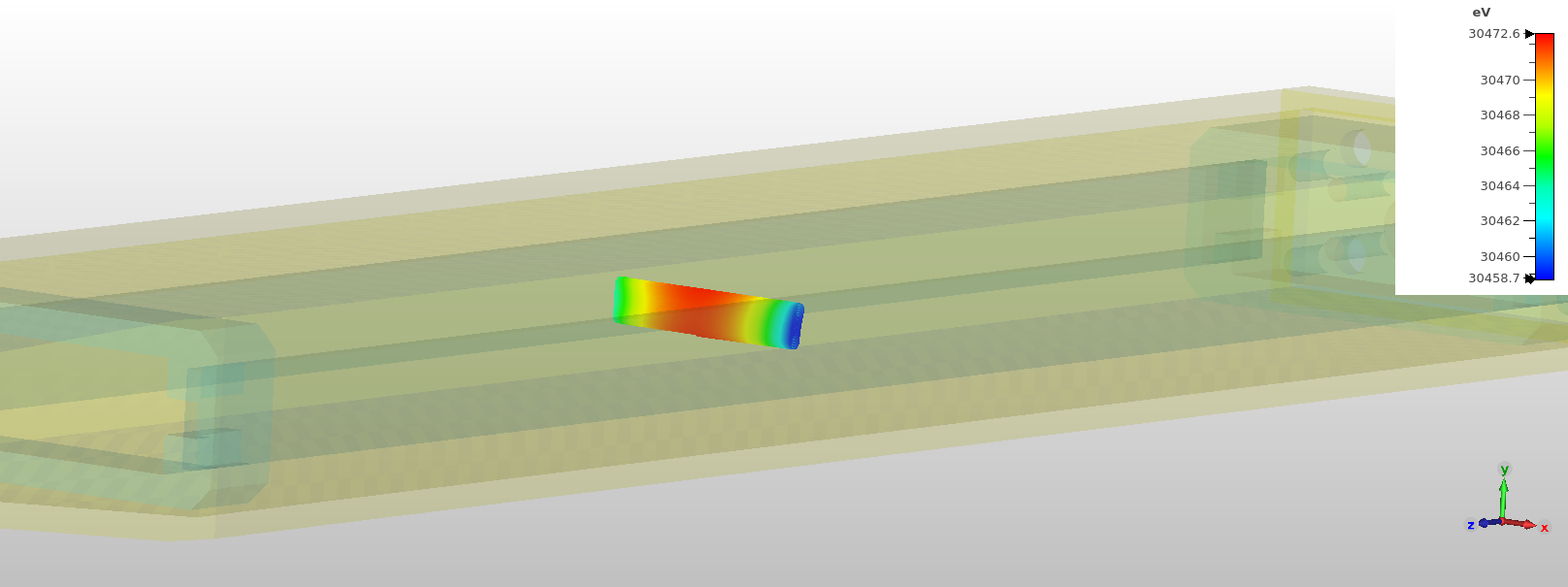}
        \caption{Trajectory simulated with the measured magnetic field. Is is possible to see that the gradient brings the electron to drift toward the Electron Trap roof during the bouncing motion.}
        %\label{fig:trajectories-unif}
    \end{subfigure}
    \caption{Electron trajectories in the case of uniform magnetic field and by using the measured magnetic field.}
    \label{fig:traiettorie_unif_vs_non_unif}
\end{figure}

From the data shown in Figure (\ref{fig:B_sim_meas}) we note two main effects: the measured magnetic field is larger than expected and non uniformity are found. An overall offset could be explained with an incorrect calibration of the probe with which the field was measured, which results in a shift of the measured frequencies. Instead, the non uniformity is more problematic because it introduces more issues. Firstly, there is a direct influence on the electron drift time, in fact the electron experiences a $\Vec{\nabla}_{z}B \times \Vec{B}$ drift that generates a $V_y$ velocity component that brings electrons to drift toward the upper floor of the Electron Trap, that is $15~mm$ wide. A comparison between an electron trajectory in a uniform magnetic field and in a field with the measured non uniformity is shown in Figure (\ref{fig:traiettorie_unif_vs_non_unif}).

\newpage

Another effect of the non uniformity of the magnetic field is that the frequency is changing during the motion non only for kinetic energy averaging effect, but also from the magnetic field that is different during the motion. In conclusion, because of these systematics, the drift time of the electrons is significantly reduced and the frequencies are spread out. Another source of systematics is the non perfect alignment of the electron trap with the magnetic field lines. This angular uncertainty propagates into an uncertainty in the pitch angle $\theta$ and has also an effect on the electron trajectories and in the reconstruction of $K$ and $K_{\parallel}$. In fact, the electron does not drift along the same planes like in the plot shown in Figure (\ref{fig:Tb_KL_Y}). The last source of systematics is the temperature in the experimental setup. In Section (\ref{sec:lifetime}) we have already discussed the effect of the variation of temperature in the electronics that produces a variation of the gain which affects the trigger rate. Furthermore, the temperature also affects the releasing of Krypton molecules from the cold surfaces. In fact, when the system is warmer more atoms are free-flying and so more electrons are emitted in the fiducial volume.

\subsection{Energy Resolution}

In this Section an estimate on the energy resolution achievable in the PTOLEMY's RF tracker is given, taking into account the experience gained in the Electron Trap phase. In the following, the calculation of the expected energy resolution for the RF tracker is presented by using some parameters used in the Electron Trap like the $RBW$ and the reference drift time $t_D=50~\mu s$. Because those calculations are for a future experimental setup (RF tracker), the critical elements identified in the Electron Trap, like the uniformity of the magnetic field and the round shape of the potential lines, are assumed under control. As described in Section (\ref{sec:kin_reconstruction}), the first step for the PTOLEMY CRES measurement is the measurement of the average kinetic energy $\langle K \rangle$ of the carrier frequency $f_c$ performed using the following Formula

\begin{equation}
    \langle K \rangle=m(eV)\left( \frac{1}{2\pi}\frac{|q|B}{m}\frac{1}{f_c}-1 \right)
\end{equation}

where the values $m$, $B$, $q$ and $f_c$ in parenthesis are given in MKS units while the electron mass outside of the parenthesis is given in units of $eV$. The uncertainty in the average kinetic energy is due to two components: $\sigma_{\langle K \rangle,B}$ which is the contribution to the resolution by the magnetic field uncertainty and $\sigma_{\langle K \rangle,f}$ which is the contribution due to the frequency measurement. The resolution on the average kinetic energy is then

\begin{equation}
    \sigma_{\langle K \rangle}=\sigma_{\langle K \rangle,B} \oplus \sigma_{\langle K \rangle,f}
\end{equation}

where

\begin{equation}
\begin{cases}
     \sigma_{\langle K \rangle,B}&= m(eV) \left( \frac{1}{2\pi}\frac{|q|}{m}\frac{1}{f_c} \right)\sigma_B \\
    \sigma_{\langle K \rangle,f} &= m(eV) \left | \frac{1}{2\pi}\frac{|q|B}{m}\frac{1}{f_c^2} \right|RBW
\end{cases}
\end{equation}

If we use the reference frequency resolution used in the Electron Trap of about $RBW=20~kHz$ and a field uniformity of the PTOLEMY magnet of about $\sigma_B=10^{-5}~T$ we obtain

\begin{equation}
\begin{cases}
    \sigma_{\langle K \rangle,B}&= 5.5\times 10^{5} \frac{eV}{T} \cdot \sigma_B=5.5~eV\\
     \sigma_{\langle K \rangle,f}&=131.6\times 10^{-6} \frac{eV}{Hz} \cdot RBW=2.6~eV
\end{cases}
\end{equation}

and so a resolution on the average kinetic energy of about

\begin{equation}
\sigma_{\langle K \rangle}=6.1~eV
\end{equation}
is obtained.

The next step is the measurement of the bouncing period by looking at the frequency deviation $\Delta f$ of the FM

\begin{equation}
    T_b=\frac{1}{\Delta f}
\end{equation}

that, considering an uncertainty on $\Delta f$ as

\begin{equation}
    \sigma_{\Delta f}=\sqrt{2} \cdot RBW=28.2~kHz
\end{equation}

we find that the resolution on the bouncing period is

\begin{equation}
    \sigma_{T_b}=\left| \frac{1}{\Delta f^2} \right|\sigma_{\Delta f}
\end{equation}

To get an estimate on $\sigma_{T_b}$ we can take as a reference the electron simulated in Section (\ref{sec:montecarlo}). In this case $\Delta f=\mathcal{O}(80~MHz)$ and we get

\begin{equation}
    \sigma_{T_b} \sim4.4\times 10^{-3}~ns
\end{equation}

where $T_b= \mathcal{O}(10 ~ns) $. As described in Section (\ref{sec:kin_reconstruction}), at this stage it is possible to estimate $K_{\parallel}$ from the plot in Figure (\ref{fig:Tb_KL_Y}) in which $\sigma_{K_{\parallel}}$ can be estimated numerically with $|\Delta K_{\parallel}/\Delta T_b|=|(10-0)/(13-15)|\simeq5~\frac{eV}{ns}$, which gives

\begin{equation}
    \sigma_{K_{\parallel}}=\left| \frac{\Delta K_{\parallel}}{\Delta T_b} \right| \sigma_{T_b}\sim0.02~eV
\end{equation}

It is now possible to reconstruct the kinetic energy at the bottom of the potential well starting from the calculation of $\langle K_{\parallel} \rangle$ using Equation (\ref{eq:avg_K_par}). The resolution of the average longitudinal kinetic energy depends on two factors: the bouncing period and the $K_{\parallel}$

\begin{equation}
    \sigma_{\langle K_{\parallel} \rangle}=\sigma_{\langle K_{\parallel} \rangle,T_b} \oplus \sigma_{\langle K_{\parallel} \rangle,K_{\parallel}}
\end{equation}

where the two contributions are calculated as shown below

\begin{equation}
\begin{cases}
    \sigma_{\langle K_{\parallel} \rangle,T_b}=\sqrt{\frac{m}{2}}\int_{-X_b}^{X_b} \sqrt{K_{\parallel}-V(x)}dx \cdot\left|\frac{1}{T_b^2}\right| \sigma_{T_b} \\
    
    \sigma_{\langle K_{\parallel} \rangle,K_{\parallel}}=\frac{1}{T_b}\sqrt{\frac{m}{2}}\left| \int_{-X_b}^{X_b} \frac{1}{2\sqrt{K_{\parallel}-V(x)}}dx \right| \sigma_{K_{\parallel}}
    \end{cases}
\end{equation}

\begin{comment}
dove 

\begin{equation}
    \left|\frac{d<K_{\parallel}>}{dK_{\parallel}}\right|=\frac{1}{T_b}\sqrt{\frac{m_{kg}}{2}}\left| \int_{-X_b}^{X_b} \frac{1}{2\sqrt{K_{\parallel}-V(x)}}dx \right|
\end{equation}
\end{comment}

The last step is the calculation of $K_{\perp}$ using Equation (\ref{eq:K_perp_recon}) and then it is possible to get $K$ using Equation (\ref{eq:K_recon}) which has an uncertainty of

\begin{equation}
    \sigma_K=\sigma_{K_{\perp}} \oplus  \sigma_{K_{\parallel}}
\end{equation}

This resolution $\sigma_K$ must be compared with the energy window $\Delta E$, before the Tritium endpoint, that we want to investigate in the RF tracker. 
In fact, this uncertainty has an effect on the prediction of the number of events in the ROI $\Delta E$ for physics measurements.
%This comparison is important to understand the capability of the RF measurement with respect to the expected event rate.
Considering that the rate $R$ in the final energy window $\Delta E$ before the end point is

\begin{equation}
    R \propto \left( \frac{\Delta E}{Q_{\beta}} \right)^3A[g]
\end{equation}

and considering as region of interest $\Delta E=10~eV$ and the\footnote{See the sensitivity plot shown in Figure (\ref{fig:PTOLEMY_sensitivity}).} instrumented Tritium mass of $1~\mu s$ required to have sensitivity for neutrino mass, we obtain

\begin{equation}
    R \sim 1.5\times 10^{-10}A[1~\mu g]=1.5\times 10^{-10}\cdot 3.6\times10^8=5.4\times 10^{-2}~Hz
\end{equation}

which corresponds to 1 event every 20 seconds. The inverse of rate $R$ is much larger than the reference drift time needed for RF measurement, which it corresponds to $t_D=50~\mu s$.
\newpage
The last topic on the energy resolution and systematics of RF measurements for the PTOLEMY detector is the measurement of the energy loss due to RF emission $E_{RF,corr}$, which is obtainable by integrating the signal power over time. In Equation (\ref{eq:PTOLEMY_initial_energy_reconstruction}) has been shown that an estimate of $E_{RF,corr}$ is needed for the reconstruction of electron energy at the target.
As shown in Figure (\ref{fig:calib_P_t}), the power profile of a $50~\mu s$ drift time signal has a triangular shape for signals with $t_D \leq 1/RBW$, which is due to the increasing (decreasing) signal integration with the sliding FFT, while signals with larger drift time are expected to appear like a trapezoid. The measurement of the energy loss can be done with the following steps:

\begin{enumerate}
    \item Measure the kinetic energy $K$ and the transverse kinetic energy $K_{\perp}$ from the RF pattern
    \item Measure the drift time $t_D$ 
\end{enumerate}

and then evaluate $E_{RF,corr}$ 

\begin{equation}
E_{RF,corr}=P(K,K_{\perp})\cdot t_D
\label{eq:E_RF_corr}
\end{equation}

where $P$ is calculated using Equation (\ref{eq:P}). The critical factor for energy loss measurement is the correct reconstruction of the drift time; this can be improved with a future dedicated calibration campaign by including the possible electron drift in the high field region outside the RF Tracker. Further improvements would be possible by implementing machine learning for the reconstruction of the electron drift time.

%% file: sections/6Conclusions.tex
\chapter*{Conclusions} \label{chap:Conclusions}

The topic of this work is the description of the measurement of the RF signal generated by a single electron during cyclotron motion in the PTOLEMY experiment, and on the design and construction of the Electron Trap, which serves as a test setup for detecting electron RF emission.  
The development of the Electron Trap has been crucial for testing electrons with different motion parameters. It has been possible to modify the pitch angle acceptance $\Delta \theta$, 
by tuning the bouncing potential, thereby enabling the study of different bouncing periods and low-power signals. In addition, drift electrodes were used to accelerate electrons along the Z direction, which was intended to probe the highest velocities achievable for RF measurements. However, due to power losses, in the Electron Trap, this high-velocity test was ultimately not pursued.  
Overall, the Electron Trap design ended up being very valuable for evaluating the performance requirements of future RF tracker, such as maximum and minimum acceptable drift time ($\mathcal{O}(50 ~\mu s )$), if one consider the PTOLEMY measuring process.\\
Thanks to the simulation studies, it has been possible to predict the characteristics of electron trajectories, primarily to investigate the electron RF pattern. These studies allowed to elaborate a possible signal model that makes it possible to deduce kinematic variables of the electrons from the pattern of the FFT of the RF signal of single electron. Among the various approaches considered, the sideband method has been chosen as the baseline option. Since this method relies on frequency measurements, it can be easily implemented on an FPGA to generate a real time trigger signal for the filter.\\
Following the design and simulation phases, data were collected. The first hint, even without an electron's signal evidence claim, of RF emission detection was obtained through the measurement of $\tau_{Kr}$, derived from the decay of the trigger rate. These measurements provide the first hint, within the PTOLEMY framework, of RF detection from a single electron undergoing cyclotron, bouncing, and drift motion, thereby demonstrating the feasibility of future PTOLEMY RF Tracker measurements.
The capabilities of the design have been validated through simulations, with particular emphasis on tuning the drift time for effective electron tagging and on clustering the frequency distribution. However, non-uniformity in the magnetic field compromised the stability of the electron motion, shortening the drift time and increasing the magnetic field variations experienced during the motion. In summary, a satisfactory result was achieved in terms of trigger rate, but challenges were encountered in reproducibility and signal tagging due to magnetic field map non-uniformity, misalignment, and temperature effects.
Because of these systematics no evidence of electron signal has been claimed.
Nonetheless, substantial experience has been acquired in detecting electron RF signals and refining the measurements, especially with regard to systematic effects. The future RF tracker, which will build on the Electron Trap experience, will be specifically designed to reduce, as much as possible, the effect of the systematics, just mentioned.\\
The measurements and design of the Electron Trap represent an original work when compared to the most significant results in the field of CRES measurement namely, the Phase 1 PROJECT-8 demonstrator. The key distinction lies in the measurement goals: PROJECT-8 is designed for precision spectroscopic measurements, whereas PTOLEMY only requires a relatively coarse and rapid measurement to trigger the dynamic filter of the PTOLEMY detector. The Electron Trap data provide strong evidence for the feasibility of such measurements, particularly concerning the drift time required to perform them.
The electron motion in PROJECT-8 and in the PTOLEMY Electron Trap differs, although the approach to determining the electron kinematic variables is similar. The key distinction is that, in PTOLEMY, electrons are emitted from a fixed potential and their trajectories are controlled by electrostatic potentials, because in the PROJECT-8 the atoms are in a gaseous form thus their position is unknown and so the potential of the initial position of the electron. Thanks to the different approach of the experiment, the PTOLEMY detector is capable to infer the electron energy using the reconstruction method described in Chapter (\ref{chap:Electron_Trap}), without the complications of the potential well effects and with the flexibility to shape the potential profiles. Furthermore, the distinctive PTOLEMY injection scheme allows the electron pitch angle, at the RF tracker, to be compressed, enabling a detector with large acceptance and tunable electron drift parameters by means of the applied potentials.\\
In summary, this work serves as a foundation for the development of the PTOLEMY RF tracker, and improved performance is expected once the dedicated PTOLEMY magnet is installed at LNGS, scheduled for delivery in 2026.